\documentclass[10pt,twocolumn,showpacs,amsmath,amssymb,aps,prx,longbibliography,superscriptaddress]{revtex4-2}
\usepackage{times}
\usepackage[unicode=true,
 bookmarks=true,bookmarksnumbered=true,bookmarksopen=false,
 breaklinks=true,pdfborder={0 0 0},backref=false,colorlinks=true]
 {hyperref}
\usepackage{graphicx,amsfonts,amssymb,amsmath,hypcap,enumerate}
\usepackage{enumitem}
\usepackage{braket}
\usepackage{lipsum} 
\usepackage{amsthm}
\usepackage{multirow}
\usepackage{graphicx}
\usepackage{dcolumn}   %
\usepackage{bm}        %
\usepackage{amssymb}   %
\usepackage{amsmath}
\usepackage{braket}
\usepackage{epstopdf}
\usepackage{color}
\usepackage{subfigure}
\usepackage{diagbox}
\usepackage{slashbox} 
\usepackage{makecell}
\usepackage{bbold}
\usepackage{array}
\usepackage{longtable}
\usepackage[utf8]{inputenc}
\usepackage{mathtools}
\usepackage{overpic}

\usepackage{tabularx}
\newcolumntype{Y}{>{\centering\arraybackslash}X}
\usepackage[dvipsnames]{xcolor}
\usepackage{tikz}
\usetikzlibrary{arrows,
                chains,
                positioning,
                shapes.geometric,
                }
\tikzstyle{line}      = [draw, -latex']

\DeclareMathAlphabet{\mathitb}{OT1}{cmr}{bx}{sl}
\usepackage{subfiles}

\usepackage{pifont}
\newcommand{\cmark}{\ding{51}}%
\newcommand{\xmark}{\ding{55}}%

\usepackage{xpatch}

\hypersetup{
    linkcolor=red,          %
    citecolor=blue,        %
    filecolor=magenta,      %
    urlcolor=magenta           %
}

\makeatletter
\patchcmd{\@ssect@ltx}
    {\addcontentsline{toc}{#1}{\protect\numberline{}#8}}
    {}
    {}
    {}
\makeatother

\begin{document}
\def\bluee#1{\textcolor{blue}{#1}}
\def\blue#1{#1}
\def\red#1{\textcolor{red}{#1}}
\def\orange#1{\textcolor{orange}{#1}}
\def\R{R}
\def\r{\boldsymbol{r}}
\def\q{\boldsymbol{q}}
\def\t{\boldsymbol{t}}
\def\k{\boldsymbol{k}}
\def\A{\boldsymbol{A}}
\def\AA{\mathcal{A}}
\def\BB{\mathcal{B}}
\def\a{\boldsymbol{a}}
\def\p{\prime}
\def\b{\boldsymbol{b}}
\def\n{\boldsymbol{n}}
\def\S{\boldsymbol{S}}
\def\v{\boldsymbol{v}}
\def\V{\mathcal{V}}
\def\H{\mathcal{H}}
\def\z{\boldsymbol{z}}
\def\U{U}
\def\0{\boldsymbol{0}}
\def\T{\mathcal{T}}
\def\vS{\vec{S}}
\def\tS{\tilde{\S}}
\def\p{\prime}
\def\hz{{\boldsymbol{z}}}
\def\hx{{\boldsymbol{x}}}
\def\hy{{\boldsymbol{y}}}

\def\kk{\tilde{\bm k}}
\def\qq{\tilde{\bm q}}
\def\tk{\tilde{k}}
\def\tq{\tilde{q}}
\def\mG{\mathcal{G}}
\def\mS{\mathcal{S}}
\def\mU{\mathcal{U}}
\def\mD{\mathcal{D}}
\def\MAGNDATA{\href{http://webbdcrista1.ehu.es/magndata/}{MAGNDATA}}

\def\pare#1{\left( #1 \right)}
\def\brak#1{\left[#1\right]}
\def\brace#1{\left\{#1\right\}}

\def\AppSecSSGTable{Appendix~\ref{sec:SSG-table}}
\def\AppSecSSGTableTab{Appendix~\ref{sec:SSG-table}}
\def\AppSecSSGTableTabb{Appendix~\ref{sec:SSG-table}}
\def\AppSecMaterialTable{Appendix~\ref{sec:material-table}}
\def\AppIrrep{Appendix~\ref{app_irrep_sg}}
\def\AppEquivRep{Appendix~\ref{app sec: equiv rep}}
\def\AppSNF{Appendix~\ref{App Sec: d_M}}
\def\AppSecNoncommutingSBZ{Appendix~\ref{App Sec: d_M}}
\def\AppC{Appendix~\ref{App Sec: d_M}}
\def\AppD{Appendix~\ref{Iden_SSG}}
\def\AppE{Appendix~\ref{App Sec: Dirac TI}}
\def\FigMnGeSOC{\ref{Fig: app Mn3Ge SOC} }

\newcommand{\SZD}[1]{\textcolor{red}{(SZD: #1)}}

\newcommand{\remove}[1]{\textcolor{red}{\sout{#1}}}
\newcommand{\add}[1]{\textcolor{blue}{\uwave{#1}}}

\newcommand{\RS}[1]{\textcolor{magenta}{#1}}

\title{Spin Space Groups: Full Classification and Applications}

\author{Zhenyu Xiao}
\thanks{These authors contribute equally to this work.}
\affiliation{International Center for Quantum Materials, School of Physics, Peking University, Beijing 100871, China}

\author{Jianzhou Zhao}
\thanks{These authors contribute equally to this work.}
\affiliation{Co-Innovation Center for New Energetic Materials, Southwest University of Science and Technology, Mianyang 621010, China}

\author{Yanqi Li}
\affiliation{International Center for Quantum Materials, School of Physics, Peking University, Beijing 100871, China}

\author{Ryuichi Shindou}
\affiliation{International Center for Quantum Materials, School of Physics, Peking University, Beijing 100871, China}

\author{Zhi-Da Song}
\email{songzd@pku.edu.cn}
\affiliation{International Center for Quantum Materials, School of Physics, Peking University, Beijing 100871, China}
\affiliation{Hefei National Laboratory, Hefei 230088, China}
\affiliation{Collaborative Innovation Center of Quantum Matter, Beijing 100871, China}

\date{\today}
\begin{abstract}
In this work, we exhaust all the spin-space symmetries, which fully characterize collinear, non-collinear, commensurate spiral, as well as incommensurate spiral magnetism, {\it etc.}, and investigate enriched features of electronic bands that respect these symmetries.
We achieve this by systematically classifying the so-called spin space groups (SSGs) - joint symmetry groups of spatial and spin operations that leave the magnetic structure unchanged.
Generally speaking, they are accurate (approximate) symmetries in systems where spin-orbit coupling (SOC) is negligible (finite but weaker than the interested energy scale), but we also show that specific SSGs could remain valid even in the presence of strong SOC.
In recent years, SSG has played increasingly pivotal roles in various fields such as alter-magnetism, topological electronic states, and topological magnon, {\it etc}.
However, due to its complexity, a complete SSG classification has not been completed up to now.
By representing the SSGs as O($N$) representations, we - for the first time - obtain the complete classifications of 1421, 9542, and 56512 distinct SSGs for collinear ($N=1$), coplanar ($N=2$), and non-coplanar ($N=3$) magnetism, respectively.
SSG not only fully characterizes the symmetry of spin d.o.f., but also gives rise to exotic electronic states, which, in general, form projective representations of magnetic space groups (MSGs).
Surprisingly, electronic bands in SSGs exhibit features never seen in MSGs, such as
(i) nonsymmorphic SSG Brillouin zone (BZ), where SSG operations behave as glide or screw when act on momentum, 
(ii) effective $\pi$-flux, where translation operators anti-commute with each other and yield duplicate bands, 
(iii) higher-dimensional representations unexplained by MSGs,
and (iv) unconventional spin texture on Fermi surface, which is completely determined by SSG, independent of Hamiltonian details. 
To apply our theory, we identify the SSG for each of the \blue{1595} published magnetic structures in the MAGNDATA database on the Bilbao Crystallographic Server. 
Material examples exhibiting the novel features (i)-(iv) are discussed with emphasis.
We also investigate new types of SSG-protected topological electronic states that are unprecedented in MSGs.
In particular, we propose a 3D $\mathbb{Z}_2$ topological insulator (TI) state with a four-fold degenerate Dirac point on the surface, and a new scenario of anomalous $\mathbb{Z}_2$ helical states that appear on magnetic domain walls.
\end{abstract}
\maketitle

\tableofcontents

\section{Introduction}

Symmetry is one of the central concepts in modern physics.
In condensed matter, symmetry plays determining roles in phase transitions, low-energy excitations, transport in disordered systems, {\it etc.}, allowing physicists to qualitatively understand a realistic system without knowing microscopic details.
One cannot overemphasize the importance of symmetry in physics.
For more than one hundred years, the 230 space groups have been the standard mathematical description of symmetries in solid material crystallography.
For magnetic materials, the complete symmetry theory was generally believed to be the 1651 magnetic space groups (MSGs), where a symmetry operation could be a pure spatial operation or a combination of the time-reversal symmetry (TRS) and a spatial operation.
In the 1960's, people realized that Heisenberg models with negligible spin-orbit coupling (SOC) enjoy higher symmetries, namely spin space groups (SSGs) \cite{brinkman66}.
In an SSG, an operation can be a combination of a spatial operation and an arbitrary spin rotation that is compatible with the group structure. 
Such symmetries are found crucial to describe the spin waves correctly. 
Later, the spin point groups, where the spatial operations are limited to point group operations, were fully classified by Litvin \cite{litvin74,litvin1977spin}.
However, due to the complexity of \blue{joining} spatial and spin operations, a full classification of SSGs has not been completed up to now. 

\begin{figure*}[tb]
    \centering
   \includegraphics[width=1 \linewidth]{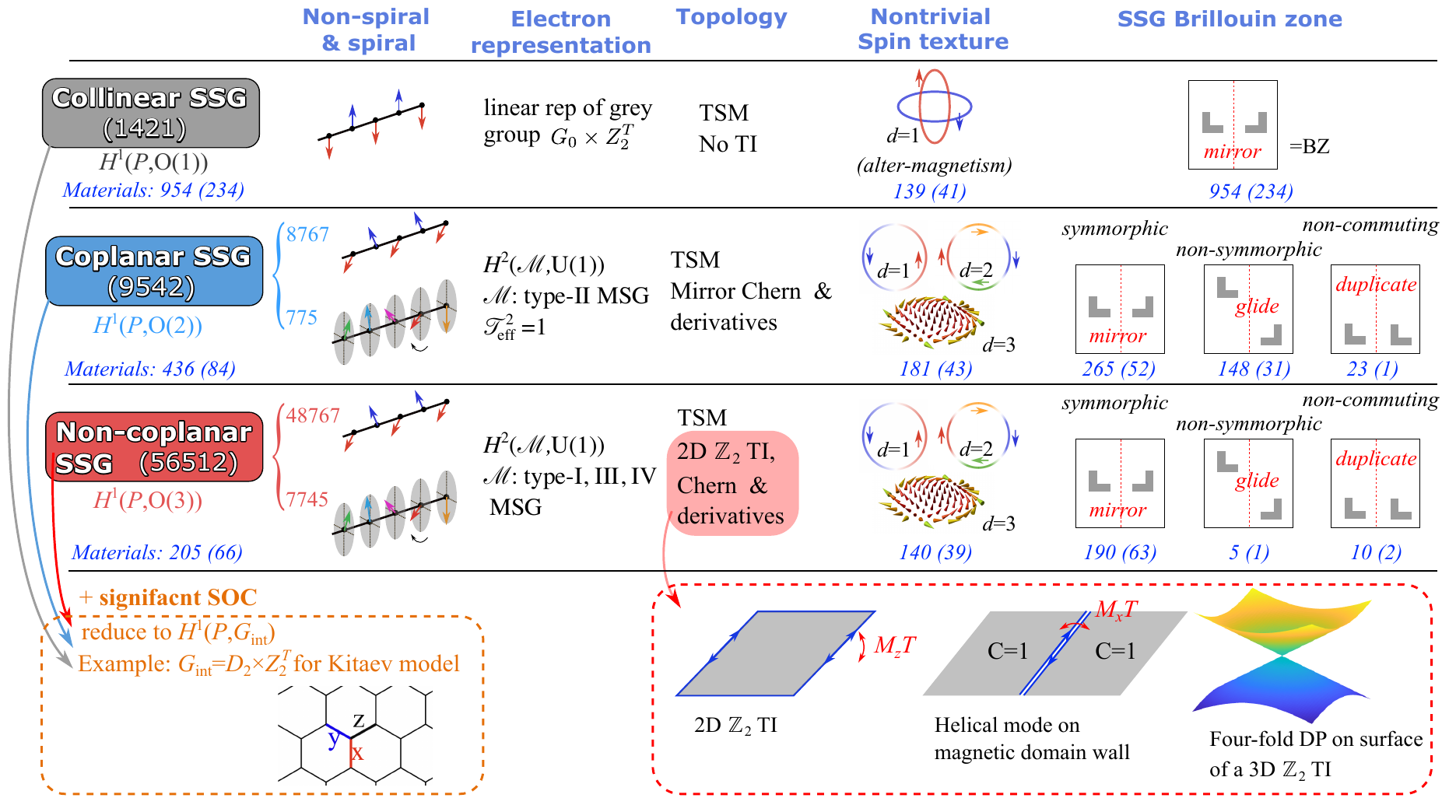}
    \caption[]{Summary of results. 
    We have obtained all 1421, 9542, and 56512 distinct spin space groups (SSGs) for collinear ($N=1$), coplanar ($N=2$), and non-coplanar ($N=3$) magnetic structures by enumerating O($N$) representations.
    (\textbf{Non-spiral \& spiral}) All the collinear SSGs describe commensurate magnetic structures, whereas 775 (7745) of the coplanar (non-coplanar) SSGs describe spiral magnetic structures with either commensurate or incommensurate spiral angles. 
    (\textbf{Electron Representation}) In collinear magnetism, electronic Bloch states form a linear representation of a magnetic space group (MSG) of type II (grey group). In coplanar (non-coplanar) magnetism, they form projective representations of MSGs $\mathcal{M}$ of type II (I, III, and IV), classified by group cohomology $H^2(\mathcal{M},U(1))$. 
    (\textbf{Topology}) We show possible electronic topological states for different SSGs, where TSM denotes topological semi-metal. We provide concrete models of topological states, {\it e.g.,} a four-fold degenerate surface Dirac fermion, in non-coplanar magnetism. 
    (\textbf{Spin texture}) SSGs determine the dimension $d$ of the span of spin expectations on Fermi surfaces, and, if $d>0$, whether the spin textures are nontrivial. 
    (\textbf{SSG Brillouin zone (SBZ)})  The action of SSG translations on \blue{SSG} Bloch states might anti-commute, resulting in a non-commuting SBZ with a duplicated band structure.
    When the SSG translations commute, the SBZ can be symmorphic or nonsymmorphic, and SSG momentum in the latter is transformed nonsymmorphically under SSG symmetries. 
    We also identify the SSGs of experimental materials. We list the (blue and italic) number of magnetic materials that are subject to the category at the bottom of each cell, and in the parentheses is the number of materials constituent of light elements from the first four periods. Finally, it is notable that some SSGs can be applied to systems with significant SOC, such as symmetry-breaking phases in the Kitaev model.  
    }
    \label{Fig: summary} 
\end{figure*}

In recent years, SSG has drawn growing attention because of their applications in magnetic materials with negligible or weak SOC, such as magnetic topological electronic states \cite{liu2022spin,yang2021symmetry-invariants,guo_eightfold_2021,ghorashi23}, topological magnon bands \cite{shindou_2013_topological_magnon,zhang_2013_topological_magnon,mook_2014_edge_states,chisnell_2015_topological_magnon,li_dirac_2017,yao_topological_2018,corticelli_spin-space_2022}, unconventional spin-momentum locking without SOC~\cite{yuan_giant_2020,smejkal_crystal_2020,reichlova_macroscopic_2021,ma2021}, which was later realized as a common feature of the alter-magnetism~\cite{smejkal_emerging_2022,cheong_magnetic_2022,bose_tilted_2022,feng_anomalous_2022,karube_observation_2022,bai_observation_2022,smejkal_beyond_2022,mazin_editorial_2022}, {\it etc.}
To be specific, SSGs can protect: 
(i) Exotic nodal-line/point semi-metals, {\it e.g.,} twelve-fold degenerate fermion in three dimensions \cite{yang2021symmetry-invariants} and eight-fold degenerate fermion in two dimensions \cite{guo_eightfold_2021}, that are disallowed by conventional symmetries \cite{bradlyn_beyond_2016};
(ii) Exotic topological insulators (TIs), {\it e.g.,} SOC-free and TRS-free two-dimensional $\mathbb{Z}_2$ TIs~\cite{liu2022spin}, that are disallowed by conventional symmetries~
\cite{kitaev_periodic_2009,ryu_topological_2010,Hasan2010_rmp,Qi2011_rmp,Chiu2016_rmp}; 
(iii) Exotic nodal magnon bands, {\it e.g.,} Dirac points with $\mathbb{Z}$-valued monopole charges \cite{li_dirac_2017}, which will evolve into nodal lines with $\mathbb{Z}_2$-monopole charges \cite{fang_topological_2015,ahn_band_2018} if weak SOC is considered, as confirmed by the experiment \cite{yao_topological_2018}.
SSGs are also crucial for understanding the so-called alter-magnetism \cite{smejkal_emerging_2022,mazin_editorial_2022} where anti-parallel spin momenta at two sublattices compensate each other exactly (as in anti-ferromagnetism) but the two sublattices are not related by inversion or translation.
Alter-magnetism can exhibit unconventional, {\it e.g.,} $d$-, $g$-, $i$-wave-like, spin-momentum locking \cite{yuan_giant_2020,smejkal_beyond_2022}, which can lead to a surprisingly large anomalous Hall conductivity when a weak SOC is introduced \cite{smejkal_crystal_2020,reichlova_macroscopic_2021}, as confirmed by experiments \cite{bose_tilted_2022,feng_anomalous_2022,karube_observation_2022,bai_observation_2022}. 
Unconventional momentum-dependent spin polarization also exists in non-collinear magnetism without SOC~\cite{yuan_prediction_2021, ma2021}.  
Therefore, SSGs not only give rise to theoretically exotic states but also have a strong relevance in realistic magnetic materials. 
It provides a precise description of magnetic materials with negligible SOC and a good starting point to understand magnetic materials with weak SOC. 
However, due to lacking a complete classification of SSGs, most discussions are limited to spin point groups or several simple SSGs.

In this work, by reducing the group classification problem to a representation problem, we obtain the full classification of SSGs and find that there exist 67475 inequivalent SSGs in total (see \AppSecSSGTable). 
These SSGs fully characterize collinear magnetism (including the ferro-, ferri-, anti-ferro-, and alter-magnetisms), coplanar magnetism, non-coplanar magnetism, and spiral magnetism (including commensurate and incommensurate structures).

The complete classification reveals additional novel features of magnetic states beyond those found in previous works. 
To investigate electron bands using SSGs, we generalize crystal momentum to SSG momentum $\kk$ and introduce a concept of SSG Brillouin zone (SBZ) -- the reciprocal space formed by SSG momenta.
Due to enriched symmetry algebra, the action of SSG translations on electronic states may exhibit anti-commutation.
An SBZ is said non-commuting if the SSG translations have this property, and commuting otherwise. 
Commuting SBZs can be further classified into symmorphic SBZs and nonsymmorphic SBZs, wherein the latter exhibits non-trivial transformations of SSG momentum. 
In a symmorphic SBZ, the momentum transforms normally under an SSG operation $g$, {\it i.e.,} $\kk \to \kk^{\prime} = s_g R_g \kk$, with $R_g$ being point group part of $g$ and $s_g$ being $1$ ($-1$) for unitary (anti-unitary) $g$.   
In a nonsymmorphic SBZ, $\kk$ transforms to $s_g(R_g \kk + \qq_g)$, where $\qq_g$ is {\it not} a reciprocal lattice vector of SBZ, 
but a ``fractional translation'' in momentum space. 
In Sec~\ref{subsec: rep PN SSGs} and Sec~\ref{subsec: Materials}, we will discuss the non-commuting and nonsymmorphic SBZ in more detail and present several theoretical and material examples. 
We identify whether the SBZ is non-commuting, nonsymmorphic for each of the 67475 SSGs (\AppSecSSGTableTab). 
\blue{We note that while anti-commuting translations and nonsymmorphic transformation of momentum~\cite{zhang2023c} can arise in systems subjected to external magnetic flux, these exotic behaviors studied in this work are intrinsically derived from the magnetic structures.}

Another advancement from the SSG classification is a systematic classification of spin textures of \blue{SSG} Bloch states in the momentum space. 
Spin expectation of SSG Bloch states is transformed under SSG operations, and the transformation realizes a representation of the SSG.
If the realized representation is non-trivial, the spin polarization on the Fermi surface must be momentum-dependent and generally cannot be described by MSGs.
This description not only fully characterizes the spin textures in alter-magnetism~\cite{yuan_giant_2020,smejkal_beyond_2022} but also in non-collinear magnetism~\cite{yuan_prediction_2021}, which can be $p$-, $d$-, $f$- $\cdots$ -wave-like.
We explicitly derived the representations formed by spin textures in {\it all} SSGs that have symmorphic SBZs (\AppSecSSGTableTab).

To apply our theory, we identify the SSGs for the \blue{1595} published experimental magnetic structures in the \href{http://webbdcrista1.ehu.es/magndata/}{MAGNDATA} database \cite{gallego_magndata_2016,gallego_magndata_2016-1} on the \href{https://www.cryst.ehu.es/}{Bilbao Crystallographic Server}, and tabulate the results in \AppSecMaterialTable. 
For every material, we further determine whether it has a non-commuting, nonsymmorphic, or symmorphic SBZ and whether it possesses a non-trivial spin texture.
Remarkably, these materials exhibit all the aforementioned exotic features of SSG. 
Among them, 33 materials have non-commuting SBZs; 153 materials have nonsymmorphic SBZs; \blue{460} of \blue{1409} remaining materials with symmorphic SBZs possess non-trivial spin textures in the momentum space.
Within this group of \blue{460} materials, \blue{139} are collinear and fall into the category of alter-magnetism.
We believe that these useful information will lead to further research for new physics and potential applications in fields like spintronics.
To illustrate these concepts, we perform first-principle calculations on representative examples of materials.

SSGs also protect novel electron topological states that are absent in conventional space groups \cite{bradlyn_topological_2017,po_symmetry-based_2017,kruthoff_topological_2017} and MSGs \cite{zhang2015_prb,dong_classification_2016,Shiozaki2017_prb,xu20,elcoro21,peng_topological_2022}. In Sec.~\ref{sec: topo phase}, we present three examples of SSG-protected topological states. 
The first example is a physical realization of a two-dimensional SOC-free and TRS-free $\mathbb{Z}_2$ TI.
The hopping Hamiltonian of this model is real, and the topology is introduced by a noncollinear magnetic order. 
The second example is a three-dimensional SOC-free and TRS-free $\mathbb{Z}_2$ TI possessing a four-fold Dirac point on surfaces. 
The third example is a two-dimensional system with a protected helical mode along the domain wall between two magnetic domains with the same topological invariant.
To our knowledge, the latter two states have never been discussed in conventional symmetry groups or SSGs.
In particular, the third example demonstrates a new scenario of topological states - topological gapless domain wall - that is unseen in previously known free-fermion states (in the absence of chiral and particle-hole symmetries). 
These examples should be a tip of the iceberg of the unexplored fruitful topological states in SSGs. 

We summarize the main results of this work in Fig.~\ref{Fig: summary}, and the rest of this paper is organized as the followings. 
In Sec.~\ref{Sec: class}, we explain why O($N$) representations of space groups can classify all SSGs (see Sec.~\ref{class_general}).
We also describe a method of constructing all O($N$) representations (Sec.~\ref{sec O3_Rep}) and criteria for determining whether two representations yield physically distinct SSGs (Sec.~\ref{distinct_O3}). 
We use O(3) representations of the space group $P3$ as an example of the classification (Sec.~\ref{sec:example-P3}).
We summarize basic information of SSGs in Sec.~\ref{sum_class} and tabulate all the SSGs in \AppSecSSGTable.
In Sec.~\ref{sec:band-theory}, we investigate the representation theory of SSGs and its application in electronic states. 
We demonstrate that SSGs acting \blue{SSG} Bloch states lead to enriched symmetry algebra, necessitating the introduction of concepts of SSG momentum, SSG Brillouin zone (SBZ), and non-commuting and non-symmorphic SBZs. 
In addition, we utilize SSGs to determine the spin texture in the momentums space (see Sec.~\ref{sub sec: spin texture}). 
Sec.~\ref{sec: material example} focuses on experimental magnetic materials. 
We identify the SSGs of \blue{1595} magnetic materials in the \href{http://webbdcrista1.ehu.es/magndata/}{MAGNDATA} database.
The statistics are present in Sec.~\ref{subsec: Materials stat}, while the complete 
information is provided in \AppSecMaterialTable. 
We perform first-principle calculations on material examples exhibiting non-symmorphic SBZ, extra degeneracies, and unconventional spin texture in the momentum space. 
In Sec.~\ref{sec: topo phase}, we discuss possible electronic states protected by SSGs and construct three models of topological phases that are protected by symmetries unique in SSGs.
Sec.~\ref{Sec: summary} is devoted to discussion and summary, \blue{where we also compare our representation-based approach of classifying SSGs with Litvin's approach~\cite{litvin74,litvin1977spin}.}

\section{Classification of spin space groups}
\label{Sec: class}

\subsection{General considerations}
\label{class_general}
Without SOC, the many-body electronic Hamiltonian of a material exhibits an SU(2) spin-rotation symmetry, a spinful time-reversal symmetry $Z_2^T$, and a space group $G_{\rm latt}$ symmetry of the lattice, where the spatial operations do not act on the spin.
\blue{The group SU(2) is a double cover of SO(3), and two SU(2) operations corresponding to the same SO(3) rotation differ only by a phase of $-1$ when acting on the electronic states.
In this work, we use the SO(3) group to describe the spin-rotation symmetry of magnetic structures. 
Additionally, the electron's spin-1/2 nature manifests itself in the projective representation theory (Sec.~\ref{sec:band-theory}).}
\blue{The elements in $G_{\rm latt}$, the $\rm SO(3)$ rotation group, and $Z_2^T$ mutually commute.}
Thus, the full symmetry group is \blue{$G_{\rm latt}\times \mathrm{SO}(3) \times Z_2^T$}. 
A generic magnetic structure may individually break $G_{\rm latt}$, $\mathrm{SO(3)}$, $Z_2^T$ but preserve some joint operations. 
\blue{An SSG, denoted as $\mG$ hereafter, is defined as the symmetry group of the magnetic structure, which is a continuous or discrete subgroup of $G_{\rm latt} \times \mathrm{SO}(3) \times Z_2^T$}, depending on the magnetic order.

\begin{figure*}
    \centering
\tikzstyle{line}      = [draw, -latex']
\begin{tikzpicture}[
    node distance = 6mm and 9mm,
      start chain = going right,
 disc/.style = {shape=cylinder, draw, shape aspect=0.3,
                shape border rotate=90,
                text width=17mm, align=center, font=\linespread{0.8}\selectfont},
  mdl/.style = {shape=ellipse, aspect=2.2, draw},
  alg/.style = {draw,  rounded corners,align=center, font=\linespread{0.8}\selectfont}]
    \node (n1) [alg] {Specify the parent \\ space group of an SSG \\ $ \left( \{ XU | R | \v   \} \rightarrow  \{ R | \v \} \right)$ (Sec.~\ref{class_general}).};
    \node (n2) [alg, right=of n1]  {Induce irreps of the parent space group from\\ the little-group irreps of momenta\\ in the irreducible Brillouin zone (Sec.~\ref{sec O3_Rep}).};
    \node (n3) [alg, right=of n2]  {Construct O($N$) ($N=1,2,3$) representations\\ from the induced irreps   \\ (Table~\ref{tab:O(N)rep} and Sec.~\ref{sec O3_Rep}). };
    \node (n4) [below =of n1, xshift=3cm, alg] {Enumerate equivalence among O($N$) representations \\ including coordinate transformation of space groups,\\ and continuous deformation of spiral angles (Sec.~\ref{distinct_O3}). };
    \node (n5) [right=of n4, alg] {Utilize equivalence relations to divide\\ O($N$) representations into different classes.\\ Each class corresponds to a distinct SSG (Sec.~\ref{sum_class}).};
     
     \path [line] (n1) -- (n2);
     \path [line] (n2) -- (n3);
     \path[line] (n3)  |- ([yshift=-0.3cm, xshift=0cm]n2.south) -| (n4);
     \path[line] (n4) -- (n5);
\end{tikzpicture}
    \caption[]{Flowchart of obtaining all distinct spin space groups (SSGs) based on the representation theory of space groups. Here irrep stands for irreducible representations.   }
    \label{fig:flow chart}
\end{figure*}

To proceed, we consider a magnetic structure, \blue{characterized by local spin magnetic moments $\S(\r_i)$'s, where $\r_i$'s are the position of magnetic atoms.}
\blue{The spin rotations act on the magnetic structure as SO(3) matrices.}
Hence, for $\S(\r_i)$, a generic symmetry operation consists of a spatial operation $\{\R |\v \} \in G_{\rm latt}$, a spin rotation $\U \in {\rm SO(3)}$, and a possible time-reversal operation $T$.
It can be written as $\{ X \cdot \U  | \R |\v \}$ with $X$ being identity $I$ or $T$, and it transforms the magnetic structure $\S(\r_i)$ to
\begin{equation} 
\S'(\r_i) = s(X) \cdot \U \cdot \S(  \{\R |\v \}^{-1} \r_i)
   \label{O3_1}
\end{equation}
where $s(X)=1$ and $-1$ for $X=I$ and $T$, respectively.
The SSG consists of all such composite operations that leave the magnetic structure unchanged, {\it i.e.,}
\begin{equation}
    \mG = \brace{ \{ X \cdot \U  | \R |\v \}\ \big|\ \{ X \cdot \U  | \R |\v \} \cdot \S = \S }\ ,
\end{equation}
where $\{ X \cdot \U  | \R |\v \} \cdot \S$ is the transformed magnetic structure defined in Eq.~(\ref{O3_1}).

The spatial operations $\{\R |\v \}$ in $\mG$ form a space group
\begin{equation}
P = \brace{ \{ \R |\v \}\ \big|\ \{ X \cdot \U  | \R |\v \} \in \mG }\ ,
\end{equation}
which is named the parent space group.
It is worth mentioning that, in general, $P$ is a subgroup of the full space group $G_{\rm latt}$ of the lattice. 
A homomorphism exists from the SSG $\mG$ to its parent space group $P$, and the kernel of the homomorphism is the pure-spin-operation group $\mS$, where operations are in the form $\{ X \U |1|\0 \}$. 
$\mS$ is uniquely determined by \blue{whether} the magnetic structure is collinear, coplanar, or non-coplanar~\cite{litvin74,litvin1977spin,liu2022spin}.
Without loss of generality, we always assume that the magnetic moments are confined to the $x,y$ plane for the coplanar structures, and are oriented along the $z$ direction for the collinear structure. 
Before investigating $\mS$, let us consider the restrictions of the spin-operation part $X \cdot U$ of a general operation in $\mG$ for different arrangements.
Non-coplanar structures have no specific restrictions on $X \cdot U$. 
However, for coplanar (collinear) structures, the spin-operation part $X \cdot U$ of any operation in $\mathcal{G}$ must preserve the spin moments within the $xy$ plane (along the $z$ direction). 
Thus, for these two kinds of structures, $s(X)U$ takes a block diagonal form ${\rm diag}\left( O_{xy},O_z \right)$, where $O_{xy}$ is an O(2) matrix acting on the $x,y$ components of the spin moments, and $O_z$ is an O(1) matrix acting on the $z$ components.

For non-coplanar structures, operations in $\mS$ must leave all components of spin moments invariant, and
\begin{equation}
    \mS = \big\{ \{I|1|\0\}  \big\} \, \label{PSO0}
\end{equation}
{\it i.e.,} there is no nontrivial on-site spin symmetry, or TRS left. 
For coplanar structures, since $S_z = 0$, only $O_{xy}$ in the $s(X)U = {\rm diag}\left( O_{xy},O_z \right)$ affects the transformation of spin moments, while $O_z$ has no impact. 
$\mS$ is given by    
\begin{equation}
    \mS = \big\{ \{ \{ I| 1 | \0 \} , \, \{ T \cdot U_{\hz}(\pi)| 1 | \0 \} \big\} \,  ,
    \label{PSO2}
\end{equation}
where $\U_{\n}(\theta)$ represents a $\theta$-rotation along the direction $\n$ ($\n = \hz$), and the operation $T \cdot U_{\hz}(\pi)$ transforms a spin moment $(S_x, S_y, S_z)$ to $(S_x, S_y, - S_z)$.
In this work, sometimes we also denote the above $\mS$ as $\mS_{Z_2^T}$. 
For the collinear structure, only $O_z$ affects the transformation, and 
{\small
\begin{equation}
\mS =  \big \{ \{\U_{\hz}(\theta)|1| \0 \} \ \big| \theta \in [0,2\pi) \big \} 
    \bigcup \big \{\{T  \U_{\n_{\theta}}(\pi)  |1| \0 \} 
    \ \big| \theta \in [0,\pi) \big \}
    \label{PSO1}
\end{equation}}%
where the first term is a $\theta$-rotation along the $z$ direction, and the second term is a $\pi$-rotation along an in-plane direction $\n_{\theta}=(\cos\theta,\sin\theta,0)$ followed by a TRS.
The two types of operations leave $\S$ unchanged if $S_x=S_y=0$. 
We can equivalently write the collinear $\mS$ as 
\begin{equation} \label{eq:mS-U1-Z2T}
\mS = \mS_{Z_2^T} \ltimes \mS_{U(1)}
\end{equation}
where $\mS_{U(1)}$ contains the continuous rotation of $\mS$ [the first term of Eq.~(\ref{PSO1})] and is a normal subgroup of $\mS$, and $\mS_{Z_2^T} = \{1, T U_{\hx}(\pi) \}$.

An SSG can be decomposed to cosets with respect to the pure-spin-operation group $\mS$
\begin{equation}
    \mG = \mS g_1 + \mS g_2 + \cdots
\end{equation}
As shown in the following paragraphs, we can properly choose the coset representatives such that they commute with every element in $\mS$ and form a discrete normal subgroup $G$, which is isomorphic to the quotient group $\mG/\mS$, of $\mG$. 
Therefore, a generic SSG has the structure 
\begin{equation} \label{eq:mG=SxG}
    \mG = \mS \times G\ . 
\end{equation}
We name $G$ the quotient SSG (qSSG). For given $\mS$, the classification problem of SSGs is now reduced to the classification problem of qSSGs. 

For non-coplanar magnetic structures, $\mS$ is trivial, and the homomorphism from $\mG=G$ to $P$ is an isomorphism.
In $G$, each spatial operation $ \{\R |\v \}$ is equipped with a unique spin operation $X \U$. (Suppose both $ \{ XU |\R |\v \}$ and $ \{ X'U' |\R |\v \}$ belong to the SSG and $X'U'\neq XU$, then $\{ XU |\R |\v \}^{-1}\cdot \{ X'U' |\R |\v \} = \{X^{-1}X' U^{-1}U'| 1 | \0 \}$ is a nontrivial pure-spin-operation, contradicting with Eq.~(\ref{PSO0}).)
Therefore, $G$ realizes a mapping from $P$ to $X U \in Z_2^T\times \mathrm{SO}(3) \simeq \mathrm{O}(3)$, {\it i.e.,} a linear O(3) representation of $P$. 
Let $\rho$ and $\sigma$ be O(3) representations of $P$ realized by two SSGs. 
If $\rho$ and $\sigma$ are equivalent, an orthogonal matrix $O$ exists such that $D_{\rho}(p) = O^T D_{\sigma}(p) O$ for any $p \in P$, which means that differences between the two representations can be eliminated by a change of the spin axes.
These two SSGs are considered to be the same.
All possible non-coplanar SSGs can be obtained from the set of inequivalent O(3) representations, classified by the group cohomology $H^1(P, \mathrm{O}(3))$.

There are various possible choices of coset representatives of $\mG/\mS$ for coplanar and collinear structures. 
To provide a concrete scheme, we adopt the following approach. 
For coplanar (collinear) structures, only $O_{xy}$ ($O_z$) in a spin operation influences the transformation of spin moments. 
Therefore, we can choose the $O_z = \det (O_{xy})$ (for coplanar) or $O_{xy} = O_z \cdot 1_{2\times 2}$ (for collinear) in the spin operations of the coset representatives.
\blue{For a coplanar structure, the only nontrivial spin operation in $\mS$ [Eq.~(\ref{PSO2})] is $T U_{\hz}(\pi)$, containing time-reversal $T$.
Consequently, each spatial operation in a coplanar SSG corresponds to two spin operations: one with the time-reversal operator $T$ (anti-unitary) and another without it (unitary). 
The choice $O_z = \det (O_{xy})$ implies that $\det \left[s(X)U\right] = 1$, leading to all coset representatives being unitary and forming the unitary subgroup $G$ of the SSG $\mG$.
As the corresponding O(3) matrix of $T U_{\hz}(\pi)$ is ${\rm diag}(1_{2\times2}, -1)$, the spin operations ${\rm diag}(O_{xy},O_z)$ of the SSG representatives commute with $\mS$.}
\blue{
For a collinear structure, the choice $O_{xy} = O_z \cdot 1_{2\times 2}$ implies that the spin operations of coset representatives are either $I$ or $T$. 
These spin operations constitute a group and commute with all spin operations.
Thus, the coset representatives form a group $G$ and commute all operations in $\mS$.}
{We also note that qSSG $G$ for a collinear SSG is the same as a magnetic space group of type I, III, or IV, where each spatial operation is uniquely accompanied by $I$ or $T$. }
Following the same derivation as in the non-coplanar case, the qSSG $G$ realizes a linear O(2) (O(1)) representation of $P$, and is classified by $H^1(P, \mathrm{O}(2))$ (for coplanar), or $H^1(P, \mathrm{O}(1))$ (for collinear).

In summary, the SSGs for non-coplanar, coplanar, and collinear magnetic structures can be classified by the O(3), O(2), and O(1) representations of their parent space group, respectively.
Before presenting our classification scheme of SSGs based on the representation theory of space groups in detail, we utilize a flow chart to depict this process (Fig.~\ref{fig:flow chart}).
In the subsequent section, we will outline the initial steps of the flow chart, which involve obtaining all distinct O(3) representations of a given space group and obtaining all O(2) and O(1) representations as byproducts.

\subsection{O(3), O(2), and O(1) representations}
\label{sec O3_Rep}

\begin{table}[]
\centering
\caption[Sixteen types of nontrivial O($n$) ($n = 0, 1,2,3$) representations.]{Sixteen types of nontrivial O($n$) ($n = 0, 1,2,3$) representations, which decompose into non-identity real irreps and pairs of complex irreps. $n=0$ corresponds to the identity representation. The classification is based on the nature of constituent irreps. 
    A non-trivial $d$-dimensional real ($s=r$) or complex ($s = c$) irrep induced from $\rho_{\k}$ of the little group $P^{\k}$ is written as $[\rho_{\k} \uparrow P]^s_d$.
    $[\rho_{\k}^* \uparrow P]^c_d$ represent the complex conjugate of $[\rho_{\k} \uparrow P]^c_d$. 
    A generic O($N$) ($N=1,2,3$) representation consists of a non-trivial O($n$) ($n=1\cdots N$) representation and $(N-n)$ number of identity representations. 
    The column ``$|P|/|P^{\k}|$'' specifies the order of $\k$-star.
    The columns ``TRS'' and ``HSP'' specify whether $\k$ is a TRS-invariant momentum and an HSP, respectively. $\k$ being an HSP means that $\k$ is TRS-invariant or the fixed point manifold of $P^{\k}$ is point-like.
    The third row indicates that a type-III nontrivial O(2) representation consists of two independent type-II nontrivial O(1) representations. 
    The ninth row indicates that a type-$A$ ($A=$IX-XIV) nontrivial O(3) representation consists of a type-II nontrivial O(1) representation and a type-$(A-6)$ nontrivial O(2) representation. 
    }
\begin{tabular}{c|c|c|c|c|c}
\hline\hline
$n$ & Type &  Irreps & $|P|/|P^{\k}|$ & TRS & HSP \\
\hline
0 & I    &  1    &   1    & \cmark & \cmark \\
\hline
1& II   & $[\rho_{\k} \uparrow P]^r_1 $ & 1 & \cmark  & \cmark \\
\hline
\multirow{6}{*}{2} & III  & II $\oplus$ II & - & - & - \\
& IV   & $[\rho_{\k} \uparrow P]^c_1 \oplus [\rho_{\k}^* \uparrow P]^c_1 $ & 1 & \cmark or \xmark & \cmark \\
& V   & $[\rho_{\k} \uparrow P]^c_1 \oplus [\rho_{\k}^* \uparrow P]^c_1 $ & 1 & \xmark & \xmark \\
& VI   & $[\rho_{\k} \uparrow P]^r_2 $ & 1 & \cmark & \cmark \\
& VII   & $[\rho_{\k} \uparrow P]^r_2 $ & 2 & \cmark or \xmark & \cmark \\
& VIII   & $[\rho_{\k} \uparrow P]^r_2 $ & 2 & \xmark & \xmark \\
\hline
\multirow{3}{*}{3}
& IX-XIV & II $\oplus$ (III-VIII) & - & - & - \\
& XV & $[\rho_{\k} \uparrow P]^r_3 $ & 1 & \cmark & \cmark \\
& XVI & $[\rho_{\k} \uparrow P]^r_3 $ & 3 & \cmark & \cmark  \\
\hline\hline
\end{tabular}
\label{tab:O(N)rep}
\end{table}

A generic O(3) representation consists of a nontrivial O($n$) ($n=0,1,2,3$) representation and $(3-n)$ identity representations.
Here $n=0$ means that the O(3) representation is an identity representation.
The nontrivial O($n$) representation is equivalent to an $n$-dimensional real representation that is decomposed into a direct sum of non-identity real irreducible representation (irrep), conjugate pairs of complex irreps (see \AppIrrep\ for more details). 
Therefore, in order to enumerate all the distinct O(3) representations, we only need to find all possible combinations of the irreps of the parent space group $P$. 
All irreps of $P$ can be induced from allowable irreps of the little groups $P^{\k}$ of $\k$ vectors in the {irreducible} BZ~\cite{bradley10}.
\blue{An irrep $\rho_{\k}$ of $P^{\k}$ at $\k$ is considered ``allowable" if the momentum of the irrep's basis is $\k$, requiring} 
that $D_{\rho_{\k}} (\{ I | \t\}) = e^{ i \k \cdot \t} 1_{r \times r}$, where $D_{\rho_{\k}}(\cdot)$ is the representation matrix and $r$ is the dimension of $\rho_{\k}$. 
Hereafter, the notation $\rho_{\k}$ always refers to an allowable irrep of $P^{\k}$.
The construction of $\rho_{\k}$ is detailed in \AppIrrep, and explicit forms of all irreps of all space groups can be obtained from the \href{https://www.cryst.ehu.es/cgi-bin/cryst/programs/representations.pl?tipogrupo=dbg}{Representations DSG} program \cite{Elcoro17} on the Bilbao Crystallographic Server. 
The induced irrep of $P$, $\rho_{\k} \uparrow P$, is supported by the basis of $\rho_{\k}$ and the rotated bases at %
the star of $\k$, $\brace{R_g \k \ |\  g \in P / P^{\k} }$, implying that its dimension is $|P|/|P^{\k}|$ times the dimension of $\rho_{\k}$, where $|\cdot |$ denotes the order of a group.
If $\k$ is not TRS-invariant, $\rho_{\k} \uparrow P$ must be a complex irrep, and its complex conjugation, $\rho_{\k}^* \uparrow P$, is induced from an irrep $\rho_{\k}^*$ of $P^{-\k}$.
If $\k$ is a TRS-invariant momentum, $\rho_{\k} \uparrow P$ can be a real, complex, or pseudo-real irrep of $P$.
As O(3) representations are constructed from irreps with dimensions smaller than or equal to three, we only need to consider TRS-invariant $\k$'s with $|P^{\k}| = |P|, \frac12 |P|, \frac13 |P|$ and non-TRS-invariant $\k$'s with $P^{\k}=P$. 
Notably, the dimension of a pseudo-real irrep must be even (see \AppIrrep), and hence a pair of pseudo-real irreps has a minimum dimension of four, and it is unnecessary to consider them when studying O(3) representations.

A space group has a finite number of high-symmetry points (HSPs) $\k$ in the BZ.
Here $\k$ being an HSP means that $\k$ is TRS-invariant or the little group of $\k$'s neighborhood is smaller than $P^{\k}$.
Hence, it is direct to enumerate the finite number of irreps induced from these $\k$'s.  
On the other hand, a space group also has high-symmetry lines and planes of momenta, and generic momenta in the irreducible BZ.
Different $\k$ in these regions induce different irreps of the space group. 
To handle the infinite number of irreps induced by $\k$ in these regions, 
we regard that an irrep induced by $\k_1$ and that by $\k_2$ belong to the same class,
if $P^{\k_1}=P^{\k_2}\equiv P^{\k}$, $\k_1$ and $\k_2$ are within the fixed-point manifold of $P^{\k}$, and if the two irreps can be continuously deformed to each other as $\k_{1,2}$ move continuously within the manifold.
SSGs described by O(3) representations induced from the irreps in the same class will be treated as the same. 
With these in mind, we only need to consider a finite number of irreps for each space group. 
It is worth noting that this scheme can include the SSGs describing incommensurate magnetic structures. 
If an O(3) representation consists of irreps induced from $\k$'s at high symmetry lines, planes, or generic momenta, the corresponding SSG describes both commensurate and incommensurate magnetic structures, depending on whether $\k$'s are rational or not.

We classify the O(3) representations into sixteen different types, as summarized in Table~\ref{tab:O(N)rep}. 
The $n=0,1,2,3$ blocks in Table~\ref{tab:O(N)rep} correspond to the nontrivial O($n$) representations, respectively. 
The nontrivial O($n$) representations in each block are further classified into different types based on the little group irreps $\rho_{\k}$ from which they are induced. 
In the following, we use the notation $[\rho_{\k} \uparrow P]^{s}_d$ to represent a $d$-dimensional real ($s=r$) or complex ($s=c$) irrep of the space group $P$. 
First, we have the 3D identity representation:
\begin{itemize}[leftmargin=0.5cm]
\item Type I: A 3D identity representation. 
\end{itemize}
Second, we have only one type of the $n=1$ O(3) representations:
\begin{itemize}[leftmargin=0.4cm]
\item Type II: A direct sum of two trivial irreps and a 1D non-trivial real irrep, $[\rho_{\k} \uparrow P]^{r}_1$, where $P^{\k} = P$ (otherwise the dimension is larger than one) and $\k$ must be TRS-invariant (otherwise the irrep cannot be real).
\end{itemize}
Third, the $n=2$ O(3) representations are divided into six types:
\begin{itemize}[leftmargin=0.5cm]
\item Type III: A direct sum of one trivial irrep and two non-trivial 1D real irreps, $[\rho_{\k} \uparrow P]^{r}_1 \oplus [\rho_{\k'}' \uparrow P]^{r}_1$, where both $\k,\k'$ are TRS-invariant and $P^{\k}=P^{\k'}=P$. 
\item Types IV-V: A direct sum of one trivial irrep and a pair of 1D complex irreps, $[\rho_{\k} \uparrow P]^{c}_1 \oplus [\rho_{\k}^* \uparrow P]^{c}_1$, where $\k$ is not necessarily TRS-invariant and $P^{\k} = P$.
The O(3) representation is type IV if $\k$ is an HSP and is type V otherwise.
We distinguish the two types because the former SSGs describe only commensurate magnetic structures, whereas the latter describes both incommensurate and commensurate structures. 
In the latter case, the fixed-point manifold of $P^{\k}$ can be high-symmetry lines, planes, or generic momenta in the BZ.
SSGs given by the O(3) representations from irrational (rational) $\k$ in the fixed-point manifold describe incommensurate (commensurate) magnetic structures [see Figs.~\ref{p3-ssg}(f)-(h) for examples for type-V SSGs].

\item Type VI: A direct sum of one trivial irrep and a 2D real irrep, $[\rho_{\k} \uparrow P]^r_2$, where $P^{\k}=P$, $\k$ is TRS-invariant, and $\rho_{\k}$ is a 2D real irrep. %
\item Type VII: A direct sum of one trivial irrep and a 2D real irrep, $[\rho_{\k} \uparrow P]^r_2$, where $|P^{\k}|=|P|/2$, $\k$ is an HSP.
$\k$ is not necessarily TRS-invariant. However, if it is not, the star of $\k$ must include $-\k$. Otherwise, the induced irrep cannot be real.
We distinguish type VII from type VI because it involves two propagating wave vectors. 
\item Type VIII: A direct sum of one trivial irrep and a 2D real irrep, $[\rho_{\k} \uparrow P]^r_2$, where $|P^{\k}|=|P|/2$, and $\k$ is not an HSP. The star of $\k$ must include its TRS partner $-\k$. As the wave vector $\k$ can move in the fixed point manifold of $P^{\k}$, type VIII can describe both incommensurate and commensurate magnetic structures. 
\end{itemize}
The $n=3$ O(3) representations can be divided into eight types, where the first six are direct sums of 1D irrep and 2D representations, and the latter two are 3D irreps:
\begin{itemize}[leftmargin=0.5cm]
\item Types IX-XIV: Direct sums of a 1D real irrep, $[\rho_{\k} \uparrow P]^r_1$, and a 2D real representation. The 1D real irrep has the same form as in type II, and the 2D real representations in the types IX-XIV are the same as the non-trivial parts in types III-VIII, respectively (see Table~\ref{tab:O(N)rep}). 
\item Type XV: A 3D real irrep, $[\rho_{\k} \uparrow P]^r_3$, with $|P^{\k}|=|P|$. $\k$ must be TRS-invariant; otherwise, the induced irrep cannot be real. 
\item Type XVI: A 3D real irrep, $[\rho_{\k} \uparrow P]^r_3$, with $|P^{\k}|=|P|/3$. $\k$ must be TRS-invariant; otherwise, the induced irrep cannot be real. 
\end{itemize}

Dividing the O(3) representations into the sixteen types allows us to enumerate all the representations efficiently. 
Additionally, it also automatically yields all the O(2) representations (types I-VIII) and O(1) representations (types I-II), which, according to Sec.~\ref{class_general}, classify the SSGs with coplanar and collinear magnetic structures, respectively.  

Note that two different complex irreps may induce the same real representation. 
Consider a complex irrep $\rho_{\k}$ at $\k$, its complex conjugation $\rho_{\k}^* $ must be an irrep at $-\k$, denoted as $\rho_{-\k}'$. 
Hence, there must be $[\rho_{\k}\uparrow P]^c_1 \oplus [\rho_{\k}^* \uparrow P]^c_1 = [\rho_{-\k}'\uparrow P]^c_1 \oplus [\rho_{-\k}^{\prime *} \uparrow P]^c_1$.
To avoid this redundancy, we limit $\k$ of $\rho_{\k}$, from which the O($N$) representations are induced, in the {irreducible} BZ. 

\subsection{Example: SSGs in the parent space group \texorpdfstring{$P3$}{P3}}
\label{sec:example-P3}

\begin{figure*}
     \begin{overpic}[%
    width=1\textwidth]{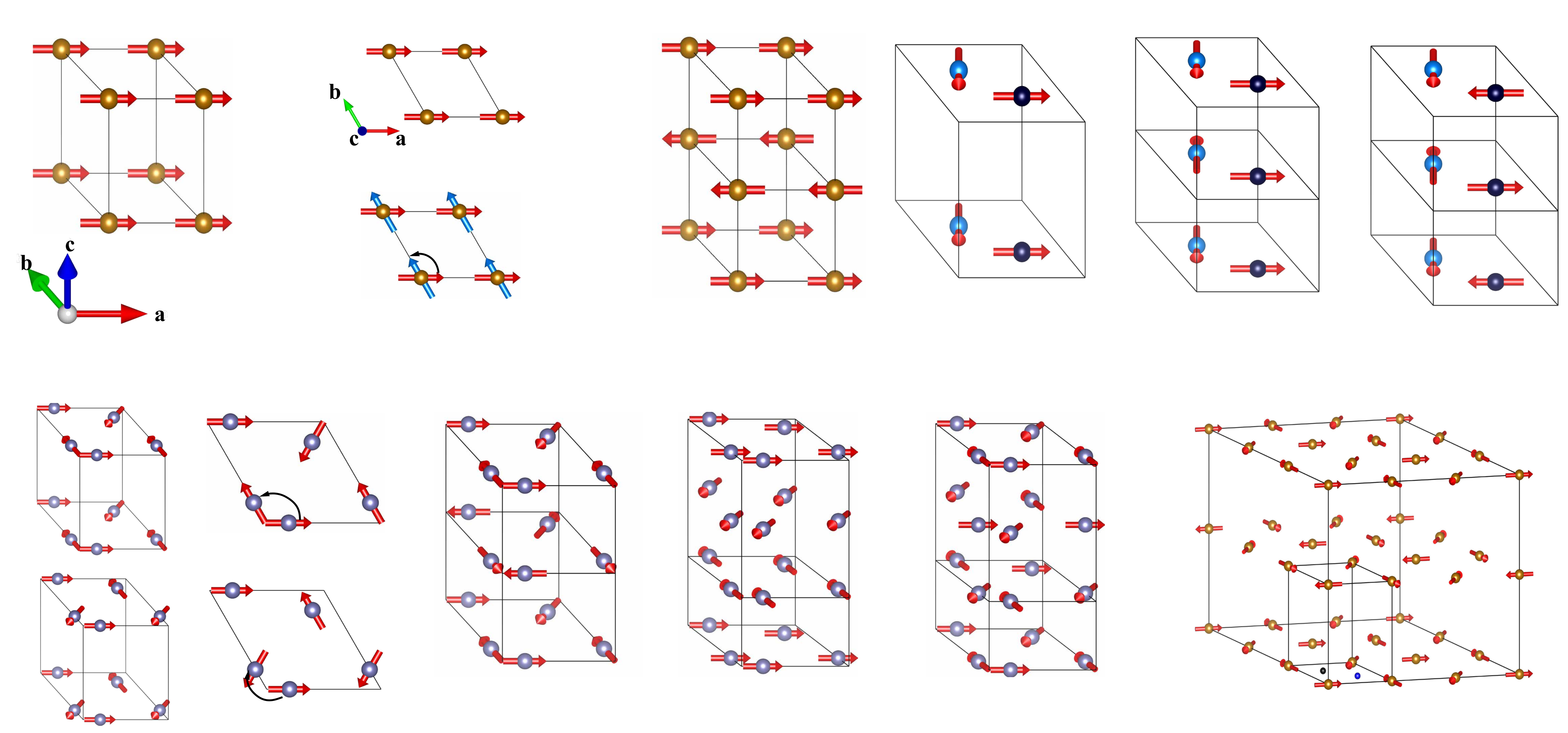}
     \put(0,46){(a) L143.1.1 GM1}
     \put(42,46){(c) L143.2.1 A1}
     \put(56,46){(d) P143.1.1 GM1}
     \put(71,46){(e) P143.2.1 A1}
     \put(87,46){(f) P143.3.1 A1$\oplus$A1}
     \put(0,23){(g) P143.4.1 GM2GM3}
     \put(26,23){(h) P143.4.2 A2A3}
     \put(42,23){(i) P143.5.1 DT1DU1}
     \put(60,23){(j) P143.5.2 DT2DU2}
     \put(80,23){(k) P143.5.3 P1PC1}
     \put(16, 46){(b) $C_{3z}$ in SSG and MSG}
     \put(17.5,44){(b1)}
     \put(17.5,35){(b2)}
     \put(27,40.5){$I$}
     \put(27,30.5){$U_{\hz}(2\pi/3)$}
     \put(0,21){(g1)}
     \put(12,21){(g2)}
     \put(0,11){(g3)}
     \put(12,11){(g4)}
     \put(18,15){$U_{\hz}(2\pi/3)$}
     \put(12,1){$U_{\hz}(-2\pi/3)$}
\end{overpic}   
\caption{(a), (c) Collinear, and (d)-(k) coplanar magnetic structures for all distinct spin space groups (SSGs) whose parent space group is $P3$ (No.~143).
The notations of SSGs, which consist of one prefix letter, three indices, and O(3) representations of the space group (see Sec.~\ref{sec: nonmenclature}), are labeled at the top of each subfigure. 
Atoms in different Wyckoff positions are depicted with different colors. 
The golden atoms occupy Wyckoff position 1a [$(0,0,z)$ with $z=0$]; the light-blue atoms occupy Wyckoff position 1b [$(1/3,2/3,z)$ with $z=0$]; the dark-purple atoms occupy Wyckoff position 1c [$(2/3,1/3,z)$ with $z=0$]; the grey atoms occupy Wyckoff position 3d [$(x,y,z), (-y,x-y,z), (-x+y,-x,z)$ with $y = z = 0$].
\blue{(b) The comparison between $2\pi/3$-rotation in SSG and magnetic space group (MSG).}
\blue{(a) Collinear magnetic moments $\S(\r_i)$'s remain invariant under all spatial operations. (c) Collinear $\S(\r_i)$'s change sign under $\{1|001\}$. (d) Coplanar $\S(\r_i)$'s remain invariant under all spatial operations. (e) The SSG 143.2.1 A1 is given by the O(2) representation $[{\rm GM1} \uparrow P]_1^r\oplus[{\rm A1} \uparrow P]_1^r$, while the trivial irrep $[{\rm GM1} \uparrow P]_1^r$ is omitted in its notation. Irrep $[{\rm GM1} \uparrow P]_1^r$ implies that one component ({\it e.g.}, $x$) of coplanar $\S(\r_i)$'s is invariant under all operations. Irrep $[{\rm A1} \uparrow P]_1^r$ implies that the other component ({\it e.g.}, $y$) changes sign under $\{I | 001\}$. (f) Two irreps $[{\rm A1} \uparrow P]_1^r$'s imply that both components of coplanar $\S(\r_i)$'s change sign under $\{1 | 001\}$. (g) Coplanar $\S(\r_i)$'s are rotated by $2\pi/3$ [(g1), (g2)] or $-2\pi/3$ [(g3), (g4)] under $C_{3z}$ rotation. (h) Coplanar $\S(\r_i)$'s are rotated by $2\pi/3$ under $C_{3z}$, and change sign under $\{1|001\}$.}
\blue{(i)-(k) Spiral magnetic structures. (i) Coplanar $\S(\r_i)$'s are invariant under $C_{3z}$, and are rotated by $u \pi$ ($u = 2/3$) under $\{1|001\}$. (j) Coplanar $\S(\r_i)$'s are rotated by $2 \pi/3$ under $C_{3z}$, and are rotated by $u \pi$ ($u = 2/3$) under $\{1|001\}$. (k) Coplanar $\S(\r_i)$'s are invariant under $C_{3z}$, are rotated by $2\pi/3 $, $2 \pi/3$, and $ v \pi$ ($v = 1$) under the translations $\{1| 100\}$, $\{1 | 010\}$, and $\{1 | 001\}$, respectively.}
These crystal structures and subsequent structures were illustrated by using the VESTA software~\cite{momma_vesta_2011}.
}
\label{p3-ssg-1}
\end{figure*}

To demonstrate how we obtain all \blue{O($N$) ($N = 1,2,3$)} representations of a space group and how they classify \blue{all the collinear, coplanar, and non-coplanar SSGs}, we use magnetic structures with the parent space group $P3$ (No.~143) as examples.
\blue{After obtaining these SSGs, we clarify the SSG and magnetic unit cells in SSGs, and their relationship with the momenta of O($N$) representations.}
The group $P3$ is generated by a three-fold rotation $C_{3z} = \{3_{001}| \0 \}$, and lattice translation.
It possesses a hexagonal prism BZ.
As explained at the beginning of the last subsection, to obtain all \blue{O($N$) ($N \leq 3$)} representations, we only need to consider TRS-invariant $\k$'s with $|P^{\k}| = |P|, |P|/2, |P|/3$ and non-TRS-invariant $\k$'s with $P^{\k}=P$. 
Such momenta in {the irreducible BZ} include $\Gamma$ $(0,0,0)$, A $(0,0,\pi)$, M $(\pi,0,0)$, L $(\pi,0,\pi)$, DT $(0,0,u \pi)$, and P $(2\pi/3,2\pi/3,v \pi)$, where the component forms are given on the basis of the reciprocal lattice vectors of $P3$.
DT and P are high symmetry lines parameterized by the {continuous} variables $u,v$, respectively
\footnote{Note that point K [$\k = ( 2\pi/3,2 \pi/3,0) $] or H [$\k = (2 \pi/3,2 \pi/3,\pi) $] is not considered as an HSP because they share the same little group as any other points on the line P, and they are not TRS-invariant.}. 

The little groups of $\Gamma$, A, DT, and P equal the space group since they are invariant under the $C_{3z}$ rotation.
There are three inequivalent 1D little group irreps on each of them, labeled by S1, S2, and S3 (S =  GM, A, DT, P. Hereafter, GM means $\Gamma$.). 
The explicit representation matrices are given as
\begin{equation} 
    D_{\rm S1}(C_{3z}) = 1 , \,  D_{\rm S2}(C_{3z}) = e^{i \frac{2\pi}3} , \, D_{\rm S3}(C_{3z}) = e^{-i \frac{2\pi}3} \ .
    \label{DT_rep}
\end{equation}
For all the irreps S1, S2, S3, the representation matrix for translation $\{1|m_1,m_2,m_3\}$ is $e^{i(k_1m_1 + k_2m_2 +k_3m_3)}$, where $ (k_1,k_2,k_3)$ is the momentum of the irrep written on the basis of the reciprocal lattice, and $(m_1,m_2,m_3)$ is an integer-valued vector written on the basis of lattice vectors. 
The little group of M or L is the translation subgroup of P3. Therefore, $|P^{\rm M, L}|=|P|/3$. 
$P^{\rm M, L}$ only has one 1D irrep (M1, L1) on each of them, where the translation $\{1|m_1,m_2,m_3\}$ is represented by $e^{i(k_1m_1 + k_2m_2 +k_3m_3)}$.

\subsubsection{O(1) representations and collinear SSGs}
\blue{
We construct O(1) representations to classify collinear SSGs.
As discussed in Sec.~\ref{sec O3_Rep}, collinear SSGs are categorized under the first two of the total sixteen SSG types. 
}
The type-I collinear SSG is described by the 1D identity representation (GM1).
Hence the collinear magnetic moments $\S(\r_i)$'s are invariant under any spatial operation without spin operations, as exampled in Fig.~\ref{p3-ssg-1}(a).
This SSG is named L143.1.1 GM1, where ``L'' stands for collinear, ``143'' is the parent space group index, the first ``1'' represents type-I, the second ``1'' is the index of SSG for a given parent space group and type, and ``GM1'' represents the little group irrep from which the O(1) representation is induced (refer to Sec.~\ref{sec: nonmenclature} for the detailed naming convention of SSGs).
\blue{
Notably, although magnetic moments in Fig.~\ref{p3-ssg-1}(a) align collinearly along the $x$-axis, rotating them simultaneously to a different direction $\n$ does not alter the SSG symmetry.
On the contrary, the MSG of a structure depends on the specific orientation of $\S(\r_i)$'s.
In an MSG, a $C_{3z}$ rotation necessitates $2\pi/3$-spin rotation $U_{\hz}(2\pi/3)$ along the $z$ direction or its combination with time reversal $T$.
The latter implies the existence of $(C_{3z}T)^3 = T$, which is impossible in the presence of non-zero magnetic moments; and the former also does not preserve the magnetic structure invariant [Fig.~\ref{p3-ssg-1}(b2)].}
The type-II SSGs are described by a non-trivial real irrep, $[\rho_{\k} \uparrow P]^{r}_1$, where $\k$ is TRS-invariant and $P^{\k}=P$.
The space group $P3$ only has one nontrivial 1D real irrep - $[\mathrm{A1} \uparrow P]^{r}_1$.
Therefore, only one type-II SSG, named L143.2.1 A1, exists for the space group [Fig.~\ref{p3-ssg-1}(c)].
The 1D real irrep implies that collinear magnetic moments $\S(\r_i)$'s change sign under the translation $\{I | 0,0,1 \}$.

\subsubsection{O(2) representations and coplanar SSGs}
\blue{
We consider O(2) representations and coplanar SSGs. 
These SSGs are categorized within the first eight types in Table~\ref{tab:O(N)rep}.
The type-I and type-II O(2) representations are given by the direct sum of an identity irrep combined with type-I and type-II O(1) representations, respectively. 
Thus, $P3$ has one type-I coplanar SSG, P143.1.1 GM1 [Fig.~\ref{p3-ssg-1}(d)], and one type-II coplanar SSG, P143.2.1 A1 [Fig.~\ref{p3-ssg-1}(e)].
Here, ``P'' represents coplanar, with the rest of the naming convention similar to that of collinear SSGs. 
Note that the O(2) representation A1$\oplus$1 characterizes the SSG P143.2.1, but the trivial irrep $1$ is omitted in the notation (also refer to Sec.~\ref{sec: nonmenclature}). 
Type-III SSGs are given by $[\rho_{\k} \uparrow P]^{r}_1 \oplus [\rho_{\k'}' \uparrow P]^{r}_1$. 
Given that $P3$ possesses only a single 1D non-trivial real irrep, there is only one type-III SSG - N143.3.1 A1$\oplus$A1 [Fig.~\ref{p3-ssg-1}(f)] - given by the direct sum of two A1's.
Refer to the caption of Fig.~\ref{p3-ssg-1} for the descriptions of these SSG symmetries that are determined by the corresponding representations.}

Type-IV SSGs are described by a direct sum of a complex irrep $[\rho_{\k}\uparrow P]^{c}_1$ and its complex conjugation $[\rho_{\k}^* \uparrow P]^{c}_1$
, where $P^{\k}=P$ and $\k$ is TRS-invariant. 
Only GM and A satisfy the requirements of type IV. 
Irreps GM2 and GM3 form a complex conjugate pair, giving the SSG P143.4.1 GM2GM3. 
Irreps A2 and A3 form another complex conjugation pair, giving another SSG P143.4.2 A2A3. 
Let us first look at P143.4.1 GM2GM3. 
The representation matrix of $C_{3z}$ given by GM2$\oplus$GM3 is equivalent to the real rotation matrix 
\begin{equation}
\begin{pmatrix}
 \cos(\frac{2\pi}3) & -\zeta \sin(\frac{2\pi}3) \\
 \zeta \sin(\frac{2\pi}3) & \cos(\frac{2\pi}3)
 \label{GM2GM3}
\end{pmatrix} \, ,
\end{equation}
where \blue{$\zeta$ being either $1$ or $-1$ indicates} that \blue{the coplanar} magnetic moments are rotated by $2\pi/3$ [Fig.~\ref{p3-ssg-1}(g1)] and $-{2\pi}/3$ [Fig.~\ref{p3-ssg-1}(g3)] under the $C_{3z}$ rotation, respectively.
A magnetic structure satisfying the rotation matrix with $\zeta=1$ is also (equivalently) described by the conventional magnetic space group \blue{[Fig.~\ref{p3-ssg-1}(g2)]}, whereas a magnetic structure satisfying the rotation matrix with $\zeta=-1$ can only be described by the SSG \blue{[Fig.~\ref{p3-ssg-1}(g4)]}. %
The $\zeta=\pm1$ configurations belong to the same SSG, because in the absence of SOC, we can adopt different spin coordinates irrespective of the real space coordinate, and the two configurations are continuously connected to each other under a spin coordinate transformation. 
For example, consider a continuous rotation around the 
$x$ axis $e^{-i \theta \hat{S}_x}$ as  
the spin coordinate transformation. %
When $\theta$ changes continuously and reaches %
$\pi$, the anti-clockwise structure in the $x,y$ spin plane 
($\zeta=-1$) will be transformed into a clockwise structure 
($\zeta = +1$).
{Notably, 
besides the $\pm 2\pi/3$ rotation around the $z$ axis, 
this SSG can describe cases with the 
$\pm 2\pi/3$ rotation around a generic axis $\n$.}
\blue{
Fig.~\ref{p3-ssg-1}(h) shows the other type-IV SSG N143.4.2 A2A3 and describes its symmetry.}

The type-V SSGs are also constructed from $[\rho_{\k} \uparrow P]^{c}_1 \oplus [\rho_{\k}^* \uparrow P]^{c}_1$ %
as in type-IV, 
except that $\k$ is now not an HSP. 
$P^{\k}=P$ is satisfied by those $\k$'s on DT and P. 
Let us consider %
$\k = (0,0,u \pi)$ on the line DT and its 
irrep DT$i$ ($i = 1,2,3$). %
The complex conjugate of DT$i$ is a complex irrep with momentum at the other %
point $(0,0,-u\pi)$ on the line DT. 
To emphasize %
that these two irreps have different momenta on 
the same line, we follow the convention in \href{https://www.cryst.ehu.es/}{Bilbao Crystallographic Server} \cite{Elcoro17}, and denote the complex conjugate of DT$i$ as DU$i$, and the real representation constructed from DT$i$ and DU$i$ as DT$i$DU$i$. 
Although DT$i$DU$i$ from different momenta (\blue{different $u$}) on the line DT represent inequivalent representations, we classify them as the same SSG, as they describe the same kind of magnetic structures (see the discussion in Sec.~\ref{sec O3_Rep}).  
Similarly, the complex conjugate of irrep P$i$ ($i=1,2,3$) on the P line is denoted as PC$i$, and their direct sum is referred to as P$i$PC$i$. 
In the following, we will explain DT$i$DU$i$ ($i=1,2,3$) and 
P$i$PC$i$ ($i=1,2,3$), and demonstrate that some of 
these six describe the same SSG; some are transformed into one another under a change of the coordinate system. 
First, let us examine the representations DT$i$DU$i$ on the line DT. 
If the momentum of DT$i$ ($i=1,2,3$) is $(0,0,u\pi)$ ($|u|\le 1$), the representation matrices of translation $\{ I | 0,0,1\}$ for DT$i$DU$i(u)$ are the same for different $i$, and equivalent to the rotation matrix 
\begin{equation}
D_{\mathrm{DT}i\mathrm{DU}i(u)}\left(\{ I | 0,0,1\}\right) = 
\begin{pmatrix}
 \cos(u \pi) & -\zeta \sin( u \pi) \\
 \zeta \sin( u \pi) & \cos( u \pi)
\end{pmatrix} \, ,
\label{DT_rep1}
\end{equation}
where $\zeta =\pm 1$. %
The representation matrices of $C_{3z}$ for DT$i$DU$i(u)$ 
with different $i$ are given by 
\begin{equation}
\begin{aligned}
    D_{\rm DT1DU1(u)}\left(C_{3z}\right) &  =  
\begin{pmatrix}
1 & 0 \\
 0& 1
\end{pmatrix} , \\
    D_{\rm DT2DU2(u)} \left(C_{3z}\right)&  =  
\begin{pmatrix}
 \cos(\frac{2\pi}{3}) & -\zeta \sin( \frac{2\pi}{3}) \\
 \zeta \sin( \frac{2\pi}{3}) & \cos(\frac{2\pi}{3})
\end{pmatrix} \, , \\
    D_{\rm DT3DU3(u)} \left(C_{3z}\right)&  =  
\begin{pmatrix}
 \cos(\frac{2\pi}{3}) & \zeta \sin( \frac{2\pi}{3}) \\
 -\zeta \sin( \frac{2\pi}{3}) & \cos(\frac{2\pi}{3})
\end{pmatrix} \, , \\
\end{aligned}
\label{DT_rep2}
\end{equation}
respectively. 
Notice that, for a given DT$i$DU$i$ representation, the $\zeta$'s in Eqs.~(\ref{DT_rep1}) and (\ref{DT_rep2}) must be the same because the representation matrices of translation and rotation are transformed from the two 1D complex irreps by the same unitary matrix. 
The first type-V SSG is N143.5.1 DT1DU1 \blue{[see the structure and symmetry in Fig.~\ref{p3-ssg-1}(i)].}
As discussed in the type-IV SSG, %
the two configurations from $\zeta = \pm 1$ should belong to %
the same SSG. 
Additionally, it is worth noting that %
different $u$ only gives %
different spiral angle in the magnetic structure of Fig.~\ref{p3-ssg-1}(i), and the other features of the structure are exactly the same for different $u$. This lets us regard %
that different $u$'s on the line DT give %
the same SSG.

Next, let us consider DT2DU2 and DT3DU3. 
For DT2 at $\k = (0,0,u\pi)$ and $\zeta = \pm1$ in Eqs.~(\ref{DT_rep1}) and (\ref{DT_rep2}), DT2DU2 describes a magnetic structure where two components ({\it e.g.,} $x,y$) of the magnetic moments are rotated by $\pm 2\pi/3$ under $C_{3z}$ and rotated by $\pm u\pi$ under $\{1 | 0,0,1\}$ respectively [see Fig.~\ref{p3-ssg-1}(j)].  
Meanwhile, for DT3 at $\k = (0,0,-u\pi)$ and $\zeta = \mp 1$ in Eqs.~(\ref{DT_rep1}) and (\ref{DT_rep2}), DT3DU3 give the same representation matrices.
It seems that DT2DU2 and DT3DU3 are the same representation. 
However, given that we only consider $\rho_{\k}$ in the irreducible BZ, DT2DU2 and DT3DU3 are still inequivalent representations because DT2DU2 induced from $u\in(0,1)$ (within the irreducible BZ) is only equivalent to DT3DU3 induced from $u\in (-1,0)$ (outside the irreducible BZ). 
Nevertheless, the two representations are continuously connected as $u$ approaches zero. 
Physically, if the spin axes are properly chosen, DT2DU2 ($\zeta=1$) and DT3DU3 ($\zeta=-1$) describe similar magnetic structures with spiral angles $u\pi \in (0,\pi)$ and $u\pi\in (\pi,2\pi)$, respectively. 
In this work, we will identify them as in the same class, as the spiral angles can continuously change. 
Such equivalences will be systematically addressed in Sec.~\ref{distinct_O3}. 
We name the SSG as P143.5.2 DT2DU2.

Lastly, we consider irreps P$i$ ($i = 1,2,3$) from the momentum $\k = (2 \pi/3,2 \pi/3,v\pi)$ and their complex conjugations PC$i$. 
For each P$i$PC$i$ representation, the coplanar magnetic moments are rotated by $2\pi/3 $, $2 \pi/3$, and $ v \pi$ under the translations $\{1| 1,0,0\}$, $\{1 | 0,1,0\}$, and $\{1 | 0,0,1\}$, respectively. 
Note that the choice of rotating
$\S(\r_i)$'s in either an anti-clockwise or clockwise direction is arbitrary, as discussed around Eqs.~(\ref{GM2GM3}) and (\ref{DT_rep1}). 
Here we choose to rotate them in the anti-clockwise direction without loss of generality. 
P$i$PC$i$ ($i=1,2,3$) also requires that $\S(\r_i)$'s are rotated by $(i-1)2 \pi/3$ under the $C_{3z}$ rotation around the origin.
Although the rotation angles are different for different $i$, %
magnetic structures for P$i$PC$i$ ($i=1,2,3$) 
are indeed equivalent; they are transformed to each other by %
shifts of the %
origin.
To be concrete, let us begin with %
a magnetic structure  %
for P1PC1 [Fig.~\ref{p3-ssg-1}(k)], where the magnetic moments are 
invariant under $C_{3z}$ around the golden point in Fig.~\ref{p3-ssg-1}(k).  
By shifting the origin %
to the blue (black) point in Fig.~\ref{p3-ssg-1}(k), 
one can see that the magnetic moments in the same magnetic structure 
will be rotated by $2\pi/3$ ($-2\pi/3$) under $C_{3z}$ around the new origin.
Therefore, all P$i$PC$i$ representations ($i = 1,2,3$) describe 
the same SSG, and we name it N143.5.3 P1PC1 in this work. 
The equivalence among P$i$PC$i$ representations can be  
shown by %
transformations of representations associated with %
the lattice coordinate transformation, which will be systematically 
discussed in Sec.~\ref{distinct_O3}. 
For $P3$, the type-VI SSG does not exist;  
$P^{\k}=P$ at TRS-invariant $\k$ requires that 
$k$ must be $\Gamma$ or A, while these two points do not have 2D 
irreps. 
Type-VII and VIII SSGs do not exist for the parent space group $P3$ either, %
because there is no $\k$ with $|P^{\k}|=|P|/2$. 

\subsubsection{O(3) representations and non-coplanar SSGs}

\begin{figure*}
    \centering
    \begin{overpic}[%
    width=1\textwidth]{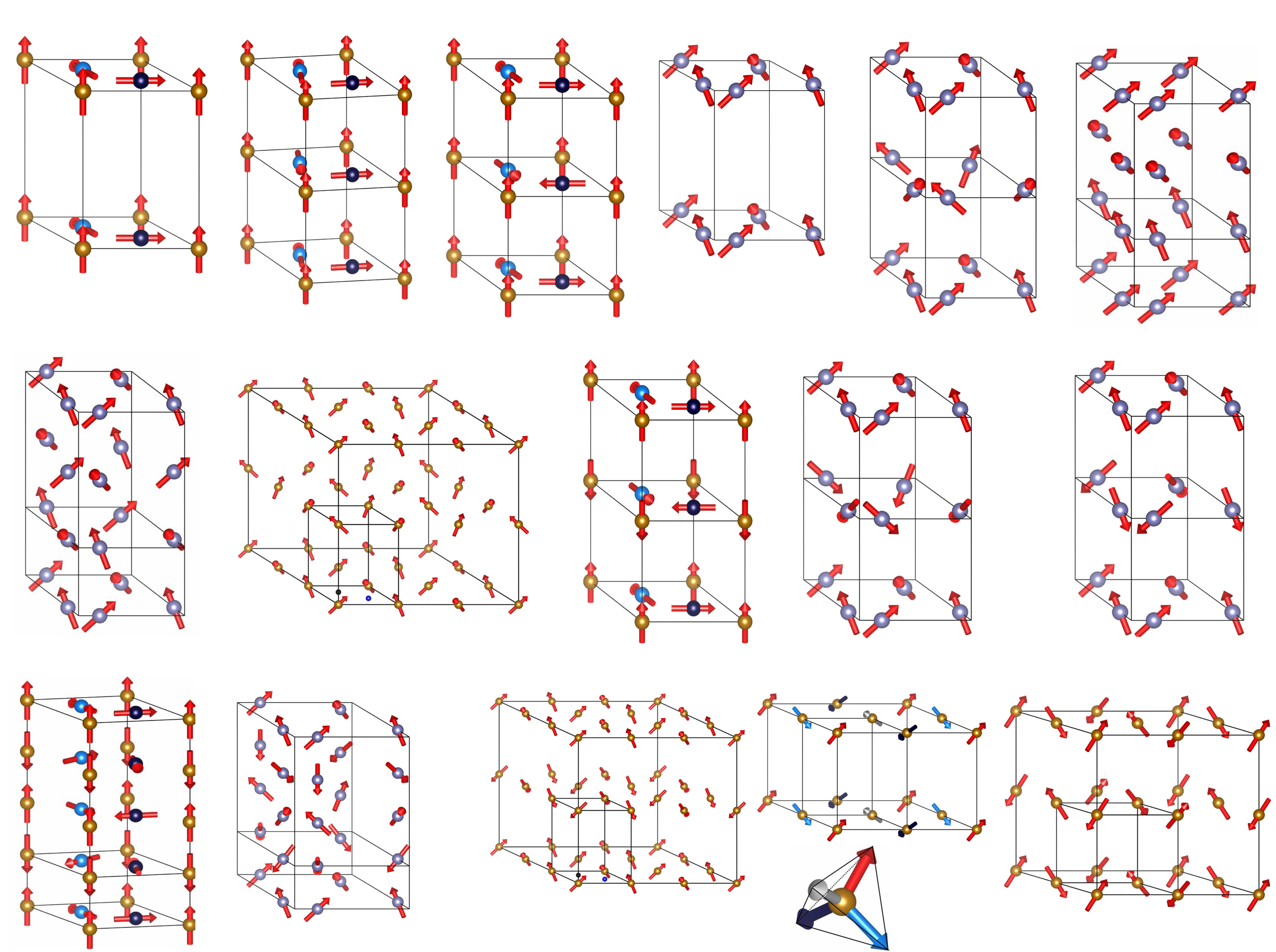}
    \put(-1,72.5){\footnotesize (a) N143.1.1 GM1}
    \put(19,72.5){\footnotesize (b) N143.2.1 A1}
    \put(34,72.5){\footnotesize (c) N143.3.1 A1$\oplus$A1}
    \put(50,72.5){\footnotesize (d) N143.4.1 GM2GM3}
    \put(69,72.5){\footnotesize (e) N143.4.2 A2A3}
    \put(84,72.5){\footnotesize (f) N143.5.1 DT1DU1}
    \put(-1,48){\footnotesize (g) N143.5.2 DT2DU2}
    \put(19,48){\footnotesize (h) N143.5.3 P1PC1}
    \put(42,48){\parbox{0.2 \linewidth}{ 
    \footnotesize(i) N143.9.1  A1$\oplus$A1$\oplus$A1
   }}
 \put(62,48){ \footnotesize
 (j) N143.10.1 A1$\oplus$GM2GM3
   }
\put(83,48){\parbox{0.2 \linewidth}{ 
    \footnotesize (k) N143.10.2 A1$\oplus$A2A3
   }}
    \put(-1,22){\footnotesize (l) N143.11.1 A1$\oplus$DT1DU1}
    \put(18,22){\footnotesize (m) N143.11.2 A1$\oplus$DT2DU2}
    \put(38,22){\footnotesize (n) N143.11.3 A1$\oplus$P1PC1}
    \put(60,22){\footnotesize (o) N143.16.1 M1}
    \put(80,22){\footnotesize (p) N143.16.2 L1}
  \end{overpic}
\caption{Non-coplanar magnetic structures for all distinct spin space groups (SSGs) whose parent space group is $P3$ (No.~143). The notations of SSGs, which consist of one prefix letter, three indices, and O(3) representations of the space group (see Sec.~\ref{sec: nonmenclature}), are labeled at the top of each subfigure. 
Atoms in different Wyckoff positions are depicted with different colors, \blue{with the same rule as Fig.~\ref{p3-ssg-1}.}
(f)-(h) and (l)-(n) display spiral magnetic structures, with (f), (g) having $\k$ on line DT chosen as $(0,0,2\pi/3)$, (l), (m) having $\k$ on line DT chosen as $(0,0,\pi/2)$, and (h), (n) having $\k$ on line P chosen as $(2\pi/3,2\pi/3,\pi)$. (h), (n) the coordinates of the black and blue points are $(2/3,1/3,0)$ and $(1/3,2/3,0)$, respectively.}
\label{p3-ssg}
\end{figure*}

\blue{We construct O(3) representations that classify non-coplanar SSGs.
An O(3) representation of types I-VIII consists of an O(2) representation in the same type and a trivial irrep.
Figs.~\ref{p3-ssg}(a)-(h) show magnetic structures conforming to these SSGs, sharing the same number of coplanar SSGs.
The O(2) representation indicates that two components ({\it e.g.}, $x,y$) of the magnetic moments transform as those in coplanar SSGs, while the trivial irrep indicates that the remaining component ({\it e.g.}, $z$) is invariant under all spatial operations.}
Types IX-XIV SSG are described by direct sums of a non-trivial 1D real irrep and a 2D non-trivial real representation. 
Here, the 1D real irreps are the same as those in type II, and the 2D real representations are the same as those in types III-VIII, respectively. 
Having only types III-V SSGs, $P3$ has only types IX, X, and XI SSGs. 
Since $P3$ possesses only one non-trivial 1D real irrep A1, there exists a one-to-one correspondence between SSGs of types IX-XI [Fig.~\ref{p3-ssg}(i)-(n)] and SSGs of types III-V [Fig.~\ref{p3-ssg}(c)-(h)]. 
The additional A1 representation in types IX-XI SSG makes one component ({\it e.g.}, $z$) of the magnetic moments change sign under a lattice translation along the $z$ axis, whereas the other two components ({\it e.g.}, $x,y$) transform identically as those in types III-V SSG discussed so far.

Type-XV SSGs, characterized by $[\rho_{\k} \uparrow P]^r_3$ with $|P^{\k}|=|P|$ and $\k$ being TRS-invariant, do not exist for $P3$. Two TRS-invariant $\k$ points, A and $\Gamma$, satisfy $|P^{\k}|=|P|$, but neither of them possesses a 3D irrep.

Type-XVI SSGs are characterized by a 3D real irrep $[\rho_{\k} \uparrow P]^r_3$ with $|P^{\k}|=|P|/3$ and TRS-invariant $\k$. %
M and L satisfy the requirements, and each possesses only one irrep. 
Hence, $P3$ has only two type-XVI SSGs, namely N143.16.1 M1 [Fig.~\ref{p3-ssg}(o)] and N143.16.2 L1 [Fig.~\ref{p3-ssg}(p)]. 
For the SSGs N143.16.1 M1 and N143.16.2 L1, the O(3) representation matrices of the lattice translation $\{ 1|m_1,m_2,m_3 \}$ are given by 
\begin{equation}
     {\rm diag} \left( (-1)^{m_1},(-1)^{m_2},(-1)^{m_1 + m_2}\right) \, ,
     \label{M1_gen}
\end{equation}
and 
\begin{equation}
     {\rm diag} \left( (-1)^{m_1 + m_3},(-1)^{m_2 + m_3},(-1)^{m_1 + m_2 + m_3}\right) \, ,
     \label{L1_gen}
\end{equation}
respectively.
Here the basis of the O(3) matrices is composed of three inequivalent TRS-invariant points.
For N143.16.1 M1, the points are $(\pi,0,0)$, $(0,\pi,0)$, $(\pi,\pi,0)$, while for N143.16.2 L1, they are $(\pi,0,\pi)$, $(0,\pi,\pi)$, $(\pi,\pi,\pi)$. %
The O(3) representation matrix of the $C_{3z}$ rotation is 
given by the same matrix for N143.16.1 M1 and N143.16.2 L1, 
\begin{equation}
\begin{pmatrix}
    0 & 1 & 0 \\
    0 & 0 & 1 \\
    1 & 0 & 0 \\ 
\end{pmatrix} \, .
\label{L1_gen2}
\end{equation}
\blue{
For the magnetic structure with SSG N143.16.1 M1, the magnetic moments are rotated by $2\pi/3$ along the $111$-direction [Eq.~(\ref{L1_gen2})] under the $C_{3z}$ rotation; and are rotated by $\pi$ along the $y$ and $x$ directions [Eq.~(\ref{M1_gen})] under $\{1 | 1,0,0 \}$ and $\{1| 0,1,0\}$, respectively.}
\blue{
Therefore, the spin operations in this SSG constitute the chiral tetrahedral group T.
Required by the symmetries, the magnetic moment at the $C_{3z}$-invariant origin aligns with the rotation axes of $2\pi/3$-spin-rotation ({\it i.e.,} $111$-direction).
Under $\{1|1,0,0\}$, $\{1|0,1,0\}$, and $\{1|1,1,0\}$, this magnetic moment is transformed into different directions.
Thus, in the magnetic unit cell, four distinct magnetic moments exist, presented by different colors in Fig.~\ref{p3-ssg}(o).
Notably, the magnetic moments point to the four vertexes of a tetrahedron [Fig.~\ref{p3-ssg}(o)], manifesting the tetrahedral group structure of spin operations.
For SSG N143.16.2 L1, the magnetic structure on the $z = 0$ plane is identical to those for SSG N143.16.1 M1 [Fig.~\ref{p3-ssg}(p)]. 
However, these magnetic moments are reversed under $\{1 | 0,0,1\}$.
Thus, the spin operations in this SSG constitute the achiral tetrahedral group T$_{\rm d}$. We also}
note that SSG N143.16.1 M1 possesses an interesting property: there exist two anti-commuting translation operations for the electronic Hamiltonian described by this SSG (see Sec.~\ref{subsec: rep PN SSGs}). 

Magnetic structures presented in Figs.~\ref{p3-ssg-1} and \ref{p3-ssg} are merely examples illustrating the SSGs.
\blue{For collinear magnetic structures [Figs.~\ref{p3-ssg-1}(a), (c)], we choose magnetic moments $\S(\r_i)$'s aligned along the $x$ direction.
For coplanar magnetic structures [Figs.~\ref{p3-ssg-1}(d)-(k)], we choose $\S(\r_i)$'s aligned within the $x,y$ plane.
For non-coplanar magnetic structures in Figs.~\ref{p3-ssg}(a)-(h), we choose the $z$ component of $\S(\r_i)$'s to be transformed according to the 1D identity irrep; for those in Figs.~\ref{p3-ssg}(i)-(n), we choose the $z$ component to be transformed according to the 1D nontrivial real irrep.
}
The particular choice of \blue{the spin orientation and} the spin coordinate system is only for the sake of clarity in the illustration. 
The same SSGs still apply if all the magnetic moments are simultaneously rotated by an arbitrary O(3) matrix $O$, as we discussed for the SSG N143.4.1 GM2GM3.

\blue{\subsubsection{SSG unit cell and magnetic unit cell}}
\blue{The SSG unit cell of an SSG $\mG$ refers to the unit cell of its parent space group $P$, and is associated with the translation subgroup $ T = \left\{ \{ 1| \v_g \} | R_g = 1,  g \in \mG \right\}$ of $P$.   
In Figs.~\ref{p3-ssg-1} and \ref{p3-ssg}, the smaller cells enclosed by the solid black line are the SSG unit cells.  
The magnetic unit cell, also known as magnetic supercell, is associated with the pure translation subgroup $T_M$, given as 
\begin{equation} \label{eq: T_M}
    T_M= \brace{ \brace{ I| 1 | \v }| \brace{ I| 1 | \v } \in \mG} \, .
\end{equation}
Here the subscript ``$M$'' denotes MSG because $T_M$ is also a subgroup of an MSG.
}

\blue{The magnetic unit cell and $T_M$ are uniquely determined by the momenta of the $O(N)$ representation.
For a non-coplanar SSG, the O(3) representation matrix of the translation $\{ 1| \v \} \in T$ is equivalent to the matrix $O_{\v} = {\rm diag} (e^{i \k_1 \cdot \v }, e^{i \k_2 \cdot \v }, e^{i \k_3 \cdot \v }  )$ up to a unitary transformation, where
$ \k_i$'s ($i = 1,2,3$) are the momenta of the O(3) representation. 
Note that $ e^{i \k_i \cdot \v}$'s are either all real or consist of a conjugate pair and a real number, required by $O_{\v}$ being equivalent to an O(3) matrix.
All $\{ 1| \v \} \in T $ such that $O_{\v} = I$ constitute $T_M$.}

\blue{Let us consider the SSG N143.5.1 DT1DU1 [see Fig.~\ref{p3-ssg}(f)] as an example. 
The SSG is given by the direct sum of a complex irrep DT1, its complex conjugate DU1, and an identity irrep, the momenta of which are $(0,0, u \pi)$ with $u \notin \mathbb{Z}$, $(0,0, -u \pi)$, and ${\bm 0}$, respectively. 
Thus, the representation matrix $O_{\v}$ for $\{1 | \v \}$ with $\v = \sum_{i = 1}^3 m_i \a_i$ ($m_i \in \mathbb{Z}$) is equivalent to ${\rm diag} (e^{i u \pi m_3 }, e^{-i u \pi m_3 }, 1 )$. 
The condition $O_{\v} = I$ requires that $\frac{1}{2}u m_3 \in  \mathbb{Z} $ and have no constraints on $m_1,m_2$. 
For the structure in Fig.~\ref{p3-ssg}(f), $u = 2/3$, and hence $m_3$ is required to be a multiple of $3$. 
The lattice vectors of the magnetic unit cell are $\a_1,\a_2$, and $3 \a_2$, which is three times as large as the SSG unit cell. 
For more generic $u$, if $\frac{1}{2}u$ is a rational number $p/q$ with $p,q$ being coprime integers, the magnetic unit cell is $q$ times as large as the SSG unit cell.
If $\frac{1}{2}u$ is irrational, only $m_3 = 0$ satisfy the requirement. 
An irrational $u$ corresponds to the incommensurate magnetic structure, where the size of the magnetic unit cell is infinite. 
As also discussed earlier in this section, although the size of a magnetic unit cell depends on $u$, DT1DU1 with different $u$ correspond to the same SSG in our classification.}
\blue{The magnetic unit cell in generic non-coplanar SSGs can be determined similarly, and for coplanar and collinear SSGs, $O_{\v}$ should be a $2\times 2$, $1\times 1$ identity matrix, respectively.}

\blue{
The magnetic moments in the equivalent positions of two different SSG unit cells are not necessarily identical. 
The magnetic moment $\S(\r_i + \v)$ at position $\r_i + \v$ ($\{ 1| \v\} \in T_M$) satisfies that $\S(\r_i + \v) = \mathcal{U} O_{\v} \mathcal{U}^{-1} \S(\r_i)$, where $\mathcal{U}$ is a unitary matrix transforming the diagonal matrix $O_{\v}$ to a O(3) matrix. 
Thus, the momenta of O($N$) representations give the magnetic propagation vectors. 
The O($N$) representations with multiple non-zero momenta correspond to SSGs describing the multiple-$Q$ structure.
For example, the SSG N143.5.1 DT1DU1 have two opposite propagation vectors $(0,0, \pm u\pi)$. 
Besides these two vectors, the third propagation vector of the SSG N143.11.1 A1$\oplus$DT1DU1 is $(0,0,\pi)$.    
}

\subsection{Equivalence in spin space groups}
\label{distinct_O3} 

In the last subsection, we have often seen that  
some inequivalent O($N$) ($N=3,2,1$) representations %
correspond to the physically same SSG. 
In this subsection, we introduce the concept of O($N$) 
representation {\it class}, which %
define the physically distinct SSGs uniquely and in a mutually exclusive way. %
Two O($N$) representations are categorized into the same class if and only if they satisfy 
either one %
of the following three conditions: (i) They are equivalent representations; 
(ii) They change into each other upon a coordinate transformation in real space;  
(iii) They are induced from momenta that share the same little group, and they 
are continuously connected. 
In the following, we explain these conditions in detail.

\paragraph*{(i) Coordinate transformation in spin space.}
Equivalent O($N$) representations are in the same class. This equivalence arises from the freedom of the %
spin coordinates; in the SSG, spin rotation and rotation in real space are decoupled.
To be more specific, two O($N$) representations, $\rho$ and $\sigma$, of the parent space group $P$ are equivalent, if %
the representation matrices of $\rho$ and $\sigma$, $D_{\rho}(p)$ and $D_{\sigma}(p)$, are transformed to each other by an orthogonal matrix $O$ for any elements $p$ in the parent space group; $D_{\sigma}(p) = O^T D_{\rho}(p) O$, $\forall p \in P$. %
This means that $\rho$ and $\sigma$ correspond to the same spin operations in different coordination of the spin axes.

\paragraph*{(ii) Coordinate transformation in real space.}
If an O($N$) representation $\rho$ is transformed into another $\sigma$ upon allowable coordinate transformation in real space%
, $\rho$ and $\sigma$ belong to the same O($N$) representation %
class.
A coordinate transformation $\{ V | \t \}$ for a space group $P$ involves a change of %
the axes by a matrix $V$, and a shift %
of the origin by a vector $\t$. 
An allowable $\{ V  | \t \}$ generates an automorphism in the space group $P$ 
\begin{equation} 
    \{ \R | \v \} \rightarrow \{ V  | \t \}^{-1} \{ \R | \v \} \{ V | \t \} \in P , \; \forall \{ \R | \v \} \in P  . %
    \label{coord_group}
\end{equation}
The group structure is invariant under the allowable transformation $\{V|\t\}$. %
Especially, for $R=1$, %
$\{ V  | \t \}^{-1} \{ 1 | \v \} \{ V | \t \} = \{1 | V^{-1}\v \} \in P$ requires %
that the Bravais lattice is invariant %
by $\{V|\t\}$. The invariance implies that %
for lattice basis %
$\a_i$ ($i=1,2,3$), $V \a_i$ can be expanded in terms of 
the lattice basis with integer coefficients,
$V \a_i = \sum_{j=1}^3 \a_j M_{ji}$ with a unimodular integer matrix $M$ ($M_{ij} \in \mathbb{Z}$ and $\det M=\pm1$). 
Here $M$ is the representation matrix of $V$ on the basis $\a_i$.
We require that the chirality of the unit cell is preserved by $\{V|\t\}$, resulting in $\det V = 1$. 
Thanks to the automorphism, an allowable $\{V | \t\}$ transforms an irrep $\rho$ 
into another irrep $\sigma$ according to %
\begin{equation}
    D_{\sigma}(  \{ \R | \v \} ) = D_{\rho}( \{ V  | \t \} \{ \R | \v \} \{ V | \t \}^{-1} ) \, .
    \label{tran_rep}
\end{equation}

The transformation of an O($N$) representation under $\{V|\t\}$ is fully determined by the transformations of each irrep that constitutes the O($N$) representation. 
Notably, the type of an O($N$) representation listed in Sec.~\ref{sec O3_Rep} is invariant under an allowable transformation $\{V | \t\}$. 
This is because the type of the 
O($N$) representation %
is determined only by the order and fixed points of $P^{\k}$, and 
the TRS-invariance of $\k$, all of which remain invariant under an allowable transformation. Under the transformation $\{V | \t\}$, the original little group 
$P^{\k}$ %
is transformed into $ P^{V^{-1}\k} =  \{ V  | \t \}^{-1} P^{\k} \{V|\t\}$, which is isomorphic to $P^{\k}$, and hence it has the same order and same type of fixed-points region (point-like, line-like, {\it etc.}). Note also that if $\k$ is TRS-invariant, %
$V^{-1} \k$ must also be TRS-invariant. This is because $V$ is represented by 
a unimodular integer matrix on the basis of the reciprocal lattice vectors 
as well as the lattice vectors. Therefore, the transformed O($N$) representation 
must have the same type as the original one.

Every space group generally has an infinite number of allowable transformations, and the allowable transformations form a group. 
Therefore, we have only to consider the transformations of the constituent irreps under 
the generators of the group. 
For example, consider the parent space group $P3$. To keep the hexagonal lattice of $P3$ unchanged, $V$ must be an element of the point group of the hexagonal lattice, $D_{6h}$.
Given that $V$ must be a proper operation ($\det M=1$), allowable $V$'s form the point group $D_6$. The shift vector $\t$ is also subject to symmetry constraints.
For $P3$, the shifted origin must be $C_{3z}$ invariant, allowable $\t$ must have the form of $(m_1,m_2,z)$, $(m_1 + 1/3, m_2  + 2/3,z)$, or $(m_1 + 2/3, m_2  + 1/3,z)$ with $m_1, m_2 \in \mathbb{Z}$ and $z \in \mathbb{R}$.
In summary, the group formed by the allowable transformations for $P3$ is 
generated by $\{ 6_{001}|\0 \}$, $\{2_{100} | \0\}$, $\{1|1/3,2/3, 0\}$, 
$\{1|2/3,1/3,0\}$, $\{1|0,0,z\}$, where rotations $6_{001}$ and $2_{100}$ 
are the generators of the $D_6$ group.

Let us next explain how %
all type-V O(3) representations of $P3$, 
DT$i$DU$i$ and P$i$PC$i$ for $i=1,2,3$, are transformed 
under the generators of allowable coordinate transformations for $P3$.
According to Eq.~(\ref{tran_rep}), the momentum $\k$ of an irrep 
is transformed to $V^{-1}\k$ under $\{V | \t\}$,
while any $V$ in the group  
can not transform the line DT to P or vice versa. This shows  
that DT$i$DU$i$ and P$j$PC$j$ lead to distinct representation classes 
respectively. 
\AppEquivRep\ shows that all coordinate transformations leave DT1 invariant; $\{2_{100} | \0\}$ transforms $C_{3z}$ to $C_{3z}^{-1}$, and hence 
transforms irrep DT2 into DT3 [Eq.~(\ref{DT_rep2})]. 
The transformation between DT2 and DT3 is consistent with the previous discussion in Sec.~\ref{sec:example-P3}; DT2DU2 and DT3DU3 represent the same magnetic structure under a coordinate transformation and hence belong to the same class (Sec.~\ref{sec:example-P3}). %
(One should notice that DT2DU2 and DT3DU3 are also subject to the equivalence-(iii), as explained in the second paragraph below.)
\AppEquivRep\ %
also demonstrates that under the action of $\{1 |2/3, 1/3, 0 \}$ 
$\left(\{1 |1/3, 2/3, 0 \}\right)$, the irrep P1 is transformed into P2 (P3). 
The transformation suggests that P1PC1, P2PC2, and P3PC3 all lead to 
the same SSG N143.5.3 P1PC1. In fact, the equivalence among P1, P2, and P3 
is also consistent with the observation in the real space (Sec.~\ref{sec:example-P3}).

\paragraph*{(iii) Continuously connected wave-vector.}
The third equivalence relation for the O($N$) representation 
classification applies to representations induced from non-HSP $\k$'s. 
Such O($N$) representations include type-V O(3) and O(2) ($[\rho_{\k}\uparrow P]^{c}_1 \oplus [\rho_{\k}^* \uparrow P]^{c}_1$), type-VIII O(3) and O(2) ($[\rho_{\k}\uparrow G]^{r}_2$), type-XI O(3) ($[\rho_{\k}\uparrow P]^{c}_1 \oplus [\rho_{\k}^* \uparrow P]^{c}_1 \oplus [\rho'_{\boldsymbol{p}}\uparrow P]^{r}_1$), and type-XIV O(3) ($[\rho_{\k}\uparrow G]^{r}_2 \oplus [\rho'_{\boldsymbol{p}}\uparrow P]^{r}_1$) representations. %
As $\k$ is not an HSP, %
the fixed point manifold of $P^{\k}$ has a dimension larger than 0, and $\k$ is not TRS-invariant. Consider another momentum $\k^{\prime}$ within the same 
fixed-point manifold and compare two O($N$) representations 
induced from $\k$ and $\k^{\prime}$. The third equivalence claims that 
the two representations belong to the same representation class if they can 
be continuously deformed to each other by a continuous change of the momentum 
between $\k$ and $\k^{\prime}$ in the manifold. 
Notice that any O(1) representation is free from %
this equivalence relation because an O(1) representation must always be induced from a TRS-invariant momentum (an HSP). 
Without this equivalence relation, the number of distinct SSGs (representation classes) would be infinite; different momenta in the BZ could define different SSGs. %
The equivalence relation is crucial for a meaningful classification of SSGs. 

We have seen two examples of the third equivalence relation %
in Sec.~\ref{sec:example-P3}, N143.5.1 DT1DU1 and N143.5.2 DT2DU2.
Thereby, we considered that DT1DU1 induced from different momenta on the line DT belong to 
the same class, and showed that all of them describe the same magnetic spiral structure with different spiral angles.
For N143.5.2 DT2DU2, we found two different 2D real representations, {\it i.e.}, DT2DU2 and DT3DU3 defined after Eq.~(\ref{DT_rep2}), and showed that they can be continuously connected to each other, describing the same kind of a magnetic structure. 
(One should notice that DT2DU2 and DT3DU3 are also subject to the equivalence-(ii), as explained in the second paragraph above.)
These observations justify the third equivalence relation.

The third equivalence becomes more intricate when the parent space group is nonsymmorphic. 
To illustrate this point, let us consider the parent space group $P2_1$ (No.~4), which is generated by lattice translations $\{1|1,0,0\}$, $\{1|0,1,0\}$,$\{1|0,0,1\}$, and the screw rotation $\{2_{010}|0,1/2,0 \}$. We consider irreps on the high-symmetry line LD $(0,u\pi,0),\, (u \notin \mathbb{Z})$. 
At every $\k = (0,  u \pi,0 )$, there exist two complex irreps ${\rm LD1}(u)$ and ${\rm LD2}(u)$.
The representation matrices of translations and the screw are $ D_{ {\rm LD1}(u)} (\{ 1 | \v \}) =  D_{ {\rm LD2}(u)} (\{ I | {\bm v} \}) = e^{i \pi u m_2 } $ and
\begin{equation}
    \begin{aligned}
        D_{ {\rm LD1}(u)} (\{ 2_{010}| 0,1/2,0 \}) & = e^{{\rm i} \frac{\pi}2  u} \\
        D_{ {\rm LD2}(u)} (\{ 2_{010}| 0,1/2,0 \}) & = e^{{\rm i} \frac{\pi}2 (u+2)} \,  ,
    \end{aligned}
\end{equation}
respectively. Here $\v = (m_1,m_2,m_3) \in \mathbb{Z}^3$ is the translation vector. 
Note that ${\rm LD1}(u )$ and ${\rm LD2}(u )$ are deformed to each other as $u$ increases by two, so they are in the same representation class according to 
the third equivalence relation. %
Physically, this equivalence means that two magnetic structures described by the two irreps (plus their complex conjugations) can be deformed to each other by a continuous change of the spiral wave vector. %
\AppEquivRep\ provides more technical details regarding the 
application of the third equivalence relation.

\subsection{Summary of the full classification} 
\label{sum_class}

As detailed in Sec.~\ref{sec O3_Rep}, for each parent space group,  
we can construct all the O($N$) ($N=3,2,1$) representations from the irreps $\rho_{\k}$, where the irreps $\rho_{\k}$ have been exhaustively tabulated on the \href{https://www.cryst.ehu.es/}{Bilbao Crystallographic Server} \cite{Elcoro17}. 
Thanks to the third equivalence relation %
in Sec.~\ref{distinct_O3}, each connected region 
in the BZ that shares the same little group 
$P^{\k}$, {\it e.g.,} high symmetry lines, planes, or the asymmetric unit, is 
represented by one $\k$ in the region. %
Under the modulo of the three types of equivalence relations 
in Sec.~\ref{distinct_O3}, we obtain all the distinct SSGs. 
As summarized in Table~\ref{num_ssg}, for $N=$3 (non-coplanar), 2 (coplanar), and 1 (collinear), we obtain 56512, 9542, and 1421 SSGs, respectively. We explicitly tabulate all the SSGs in \AppSecSSGTable. 

\blue{}

It is worth noting that the number of collinear SSGs (1421) equals the number of MSGs of types I, III, and IV. As established in Sec.~\ref{class_general}, we require the spin operations in a qSSG $G$ to be either time-reversal ($T$) or identity ($I$), and hence, $G$ is exactly the same as a magnetic space group of type I, III, or IV. The consistency in the number of groups confirms the validity of our method based on the representation theory.

\subsection{\blue{Nomenclature of SSGs}}
\label{sec: nonmenclature}
We label an SSG with one letter and three indices, %
$\alpha\mathcal{I.J.K}$. The prefix letter $\alpha=$~L, P, N refers to %
collinear, coplanar, and non-coplanar magnetic structures for 
the SSG. The SSG with these three kinds of magnetic structures  
are defined by O(1), O(2), and O(3) representations, respectively. 
The first index $\mathcal{I}$, ranging from 1 to 230, specifies the parent space group. 
The second index $\mathcal{J}$ specifies the type of O($N$) representations (Table~\ref{tab:O(N)rep}). For $N=1,2,3$, $\mathcal{J}$ ranges from 1 to 2, 8, and 16, respectively. 
For a given parent space group and a given O($N$) type, the third index $\mathcal{K}$ 
specifies %
distinct O($N$) representation classes. 
For clarity, we {always} indicate the constituent irreps of the O($N$) representation after the third index $\mathcal{K}$. 
For examples, the type-I non-coplanar SSG given by the identity representation [Fig.~\ref{p3-ssg}(a)] with the parent space group $P3$ (No.~143) is named N143.1.1 GM1, where GM1 refers to the identity representation; the type-III non-coplanar SSG given by the O(3) representation A1$\oplus$A1$\oplus$1 [Fig.~\ref{p3-ssg}(c)] is named N143.3.1 A1$\oplus$A1, where the identity irrep ($\oplus$1) is omitted for simplicity. In the nomenclature, the identity irrep will always be omitted except for the type-I SSGs. 
To avoid ambiguity, if several inequivalent O($N$) representations correspond 
to the same SSG, we consistently choose one of them, %
where the explicit choice for each SSG can be found in \AppSecSSGTable.
For example, the type-V SSG shown in Fig.~\ref{p3-ssg}(h) is always named as N143.5.3 P1PC1 but not N143.5.3 P$i$PC$i$ ($i=2,3$), though these representations can be transformed into each other under a coordinate transformation in real space.

It should be emphasized that SSGs with the same indices $\mathcal{I.J.K}$ but different prefixes $\alpha$ are different. 
For instance, the SSG N143.2.1 A1 characterizes non-coplanar magnetic structures [Fig.~\ref{p3-ssg}(b)], where the spin rotations assigned to the spatial operations are given by the O(3) representation A1$\oplus$1$\oplus$1. 
\blue{On the contrary, the SSG P143.2.1 A1, referred to as $\mG$, characterizes coplanar magnetic structures [Fig.~\ref{p3-ssg-1}(e)], and has a non-trivial pure-spin-operation subgroup $\mS$ [Eq.~(\ref{PSO2})].
}
In its qSSG $G\simeq \mG/S$ [Eq.~(\ref{eq:mG=SxG})], spin rotations assigned to the spatial operations are given by the O(2) representation A1$\oplus$1 that only acts in the $x,y$ subspace of the spin space.

\begin{table}[t]
    \caption[The numbers of SSGs for collinear, co-planar, and non-co-planar magnetic structures.]{The numbers of SSGs for collinear, co-planar, and non-co-planar magnetic structures. The table also shows the statistics 
    of SSGs according to the type of momenta $\k$ whose irreps $\rho_{\k}$ constitute the O($N$) representations ($N =1,2,3$). %
    ``$(0,0,0)$'' means that the representations are induced from irreps at the origin of the BZ. 
    In the corresponding SSGs,  %
    translation operations are always accompanied with the identity spin operation. 
    ``HSP'' means that the representations are induced by high-symmetry points (HSPs), some of which are not the origin of the BZ. 
    The corresponding SSGs describe commensurate magnetic structures. 
    ``Non-HSP'' means that the representations consist of irreps induced by non-HSP momenta. 
    The corresponding SSGs can describe in-commensurate magnetic structures.}
    \begin{tabular}[c]{c|ccc}
        \hline \hline
 Type of momenta $\k$ %
 & Collinear  & Coplanar  & Non-coplanar \\
        \hline
$(0,0,0)$ & 904  & 3019 & 8505  \\
HSP & 517  & 5748 & 40262 \\
Non-HSP & --    & 775  & 7745  \\
\hline
Total & 1421 & 9542 & 56512 \\
\hline \hline
    \end{tabular}
    \label{num_ssg}
\end{table}

\section{Electronic band theory in SSGs} 
\label{sec:band-theory}

In this section, we explore SSG applications to electronic Hamiltonians. 
We will demonstrate how symmetry algebra, characterized by projective representations, impacts the Bloch states.
A generic mean-field electronic Hamiltonian in a magnetic material with negligible SOC is given by
\begin{equation}
    \H = \frac{{\hat{ \bm  p}}^2}{2m} \sigma_0 + V(\r) \sigma_0 + J \S(\r) \cdot {\bm \sigma} \, , \label{electron-Hami} 
\end{equation}
where $\S(\r)$ is the magnetic moment density at $\r$; ${\bm \sigma} = (\sigma_x, \sigma_y,\sigma_z)$ is the Pauli matrices for the spin-1/2 operator; $J$ is the coupling strength between them.
The Hamiltonian $\H$ respects the symmetry of the SSG $\mG$, if $V(\r)$ and $\S(\r)$ are invariant under the SSG operations, {\it i.e.},
{\small
\begin{equation}
    V(\r) = V( \{ \R | \v  \}^{-1} \r),\quad \S(\r) = s(X) \U \S( \{ \R | \v  \}^{-1}\r) \, ,
    \label{Eq-20}
\end{equation} } 
for any $\{ X U | \R | \v  \} \in \mG$.

\subsection{Projective representation}
\label{generic_H}
The symmetry operators acting on electronic states form a {\it projective} representation of SSG.
The adjoint representation of SU(2) group are given by SO(3) matrices: $\hat{U} (\mathbf{d}\cdot \boldsymbol{\sigma}) \hat{U}^{-1} = (U\mathbf{d})\cdot \boldsymbol{\sigma}$, where $\hat{U}$ is the SU(2) matrix corresponding to $U$.
For a $\theta$-rotation $U$ along a direction $\boldsymbol{n}$, %
$\hat{U}$ can be given by either $e^{-i\frac{\theta}2\boldsymbol{n}\cdot\boldsymbol{\sigma}}$ or $ - e^{-i\frac{\theta}2\boldsymbol{n}\cdot\boldsymbol{\sigma}}$. This can be regarded as a one-to-two mapping between 
$U$ and $\hat{U}$. Similarly, one also has a mapping for the time-reversal operation, $\hat{X} \boldsymbol{\sigma} \hat{X}^{-1} = s(X) \boldsymbol{\sigma}$. Here $\hat{X}=i\sigma_y \mathcal{K}$ with $\mathcal{K}$ being complex conjugate if $X = T$, and $\hat{X} = \sigma_0$ 
if $X = I$.  
In terms of $\hat{U}$ and $\hat{X}$ thus defined,  
a two-component fermion wave function 
$\psi(\r) = (\psi_{\uparrow}(\r), \psi_{\downarrow}(\r))^T$ is 
transformed under generic 
$g=\{X_g U_g| R_g | \t_g \} \in \mG$ as 
\begin{equation}
(\hat{g}\cdot \psi) (\r) \equiv \hat{X}_g \hat{U}_g \psi (\r')
\end{equation}
where $\r' = R_g^{-1} (\r - \t_g)$, and $\hat{X}_g$ and $\hat{U}_g$ are obtained from $X_g$ and $U_g$ by the mapping, respectively.
This action on the wave functions verifies that $\hat{g} \H \hat{g}^{-1}$ is transformed according to the right-hand sides of Eq.~(\ref{Eq-20}).
As the mapping from $U$ to $\hat{U}$ is one-to-two, we should specify an (arbitrarily chosen) sign of $\hat{g} \equiv \hat{X}_g \hat{U}_g$ for every $g$. 
One can verify that, for $g_1g_2=g_3$, we have 
\begin{equation} \label{eq:w2}
    \hat{g}_1 \cdot \hat{g}_2 = \omega_2(g_1,g_2) \hat{g}_3\ ,
\end{equation}
where $\omega_2 (g_1, g_2)$ is referred to as the {\it factor system} of the projective representation. 
For example, if we choose $\hat{U}_{g_i} = e^{-i\frac{\theta_i}2\boldsymbol{n_i}\cdot\boldsymbol{\sigma}}$ for all $g_{1,2,3}$,  where $\boldsymbol{n}_{i}$ and $\theta_i$ are the rotation axis and angle of $U_{g_i}$, respectively, then $\omega_2(g_1,g_2)$ can be determined as $e^{-i \frac{\theta_3}2 \boldsymbol{n}_3\cdot \boldsymbol{\sigma} } \cdot e^{i \frac{\theta_2}2 \boldsymbol{n}_2\cdot \boldsymbol{\sigma} } \cdot e^{i \frac{\theta_1}2 \boldsymbol{n}_1\cdot \boldsymbol{\sigma} } \cdot \hat{X}_{g_3} \cdot \hat{X}_{g_2}^{-1} \cdot \hat{X}_{g_1}^{-1} =\pm 1$. 

A different choice of the signs of $\hat{U}_g$ yields an equivalent factor system. 
In general, we can further assign an additional U(1) factor for each $\hat{g}$ without changing the algebra of the symmetry group. %
By a change of $\hat{g}_i$ in Eq.~(\ref{eq:w2}) into $e^{i\alpha_i} \hat{g}_i$, the factor system will transform to 
\begin{equation}
    \omega_2(g_1,g_2) \rightarrow e^{i (\alpha_1 + \zeta \alpha_2 -\alpha_3)} \omega_2(g_1,g_2) \, 
\end{equation}
where $\zeta=1$ ($-1$) if $g_1$ is unitary (anti-unitary). 
Two factor systems related by such a U(1) gauge transformation are considered to be equivalent. 
For each SSG, we can determine its \blue{unique} factor system based on the spin operations $\hat{g} \equiv \hat{X}_g \hat{U}_g$, and study the electronic band theory under the determined factor system.
\blue{
On the other hand, non-collinear SSGs are isomorphic to MSGs (see Sec.~\ref{subsec: rep PN SSGs}), and two distinct SSGs isomorphic to the same $\mathcal{M}$ might realize its inequivalent factor systems.}

In a conventional MSG, the SOC term being invariant requires that $U_g \equiv \det(R_g)\cdot R_g$.
Thus $\hat{g}$'s also form a projective representation of the parent space group, which is usually referred to as the double magnetic space group. 
A crucial feature of SSG is that the factor system can be inequivalent to those of MSGs because, in the absence of SOC, $U_g$ can be different from $\det(R_g)\cdot R_g$ and hence giving 
different signs to $\omega_2$. 
In the following, we present an example of $\omega_2$ that cannot be realized in MSGs.

Let us consider the SSG N143.16.1 M1 [Fig.~\ref{p3-ssg}(o)]. 
The transformation of the magnetic moments under lattice translation $\{ 1|m_1,m_2,m_3 \}$ is given by Eq.~(\ref{M1_gen}).
Specifically, the lattice translation $t_3 = \{1|0,0,1\}$ is associated with an identity spin operation, while $t_1 = \{1|1,0,0\}$, $ t_2 = \{1|0,1,0\}$, and $t_{12} = \{1|1,1,0 \}$ should be accompanied by $\pi$ spin rotation along the $y$, $x$, and $z$ directions, respectively.  
We can choose the corresponding SU(2) spin rotation matrices of $t_1$, $t_2$, and $t_{12}$ as $\hat{U}_{t_1} = i \sigma_y$, $\hat{U}_{t_2} = i \sigma_x$, and $\hat{U}_{t_{12}} = i \sigma_z$, respectively. 
Then it is direct to verify that 
\begin{equation}
\hat{t}_1 \cdot \hat{t}_2 = - \hat{t}_2 \cdot \hat{t}_1 = \hat{t}_{12} \ . 
\end{equation}
The translation operators $\hat{t}_{1,2}$ anti-commute with each other as if they were magnetic translations encompassing a plaquette with 
$\pi$-flux. As will be discussed in the next subsection, such translation operators lead to a non-commuting BZ. This feature is unique to SSG and can never be realized in conventional space groups or MSGs.

\subsection{Collinear SSGs}
\label{bloch_s}

The representation theory of Bloch states in collinear SSGs is simpler than that in generic non-collinear SSGs. 
We will show in this subsection that they are effectively described by {\it single-valued grey} space groups. 
In our construction, the spin operations in a collinear qSSG $G$ are either $I$ or $T$, and operations with identity spin operation ($I$) form a subgroup $G_0$.
For type-I SSGs that are described by the identity representation of the parent space group $P$, $G=G_0$, and a collinear SSG $\mG$ satisfies that [Eq.~(\ref{eq:mS-U1-Z2T})]
\begin{equation} 
\label{eq:collinear-SSG-decompose2}
\mG = G_0 \times [\mS_{Z_2^T} \ltimes \mS_{U(1)}]\ .
\end{equation}
For type-II SSGs that are described by nontrivial O(1) representations of $P$, $|G_0| = 1/2 |G|$, and $\mG$ can be decomposed as [Eq.~(\ref{eq:mS-U1-Z2T})]
\begin{equation} \label{eq:collinear-SSG-decompose}
\mG = (G_0 + h\cdot G_0) \times [\mS_{Z_2^T} \ltimes \mS_{U(1)}] \ ,
\end{equation}
where $h \in \left( G-G_0 \right)$ is a spatial operation accompanied by $T$. 
$\mS_{Z_2^T}$ is a $Z_2$ group generated by $T U_{\hx}(\pi)$, and $\mS_{U(1)}$ is the continuous spin-rotation (along the $z$ direction) group. 
$\mS_{U(1)}$ allows the block diagonalization of the electronic Hamiltonian $\H$ into spin-up ($\H_{\uparrow}$) and spin-down ($\H_{\downarrow}$) sectors. 
In both types of collinear SSGs, the little group of both $\H_{\uparrow}$ and $\H_{\downarrow}$ is $G_0 \times [\mS_{Z_2^T} \ltimes \mS_{U(1)}]$. 
The band structure of $\H_{\uparrow}$ or $\H_{\downarrow}$ is fully characterized by the discrete subgroup $G_0 \times \mS_{Z_2^T}$, because the action of $\mS_{U(1)}$ on different states in a given spin sector is the same and proportional to an identity matrix. 
As $G_0$ only consists of spatial operations with identity spin operation, the factor system of $G_0$ is trivial, {\it i.e.}, $\omega_2(g_1,g_2) = 1$ ($\forall g_{1,2} \in G_0$). 
${\mS}_{Z_2^T}$ is generated by $- i\sigma_x i\sigma_y \mathcal{K} = i\sigma_z \mathcal{K}$ - a time-reversal operator (acting on fermions) that squares to 1, and operations in $G_0$ and $\mS_{Z_2^T}$ commute. 
These imply that electronic bands in each spin sector respect a {single-valued grey} group ${G}_0 \times {\mS}_{Z_2^T}$ as if $\H_{\uparrow}$ and $\H_{\downarrow}$ were in a nonmagnetic material belonging to the space group $G_0$.

The SSG of a material uniquely determines whether its electronic bands are spin split.
In a type-I collinear SSG [Eq.~(\ref{eq:collinear-SSG-decompose2})], net spin polarization is allowed, and the spin splitting is generally nonzero. 
In a type-II collinear SSG [Eq.~(\ref{eq:collinear-SSG-decompose})], operations in $h \cdot G_0 \times \mS_{Z_2^T}$ flip the spin.
If $h \cdot G_0$ contains an inversion operation combined with time reversal ($\mathcal{PT}$ operation), this spin-flipping and momentum-preserving operation implies $E_{n \uparrow} (\k) = E_{n \downarrow} (\k)$. 
Here $E_{ns}(\k)$ is the $n$th energy band in the $s$-spin sector. 
If $h \cdot G_0$ contains a translation followed by time reversal, this spin-flipping and momentum-flipping operation implies $E_{n \uparrow} (\k) = E_{n \downarrow} (-\k) = E_{n \downarrow} (\k)$, where the second equation is due to the $\mS_{Z_2^T}$ symmetry. 
Therefore, spin splitting is forbidden at every momentum for materials where the spin-up and spin-down atoms are related by inversion or translation~\cite{smejkal_beyond_2022}. %
Other types of spin-flipping operations, {\it i.e.}, rotation, mirror, screw, glide, allow spin splittings at a generic momentum. 
In Sec.~\ref{sub sec: spin texture}, we will discuss the symmetry of the spin texture of electronic bands in more detail.

\subsection{Non-collinear SSGs}
\label{subsec: rep PN SSGs}

Coplanar or non-coplanar SSGs do not have spin U(1) symmetry, and hence it is necessary to analyze the symmetries of the total Hamiltonian $\H$.
A coplanar SSG $\mG = G \times \mS_{Z_2^T}$ [Eq.~(\ref{eq:mG=SxG})] is a discrete group.
Here $G$ is the unitary qSSG that is isomorphic to a space group, and $\mS_{Z_2^T}$ [Eq.~(\ref{PSO2})] is a $Z_2$ group generated by $T U_{\hz}(\pi)$.
Thus, it is isomorphic to a grey space group (or type-II MSG). 
In a non-coplanar SSG, each spatial operation corresponds to a unique spin operation.
Consequently, a non-coplanar SSG is a discrete group and is isomorphic to a type-I, III, or IV MSG, where anti-unitary (unitary) operations in an SSG are mapped to anti-unitary (unitary) ones in the MSG. 
As explained in Sec.~\ref{generic_H}, when the symmetry operations $g\in \mG$ act on the fermionic degrees of freedom, they form a projective representation of $\mG$, or, isomorphically, a projective representation of $\mathcal{M}$.
Thus, the algebra of the symmetry operators $\hat{g}$ in a non-collinear SSG is fully characterized by the factor system $\omega_2$ that belongs to the second cohomology group 
\begin{equation}
  \omega_2 \; \in \; H^2(\mathcal{M}, \mathrm{U}(1))\ .
\end{equation}
Enumerating all irreps of $\mathcal{M}$ with inequivalent $\omega_2$'s that can be realized by SSGs complete the representation theory of SSGs, which is beyond the scope of the current work, and we leave it for future studies. 

In the following, we investigate
\blue{a generalization of Bloch states} in non-collinear SSGs.
Let us consider the unitary translation subgroup $T_U$ of the SSG $\mG$
\begin{equation} \label{eq: T_U}
    T_U = \brace{ \brace{ \U| 1 | \v }| \brace{ \U| 1 | \v } \in \mG, \det \U =1} \, .
\end{equation}
The generators of $T_U$ are joint operations of spin-rotations and translations 
\begin{equation}
    t_i = \{ \U_{t_i}| 1 | \a_i \}\, (i = 1,2,3) \, ,
\end{equation}
where $\a_i$ are primitive lattice vectors, and $\U_i \in \rm SO(3)$. 
It is noteworthy that $ U_{t_i} U_{t_j} = U_{t_j} U_{t_i}$ since both must equal the spin-rotation associated with the spatial translation $\{1|\a_i + \a_j\}$, which is unique by definition. 
The corresponding translation operators acting on fermion wave functions can be written as 
\begin{equation}
    \hat{t}_i = \{ \hat{U}_{t_i}| 1 | \a_i \}\, (i = 1,2,3) \,
\end{equation}
with $\hat{U}_{t_i}$ being the SU(2) representation matrix (with an arbitrarily chosen sign as explained in Sec.~\ref{generic_H}) of the SO(3) rotation $U_i$. 
As SU(2) is a double-covering of SO(3), even though $\hat{U}_{t_i}\hat{U}_{t_j}$ and $\hat{U}_{t_j}\hat{U}_{t_i}$ correspond to the same SO(3) rotation, they may differ by a minus sign, {\it i.e.,} $ \hat{U}_{t_i}\hat{U}_{t_j} = \zeta \hat{U}_{t_j}\hat{U}_{t_i}$ with $\zeta=\pm1$. 
Therefore, in contrast to ordinary MSGs, the translation generators $\hat{t}_i$ in a non-collinear SSG - when acting on fermions - do not necessarily commute with each other. 
In general, there is $\hat{t}_i \hat{t}_j = \zeta \hat{t}_j \hat{t}_i$, with $\zeta = \hat{U}_{t_i}\hat{U}_{t_j} \hat{U}_{t_i}^\dagger \hat{U}_{t_j}^\dagger$.

\subsubsection{Symmorphic and nonsymmorphic SBZ}
\label{subsubsec symmorphic SBZ}

When $[\hat{t}_i, \hat{t}_j]=0$ ($i,j=1,2,3$), eigenstates of $\H$ follow the Bloch theorem - they are common eigenstates of $\H$ and $\hat{t}_{1,2,3}$. 
To be specific, %
\blue{a common eigenstate} $\ket{\psi(\kk)}$ with wave vector $\kk$ satisfies 
\begin{equation}
     \hat{t}_{i}  \ket{\psi(\kk)} = e^{ i \tk_i }   \ket{\psi(\kk)} \, (i = 1,2,3) \, ,
     \label{Bloch1}
\end{equation}
where $\kk = \frac{1}{2\pi} \sum_{i} \tk_i \b_i$, $\tk_i =  \kk \cdot \a_i \in [0,2\pi)$, and $\b_i$'s are the reciprocal lattice vectors, with $\a_i \cdot \b_j = 2\pi\delta_{ij}$. 
Hereafter, we refer to $\kk$ as the {\it SSG momentum}, \blue{and refer to $\ket{\psi(\kk)}$ as the {\it SSG Bloch state}.} 
\blue{The SSG momentum $\kk$ and SSG Bloch state $\ket{\psi(\kk)}$ are different from crystal momentum $\k$ and traditional Bloch state $\ket{\psi(\k)}$. The latter pair is defined as the common eigenstate and eigenvalues of pure spatial translations [Eq.~(\ref{eq: T_M})].}
Note that to distinguish SSG momentum $\kk$ from crystal momentum $\k$, we adopt different notations.
If all $\hat{t}_i$'s are pure translations, SSG and crystal momenta coincide, we use $\k$ to denote both.
We refer to the reciprocal space formed by the SSG momenta as the SSG Brillioun zone (SBZ) to distinguish it from the \blue{magnetic BZ} formed by the crystal momenta. 
SBZ is expanded by $\b_i$ ($i=1,2,3$), while the magnetic BZ is a fraction of SBZ (see an example in the last of this subsection) unless all $\hat{t}_i$'s are pure translations.

An SSG momentum $\kk$ may transform {\it nonsymmorphically} under SSG symmetry operations, {\it i.e.} $\nexists \kk \in$ SBZ that is invariant under all the SSG symmetries, which is crucially different from a crystal momentum $\k$.
To see this, we consider the conjugate operation of $\hat{t}_i=\{\hat{U}_{t_i}|1|\a_i\}$ under a generic SSG operation $\hat{g} = \{ \hat{X}_g \hat{U}_g | R_g | \v_g \}$ 
\begin{equation}
\hat{g}^{-1} \hat{t}_i \hat{g}  
= e^{i 2 \pi \tq_i(g)} \widehat{g^{-1} t_i g}\ ,
\label{q_g_1}
\end{equation}
where $\tq_i(g)$ is determined by
\begin{equation}
     \hat{U}^{-1}_g  \hat{X}_g^{-1} \hat{U}_{t_{i}} \hat{X}_g \hat{U}_g  = e^{i 2\pi \tq_i(g)} \hat{U}_{ R_g^{-1} \a_i }  \ .
    \label{q_g}
\end{equation}
Here $\hat{U}_{ R_g^{-1} \a_i }$ denotes the SU(2) spin rotation matrix for $\{1 |  R_g^{-1} \a_i \}$ and should be chosen consistently with $\hat{U}_{t_i}$: $\hat{U}_{ R_g^{-1} \a_i } = \prod_{i=1}^3 \hat{U}_{t_i}^{m_i}$ if $ R_g^{-1} \a_i = \sum_{i}^3 m_i \a_i$. 
One can directly verify that $\hat{g} \ket{\psi(\kk)}$ is an eigenstate of $\hat{t}_{i}$ with the eigenvalue $e^{ i s_g (R_g \kk \cdot  \a_i + \tq_i(g))}$, where $s_g=1$ ($-1$) for unitary (anti-unitary) $g$.
Hence, the SSG momentum $\kk$ is transformed into $s_g \left( R_g \kk +  \qq_g \right)$ by $\hat{g}$, with $\qq_g = \sum_{i} \tq_i(g) \b_i$.
In a generic SSG with a non-trivial factor system, 
$\qq_g$ is {\it not} necessarily a reciprocal lattice vector, %
where $\hat{g}$ %
acts as a screw or glide on the SSG momentum. On the other hand, 
$2 \qq$ must be a reciprocal lattice vector, 
because $\hat{U}_g$ are SU(2) matrices, requiring %
$ e^{i 2 \pi \tq_i(g)} = \pm 1$.

Due to an arbitrary choice of the origin of the SBZ, a fractional $\qq_g$ does not necessarily mean the nonsymmorphic action of $g$ in the SBZ. 
Some fractional $\qq_g$'s can be made to zero or the reciprocal lattice vectors by a gauge transformation of $\hat{t}_i$. 
Let us consider the gauge transformation $\hat{t}_i \to e^{i 2 \pi \theta_i} \hat{t}_i$ that shifts the SSG momentum $\kk$ to $\kk' = \kk + \boldsymbol{\theta}$, where ${\bm \theta} = \sum_{i = 1}^3 \theta_i \b_i$. 
\blue{Owing to the flexibility in selecting the origin, $e^{i 2 \pi \theta_i}$ is not limited to $\pm 1$ but can be a generic U(1)-valued complex number. 
This extension also facilitates the transformation process between SSG momentum and crystal momentum (see the discussions in the context of the band of CoSO$_4$ in Sec.~\ref{subsec: Materials}).}
The shifted SSG momentum $\kk'$ is transformed by $\hat{g}$ into $s_g R_g (\kk' - {\bm \theta})+ s_g \qq_g + {\bm \theta}$. %
Hence, the fractional momentum transfer $\qq_g$ becomes $\qq_g + s_g \boldsymbol{\theta} - R_g \boldsymbol{\theta}$ in the shifted SBZ. 
If there exists such a gauge $\bm \theta$ satisfying 
\begin{equation} \label{eq:symmorphic-gauage}
    -s_g{\bm \theta} + R_g {\bm \theta} \equiv   \qq_g 
    \quad \forall g \in \mG \, ,
\end{equation}
for all $g \in \mG$, all the transfer $\qq_g$ can be eliminated by the gauge transformation.
Here the symbol ``$\equiv$'' means that SSG momenta on the left-hand and right-hand sides differ only by a reciprocal lattice vector. 
If there exists no such $\boldsymbol{\theta}$ that satisfies Eq.~(\ref{eq:symmorphic-gauage}) for all $g$, the SBZ is nonsymmorphic. 
For each SSG listed in \AppSecSSGTableTabb, we identify whether its SBZ is symmorphic or nonsymmorphic, using an automatic algorithm detailed in \AppSNF.

\begin{figure}[tb]
    \centering
   \includegraphics[width=1 \linewidth]{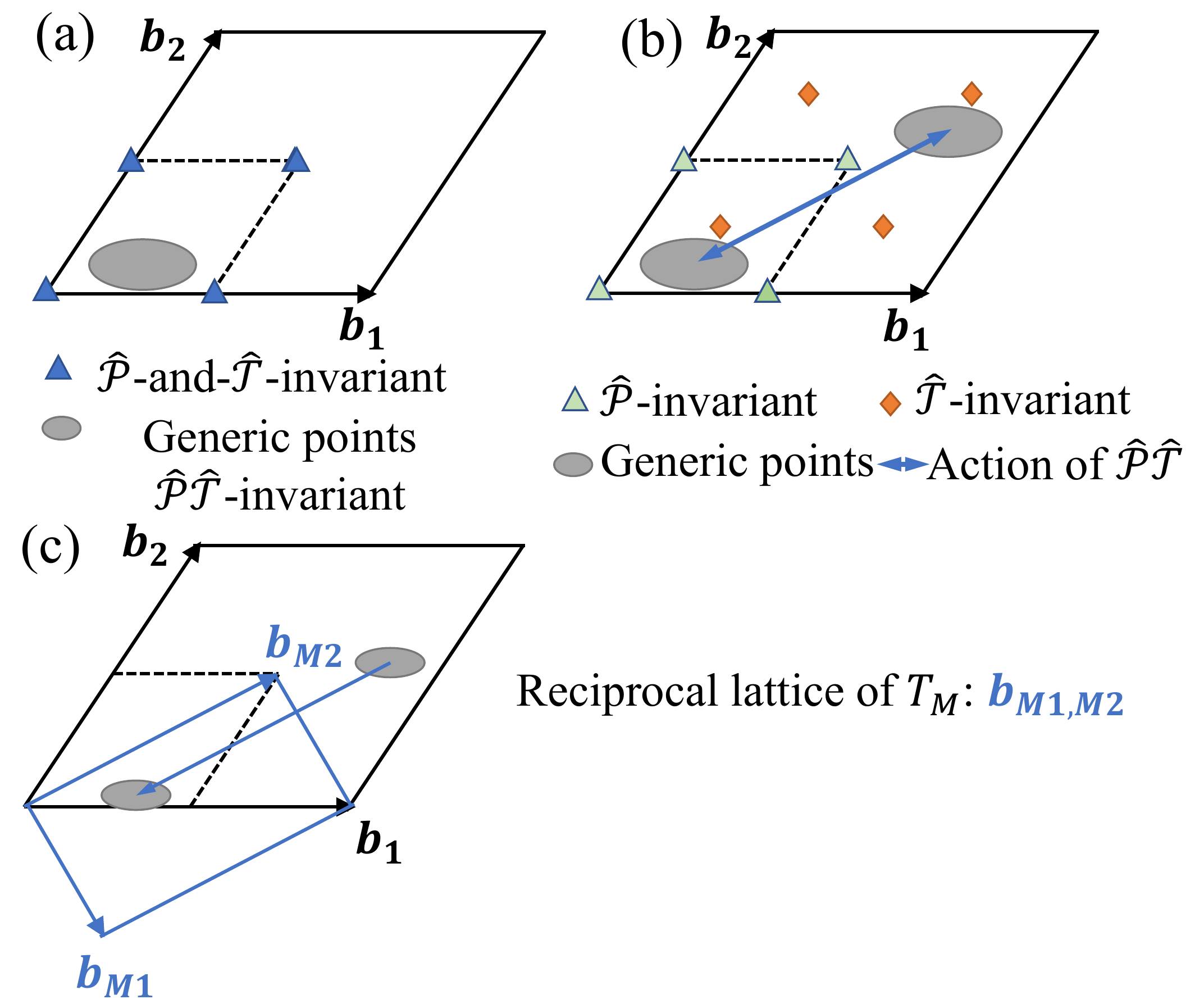}
    \caption[]{SSG Brillouin zone (SBZ) subtended by the reciprocal basis $\b_i$ ($i=1,2,3$). %
    For the clarity of demonstration, the 3D SBZ is projected into $\b_1$, $\b_2$ plane. 
    (a) SBZ of SSG P2.3.3 $ \rm R1^+ \oplus R1^+$ is symmorphic,  
    and (b) SBZ P2.3.2 $ \rm GM1^- \oplus R1^+$ is nonsymmorphic. 
    Symmetries of SSG momenta %
    are demonstrated by $\mathcal{P}$- and $\mathcal{T}$-invariant SSG momentum points in the SBZs. 
    (c) The SBZ (region enclosed by black vectors) and magnetic BZ (region enclosed by blue vectors) of SSG P2.3.2 $ \rm GM1^- \oplus R1^+$. Two grey regions that are related by $\hat{\mathcal{P}}\hat{\T}$ symmetry are inequivalent in the SBZ but are equivalent in the magnetic BZ.}
    \label{Fig: SBZ and MBZ}
\end{figure}

Here we use two SSGs with the same parent space group $P\bar1$ (No. 2) to illustrate the symmorphic and nonsymmorphic SBZs, respectively. 
$P\bar1$ is generated by translations $\{1|\a_i\}$ ($i=1,2,3$), with $a_{i}$ being lattice vectors, and the inversion $\{\bar 1|0\}$. 
Let us first consider a coplanar SSG, P2.3.4 $ \rm R1^+ \oplus R1^-$, which is induced from the even ($\mathrm{R1}^+$) and odd ($\mathrm{R1}^-$) irreps at the TRS-invariant momentum R $(\pi,\pi,\pi)$. 
Following the argument in Sec.~\ref{sec:example-P3}, we can obtain the generators of this SSG: translations $t_i = \{ U_{\hz}(\pi)| 1 | \a_i \}$, the inversion $\mathcal{P} =\{ U_{\hy}(\pi) | \bar{1} | \0 \}$, and an effective time reversal $\T= \{T U_{\hz}(\pi)| I | \0 \}$, which is the generator of the pure-spin-operation group $\mS_{Z_2^T}$ for coplanar structures [Eq.~(\ref{PSO2})].
We can choose $\hat{U}_{t_i}=i\sigma_z$, $\hat{X}_{\T} \hat{U}_{\T} = i\sigma_x \mathcal{K}$,  $ \hat{U}_{\mathcal{P}} = i \sigma_y $. 
Using Eq.~(\ref{q_g}), we find that $\qq_{\T} =  \qq_{\mathcal{P}} = \qq_{\mathcal{P}\mathcal{T}}= \0$, and hence its SBZ is symmorphic. 
On the other hand, another coplanar SSG with $P\bar1$, P2.3.2 $ \rm GM1^- \oplus R1^+$, gives an example of nonsymmorphic SBZ. 
It is generated by $t_i =  \{ U_{\hx}(\pi)| 1 | \a_i \}$, and the same $\mathcal{P}$, $\T$ as in P2.3.4 $ \rm R1^+ \oplus R1^-$. %
We choose $\hat{U}_{t_i } = i \sigma_x$, $\hat{X}_{\T} \hat{U}_{\T} = i\sigma_x \mathcal{K}$, $\hat{U}_{\mathcal{P}} = i \sigma_y$, and find that $\qq_{\mathcal{P}} = \0$ and $ \qq_{\T} = \qq_{\mathcal{P}\mathcal{T}}= \frac{1}{2}(\b_1 + \b_2 + \b_3)$ in P2.3.2 $ \rm GM1^- \oplus R1^+$. Although we can gauge eliminate $\qq_{\T}$ by shifting the origin of SBZ, the action of $\hat{\mathcal{P}}\hat{\T}$ 
on SSG momenta always induces the fractional translation, irrespective of the choice of origin.
The SBZ of P2.3.2 $ \rm GM1^- \oplus R1^+$ is hence nonsymmorphic.
We compare the SBZs of the two SSGs in Figs.~\ref{Fig: SBZ and MBZ}(a) and (b). As shown in Fig.~\ref{Fig: SBZ and MBZ}(b), in the nonsymmorphic SBZ, TRS-invariant and inversion-invariant momenta do not coincide, and no momenta have the $\hat{\mathcal{P}}\hat{\T}$ symmetry.

A non-symmorphic SBZ generally leads to extra degeneracy of energy bands in the magnetic BZ. 
In the following, we compare a description of electron bands by the 
SSG momentum with the traditional description by the crystal momentum.
Crystal momentum is related to the pure-translation subgroup $T_M$ of the SSG $\mG$ [Eq.~(\ref{eq: T_M})].
$T_M$ is an invariant subgroup of the SSG $\mG$ and a subgroup of $T_U$ [Eq.~(\ref{eq: T_U})].
Therefore, an SSG Bloch state defined in Eq.~(\ref{Bloch1}) is also a common eigenstate of operations in $T_M$.
In SSGs with nonsymmorphic SBZ, $|T_M|$ is a fraction of $|T_U|$. 
This implies that the first BZ for the crystal momentum is smaller 
than that for the SSG momentum (SBZ).  
The energy bands in the first %
BZ can be obtained by folding those in the SBZ. 
To be concrete, let us consider the example of %
P2.3.2 $ \rm GM1^- \oplus R1^+$. %
$T_M$ of this SSG can be generated by $\{I| 1 | \a_{Mi} \}$ ($i = 1,2,3$) with $\a_{M1} = \a_1 - \a_2$, $\a_{M2} =  \a_1 + \a_2$, and $\a_{M3} = \a_2 + \a_3$, 
becuase $(U_{\hx}(\pi))^2=1$. 
Their reciprocal lattice vectors are given as $\b_{M1} = \frac{1}{2}(\b_1 - \b_2 + \b_3)$, $\b_{M2} = \frac{1}{2}(\b_1 + \b_2 -\b_3)$, and $\b_{M3} = \b_3$. 
The volume enclosed by $\b_{Mi}$'s is half of that enclosed by $\b_i$'s. 
As shown before, $\hat{\mathcal{P}}\hat{\T}$ in this SSG will transform an SSG momentum $\kk$ to $\kk + \qq_{\mathcal{P}\T}$, indicating %
that the two corresponding SSG Bloch states share the same energy. 
Meanwhile, $\qq_{\mathcal{P}\T} = \b_{M2} + \b_{M3}$. 
Thus, if we fold the SBZ to the first %
BZ, these two SSG momenta will be considered equivalent [Fig.~\ref{Fig: SBZ and MBZ}(c)], and the energy bands shown in the magnetic BZ are at least double-degenerate at every momentum. 
A similar analysis can be successfully applied to explanations of extra degeneracy in the energy bands of CuSO$_4$
obtained from the first-principle calculation (see Sec.~\ref{subsec: Materials}).

\subsubsection{Non-commuting SBZ}
\label{sec: non-commuting SBZ}

Suppose that the electronic Hamiltonian $\H$ respects both $\hat{t}_{1,2}$, 
while $\hat{t}_1$ and $\hat{t}_2$ anti-commute: $\{\hat{t}_1, \hat{t}_2\}=0$. 
Then, eigenstates of $\H$ cannot be labeled by eigenvalues of $\hat{t}_{1,2}$ simultaneously because $\hat{t}_1$ and $\hat{t}_2$ do not share a common set of eigenstates.
In this case, the SBZ possesses a {\it non-commuting} nature (non-commuting SBZ).
In this section, we will use SSG N143.16.1 M1 [Fig.~\ref{p3-ssg}(o)] as an example to demonstrate several exotic features of the non-commuting SBZ. 

As discussed in Sec.~\ref{generic_H}, in SSG N143.16.1 M1 $\{\hat{t}_1,\hat{t}_2\}=0$ and $[\hat{t}_{1,2}, \hat{t}_3]=0$.
We can define \blue{SSG Bloch states} in terms of the eigenvalues of the commuting operators $\hat{t}_1$, $\hat{t}_2^2$, and $\hat{t}_3$, and introduce %
a folded SBZ spanned by $\b_1$, $\frac12\b_2$, and $\b_3$. 
The \blue{SSG Bloch state} satisfies
\begin{align} \label{Eq: Bloch noncommuting}
\hat{t}_{1,3} \ket{\psi(\kk)} = e^{i\tk_{1,3}} \ket{\psi(\kk)},\quad
\hat{t}_{2}^2 \ket{\psi(\kk)} = e^{i\tk_2} \ket{\psi(\kk)}
\end{align}
where $\kk = \frac{1}{2\pi}\left(\tk_1 \b_1 + \tk_2 \frac12 \b_2 + \tk_3 \b_3\right)$, with $\tk_{1,3} = \kk \cdot \a_{1,3} \in [0,2\pi)$ and $\tk_2 = 2\kk \cdot \a_2 \in [0,2\pi)$. 
The anti-commutation $\{\hat{t}_1, \hat{t}_2\}=0$ implies extra degeneracy in the folded SBZ.
Suppose $\ket{\psi(\kk)}$ is an eigenstate with an energy $E(\kk)$. We construct the state $\hat{t}_2\ket{\psi(\kk)}$.
As $\hat{t}_2$ commutes with $\H$, $\hat{t}_2\ket{\psi(\kk)}$ %
has the same energy. 
As $\{\hat{t}_1,\hat{t}_2\}=0$, there is $\hat{t}_1 \hat{t}_2\ket{\psi(\kk)} = - \hat{t}_2 \hat{t}_1 \ket{\psi(\kk)} = e^{i(\kk+\frac12\b_1) \cdot \a_1} \hat{t}_2 \ket{\psi(\kk)} $. In other words, $\hat{t}_2 \ket{\psi(\kk)}$ is an eigenstate with energy $E(\kk)$ and the SSG momentum $\kk+\frac12\b_1$. 
Therefore, the energy bands at $\kk$ are always identical to those at $\kk+\frac12 \b_1$. 
This degeneracy is well known in $\pi$-flux models.
Unlike the $\pi$-flux models, the electron Hamiltonian [Eq.~(\ref{electron-Hami})] has no external magnetic field, and the degeneracy emerges from the non-collinear magnetism.
Since $\hat{t}_2$ rotates spin along the $x$-direction, the degenerate states at $\kk$ and $\kk+\frac12 \b_1$ have opposite spin expectation values in the $y,z$ plane.

The shape of the SBZ is $C_{3z}$-asymmetric:
$\frac{1}{2}\b_2$ is a reciprocal lattice vector in the reduced SBZ, but $R_{\hz}(2 \pi/3)\frac{1}{2}\b_1 = -\frac{1}{2}(\b_1 + \b_2)$ is not, where $R_{\hz}(2 \pi/3)$ denotes $2\pi/3$-rotation along the $z$ axis.
However, counter-intuitively, the $C_{3z}$ symmetry is preserved, leading to extra degeneracies. 
We find that while $\hat{C}_{3z} \ket{\psi(\kk)}$ is not an \blue{SSG Bloch state}, $\hat{C}_{3z} \ket{\phi_{+}}$ $\left( \hat{C}_{3z} \ket{\phi_{-}}\right)$ is an \blue{SSG Bloch state} with energy $E(\kk)$ and SSG momentum $R_{\hz}({2 \pi}/{3}) \kk $ $\left( \R_{\hz}({2 \pi}/{3}) \kk  + \frac{1}{2}\b_1 \right)$, where $\ket{\phi_{\pm}}$'s are properly chosen linear combinations of $\ket{\psi(\kk)}$ and  $ \hat{t}_2 \ket{\psi(\kk)}$.  
Note that $\hat{C}_{3z} \ket{\phi_{\pm}}$ are also related by $\hat{t}_2$.
The details of derivation can be found in \AppSecNoncommutingSBZ, and a gauge transformation on $\hat{t}_{1,2}$ is applied to simplify the expression of the transformation.
Simply speaking, $\hat{C}_{3z}$ does not transform a single SSG momentum to another one but transforms a pair to another, requiring bands at these $C_{3z}$-related pairs to share the same energy.

\subsection{Spin texture in the momentum space}
\label{sub sec: spin texture}

In this subsection, we study symmetries of spin texture $\tS(\kk)$ in a certain equal-energy surface ({\it e.g.}, Fermi surface) in the SBZ. 
Here $\tS(\kk)$ is an expectation value of the spin operator with respect to an \blue{SSG Bloch state} at $\kk$. 
In the presence of energy degeneracy at $\kk$, $\tS(\kk)$ is a sum of the spin expectation values of the degenerate SSG Bloch states.
SSGs completely determine whether a nonzero spin texture $\tS(\kk)$ is allowed and how it transforms under symmetries. %
We first consider $\tS(\kk)$ in those SSGs where unitary translation generators $\hat{t}_{1,2,3}$ all commute with each other. 
An SSG momentum $\kk$ transforms into $s_g (R_g \kk + \qq_g)$ under an SSG operation $g = \{ X_gU_g |R_g| \v_g \}$, where $s_g = 1$ ($-1$) for $X_g = I$ ($T$). 
Meanwhile, the spin expectation transforms according to $X_g U_g$.
Hence $\tS(\kk)$ satisfies the symmetry constraint  
\begin{equation} \label{eq:Sk-constraint}
    \tS\left( s_g (R_g \kk + \qq_g )\right) = s_gU_g \tS(\kk). 
\end{equation}

We use symmetries that leave generic $\kk \in $ SBZ invariant to determine the dimension $d_{\rm SBZ}$ of the symmetry-allowed span of $\tS(\kk)$ in the momentum space.
$d_{\rm SBZ}$ represents non-coplanar ($d_{\rm SBZ}=3$), coplanar ($d_{\rm SBZ}=2$), collinear ($d_{\rm SBZ}=1$) distribution of $\tS(\kk)$ in the SSG momentum space. 
$d_{\rm SBZ}=0$ means $\tS(\kk)=0$. 
The relevant symmetries are the spin-U(1) rotation (in collinear SSGs), spin-rotation translations, and space-time inversion ($\mathcal{PT}$) with spin rotation. 
For collinear SSGs, $d_{\rm SBZ}\le 1$ due to the spin-U(1) symmetry. 
As discussed in Sec.~\ref{bloch_s}, $d_{\rm SBZ}=0$ if the spin-up and spin-down atoms in real space are related by $\mathcal{P}\T$ operation or translation followed by time reversal, and $d_{\rm SBZ}=1$ otherwise (ferromagnetism or alter-magnetism).
For non-collinear SSGs, symmetries leaving each $\kk$ invariant can only be unitary translations, generated by $\hat{t}_{1,2,3}$, and spin-rotation $\mathcal{PT}$ operation, denoted as $\hat{\mathcal{P}}\hat{\T}$. 
If $\hat{\mathcal{P}}\hat{\T}$ is present and $\qq_{\mathcal{P}{\T}}=0$, $d_{\rm SBZ}$ is given by the dimension of a subspace in the spin space satisfying Eq.~(\ref{eq:Sk-constraint}), {\it i.e.}, $d_{\rm SBZ}=\mathrm{dim}[ \bigcap_{i=1}^3 \ker( \U_{t_i} - I) \bigcap \ker(\U_{\mathcal{PT}} + I )]$.
Otherwise, $d_{\rm SBZ}=\mathrm{dim}[ \bigcap_{i=1}^3 \ker( \U_{t_i} - I)]$.

The group structure of the little group of generic SSG momenta gives certain constraints on the forms of $\U_{t_i}$, $\U_{\mathcal{PT}}$, and their relationship, which allows us to obtain $d_{\rm SBZ}$ with a simple rule as summarized in Table~\ref{tab d_M}. 
First, we consider cases without $\mathcal{PT}$ symmetry. 
If all $U_{t_i}$'s equal the identity $I$, no constraint exists and $d_{\rm SBZ} = 3$.
If some of them do not equal $I$, to satisfy $[\hat{t_i},\hat{t_j}] = 0$, they must share a same rotation axis, leading to $d_{\rm SBZ} = 1$, $\tS(\kk)$ is non-zero along the rotation axis (see \AppC). 
Next, we consider the case with $\mathcal{PT}$ symmetry and $U_{t_i} = I$ ($i= 1,2,3$), implying that $\qq_{\mathcal{P}{\T}}=\0$. 
Since the square of $\mathcal{PT}$ operation should be the identity, $U_{\mathcal{PT}}^2 = I$, and $U_{\mathcal{PT}}$ can be either $I$ or a $\pi$-rotation $U_{\bm m}(\pi)$, which lead to $d_{\rm SBZ} = 0$ and $2$, respectively.
When both $\mathcal{PT}$ symmetry and a non-trivial $U_{t_i}$ are present, 
$d_{\rm SBZ}$ apparently depends on whether $\qq_{\mathcal{PT}}$ is zero or not.    
Interestingly, in \AppC, we find that $d_{\rm SBZ}$ is always $1$ if there exists a $U_{t_i} \neq I$ independent of the form of $\mathcal{PT}$.

Notably, $d_{\rm SBZ}$ does not necessarily coincide with the dimension $N$ of the span of magnetic moments $\S(\r)$ in the real space.
In fact, in either non-coplanar or coplanar SSGs, $d_{\rm SBZ}$ can be 0, 1, 2, 3. 
\blue{
As discussed before, in the absence of $\mathcal{PT}$ operation and with $U_{t_i} = I$ in a coplanar SSG, $d_{\rm SBZ}$ can indeed be $3$, due to the lack of local constraints on $\tS(\tk)$.
To substantiate this somewhat counter-intuitive scenario, we present an electronic model with SSG P143.1.1 GM1, exhibiting $d_{\rm SBZ} = 3$ in Appendix~\ref{App Sec: d_M}.
In contrast, in some non-coplanar magnetic structures, $d_{\rm SBZ}$ can be $1$.
}
For example, in the SSG N143.3.1 A1$\oplus$A1 [Fig.~\ref{p3-ssg}(c)], the translation along $\a_3$ is accompanied by spin rotation $\U_{\t_3} = U_{\hz}(\pi)$.
Thus, $\tS(\kk)$ of an \blue{SSG} Bloch state at generic $\kk$, which is an eigenstate of $\hat{t}_3=\{ i \sigma_z| 1 | \a_3 \}$, can only have a non-zero component 
only along the $z$-direction, leading to $d_{\rm SBZ}=1$.

\begin{table}[tb]
    \centering
\caption[The dimensions $d_{\rm SBZ}$/$d_{\rm BZ}$ of the span of spin texture $\tS(\kk)$/$\vS(\k)$ in the SBZ/BZ of non-collinear SSGs.]{The dimensions $d_{\rm SBZ}$/$d_{\rm BZ}$ of the span of spin texture $\tS(\kk)$/$\vS(\k)$ in the SBZ/BZ of non-collinear SSGs.
Columns and rows specify spin-rotations that accompany translations ($t_i$) and space-time reversion $\mathcal{P}\T$, respectively. %
$\{U_{t_i}\} = \{U_{\n}(\theta_i)\}$ means that all the translations are accompanied by spin rotations along the same axis $\n$ by $\theta_i$ ($i=1,2,3$), and at least one $\theta_i$ is nonzero. 
 }
\begin{tabular}{c | c c c}
  \hline \hline
  \backslashbox{$U_{\mathcal{PT}}$ }{ $\{U_{t_i}\}$ } & Identity & $ \{U_{\bm n} (\theta_i) \}$ 
  & Non-commuting \\
  \hline
  Absent & 3/3 & 1/1  & 1/0\\ 
  Identity & 0/0 & 1/0 & 1/0\\
  \multirow{2}{*}{$ U_{\bm m}(\pi)$}
    & \multirow{2}{*}{2/2} & 1/1 (${\bm m}\perp {\bm n}$)  & \multirow{2}{*}{1/0}\\
    &     & 1/0 (${\bm m}\parallel {\bm n}$) \\ 
  \hline \hline
\end{tabular}
\label{tab d_M}
\end{table}

If the SBZ is symmorphic, {\it i.e.}, $\qq_{g}$ in Eq.~(\ref{eq:Sk-constraint}) can be eliminated by a certain gauge [Eq.~(\ref{eq:symmorphic-gauage})], the symmetry of $\tS(\kk)$ can be characterized by a
$d_{\rm SBZ}$-dimensional representation $\tilde{\rho}$ of a point group $\tilde{P} = \{s_g R_g | g\in {\mG}\}$. 
Non-identity representation $\tilde{\rho}$ implies the spin texture on the Fermi surface forming a nontrivial pattern, as exampled in Sec.~\ref{sec: material example}. 
In \AppSecMaterialTable, we tabulate the $\tilde{\rho}$ and $\tilde{P}$ for all the SSGs.

\blue{We define the spin texture $\vS(\k)$ in the magnetic BZ ($\k$ denotes crystal momentum)}, following the same approach as for $\tS(\kk)$.
Note that we use different notations to distinguish spin textures in the magnetic BZ and SBZ. In cases where BZ is identical to SBZ, we use the notation $\vS(\k)$.
$\vS(\k)$ is potentially more feasible for experimental measurement and is relevant to first-principle calculations, which are generally performed in the magnetic BZ.
We now investigate the dimension $d_{\rm BZ}$ of the span of $\vS(\k)$. 
If $\hat{t}_i=\{\sigma_0|1|\a_i\}$, the magnetic BZ is identical to SBZ, and $d_{\rm BZ} = d_{\rm SBZ}$.
If $\hat{t}_i=\{\hat{U}_{\n}(\theta_i)|1|\a_i\}$, the magnetic BZ is given by the folded commuting BZ. 
As $d_{\rm SBZ}=1$ in this case, $d_{\rm BZ}$ equals 0 if the $\vS(\kk)$'s at momenta folded to the same $\k$ always cancel each other and 1 otherwise. 
The only symmetry that leads to the cancellation at every $\k$ is $\mathcal{P}{\T}$.
One can verify that $U_{\mathcal{P}\T}=I$ or $U_{\bm m}(\pi)$ with ${\bm m} \parallel {\bm n}$ result in the cancellation. 
We summarize $d_{\rm BZ}$ in Table~\ref{tab d_M}.
Additionally, it is important to note that crystal momentum $\k$ always transforms symmorphically.
In SSGs with symmorphic SBZs but some of $U_{t_i} \neq I$, $\vS(\k)$ does not equal $\tS(\kk)$. 
However, SSG operations impose the same forms of constraints [Eq.~(\ref{eq:Sk-constraint}) with $\qq_g \equiv \0 $] on them. It implies that $d_{\rm BZ} = d_{\rm SBZ}$, and $\vS(\k)$ realizes the same representation $\tilde{\rho}$ of the point group $\tilde{P} = \{s_g R_g | g\in {\mG}\}$ as $\tS(\k)$.

Finally, we comment on some universal properties of spin textures of SSGs with the non-commuting SBZ.
In the reduced SBZ, all translation operations commute with each other, while some of them are still accompanied by non-trivial spin rotations (see Sec.~\ref{sec: non-commuting SBZ} for an example).
A similar analysis as in the commuting SBZ (see \AppC) shows that $d_{\rm SBZ}$ is always $1$. 
On the other hand, $d_{\rm BZ}\equiv 0$, because $\{ \hat{t}_i, \hat{t}_j\} = 0$ requires that $U_{t_i}$ and $U_{t_j}$ must be $\pi$-rotation along two perpendicular directions.

\section{Materials with SSG symmetries}
\label{sec: material example}

\subsection{Identifying SSGs for 1595 magnetic materials}
\label{subsec: Materials stat}
We identify the SSGs for {\it all} the \blue{1595} published experimental magnetic structures (materials with non-integer site occupation numbers are excluded) in the \href{http://webbdcrista1.ehu.es/magndata/}{MAGNDATA} database~\cite{gallego_magndata_2016,gallego_magndata_2016-1} on the Bilbao Crystallographic Server. 
The identification algorithm and the SSGs of all these materials are provided in Appendices~\ref{app: material example} and \ref{app sec ssg materials}, respectively.
Out of the 1595 structures, we find 242 distinct collinear SSGs, 183 distinct coplanar SSGs, and 106 distinct non-coplanar SSGs.
Among these SSGs, the most frequently occurring types are type-II (815 structures) for collinear structures, type-III (211 structures) for coplanar structures, and type-IX (69 structures) for non-coplanar structures. 
A collinear (coplanar, non-coplanar) SSG of type II (III, IX) indicates that the transformation of each component of the spin moments is described by an independent real 1D irrep.  
Table~\ref{tab num_material} shows the statistics of the features of these materials determined by SSGs.
Additionally, for reference, we provide statistics limited to materials with light elements (shown in parentheses), where all constituent elements are from the first four periods of the periodic table, and SOC is generally weak.
Remarkably, all the exotic features of SSGs that we have studied, such as non-commuting SBZ, nonsymmorphic SBZ, and non-trivial spin textures, occur in these published experimental materials.
In the subsequent sections, we present the results from first-principle calculations performed on selected materials as illustrative examples.

\begin{table}[h]
    \caption[The statistics of published magnetic materials in \href{http://webbdcrista1.ehu.es/magndata/}{MAGNDATA} database.]{The statistics of published magnetic materials in \href{http://webbdcrista1.ehu.es/magndata/}{MAGNDATA} database. Each cell contains the number of magnetic materials possessing specific features.
    The number in parentheses is the statistics limited to materials with light elements from the first four periods of the periodic table. 
    The statistics of spin textures are limited to the materials with SSGs processing symmorphic SBZ, and the ``Non-trivial $\tS(\kk)$'' means that the transformation of $\tS(\kk)$ realizes a non-trivial representation of the SSG.    %
    }
    \def\arraystretch{1.2}
    \begin{tabularx}{\columnwidth}{c|YYY}
        \hline \hline
   & Collinear  & Coplanar  & Non-coplanar \\
        \hline
Total & \blue{954} (234)  & 436 (84) & 205 (66) \\
Non-commuting SBZ  &  -&  23 (1) & 10 (2)\\
Nonsymmorphic SBZ   & -   &  148 (31) &  5 (1) \\
Symmorphic SBZ  & \blue{954} (234) & 265 (52) & 190 (63)\\
Non-trivial $\tS(\kk)$  & \blue{139} (41) & 181(43) & 140 (39)\\
\hline \hline
    \end{tabularx}
    \label{tab num_material}
\end{table}

\subsection{Material examples}
\label{subsec: Materials}

This subsection is devoted to four material examples, {\it i.e.}, coplanar CoSO$_4$, collinear FeGe$_2$, coplanar Mn$_3$Ge, non-coplanar Mn$_3$GaN, that exhibit novel electronic states or spin textures protected by SSGs. 
Three other material examples, including coplanar InMnO$_3$, non-coplanar Mn$_3$Ge, and coplanar FePO$_4$, are discussed in \AppD. 
These materials were calculated using the density functional theory (DFT), as implemented in the Vienna ab-initio Simulation Package (VASP)~\cite{Kresse1996kl,Kresse1996vk}.
The projector augmented wave (PAW) pseudo-potentials were adopted in the calculation~\cite{Kresse1999wc,Blochl1994zz}.
The generalized gradient approximation with the Perdew-Burke-Ernzerhof (PBE) realization~\cite{Perdew1996iq} was used for the exchange correlation functional.
The kinetic energy cutoff was fixed to 450 eV, which is larger than the ENMAX in the pseudo-potential files of all elements. The energy convergence criteria were set to be 10$^{-6}$ eV. The k-point mesh for the Brillouin zone integration is 10000/(number of atoms)~\cite{jain_high-throughput_2011}.

We first perform fully non-collinear magnetic structure calculations {\it without} considering the SOC effect. 
Remarkably, for three of the four materials, {\it i.e.}, collinear FeGe$_2$, coplanar Mn$_3$Ge, non-coplanar Mn$_3$GaN, the directions of magnetic moments converged to the experimental results (up to a global spin rotation).
The remaining material, coplanar CoSO$_4$, also converged to a magnetic structure that is close to the experimental structure but exhibits a slightly smaller canting angle. 
To match the experimental data for CoSO$_4$, we introduced a penalty contribution to the total energy (see \AppD\ for more details).
The good agreement between DFT results without SOC and experimental structures suggests that SOC, except for choosing a global spin coordinate, does not play a major role in determining the collinear or non-collinear magnetic structures of these materials. 

To demonstrate the perturbative nature of SOC, we compare the band structures with and without SOC of all these materials in \AppD.
We observe that SOC leads to small splittings of the band degeneracy predicted by SSG. 
Specifically, the typical splitting along high-symmetry paths is less than $5$ meV, which is considerably smaller than the typical distance between two adjacent non-degenerate energy bands in the absence of SOC.
These quasi-degeneracies could not be understood without SSGs.
We claim that SSGs provide accurate descriptions for these systems when the interested energy scale is larger than the splittings.

\subsubsection{Nonsymmorphic SBZ in \texorpdfstring{CoSO$_4$}{CoSO4}}

\begin{figure}[h]
    \centering
   \includegraphics[width=\columnwidth]{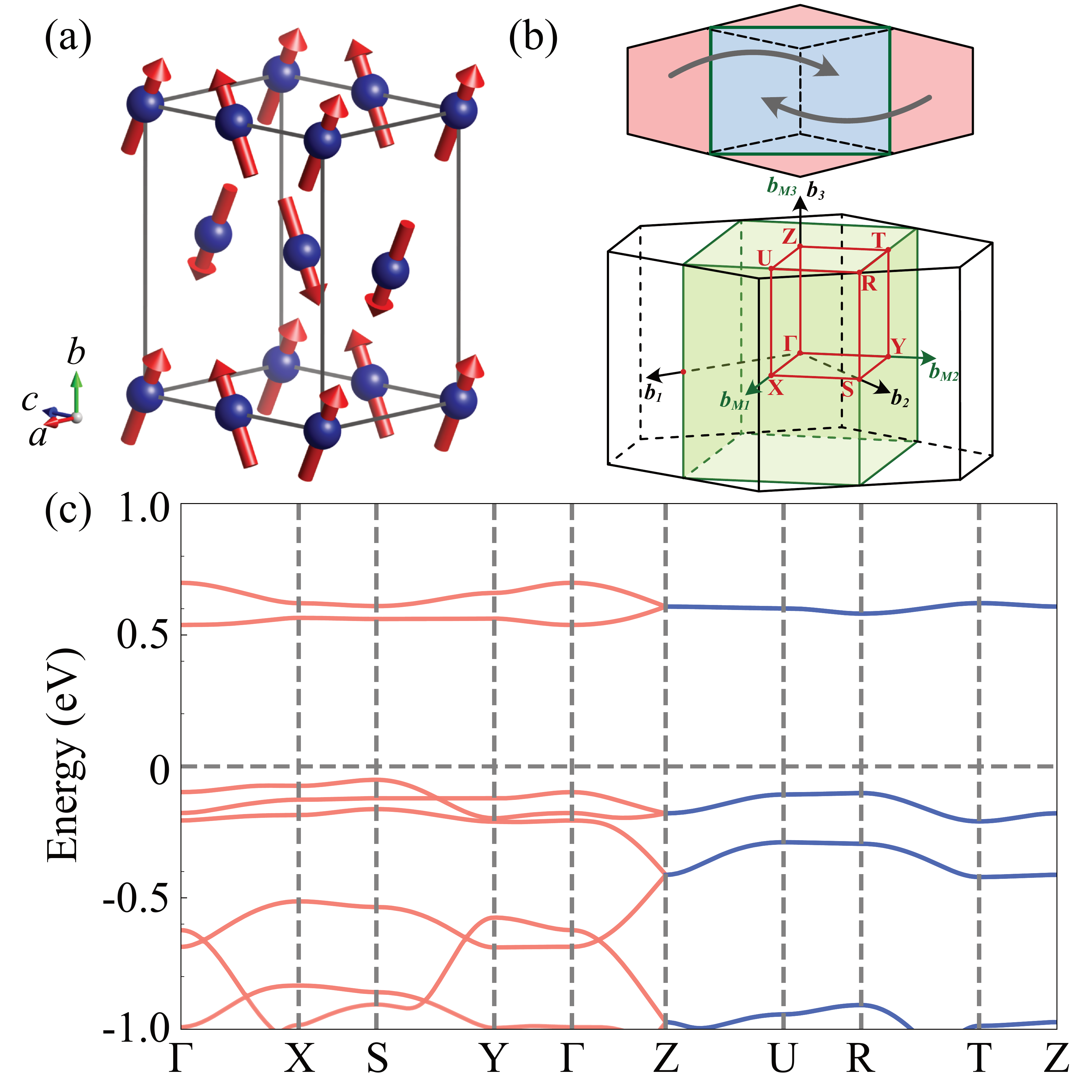}
    \caption[]{(a) The magnetic structure of CoSO$_4$ showing only the magnetic atoms (Co). (b) \blue{Bottom:} The first BZ in the SSG reciprocal lattice $\b_i$'s (black lines) and MSG reciprocal lattice $\b_{Mi}$'s (green region). \blue{Top: The folding process of the SSG BZ (the large hexagon) into the MSG BZ (the smaller rectangle), viewed from the $z$-axis perspective.} (c) The energy bands obtained from the first-principle calculation. At every $\k$, $E(\k)$ is at-least two-fold degenerate. \blue{The light red and blue lines represent the two-fold and four-fold degeneracy, respectively.}}
    \label{CoSO4_Fig}
\end{figure}

The compound CoSO$_4$ (\href{http://webbdcrista1.ehu.es/magndata/index.php?this_label=1.519}{No.~1.519} in the \href{http://webbdcrista1.ehu.es/magndata/}{MAGNDATA } database) has a base-centered orthorhombic lattice structure. 
The spin moments of CoSO$_4$ in the anti-ferromagnetic phase were determined from a neutron diffraction study~\cite{Frazer62}.
The moments lie in the $y,z$ plane and have around $\pm 25^{\circ}$ canting angles concerning the $y$ axis, and the body-centered translation relates spin moments with opposite directions [Fig.~\ref{CoSO4_Fig}(a)]. 
This magnetic structure is described by the MSG $C_Pm'cm'$ (No.~63.16.526 in the Opechowski-Guccione (OG) setting \cite{opechowski95}). 
Fig.~\ref{CoSO4_Fig}(c) shows the energy bands obtained from the first-principle calculation.
These bands are plotted in the BZ of the MSG and exhibit the following two features:
(i) The bands are at least two-fold degenerate in the whole MSG BZ. This degeneracy is protected by the MSG $\mathcal{PT}$ symmetry $\{ i \sigma_y \mathcal{K} | \bar{1} | 1/2( \a_{M1} + \a_{M2}) \}$ which squares to $-1$. Here $\a_{Mi}$'s ($i = 1,2,3$) denote the lattice vectors of the pure translation subgroup given by the conventional lattice vectors of the orthorhombic lattice (see discussion of $T_M$ in Sec.~\ref{subsubsec symmorphic SBZ}.)
(ii) The degree of degeneracy becomes four along Z-U-R-T-Z (the high symmetry points are denoted in Fig.~\ref{CoSO4_Fig}(b)). And we verified this four-fold degeneracy exists in the whole $k_z=\pi$ plane.
MSG explains the four-fold degeneracy along the high symmetry lines R-T and T-Z, which form a projective representation of the little co-group $m'mm$. 
However, MSG can only protect two-fold degeneracy and hence cannot explain the extra degeneracy at the remaining momenta in the $k_z=\pi$ plane. 
We find that the hidden SSG symmetry explains the four-fold degeneracy, as will be discussed below.

\begin{table}[t]
    \caption[The generators of the SSG P63.3.89 ${\rm Y}1^+ \oplus {\rm Y}3^+$ of coplanar magnetic material CoSO$_4$.]{The generators of the SSG P63.3.89 ${\rm Y}1^+ \oplus {\rm Y}3^+$ of coplanar magnetic material CoSO$_4$.
    Here, $\a_{1,2}$ are body-centered translation vectors.
    The last row is for the pure-spin-operation group $\mS$. 
    The last column shows how the SSG momentum is transformed under the SSG operations.
    }
    \def\arraystretch{1.2}
    \begin{tabularx}{\columnwidth}{c|ccc}
        \hline \hline
        SSG operation & \makecell[c]{Spin operations \\ on electrons} &  \makecell[c]{Transformation of \\  SSG momentum $\kk$}  \\   \hline
        $\{ U_{\hx}(\pi) |1 | \a_{1,2} \}$ &  $\sigma_x$ & $\kk$\\
        $\{I |1 | \a_{3} \}$ &  $\sigma_0$ & $\kk$\\
        $ \{ I | \bar{1} | \0 \}$  & $\sigma_0$ & $-\kk$ \\
      $ \{  U_{\hy}(\pi) |m_{001} | 1/2  \a_3 \}$ & $\sigma_y$ & $-R_{\hz}(\pi) \kk  + 1/2(\b_1 + \b_2)$  \\
      $ \{I |m_{100} | \0 \}$ & $\sigma_0$ & $-R_{\hx}(\pi) \kk $ \\
         \hline
        $\{T U_{\hx}(\pi) | 1 | \0  \}$ & $ \sigma_z {\mathcal K}$& $-\kk  + 1/2(\b_1 + \b_2)$\\ 
        \hline\hline
    \end{tabularx}
    \label{tab: CoSO4}
\end{table}

The SSG P63.3.89 ${\rm Y}1^+ \oplus {\rm Y}3^+$ with parent space group $Cmcm$ (No.~63) characterizes this magnetic structure.
Table~\ref{tab: CoSO4} shows the generators of the SSG, where $\a_i$ are lattice vectors of the unitary translation subgroup $T_U$ of SSGs.
$\a_{1,2}$ are the base-centered translation vectors, and $\a_{i}$'s are satisfies 
{\small
\begin{equation}
    \a_{1} = \frac12(\a_{M1} + \a_{M2}) , \,   \a_{2} = \frac12(\a_{M1} - \a_{M2}) , \, \a_3 = \a_{M3} \, .
\end{equation}}%
Thus, the size of the SBZ spanned by $\b_{i}$ is twice that of the magnetic BZ spanned by $\b_{Mi}$ [see Fig.~\ref{CoSO4_Fig}(b)].
With the spin operations acting on electrons given in Table~\ref{tab: CoSO4}, we determine the transformation of SSG momentum (see the last column). 
The SSG operation $\mathcal{\hat{P}\hat{T}} =  \{ \sigma_z \mathcal{K}| \bar{1}| \0 \}$ acts on SSG momentum $\kk$ as a fractional translation ($\kk \to \kk + \qq_{\mathcal{PT}}$) with $\qq_{\mathcal{PT}} = \frac12 (\b_1 + \b_2)$, and the mirror operation $M_z = \{  U_{\hy}(\pi) |m_{001} | \frac12  \a_3 \}$ acts on $\kk$ as a glide operation with $\qq_{M_z} = \frac12 (\b_1 + \b_2) $, which indicate a non-symmorphic SBZ. 
We first study the electronic bands in the SBZ. 
No SSG momentum is invariant under $\hat{P}\hat{\T}$ operation, and $\mathcal{\hat{P}\hat{T}}$ requires that $E_n(\kk) = E_n(\kk + \qq_{\mathcal{PT}})$ for all $\kk \in $ SBZ.
In addition, $\tk_z = \pi$ plane has the symmetry $\hat{C}_{2z} \hat{\T} = \{ i\sigma_x {\mathcal K}| 2_{001}| \frac12 \a_3 \}$, and $ (\hat{C}_{2z} \hat{\T})^2 \ket{\psi(\tk_x ,\tk_y, \pi)} = - \ket{\psi(\tk_x ,\tk_y ,\pi)}$.
It results in Kramer's degeneracy on that plane.
Then, we fold the SBZ to the magnetic BZ.
\blue{
We introduce a phase factor of $i$ to the spin rotation associated with the translations $\{U_{\hx}(\pi)|1 |\a_{1,2} \}$, altering the accompanying spin rotation from $-i \sigma_x$ to $\sigma_x$ (see Table~\ref{tab: CoSO4}).}
\blue{Under this gauge for SSG,} the translations along the lattice vectors of MSG are always accompanied by $\sigma_0$ spin rotation without an extra phase. 
Thus, the crystal momentum $\k$ equals the SSG momentum $\kk$ after modulo reciprocal lattice $\b_{Mi}$'s.
On the other hand, $\qq_{\mathcal{PT}} = \b_{M1}$ is a reciprocal lattice vector in the magnetic BZ.
Thus, the energy bands at a momentum $\k$ in the magnetic BZ consist of $E_n(\kk)$ and $E_n(\kk + \qq_{\mathcal{PT}})$ in the SSG BZ.
This implies that the degrees of degeneracy at $\k$ in the MSG BZ should be twice as $E_n(\kk)$ in the SSG BZ. 
After folding, every band in the $k_z = \pi$ plane is four-fold degenerate.

\subsubsection{Extra band degeneracies in \texorpdfstring{FeGe$_2$}{FeGe2}}

\begin{figure}[h]
    \centering
   \includegraphics[width=\columnwidth]{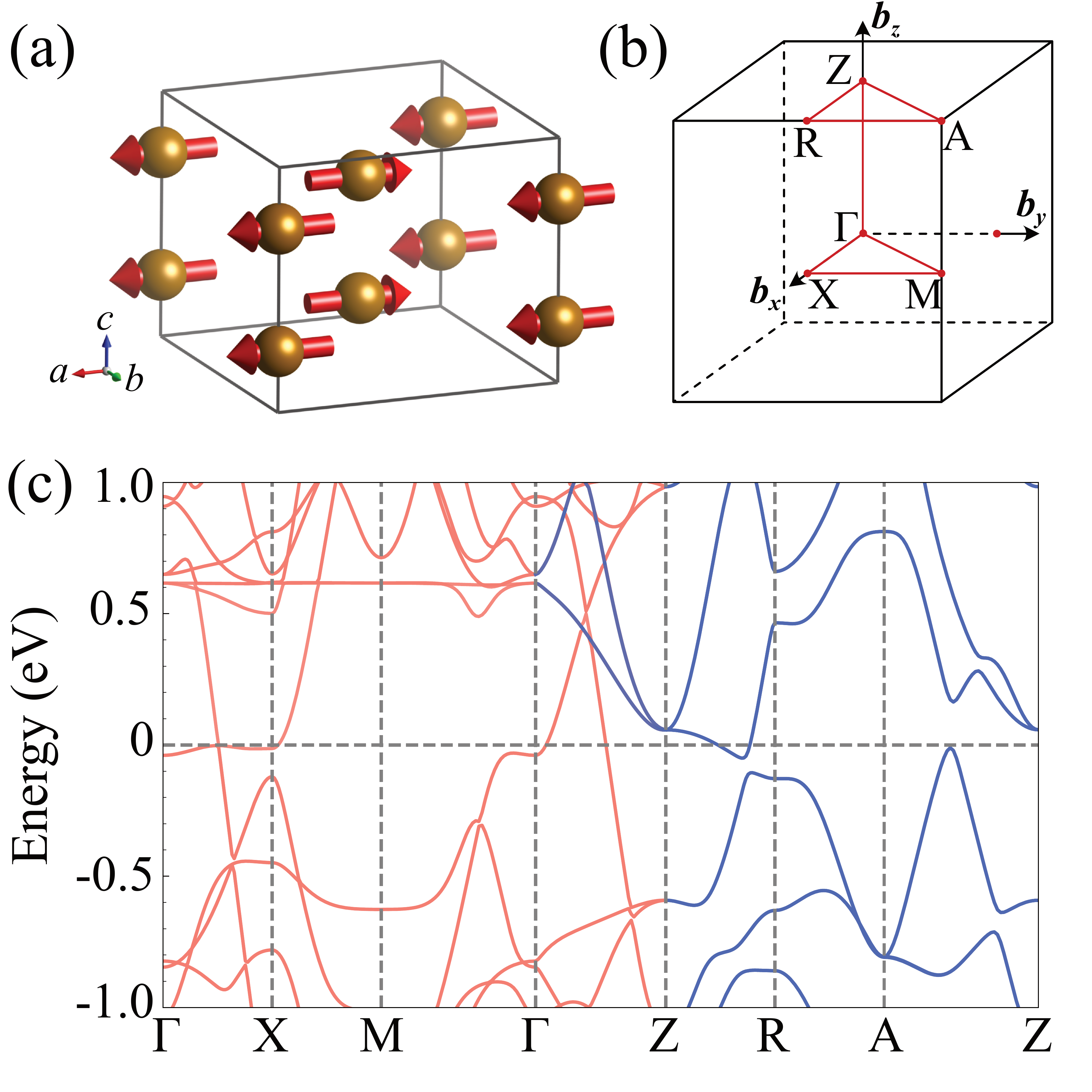}
    \caption[]{(a) The magnetic structure of FeGe$_2$ showing only the magnetic atoms (Fe). (b) The first BZ in the SSG reciprocal lattice $\b_i$'s, which coincides with MSG reciprocal lattice. (c) The energy bands obtained from the first-principle calculation. Every single line is at least two-fold degenerate. \blue{The light red and blue lines represent the two-fold and four-fold degeneracy, respectively.}}
    \label{FeGe_Fig}
\end{figure}

The anti-ferromagnetic material FeGe$_2$ (\href{http://webbdcrista1.ehu.es/magndata/index.php?this_label=1.557}{No.~1.557}) has a body-centered orthorhombic lattice structure.
A neutron diffraction experiment~\cite{murthy65} found that the spin moments are along the $x$ direction, and moments with opposite signs are related by a body-centered translation vector [Fig.~\ref{FeGe_Fig}(a)].
The structure is described by the MSG $I_Pb^{\p}am^{\p}$ (No.~72.12.641)~\footnote{
We modified the coordinate setting for FeGe$_2$ in this structure from the magnetic crystallographic information file obtained from the \href{http://webbdcrista1.ehu.es/magndata/}{MAGNDATA} database. The new coordinate setting follows the standard convention for this MSG, facilitating the analysis of MSG symmetry.}.
Fig.~\ref{FeGe_Fig}(c) shows the energy bands from the first-principle calculation.
The energy bands are always double-degenerate, which can be explained by the MSG's $\mathcal{PT}$ symmetry.
However, the four-fold degeneracy at $\Gamma$ point and path Z-R-A-Z [see their positions in Fig.~\ref{FeGe_Fig}(b)] cannot be explained by the MSG, because
the little MSG co-groups on them only have 2D coirreps (see explicit representation matrices in the \href{https://www.cryst.ehu.es/cgi-bin/cryst/programs/corepresentations.pl}{Corepresentations} tools~\cite{elcoro21,xu20}).
We can use the representation theory of collinear SSG (Sec.~\ref{bloch_s}) to explain these degeneracies.

This antiferromagnetic structure is described by the type-II collinear SSG L140.2.8 $\rm M1^+$ with parent space group $I4/mcm$ (No.~140).
Besides the pure-spin-operation group $ \mS = \mS_{U(1)} \times \mS_{Z_2^T}$, this SSG $\mG$ is generated by $\{ I| 1 |\a_i \}$ ($i = 1,2,3$), $\{I |4^+_{001} | \0 \}$, $\{ I |2_{010} | 0, 0 , 1/2 \}$, $\{I|\bar{1} | \0 \}$, $\{T| 1 | 1/2,1/2,1/2\}$.
Here $\a_i$'s denote the lattice vectors of the conventional unit cell, and the translation parts of spatial operations are written on the basis of $\a_i$'s. 
Also note that here $\mS_{Z_2^T}$ is generated by $ \{ e^{-i \frac{\pi}{2} \sigma_y} i \sigma_y \mathcal{K} |1  | \0\} = \{ \mathcal{K} | 1 | \0\}$ since spins are along the $x$ direction. 
As discussed in Sec.~\ref{bloch_s}, the Hamiltonian for the collinear magnetism can be block diagonalized into spin-up and spin-down parts. \blue{Due to the existence of a spin-flipping unitary translation $\{ -i \sigma_z|1|1/2,1/2,1/2\}$}, $E_{n \uparrow}(\k) = E_{n \downarrow}(\k )$. 
Thus, we only need to study spin-up (or spin-down) energy bands, whose symmetry is effectively described by single-valued grey group $G_0 \times \mS_{Z_2^T}$. 
Here $G_0$ is the operations in $\mG$ with identity spin operations, and $G_0 \times Z_2^T$ is identical to the grey group $P4/mcc1^{\p}$ (No.~124.2.1019).
The little co-group of $P4/mcc1^{\p}$ at $\Gamma$ point is magnetic point group $4/mmm1^{\p}$, which has eight 1D coirreps and two 2D coirreps (refer to \href{https://www.cryst.ehu.es/html/cryst/mpointrepres.html}{Bilbao Crystallographic Server} for the character table). 
Thus, $E_{n \uparrow} (\k)$ can be either two-fold degenerate or non-degenerate at $\Gamma$.
The Z-R line has a joint symmetry of mirror and time reversal, given as $ \hat{M}_z \hat{\T} =  \{  \mathcal{K}| m_{100} |0,0,1/2 \} $, and the action $ (\hat{M}_z \hat{\T})^2$ on $\ket{\psi(k_x,0,\pi)}$ equals $-1$, leading to Kramer's degeneracy of $E_{n \uparrow}(\k)$. 
Similarly, on lines R-A and A-Z, $\{  \mathcal{K}| m_{010} |0,0,1/2 \} $ and  $\{  \mathcal{K}| m_{110} |0,0,1/2 \} $ lead to Kramer's degeneracy, respectively.
The total degrees of the degeneracy of energy bands are twice as $E_{\n \uparrow}(\k)$, which can be either two or four at $\Gamma$ and always four along Z-R-A-Z. 

\blue{
The specific irreps constituted by the Bloch states at a given momentum, along with their degrees of degeneracy, are contingent upon the symmetry of the local orbitals of the atoms and the representation they form.
The Fe atoms of FeGe$_2$ are situated at the coordinates $(0,0,\frac{1}{4})$, $(0,0,\frac{3}{4})$, $(\frac{1}{2},\frac{1}{2},\frac{1}{4})$, and $(\frac{1}{2},\frac{1}{2},\frac{3}{4})$. 
Let us first consider the site-symmetry group of the spin-up (or spin-down) orbitals at $(0,0,\frac{1}{4})$.
In addition to the operations in the group $\mS_{Z_2^T}$, the symmetry group is generated by $\{I |4^+_{001} | \0 \}$, $\{ I |2_{010} | 0, 0 , -1/2 \}$, isomorphic to the magnetic point group $4221'$. 
Specifically, the orbitals corresponding to $A_{1,2}$ or $B_{1,2}$, in conjunction with the orbitals at $(0,0,\frac{3}{4})$ linked via SSG symmetry, contribute to the 1D coirreps at $\Gamma$ (see the details of the induction process, for example, in Ref.~\cite{elcoro21}). 
In contrast, the $E$-orbitals contribute to the 2D coirreps at $\Gamma$.  
We note that states $\ket{\uparrow, p_x}$ and $\ket{\uparrow, p_y }$, which remain invariant under the effective time reversal in the $\mS_{Z_2^T}$, form the coirrep $E$ of $4221'$.
The site-symmetry group in the other positions of Fe atoms is also isomorphic to $4221'$, where the orbitals contribute to the coirreps in the momentum space in a similar manner.  
For generic collinear magnetic structures, the effective symmetry group of spin-up (or spin-down) energy bands is always isomorphic to a grey group, and the effective site-symmetry group of local spin-up orbitals is isomorphic to a grey point group. Hence, the techniques of band representation of magnetic topological quantum chemistry~\cite{bradlyn_topological_2017, elcoro21} can be directly applied in the context of collinear SSGs. 
}

It is notable that applying a global spin rotation on the magnetic structure does not affect the SSG, as discussed in Sec.~\ref{sec:example-P3}, but it might enhance the MSG symmetry. 
For example, consider rotating all the magnetic moments of FeGe$_2$ [see Fig.~\ref{FeGe_Fig}(a)] by $U_{\hy}(\pi/2)$. 
After this rotation, the magnetic moments will align along the $c$ axis, the four-fold rotation axis of the crystal, thereby enhancing the MSG to $I_{\rm P}4/m^{\prime}cm$ (No.~140.11.1206).
The enhanced MSG shares the same parent space group as the SSG L140.2.8 $\rm M1^+$, but remains smaller than the SSG.
The mapping from the parent space group to the enhanced MSG is one-to-one, whereas that to the SSG is one-to-many due to the pure-spin-operation group $\mS$ (Eq.~(\ref{eq:mG=SxG})). 
By examining the coirreps of the magnetic little co-group of the enhanced MSG, we find that it can explain the four-fold degeneracies on the line R-A, but fails to explain the degeneracies at $\Gamma$ or line R-Z-A.

\subsubsection{\texorpdfstring{$E_{2g}$}{E2g} spin texture in \texorpdfstring{Mn$_3$Ge}{Mn3Ge}}
\label{sec:Mn3Ge-coplanar}

\begin{figure}[tb]
    \centering
   \includegraphics[width=1 \linewidth]{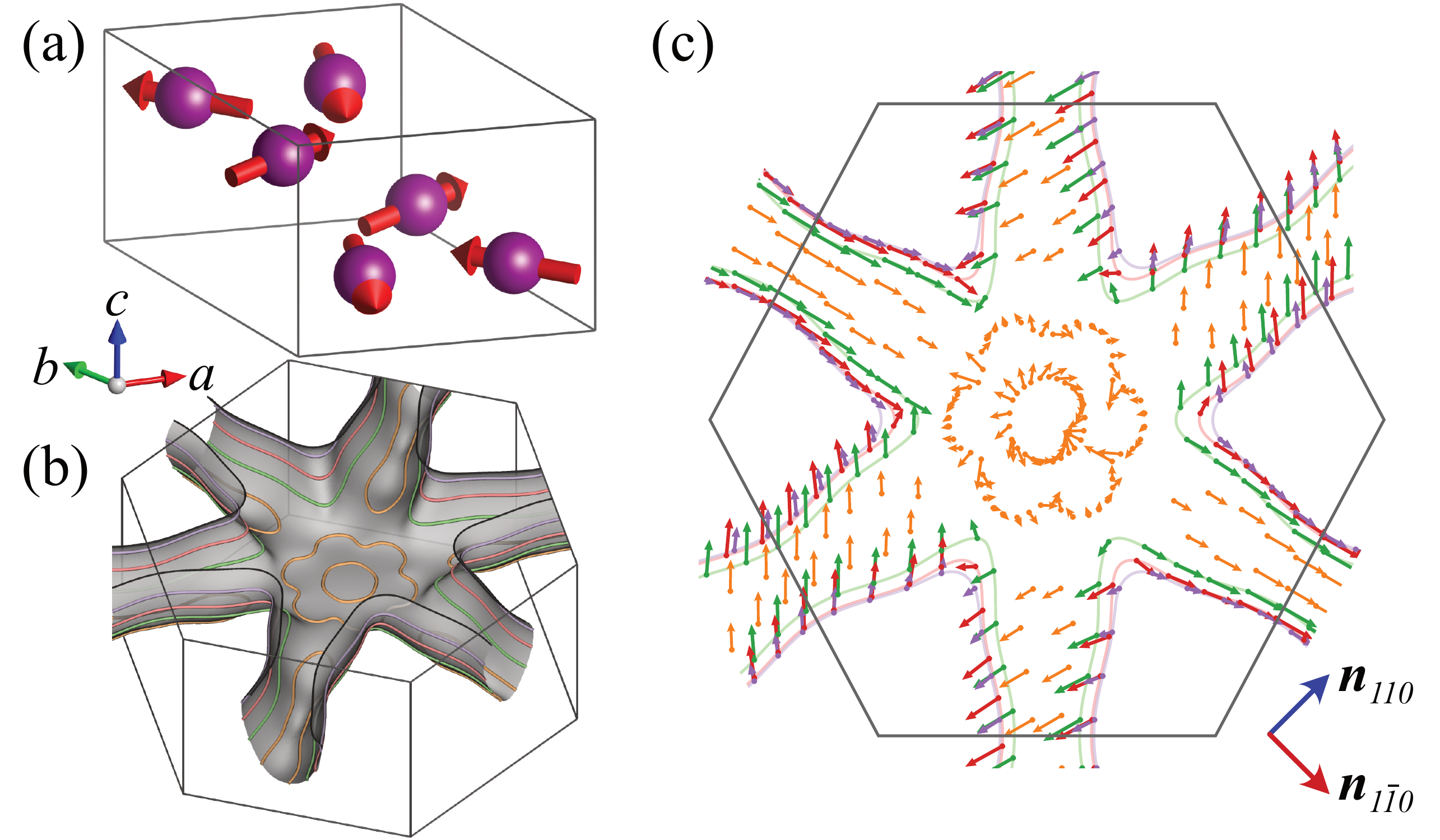}
    \caption[]{(a) The magnetic structure of Mn$_3$Ge showing only the magnetic atoms (Mn). 
    (b) The Fermi surface of Mn$_3$Ge, which is centered at A $(\k = (0,0,\pi))$ point.
    Only the lower half is shown, and the full Fermi surface is symmetric with respect to $k_z=\pi$. 
    Different colors represent equal-$k_z$ lines.
    (c) The spin texture $\vS(\k)$ on equal-$k_z$ lines labeled in (b). 
    Arrows represent the directions of $\vS(\k)$.}
    \label{Fig: Mn3Ge all}
\end{figure}

The compound Mn$_3$Ge (\href{http://webbdcrista1.ehu.es/magndata/index.php?this_label=0.377}{No.~0.377}) has a hexagonal lattice structure.
It has a coplanar ($x,y$ plane) triangular antiferromagnetic structure in the magnetic phase [Fig.~\ref{Fig: Mn3Ge all}(a)]~\cite{soh_ground-state_2020,Ajaya16,kubler2014non,chen2021anomalous}.
Its MSG is $Cm^{\prime}cm^{\prime}$ (No.~63.8.518) and its SSG is P194.6.1 GM5$^+$ with parent space group $P6_3/mmc$. 
Besides $\mS_{Z_2^T}$, the SSG is generated by $ \mathcal{P} = \{I|\bar{1}| 0, 0 , 0 \}$, $C_{6z} = \{ U_{\hz}(2\pi/3) | 6^+_{001} | 0, 0 , 1/2 \}$, $ C_{2, {1\bar{1}0}} = \{ U_{{1\bar{1}0} }(\pi) |2_{1\bar{1}0} | 0, 0 ,1/2 \}$, and pure translations $\{ I| 1 |\a_i \}$ of the hexagon lattice.
Here ${1\bar{1}0}$ represents the direction of $\a_1 - \a_2$ equivalent to the direction of $-\sqrt{3}/2 {\bm e}_x + 1/2 {\bm e}_y$.
Note that in this SSG, the 6-fold screw rotation in real space is accompanied by the 3-fold rotation in the spin space. 
Consequently, the $2\pi/3$ rotation in real space is accompanied by $-2\pi/3$-rotation in spin space, and this symmetry was noticed and studied in a 2D model of Mn$_3$Ge~\cite{liu2022spin}.

The SSG completely determines the symmetry of $\vS(\k)$. 
The $\mathcal{PT}$ operation $ \{ T U_{\hz}(\pi) | \bar{1} | \0  \}$ ($\T$ is from $\mS_{Z_2^T}$) requires that $\vS(\k)$ lies in the $x,y$ plane, as the spin moments in real space.
The operations $\mathcal{P}$, $C_{6z}$, and $C_{2,1\bar10}$ lead to a $6/mmm$ point group symmetry in the momentum space. 
Due to the $U_{\hz}(2\pi/3)$ spin rotation in $C_{6z}$, $\vS(\k)$ is rotated by $2\pi/3$ rather than $\pi/3$ under $C_{6z}$. 
Due to $\{ I | m_{001} | 0,0,1/2 \}$, which is generated by $C_{6z}^3 \cdot \mathcal{P}$ up to a lattice translation, $\vS(\k)$ is invariant under the mirror with respect to the $k_z = 0$ plane.
Due to $C_{2,1\bar10}$, which is also a MSG symmetry, $\vS(\k)$ is rotated by $U_{{1\bar{1}0}}(\pi)$ under $2_{1\bar{1}0}$.
$\vS(\k)$ subject to these constrants form the 2D real irrep $\rm E_{2g}$ of $6/mmm$.

Mn$_3$Ge with the coplanar magnetic structure has three Fermi surfaces. 
Here we focus on the Fermi surface around $k_z=\pi$ and leave discussions on the other two in \AppD. 
First, we verify that $S_z(\k)$ vanishes at every $\k$ and $\vS(\k)$ is symmetric with respect to $k_z = 0$.
Fig.~\ref{Fig: Mn3Ge all}(c) shows the spin texture on the Fermi surface viewed from the $\hz$ axis, where different colors represent equal energy contours in different $k_z$ planes. 
One can verify that $\vS(\k)$ satisfies all the symmetry constraints derived in the last paragraph.
The two-component vector $\vS(\k)$ has a non-trivial vortex configuration: As $\k$ completes an anticlockwise path that circles the $z$ axis, $\vS(\k)$ is rotated by $4\pi$. 
This implies the existence of a single vortex with charge $S_V = 2$ on the north pole of the Fermi surface, as indicated by the yellow arrows on the small circle around $k_x=k_y=0$. 
\blue{
The non-trivial $S_V$ arises from $\vS(\k)$ forming the $E_{2g}$ representation of $6/mmm$. 
Let us consider a closed path in the $\k$ space, {\it e.g.}, $(k_0 \cos(\theta), k_0 \sin(\theta),k_{z0})$ with $k_0 > 0$ and $\theta \in [0,2\pi)$. 
The vortex charge is defined as $S_V = \frac{1}{2 \pi}\int_0^{2 \pi} d \theta \frac{d}{d\theta} \arctan(\frac{S_y(\theta)}{S_x(\theta)})$, where $S_{\mu}(\theta) \equiv S_{\mu} \left( k_0 \cos(\theta), k_0 \sin(\theta),k_{z0} \right)$ ($\mu = x, y$). 
The expression can be decomposed as $S_V= \sum_{i=1}^6 S_{Vi}$, with each $S_{Vi}$ representing the integral's contribution over the interval $\theta \in [ 2\pi(i-1)/6, 2\pi i/6)$.
The SSG operation $C_{6z}$ requires that $S_{Vi} = S_{Vj}$ for any $i,j = 1,\cdots,6$.
Given the spin operation in $C_{6z}$ as $ U_{\hz} (2\pi/3)$, $\arctan({S_y(\pi/3)}/{S_x(\pi/3)}) = \arctan({S_y(0)}/{S_x(0)}) + {2\pi}/{3}$, which is also equivalently required by the $E_{2g}$ representation.
Consequently, $S_{V1} = (2\pi/3 + 2 n \pi)/2\pi$ $(n \in \mathbb{Z})$, leading to $S_V =6 S_{V1} =  6n + 2$.}
Note that this vortex differs from previously studied ones in systems where SOC plays a major role in band splitting, such as the Rashba model~\cite{Johansson16}, because the charge $S_V$ of the latter one generally can only be $1$. 
We also verify this vortex is stable under SOC (see Fig.~\FigMnGeSOC in Appendix~\ref{app sec: SOC}).
The stability of the vortex originates from its topological nature, which is characterized by the homotopy group $\pi_1(S_1)$ and will remain immune under generic weak perturbation including SOC.

\subsubsection{\texorpdfstring{$E_g$}{Eg} spin texture in \texorpdfstring{Mn$_3$GaN}{Mn3GaN}}

\begin{figure}[ht]
    \centering
    \includegraphics[width=\linewidth]{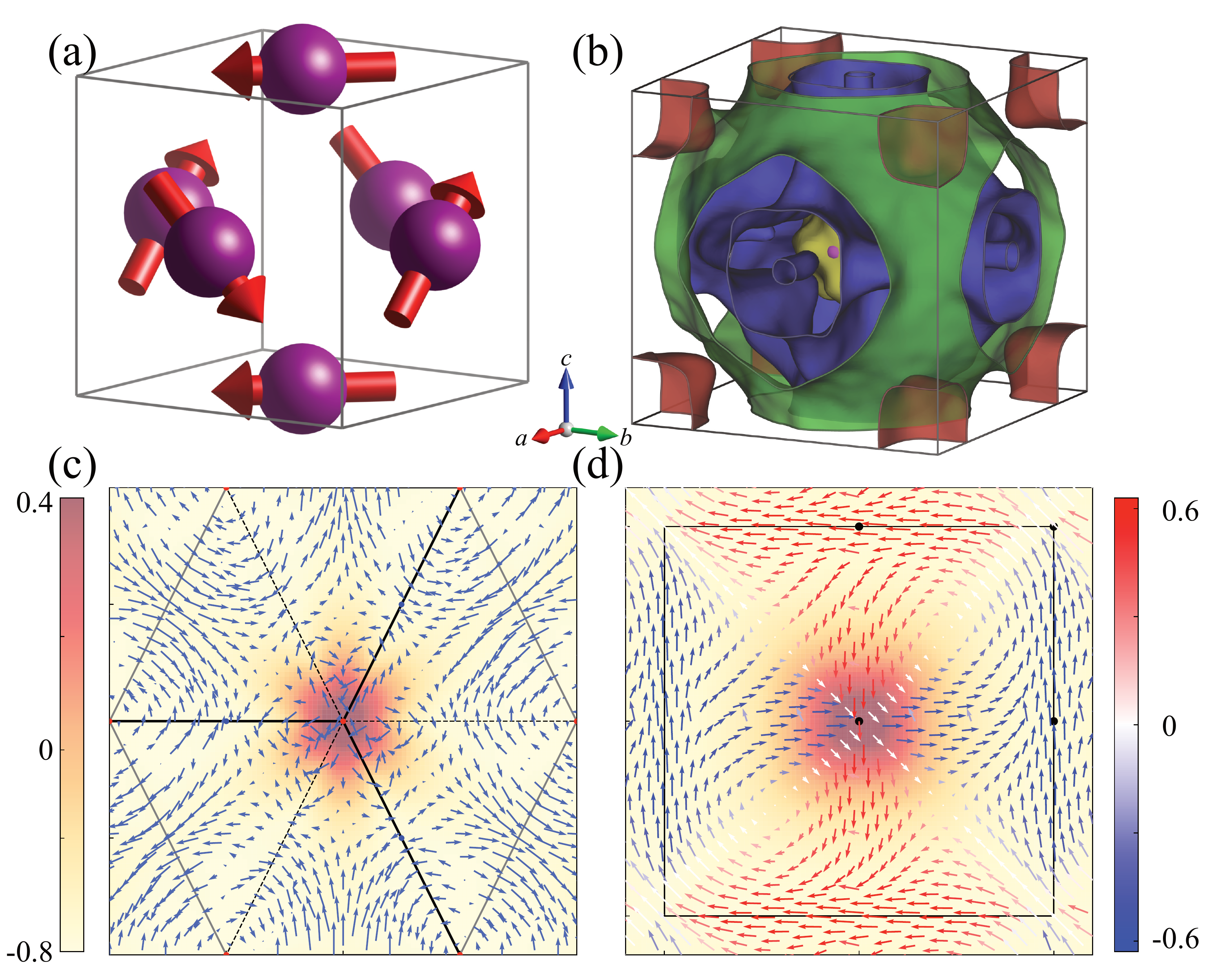}
    \caption[]{(a) Magnetic structure of Mn$_3$GaN showing only the magnetic atoms (Mn). 
    (b) The Fermi surfaces of Mn$_3$GaN.
    (c), (d) The spin texture $\vS(\k)$ on the $111$ plane ($k_x+k_y+k_z = 0$) and $001$ plane ($k_z = 0$) in the momentum space.
    The background colors denote the values of $E(\k)$ (see the left color bar).
    The outer hexagon in (c) and the square in (d) denoted the boundary of the first BZ projected to these planes.
    In (c), $\vS(\k)$'s are coplanar on the $111$ plane. 
    In (d), $\vS(\k)$'s are not confined to the $001$ plane, and the red or blue color (the right color bar) represents the value of $S_z(\k)$. 
    }
    \label{Fig: Mn3GaN all}
\end{figure}

The compound Mn$_3$GaN (\href{http://webbdcrista1.ehu.es/magndata/index.php?this_label=0.177}{No.~0.177}) has an antiperovskite crystal structure with crystalline symmetry $Pm\bar{3}m$ (No.~221). 
In the magnetic phase, the magnetic moments (Mn atoms) lie in the $111$ plane, forming a triangular antiferromagnetic structure in each $111$ cross-section \cite{fruchart_magnetic_1978}.
This structure's MSG $R\bar{3}m$ (No.~166.1.1327) is much lower than the crystalline symmetry, but its SSG P221.6.1 GM3$^+$ still enjoys the full crystalline symmetry.
This coplanar SSG is generated by $\{ I | 1 | \a_i \}$ ($i= 1,2,3$), $ \mathcal{P} = \{I | \bar{1}| \0 \}$, $C_{3,111} = \{ U_{111} (2\pi/3) |3_{111}^+ | \0 \}$, $C_{4z} = \{ U_{{1\bar{1}0}}(\pi) |4^+_{001} | \0 \}$, and $\T = \{ T U_{111}(\pi)| 1 | \0 \}$ (the generator for $\mS_{Z_2^T}$). 
One can derive $C_{2x} = \{ I | 2_{100}| \0 \}$, $C_{2,1\bar{1}0} = \{  U_{{1\bar{1}0}} | 2_{1\bar{1}0} | \0 \}$ from the generators.
Here $111$ and $1\bar10$ correspond to the directions $\mathbf{e}_x + \mathbf{e}_y + \mathbf{e}_z$ and $\mathbf{e}_x - \mathbf{e}_y$, respectively. 
Due to the $\mathcal{PT}$ symmetry, $\vS(\k)$ is within the $111$ plane. 
Following a similar analysis as in the previous sections, we find that the $\vS(\k)$ lying in 111 plane forms the $E_g$ representation of the point group $O_h$: It is parity-even, rotated by $2\pi/3$ under $C_{3,111}$, and undergoes a 2D reflection $U_{1\bar10}(\pi)$ under $C_{4z}$.

Fig.~\ref{Fig: Mn3GaN all}(b) shows the Fermi surfaces of Mn$_3$GaN obtained by the first-principle calculation. 
We verify that the $111$-component of $\vS(\k)$ at any $\k$ is always zero, consistent with the $\mathcal{PT}$ symmetry constraint.
For the clarity of demonstration, we do not show $\vS(\k)$ on the complicated 3D Fermi surfaces but plot $\vS(\k)$ of the band, which contributes to the second innermost Fermi surface shown in Fig.~\ref{Fig: Mn3GaN all}(b), on the high-symmetry planes containing $\Gamma$ point. 
The $111$ plane has the symmetry of $D_{3d}$ generated by $\mathcal{P}$, $C_{3,111}$, and $\{R_{{1\bar{1}0}}(\pi)|2_{1\bar{1}0}|\0 \}$. 
Thus $\vS(\k)$ on this plane form the reduced representation $E_g \downarrow D_{3d} = E_g$, which is even-parity, rotated by $2\pi/3$ under $C_{3,111}$, and undergoes a 2D reflection under $2_{1\bar10}$ [$1\bar10$ is the vertical direction in Fig.~\ref{Fig: Mn3GaN all}(c)]. 
The spin texture on the 111 plane can also be explained by the MSG $R\bar{3}m$ since in these aforementioned operations, spin rotations equal the spatial rotations. 
On the other hand, the pattern of $\vS(\k)$ on the $001$ plane can only be understood using SSG. 
The 001 plane has the symmetry of $D_{4h}$, and hence $\vS(\k)$ on the 001 plane forms the reduced representation $E_g \downarrow D_{4h} = {\rm A_{1g} + B_{1g}}$. 
As $(S_x, S_y, S_z)$ transforms into $(-S_y,-S_x,-S_z)$ under $C_{4z}$, and is invariant under $\mathcal{P}$ and $C_{2x}$, one can show that $S_x-S_y$ forms the $\rm A_{1g}$ representation of $D_{4h}$ and $S_x+S_y$, $S_z$ from the $\rm B_{1g}$ representation.  
The spin texture from first-principle calculation shown in Fig.~\ref{Fig: Mn3GaN all}(d) demonstrates this unconventional feature, where the $S_x(\k), S_y(\k)$ are represented by vectors and  $S_z(\k)$ is represented by the color of the vectors.

\section{Topological phases protected by SSG}
\label{sec: topo phase}

As spatial operations in SSGs are associated with spin rotations, the algebras of SSG symmetry operations, described by projective representation (Sec.~\ref{sec:band-theory}), can be different from those in the MSGs.
The enriched symmetry algebra enables novel topological states in the absence of SOC and TRS.
Here we comment on possible topological states in SSGs. For collinear SSGs, as the symmetry operators form a linear representation of grey groups, no stable TI state could be stabilized \cite{po_symmetry-based_2017,bradlyn_topological_2017}. However, various topological semi-metal can be protected as in space groups in class AI. 
For coplanar SSGs, the effective TRS $\hat{\T}_{\rm eff} = \hat{U}_{\hat{\bm x}}(\pi) \hat{\T}$, which squares to 1, forbids Chern insulators. Mirror Chern insulators can appear in coplanar SSGs where a mirror operator $\hat{M}$ satisfies $\hat{M}^2=1$ and $\{\hat{M}, \hat{\T}\}=0$. Chern numbers in the mirror-even and mirror-odd sectors must be opposite due to $\hat{\T}_{\rm eff}$. 
3D Derivatives of mirror Chern insulator can be constructed using the topological crystal approach \cite{song_topological_2017,song_topological_2019}. 
In non-coplanar SSGs, both Chern insulator \cite{ohgushi2000} and $\mathbb{Z}_2$ TI \cite{liu2022spin}, which is protected by $\hat{\T}_{\rm eff} = \hat{M} \hat{T}$ ($\hat{\T}_{\rm eff}^2=-1$), as well as their 3D derivatives \cite{song_topological_2017,song_topological_2019}, can be stabilized. 
We present three examples of topological states in non-coplanar SSGs in the following subsections.

\subsection{2D \texorpdfstring{$\mathbb{Z}_2$}{Z2} TI in the absence of SOC and TRS}
\label{2DTI}
It is widely known that the $\mathbb{Z}_2$ topological insulator (TI) is protected by the spinful TRS whose square equals $-1$, which only exists in nonmagnetic materials with SOC.
However, in 2D systems with non-collinear magnetism and negligible SOC, which seem to completely violate the conditions of the existence of TI, the SSG operation $\hat{M}_z\hat{\T} = \{ i \sigma_y \mathcal{K} |  m_{001} | \0 \}$ can serve as an effective TRS that squares to $-1$ and can give rise to a 2D magnetic $\mathbb{Z}_2$ TI \cite{liu2022spin}.  
Note that its counterpart $ \{ i \sigma_y \mathcal{K} e^{i \pi \sigma_z} |  m_{001} | \0 \}$ in MSG, which contains $\pi$-spin rotation along the $z$ direction, squares to $1$ and cannot protect a TI.

\begin{figure}[ht]
    \centering
    \includegraphics[width=\linewidth]{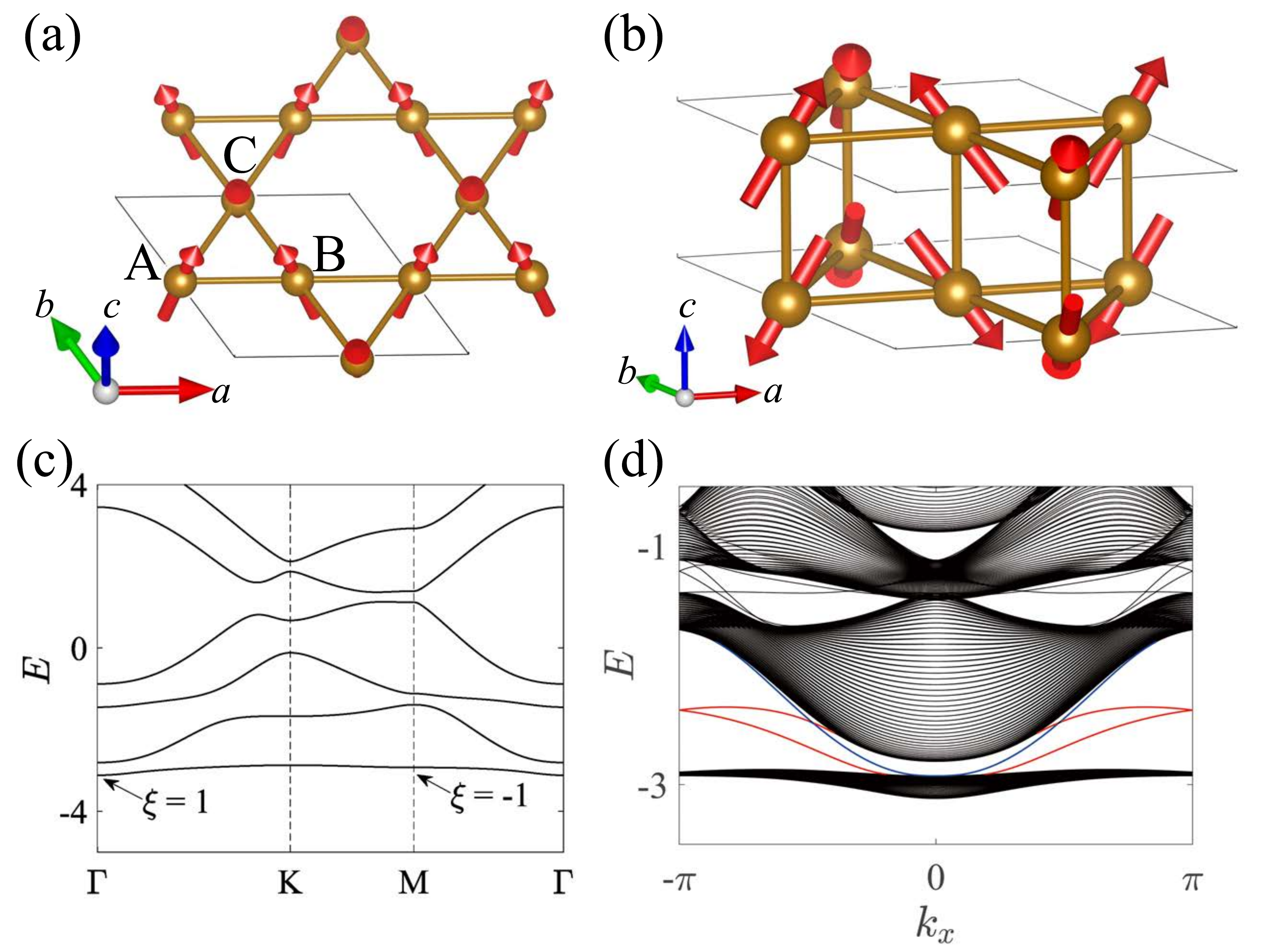}
    \caption[]{Non-coplanar magnetic structure on the (a) upper layer and (b) bilayer of the Kagome lattice. (c) Energy band of Hamiltonian $\H_{Z_2}$ [Eq.~(\ref{H_Kagome})] in the periodic boundary condition (PBC). Energy is double-degenerate in the whole Brillouin zone. (d) Energy band of $\H_{Z_2}$ under PBC along the $x$ direction and OBC along the $y$ direction with $N_y = 50$. The blue (red) line represents the helical edge mode around $y = 0$ ($y = N_y$).}
    \label{Fig: 2D TI}
\end{figure}

Here we provide a concrete model with real hoppings and non-coplanar magnetism to realize the 2D magnetic $\mathbb{Z}_2$ TI. 
We consider an A-A stacked bilayer Kagome lattice [Figs.~\ref{Fig: 2D TI}(a) and (b)]. 
The 2D crystal structure can be described by the space group $P6/mmm$ (No.~191) with lattice constant $c \to \infty$. 
In each layer, local magnetic moments $\S(\r_i)$ have a canting angle $\theta$ ($\cos\theta = |S_z|/ |\S|$), and in-plane components form an all-in-all-out spin-ice structure. 
The magnetic structure on a single-layer Kagome lattice has nonzero spin chirality and is identical to the one studied in previous work on metallic pyrochlore ferromagnets~\cite{ohgushi2000}, in which a Chern insulator phase without SOC can occur. 
The mirror operation $m_{001}$ transforms two layers into each other, and the moments of two neighboring sites in two layers are opposite to each other. 
The SSG of this magnetic structure, named N191.12.2 $\rm GM1^{-} \oplus  GM5^{-}$,  is generated by the symmetry operators (acting on fermions)
$\hat{C}_{3z} = \{ e^{-i \frac{ \pi}{3} \sigma_z} | 3_{001}^+ | \0\} $, $\hat{C}_{2z} = \{ \sigma_0 | 2_{001} | \0 \}$, $\hat{M}_z \hat{\T} = \{i \sigma_y \mathcal{K}  |  m_{001} | \0 \} $, $ \hat{C}_{2x} = \{-i \sigma_x |  2_{100} | \0 \} $, and lattice translations without spin operations.

We consider the electronic Hamiltonian with such a magnetic structure 
{\small
\begin{equation}
\begin{aligned}
        \H_{Z_2} &= \sum_{i = 1,2} \left[ t_{\|}\sum_{\langle \R ,\alpha ; \R^{\p} , \alpha^{\p} \rangle} \left( c^{\dagger }_{ \R ,\alpha,i } \sigma_0 c_{\R^{\p}  , \alpha^{\p},i} \right) + \right.
        \\ 
        & \left. J \sum_{\R ,\alpha} \S_{\R ,\alpha,i } \cdot {\bm s}_{\R,\alpha,i} \right] + 
        t_{\perp}  \sum_{\R, \alpha} \left( c^{\dagger }_{ \R ,\alpha,1 } \sigma_0 c_{ \R ,\alpha,2 } + {\rm H.c.} \right) \, ,
\end{aligned}
\label{H_Kagome}
\end{equation}}%
with
\begin{equation*}
    {\bm s}_{\R,\alpha,i }^{\mu} = \frac{1}{2}c^{\dagger }_{ \R ,\alpha,i } \sigma_{\mu}  c_{ \R ,\alpha,i } \quad (\mu = x,y,z)
\end{equation*}
being the spin operator of the electrons.  
Here $c^{\dagger }_{ \R ,\alpha,i } = (c^{\dagger }_{ \R ,\alpha,i ,\uparrow}, c^{\dagger }_{ \R ,\alpha,i, \downarrow })$ ($\alpha=A,B,C$, $i=1,2$) is a two-component spin-1/2 electron creation operator in  the unit cell $\R$, sub-lattice $\alpha $, of the $i$th layer.
The first and third terms of the Hamiltonian are intralayer and interlayer nearest-neighboring hopping, respectively. 
The hopping matrix is proportional to $\sigma_0$ because the SOC is considered negligible. 
The second term describes the on-site interaction between the conduction electron and localized magnetic moments $\S_{\R, \alpha, i}$ [Figs.~\ref{Fig: 2D TI}(a) and (b)]. 
In the following, we choose $t_{\|} = 1, t_{\perp} = 0.5, J =2$, and $|\S_{\R,\alpha,i}| = 1$ with canting angle $\theta = \pi/3$. 
The band structure is shown in Fig.~\ref{Fig: 2D TI}(c), where every band is double-degenerate, as explained in the next paragraph.

We can evaluate the $\mathbb{Z}_2$ topological invariant $\nu$ of the model by the Fu-Kane formula~\cite{fu07}.
In this system, $\hat{M}_z \hat{\T}$ plays the role of TRS, and $\hat{C}_{2z}$, which does not contain spin rotation, plays the role of inversion symmetry because it squares to 1 and commutes with $\hat{M}_z \hat{\T}$
\footnote{Note that in the 2D system with SOC, which is described by the magnetic space group which has TRS and $C_{2z}$ symmetry, the $C_{2z}$ eigenvalues of Krammer's pairs have the opposite signs. The square of $C_{2z}$ equals $-1$, so the eigenvalues of $C_{2z}$ equal $i$ or $-i$. $C_{2z}$ commutes with TRS, so after TRS, the $C_{2z}$ eigenvalue should be its complex conjugate. In such systems, a ``Fu-Kane type'' formula does not exist.}. 
The topological invariant $\nu$ is then determined by $(-1)^{\nu} = \prod_{i} \xi(\k_i)$, where $\k_i$ are TRS-invariant momenta, and $\xi(\k_i)$ is the product of the $\hat{C}_{2z}$ eigenvalues of occupied Kramers pairs at $\k_i$. 
(Notice that the joint symmetry $\hat{C}_{2z} \hat{M}_z \hat{\T}$ preserves momentum and squares to $-1$, hence it protects double-degeneracy of the energy bands.)

When the filling is $2$ per unit cell, only the lowest double-degenerate band is occupied, with $\xi(\k)=1, -1$ at $\Gamma$ and (three) $M$ points, respectively, implying that the system is topological.
We also observe the topological helical edge states on the open boundary of a cylinder geometry [Fig.~\ref{Fig: 2D TI}(d)].
\blue{The non-coplanar magnetism is essential in realizing a non-trivial $\mathbb{Z}_2$ topological invariant.
In coplanar magnetism, we can align the magnetic moments within the $x,z$ plane through a global spin rotation. 
In such a case, both the hoppings and on-site terms of the Hamiltonian become real, inevitably leading to a $\mathbb{Z}_2$-trivial state.
It is the $y$-component of $\S_{R,\alpha,i}$'s in our model that makes the Hamiltonian pseudoreal ({\it i.e.,} respecting TRS with sign $-1$) and allows a possible non-trivial $\mathbb{Z}_2$ index. 
}

\subsection{3D \texorpdfstring{$\mathbb{Z}_2$}{Z2} TI with four-fold Dirac point on surface}

\begin{figure*}[ht]
    \centering
    \includegraphics[width=\linewidth]{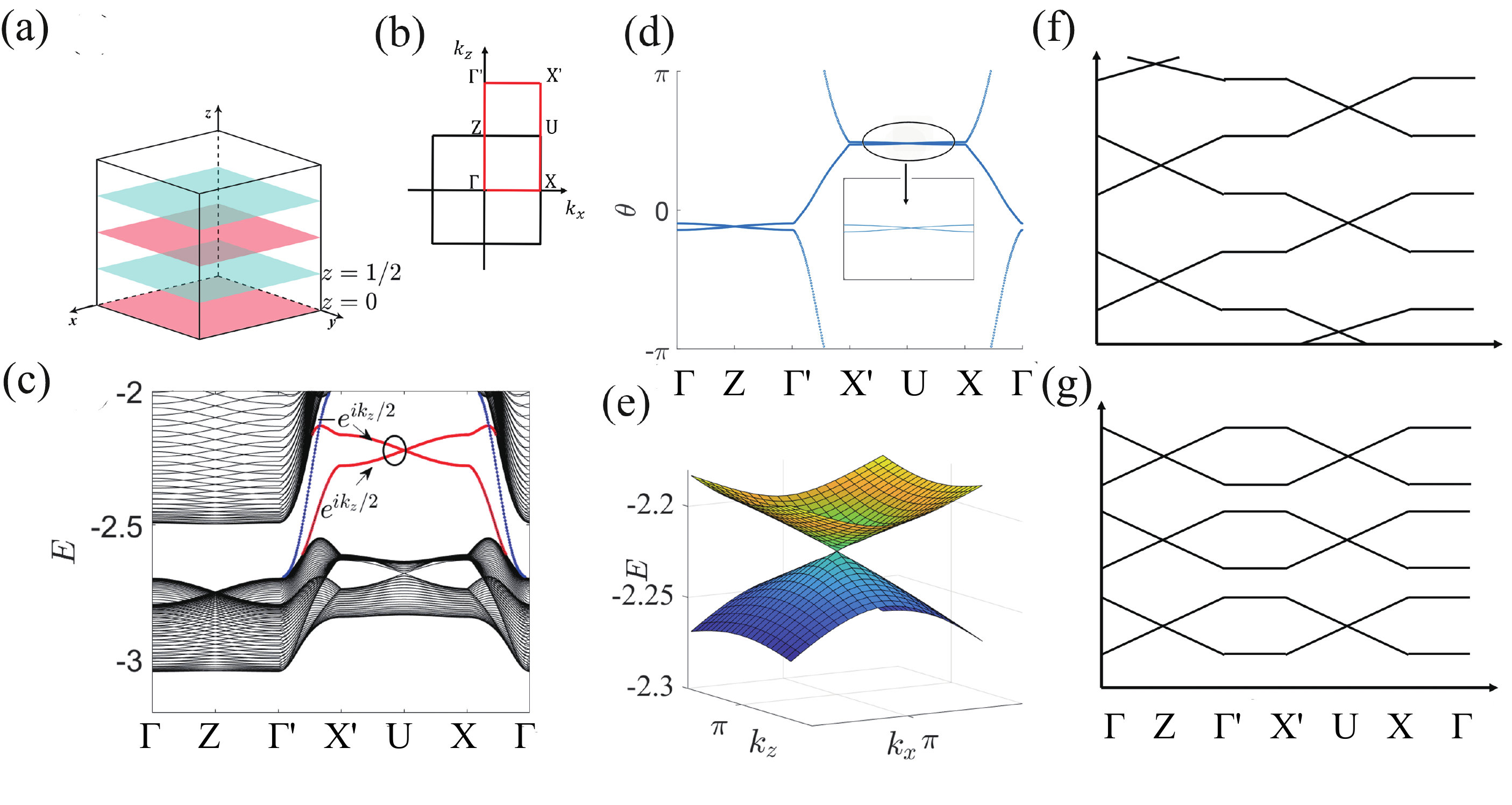}
 \caption[]{(a) Layer construction of the 3D $\mathbb{Z}_2$ TI protected by $\hat{M}_z\hat{\T} = \{i \sigma_y \mathcal{K}  |  m_{001} | \0 \}$ and $\hat{G}_x = \{ \sigma_0| m_{100} | 0,0,1/2\}$. On each pink plane and cyan plane, a $\mathbb{Z}_2$ TI protected by $\hat{M}_z\hat{\T}$ is placed.
 (b) The $010$-surface BZ of the TI.
 (c) Energy bands of the 3D $\mathbb{Z}_2$ TI [Eq.~(\ref{3DTI})] under PBC along the $x,z$ directions and OBC along the $y$ direction with $N_y = 50$. 
 The red (blue) lines represent the surface mode around $y = N_y$ ($y= 0$). 
 The glide eigenvalues of the surface states on $k_x = \pi$ are labeled. $E(k_x,k_z)$ is at least double-degenerate for any $(k_x,k_z)$.
 The position of high-symmetry points can be found in subfigure (b).
 (d) The phases $\theta_i = -{\rm Im} \ln(\lambda_i) $ of the eigenvalues $\lambda_i(k_x,k_z)$ of the Wilson loop operator $W(k_x,k_z)$ along the high-symmetry path. Note that $\theta_i$ are always double-degenerate.
 Inset: zoom-in of $\theta_i(k_x,k_y)$ along the X'-U-X.
 (e) Zoom-in of the surface Dirac point in subfigure (c).
 (f), (g) Two distinct connection scenarios of the surface bands: zigzag flow (f) and direct gaps (g). 
  }
  \label{Fig: TI 3D}
\end{figure*}

We show that an unavoidable four-fold Dirac point can be protected on the surface of a 3D $\mathbb{Z}_2$ TI protected by SSG symmetries. 
It provides an exception to the fermion-doubling theorem of 2D systems~\cite{young_dirac_2015,wieder_wallpaper_2018}. 

As shown in Fig.~\ref{Fig: TI 3D}(a), the 3D state is constructed by stacking the 2D $\mathbb{Z}_2$ TI layers protected by $ \hat{M}_z \hat{\T} = \{ i \sigma_y \mathcal{K} |  m_{001} | \0 \} $ (Sec.~\ref{2DTI}).
We first decorate the integer planes, $z=0,\pm1,\pm2\cdots$, with the 2D Hamiltonian $\H_{Z_2}(k_x,k_y)$ [Eq.~(\ref{H_Kagome})].
Different layers are related by the translation $\hat{t}_3 = \{I|1|0,0,1\}$ along the $z$ direction, and the $\mathbb{Z}_2$ topology of the layer at $z=n$ is protected by $ \hat{t}_3^{2n}\hat{M_z}\hat{\T}$. 
We then decorate the half-integer planes, $z=\pm\frac12,\pm\frac32\cdots$, with the mirror reflection of $\H_{Z_2}(k_x,k_y)$, {\it i.e.}, $D^\dagger(m_{100}) \H_{Z_2}(-k_x,k_y) D(m_{100})$, where $D_{\alpha i s, \alpha' i' s'}(m_{100}) = 
M_{\alpha,\alpha'} \delta_{ii'}\delta_{ss'}$ is the mirror representation matrix on the local orbitals. 
Here $\alpha,\alpha'=A,B,C$ represent the sub-lattice, $i,i'=1,2$ represent top and bottom sub-layers, and $s,s'=\uparrow,\downarrow$ represent the spin. 
$M$ exchanges $A,B$ sub-lattice, and its non-zero components are given as  $M_{AB} = M_{BA} = M_{CC} = 1$.
The $\mathbb{Z}_2$ topology of the layer at the half-integer position $z=n+\frac12$ is protected by $\hat{t}_3^{2n+1}\hat{M_z}\hat{\T}$. 
\blue{
The symmetry of the 3D system is determined by its constructing layers.
The layer at $z = n$ respects $\hat{C}_{3z} = \{ e^{-i \pi/3 \sigma_z}|3_{001}^+| \0 \}$ (see Sec.~\ref{2DTI}), while the layer at $z = n  + \frac{1}{2}$ respects a modified $\hat{C}_{3z}$, given as $\{ e^{-i \pi/3 \sigma_z} |m_{100}^{-1} 3_{001}^+ m_{100}^{-1}| \0 \} = \{ e^{-i \pi/3 \sigma_z} | 3_{001}^-| \0 \}$. 
The spatial rotations $3_{001}^+$ in these two layers involve opposing spin rotations, leading to the elimination of three-fold rotation symmetry in the entire 3D system. 
In contrast, the $\hat{C}_{2z}$, $\hat{M}_z \hat{\T}$, and $\hat{C}_{2x} $ symmetries are preserved in the 3D structure. 
In addition, the system respects the glide symmetry $\hat{G}_x =  \{ \sigma_0 |  m_{100} | 0,0,1/2 \} $, due to our construction.
The combination of these symmetries yields a half-lattice translation, given as $\{ \sigma_z \mathcal{K} |1|0,0,1/2\}$.
Considering all the symmetries, the 3D structure is described by the SSG N47.9.392 $\rm GM2^{-}\oplus GM2^{-} \oplus Z1^-$.
Note that the half-lattice translation in this SSG is anti-unitary and hence not relevant to the definition of \blue{SSG} Bloch states.
}

In the end, we introduce a coupling $V$ between nearby layers, and write the entire 3D Bloch Hamiltonian as 
{\small
\begin{equation}
    \H_{\rm 3D}(\k) = 
    \begin{pmatrix}
        \H_{Z_2}(k_x,k_y) & V(k_z) \\
        V^{\dagger}(k_z) & D^{\dagger}(m_{100}) \H_{Z_2}(-k_x,k_y) D(m_{100})
    \end{pmatrix} .
    \label{3DTI}
\end{equation}}%
For simplicity, we assume that $V$ couples the 1st (2nd) sub-layer of the layer at $z$ to the 2nd (1st) sub-layer of the layer at $z+\frac12$ ($z-\frac12$), and it is diagonal in the spin and sub-lattice. Then the $V$ term can be written as 
\begin{equation}
    V_{\alpha i s,\alpha'i's'}(k_z) = \begin{pmatrix}
        0 & t_{\perp}^{\p}  e^{ik_z} \\
        t_{\perp}^{\p}  & 0
    \end{pmatrix}_{ii'} \delta_{\alpha \alpha'} \delta_{ss'} \ .
\end{equation}

We choose the parameters $t_{\|} = 1, t_{\perp} = 0.3, J =2, |\S_{\R,\alpha,i}| = 1, \theta = 0.4\pi$, and $t_{\perp}^{\p} = 0.15$.
The filling is chosen as four per unit cell such that each layer is filled up to the topological gap.
We diagonalize a slab sample with PBC along the $x,z$ directions and OBC along the $y$ direction with $N_y$ ($N_y =50$) sites to study the states on the 010 surface, which preserves $\hat{G}_x$ and $\hat{M}_z\hat{\T}$ symmetry.
Note that the detailed dispersion of edge states depends on whether the interfacial atoms belong to A, B sublattices (flat edge) or C sublattice (sawtooth-shape edge) [Fig.~\ref{Fig: 2D TI}(a)]~\cite{redder16}.
For the clarity of demonstration, we focus on the right sawtooth-shape surface $y=N_y$.
The surface wave vector $(k_x,k_z)$ is invariant under the anti-unitary operation
\begin{equation}
    \hat{C}_{2y}\T = \hat{G}_x \hat{M}_z \hat{\T} = \{ i\sigma_y \mathcal{K}| 2_{010} | 0,0,1/2 \} \, .
\end{equation}
As $(\hat{C}_{2y}\hat{\T})^2=-1$, the surface band at every $\k$ is at least double-degenerate due to the Kramers' theorem. 
The two-fold bands form a single four-fold Dirac point around the high symmetry point U, as shown in Figs.~\ref{Fig: TI 3D}(c) and (e). 

The surface state manifests a $\mathbb{Z}_2$ topology of the 3D TI. 
To show this, we prove that two distinct connection scenarios exist of the surface bands; one must have an odd number of four-fold Dirac points between every two (degenerate) bands, whereas the other can be gapped. 
We first consider the bands along the $k_x=0$ line, which can be labeled by eigenvalues of $\hat{G}_x$. 
As $\hat{G}_x^2=\{I|1|0,0,1\}$, the $\hat{G}_x$ eigenvalue of a state $\ket{\psi(0,k_z)}$ can be either $e^{ik_z/2}$ or $-e^{ik_z/2}$. 
Using the relation $\hat{G}_x \hat{C}_{2y} \hat{\T} = \{I|1|0,0,1\} \hat{C}_{2y} \hat{\T} \hat{G}_x$, one can show that $\hat{C}_{2y} \hat{\T} \ket{\psi(0,k_z)}$ is another state (Kramers' theorem) that has the same $\hat{G}_x$ eigenvalue. 
Thus, every band with $k_x=0$ is two-fold degenerate and has a definite $\hat{G}_x$ eigenvalue. 
Because $\pm e^{ik_z/2}$ changes into each other as $k_z$ moves from 0 to $2\pi$, a pair of nearby bands with opposite $\hat{G}_x$ eigenvalues must evolve into each other along this path and form an odd number of crossings [Figs.~\ref{Fig: TI 3D}(f) and (g)].
The same argument applies to the high symmetry line at $k_x=\pi$. 
Now we consider the closed path $\Gamma$-Z-$\Gamma'$-X$'$-U-X-$\Gamma$ shown in Fig.~\ref{Fig: TI 3D}(b). 
There are two possible connections between $\Gamma$X: (i) the surface bands form a zigzag flow along the closed path [Figs.~\ref{Fig: TI 3D}(f)], and (ii) the surface bands have direct gaps [Figs.~\ref{Fig: TI 3D}(g)]. 
Bands in scenario (i) cannot be gapped without breaking the symmetries.

One can also define a $\mathbb{Z}_2$ topological invariant for the 3D bulk state using the Wilson loop operator $W(k_x,k_z)$ integrated along the $y$ direction.
Its spectrum satisfies the same $\hat{G}_x$ and $\hat{C}_{2y} \hat{\T}$ symmetry constraints as the surface states and hence can form a zigzag flow that indicates a nontrivial topology [Fig.~\ref{Fig: TI 3D}(d)]. 
The existence and robustness of the surface four-fold degenerate Dirac cone can also be understood by a Dirac theory, and we leave this discussion in \AppE.

It has been shown that, at specific fillings, four-fold Dirac point could exist in 2D magnetic systems with the MSG $C_P2'/m$ (No. 12.9.74) \cite{young_filling-enforced_2017} or on surface of 3D topological states with the grey MSG $Pba21'$ (No. 32.2.220) \cite{wieder_wallpaper_2018}. 
However, bands in these systems will become gapped \cite{young_filling-enforced_2017} or exhibit two double-degenerate Dirac points \cite{wieder_wallpaper_2018,fang_new_2019} if the surface band filling (for the 3D TI) or the filling (for the 2D system) is changed by two. 
Instead, the four-fold Dirac point protected by SSG is unavoidable at arbitrary even fillings [Fig.~\ref{Fig: TI 3D}(f)].

\subsection{\texorpdfstring{$\mathbb{Z}_2$}{Z2} helical mode on magnetic domain wall} 

\begin{figure}[h]
 \includegraphics[width=1 \linewidth]{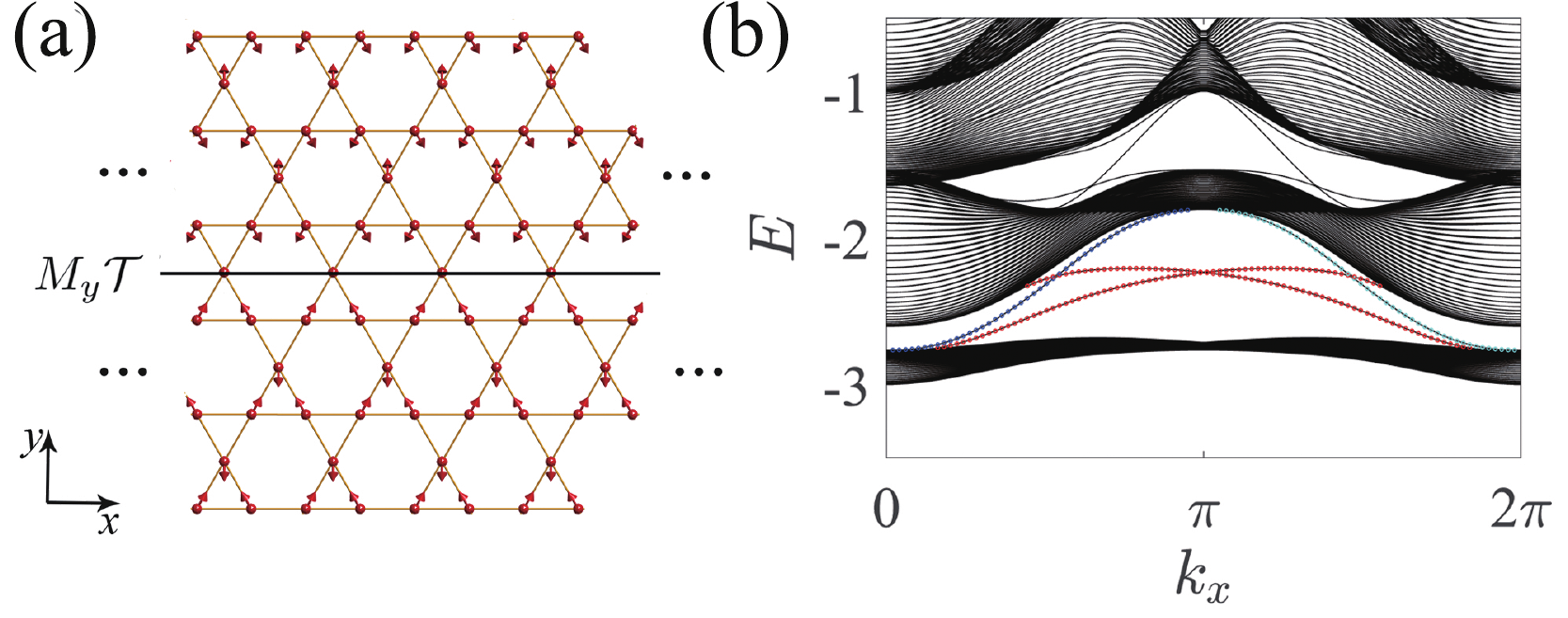}
    \caption[]{(a) Non-coplanar magnetic structure with two magnetic domains on the Kagome lattice. Two magnetic domains are related by $\{i\sigma_y \mathcal{K}|m_{010}|\0\}$, and out-of-plane canting on the upper (lower) plane is towards $z$ ($-z$) direction. (b) Energy band of $\H_{2D}$ under PBC along the $x$ direction and OBC along the $y$ direction. $N_y$ of the upper and lower domain both equal $50$. Red dots represent helical edge modes on the domain wall. The blue and cyan dots represent chiral edge modes around the upper and lower boundaries. }
    \label{Domain}
\end{figure}

Anomalous edge states are known to occur at boundaries between topologically distinct systems. Here we demonstrate a new situation where an anomalous helical state appears at a domain wall between two magnetic domains with the {\it same} Chern number. 
The helical state is protected by an SSG symmetry that relates the two domains as well as the nonzero but the same Chern number of the two domains. 
One would expect a gapped domain wall without knowing the SSG symmetry. 

We consider a single-layer Kagome lattice with two magnetic domains [Fig.~\ref{Domain}(a)]. 
The Hamiltonian has the form
\begin{equation}
\begin{aligned}
        \H_{2D} &=  t_{\|}\sum_{\langle \R ,\alpha ; \R^{\p} , \alpha^{\p} \rangle} \left( c^{\dagger }_{ \R ,\alpha } \sigma_0 c_{\R^{\p}  , \alpha^{\p}} \right) +  J \sum_{\R ,\alpha} \S_{\R ,\alpha } \cdot {\bm s}_{\R,\alpha} \, .
\end{aligned}
\label{H_Domain}
\end{equation}
In the lower plane ($y<0$), $\S_{\R,\alpha}$ are identical to those in the $i=1$ sub-system of Eq.~(\ref{H_Kagome}), which exhibits an out-of-plane canting along the $+z$ direction and forms out-in-out-out structure for the in-plane components. 
The magnetic structure in the upper plane ($y > 0$), on the other hand, is related to that of the lower plane through the SSG operation $\hat{M}_y \hat{\T} = \{i\sigma_y \mathcal{K}|m_{010}|\0\}$. The magnetic moments on the domain wall ($y=0$), which is the $\hat{M}_y \hat{\T}$-invariant line, equal zero. 

By calculating Berry curvature, we find that the Hamiltonians in both the lower and upper domains have $C=-1$ (at the filling of one electron per unit cell). 
Note that the $\hat{M}_y \hat{\T}$ symmetry promises the same Chern number of the two domains. 
We diagonalize a sample $\H_{2D}$ with OBC along the $y$ direction and PBC along the $x$ direction, in which two domains occupy the same area. 
In addition to the chiral edge modes in the upper and lower boundaries, a helical edge mode appears around the domain wall [Fig.~\ref{Domain}(b)]. 

The existence and robustness of the helical edge mode can be understood through a Dirac theory.
First, consider an infinite barrier at the domain wall. The lower and upper domains will give rise to a chiral and an anti-chiral edge mode around the domain wall, respectively. 
They do not couple with each other because of the infinite barrier.
The effective Hamiltonian can be written as $\H(k_x) = v_F k_x \sigma_z$, where $v_F$ is the Fermi velocity, and the $\hat{M}_y \hat{\T}$ symmetry is represented by $i\sigma_y \mathcal{K}$, which serves as an effective spinful TRS on the domain wall. 
Then we consider softening the barrier. 
One may expect a mass generation of the form $\sigma_y$ or $\sigma_x$. 
However, these mass terms are forbidden by the $\hat{M}_y \hat{\T}$ symmetry. 
As a result, the helical edge mode remains robust. 

\blue{Applying a generic O(3) matrix rotation to only the magnetic moments in the upper (or lower) domain would disrupt the $M_y \mathcal{T}$ symmetry, resulting in the opening of the gap of the helical mode.
However, such a configuration might be less stable than the original one.
According to the Landau–Ginzburg theory, the free energy $F$ associated with a domain is a function of the relative angle $\theta$ between the order parameters on the two sides.
For the $M_y \mathcal{T}$-symmetric domain wall, the order parameters on two sides are opposite, and hence $\theta = \pi$. 
Let us expand $F$ around $\theta = \pi$. 
As required by the symmetry, $F$ should be an even function of $(\theta - \pi)$, and at the lowest order, $F = F_0 + b(\theta - \pi)^2 $.
In the case that $b > 0$, $\theta = \pi$ is the local minimum of the free energy. 
}

\section{Summary and discussion}
\label{Sec: summary}

In this work, we completed the full classification of SSGs for the first time. 
These SSGs provide a complete mathematical description for the symmetries of all types of magnetic materials, including collinear, non-collinear, commensurate, and incommensurate spiral configurations, {\it etc.}, when the strength of SOC is weaker than the relevant energy scale.
Remarkably, we find that the classification problem can be mapped to a representation problem.
The SSGs ($\mG$) for collinear, coplanar, and non-coplanar magnetic structures can be represented by O(1), O(2), and O(3) representations of the parent space groups ($P$), respectively.
To summarize, the O($N$) representation matrix $D(p)$ of a spatial operation $p\in P$ indicates how the magnetic moments, confined within an $N$-dimensional subspace, undergo transformation after the spatial operation $p$ in such a manner that the magnetic structure remains unchanged.
We enumerate all the inequivalent O($N$) representations by exhausting all distinct combinations of real irreps of space groups.
Depending on their constituent irreps, the O($N$) representations are categorized into two (I-II), eight (I-VIII), and sixteen (I-VXI) types for $N=1,2,3$, respectively. 
It is worth mentioning that O(2) representations of types V, VIII, and O(3) representations of types V, VIII, XI, and XIV allow both incommensurate and commensurate spiral magnetism due to their inclusion of irreps induced from non-HSP momenta. Thereby, the momenta determine the spiral angles, allowing for continuous changes.
By organizing all the representations by equivalence relations associated with coordinate transformations and continuous change of the non-HSP momenta, we obtain 1421, 9542, and 56512 distinct SSGs for collinear, coplanar, and non-coplanar magnetic structures, respectively. 
For clarity and ease of reference, we have introduced a name convention $\alpha \mathcal{I.J.K}$ $\rho$ for the SSGs, where $\alpha$ indicates collinear ($\alpha = \rm L$), coplanar ($\alpha = \rm P$), or non-coplanar magnetism ($\alpha = \rm N$), $\mathcal{I}=1\cdots 230$ is the index of the parent space group $P$, $\mathcal{J}=1\cdots 16$ is the type of the representation, $\mathcal{K}$ is the additional numbering of the SSG for given $\alpha \mathcal{I.J}$, and $\rho$ specifies the constituent irreps forming the O($N$) representation (see Sec.~\ref{sec: nonmenclature}, and see Figs.~\ref{p3-ssg-1} and \ref{p3-ssg} as examples). 
A complete list of all SSGs is given in \AppSecSSGTable. 
Additionally, to facilitate future studies, we identify the SSGs of the \blue{1595} published magnetic materials in the \href{http://webbdcrista1.ehu.es/magndata/}{MAGNDATA} database \cite{gallego_magndata_2016,gallego_magndata_2016-1} on the Bilbao Crystallographic Server. 
Detailed information on these materials is given in \AppSecMaterialTable. 

We have discussed various applications of SSGs and introduced several key concepts, such as SSG momentum, SSG Brillouin zone (SBZ), symmorphic and nonsymmorphic SBZ, and non-commuting SBZ, in electronic band theory. 
In a collinear SSG, the electronic bands decouple to spin-up and spin-down sectors.
The \blue{bands} within each spin sector form a linear representation of a single-valued grey MSG $G_0 \times \mS_{Z_2^T}$, where $G_0 \subset P$ consists of pure spatial operations preserving the magnetic structure.
In anti-ferromagnetic or alter-magnetic SSGs, which are given by non-identity O(1) representations of $P$, the two spin sectors are related to each other.
In a non-collinear SSG, the electronic bands form a projective representation, characterized by factor system $\omega_2\in H^2(\mathcal{M}, \mathrm{U(1)})$ of an MSG $\mathcal{M}$ that is isomorphic to the SSG $\mG$.
$\mathcal{M}$'s are type-II, and type-I, III, IV MSGs for coplanar and non-coplanar structures, respectively. 
The projective representation allows for exotic electronic features that do not exist in conventional MSGs.
One such example is the nonsymmorphic SBZ, where an SSG operation ${g}=\{X_g U_g | R_g | \v_g\}$ transforms an SSG momentum $\kk$ to $s_g \left( R_g \kk + \qq_g \right)$ with a fractional reciprocal lattice $\qq_g$ and $s_g=1$ ($-1$) for unitary (anti-unitary) operation.
In other words, a real space rotation (screw) or mirror (glide) $g$ may behave as a screw or glide in the SSG momentum space. 
This feature will be reflected in extra degeneracy, not explained by MSG, of electronic bands in the conventional (reduced) BZ corresponding to the expanded magnetic unit cell. 
With first-principle calculations, we demonstrate that the antiferromagnetic material CoSO$_4$ exhibits this novel feature, and other materials with non-symmorphic SBZs are summarized in \AppSecMaterialTable.
Another exotic electronic feature is the effective $\pi$-flux induced by non-collinear magnetism, where the translation operators $\hat{t}_{i}$ ($i=1,2,3$) do not commute with each other, yielding the non-commuting SBZ.
We construct \blue{SSG} Bloch states and explore band theory in the folded SBZ, revealing that a single \blue{SSG} Bloch state can transform into a non-Bloch state under an SSG operation, leading to additional degeneracies.
Furthermore, SSGs enable a comprehensive characterization of the spin texture in the momentum space.
The theory of collinear SSGs fully classifies the possible types of alter-magnetism, and coplanar and non-coplanar SSGs generalize the alter-magnetism to a much boarder scope of non-collinear magnetic materials. 
Our generalizations and the identification of material candidates may boost the development of this flourishing field.

Given the enriched symmetry algebra, SSGs can protect novel topological phases.
We construct a $\mathbb{Z}_2$ TI without TRS and SOC. 
Different from previous ones~\cite{liu2022spin,herzog202}, this model does not require an external magnetic flux and hence is an intrinsic magnetic $\mathbb{Z}_2$ TI.
Moreover, we demonstrate that SSG can protect a four-fold band-crossing on the surface of a 3D $\mathbb{Z}_2$ TI and a helical edge mode between two domains with the same topological index. 
These unique features have no counterparts in the conventional magnetic topological phases.  

Our systematic investigations on SSGs may intrigue further research interests in topological states.
For example, the complete classification of SSGs can lead to the development of the complete representation theory of SSGs, as well as the topological quantum chemistry (TQC)~\cite{bradlyn_topological_2017} that can be used to diagnose topological insulators and topological semi-metals. 
Additionally, SSGs offer potential applications in magnon spectrum and magnon topological states.
Since magnons are spin-1 particles, their investigations do not require projective representations.
Thus, we can directly use the linear representation of the isomorphic MSG $\mathcal{M}$ of an SSG $\mathcal{G}$ to describe the magnon bands.
This comprehensive understanding of SSGs paves the way for generalizing the previous theory into magnon TQC~\cite{mcclarty22,Corticelli23}.

We compare our method of classifying SSGs to Litvin's approach introduced in Refs.~\cite{litvin74,litvin1977spin}. 
While these two methods appear different at first glance, they are equivalent for commensurate magnetic structures, except that we consider an additional equivalence relation between SSGs with different spiral angles that are continuously connected (Sec.~\ref{distinct_O3}).
This equivalence relation enables our theory to be applicable in incommensurate magnetic structures.
Litvin's approach involves the introduction of a normal subgroup $P_0=\{ \{R_g | \v_g \} | X_g U_g = I, g\in \mG \}$ of the parent space group $P$, consisting of pure spatial operations that leave the magnetic structure unchanged. 
Additionally, a supergroup $\mathcal{B} = \{ X_g U_g| g\in \mG \}$ of the pure-spin-operation group $\mS$ is introduced, containing the spin operation parts of all SSG operations. 
$P$ can be decomposed as cosets of $P_0$ 
\begin{equation}
    P = p_1 P_0 + p_2P_0 + \cdots + p_n P_0\ .
\end{equation}
The authors of Refs.~\cite{litvin74,litvin1977spin} proved the isomorphism $P/P_0 \simeq \mathcal{B}/\mS$. 
Distinct SSGs within the same family possess the same $P$ and $\mathcal{B}$, but differ in terms of $P_0$ and the isomorphisms between $P/P_0$ and $\mathcal{B}/\mS$.
We can show that the chosen $P_0$ and the isomorphisms can be interpreted as O($N$) representations of $P$, where $N=3,2,1$ for non-coplanar, coplanar, and collinear magnetic structures, respectively.
For non-coplanar magnetic structures, $\mS$ is trivial, and hence $\mathcal{B}=\{X_1U_1, X_2 U_2 \cdots X_n U_n\}$, as a discrete subgroup of O(3), is a point group (not necessarily a crystalline point group) represented by O(3) matrices. 
Therefore, an isomorphism between $P/P_0$ and $\mathcal{B}$, which associates each coset representative $p_i$ with an O(3) matrix $X_i U_i$, defining an O(3) representation of $P$, where $D(p) = X_i U_i$ for $p\in p_i P_0$. 
For the coplanar magnetic structures, any spin operation that leaves the moments in-plane is an O(2) rotation followed by an element in $\mS = \mS_{Z_2^T}$ [Eq.~(\ref{PSO2})], implying that $\mathcal{B}/\mS$ is a 2D point group represented by O(2) matrices.
Similarly, for the collinear magnetic structures, $\mathcal{B}/\mS$ is a subgroup of O(1).
Therefore, each isomorphism $P/P_0 \simeq \mathcal{B}/\mS$ defines an O($N$) representation of the parent space group $P$. 
The equivalence relations defined in Refs.~\cite{litvin74,litvin1977spin} are identical to the 
the first two equivalence relations among SSGs in this work (Sec.~\ref{distinct_O3}).
If two SSGs are related by a coordinate transformation in spin or real space, they are identified as equivalent. 
However, our third equivalence relation, {\it i.e.}, the equivalence between SSGs that can be deformed to each other by a continuous change of non-HSP momenta (spiral angle), is not considered in Litvin's approach. 
It is because, for commensurate magnetic structures, different spiral angles imply different magnetic unit cells and hence different $P_0$, resulting in different SSGs that are not naturally related in the language of Litvin's approach.
Considering the rational-number-valued spiral angles, Litvin's approach would give an infinite number of SSGs for commensurate magnetic structures.
Furthermore, Litvin's approach does not apply to incommensurate magnetic structures.

In the end, it is worth noting that the theory of SSG also applies to some magnetic systems with significant SOC, and the idea of SSG can be generalized to other types of symmetry breaking, such as pair-density wave states in superconductors and high-spin states in cold atom systems. 
Many intriguing physical systems have their symmetries in the form of the product between space group $G_{\rm latt}$ and 
internal symmetry group $G_{\rm int}$; $G_{\rm latt} \times G_{\rm int}$.
Here the internal symmetry group $G_{\rm int}$ can be either a discrete or a continuous group. 
For instance, the Kitaev spin model~\cite{kitaev06} and its generalizations~\cite{dong20,Jackeli09,price12,ziatdinov16,taddei23} have 
the symmetry of $G_{\rm latt} \times G_{\rm int}$ with $G_{\rm int} = D_2 \times Z_2^T \simeq Z_2^3$, due to their bond-dependent Ising couplings.
Note that the spin-spin interaction being intensively stronger along easy axes suggests a significant SOC.
A non-trivial subgroup of $G_{\rm latt} \times G_{\rm int}$ might characterize their symmetry-broken phases.
A similar analysis as for SSGs enables the classification of these symmetry groups by $H^1(P,Z_2^3)$, where $P$ is the parent space (plane) group in this case, and is generally a subgroup of $G_{\rm latt}$. 
$H^1(P,Z_2^3)$ is characterized by three independent real irreps of the group $P$.
Hence, symmetry groups given by $H^1(P,Z_2^3)$ form a {\it subset} of the SSGs with the same parent space group $P$, {\it i.e.}, type-I, II, III, and IX SSGs where only 1D real irreps are involved.
The knowledge of SSGs can be directly applied to studies of the spectral and dynamical properties in their symmetry-broken phases.  
From this illustrative example, we conclude that, in general, a subset of SSGs given by $H^1(P,\mathrm{O(3)})$ are still valid even in the presence of significant SOC - as long as the many-body Hamiltonian respects a nontrivial $G_{\rm int} \subsetneq \mathrm{O(3)}$. 
In pair-density wave states~\cite{agterberg20} and spin-$\frac32$ cold atom systems \cite{Wu2003_SO5}, $G_{\rm int}=\mathrm{U(1)}$ and SO(5), respectively, and systematic understandings of symmetries of order parameters can be achieved by a similar scheme.

{\it Note added.}
After completing the classification of SSGs, we learned that related works were carried out by Chen Fang's \cite{fang2023_ssg} and Qi-Hang Liu's \cite{liu2023_ssg} groups, \blue{employing a method similar to Litvin's approach.}

\acknowledgements
We are grateful to Jian Yang, Chen Fang, Qi-Hang Liu, Yuan Li, and Zhijun Wang for useful discussions. 
Z.-D. S. and Y.-Q. L. were supported by National Natural Science Foundation of China (General Program No. 12274005), Innovation Program for Quantum Science and Technology (No. 2021ZD0302403), National Key Research and Development Program of China (No. 2021YFA1401900). 
Z. X. and R.S. were supported by the National Basic Research Programs of China (No.~2019YFA0308401) and by National Natural Science Foundation of China (No.~11674011 and No.~12074008).
The computational work was performed on the resources of the Platform for Data-Driven Computational Materials Discovery, Songshan Lake Materials Laboratory.

\bibliography{ref}

\begin{thebibliography}{95}
\bibitem [{dat()}]{data}%
  \BibitemOpen
  \href@noop {} {}\bibinfo {note} {See the text files at [URL will be inserted
  by publisher].}\BibitemShut {Stop}%
\end{thebibliography}

\def\MAGNDATA{\href{http://webbdcrista1.ehu.es/magndata/}{MAGNDATA}}
\def\CmdDistinctO{II D in the main text}
\def\RefCoordGroup{(17) in the main text}
\def\RefTranRep{(18) in the main text}
\def\RefSecExampleP{II C in the main text}
\def\RefDTReP{(10) in the main text}
\def\RefSecSpinTexture{III D in the main text}
\def\RefSecPNSSGs{III C in the main text}
\def\RefGenericH{III A in the main text}
\def\RefPSO{(5) in the main text}
\def\RefSumClass{II E in the main text}
\def\RefSecNoncommutingSBZ{III C 2 in the main text}
\def\RefClassGeneral{II A in the main text}
\def\RefEqBlochNoncommuting{(37) in the main text}

\def\pare#1{\left( #1 \right)}
\def\brak#1{\left[#1\right]}
\def\brace#1{\left\{#1\right\}}

\clearpage

\appendix

\section{Representation theory for space groups}
\label{App_irrep}
\label{app_irrep_sg}

\subsection*{Three types of irreducible representations} 

An irreducible representation (irrep) $D(g)$ of a group $G$ must fall under one of the following categories: real irreps, complex irreps, and pseudo-real irreps. $D(g)$ is a real irrep, if $D(g)$ is equivalent to its complex conjugate, given by
\begin{equation}
    D(g)^* = \V D(g) \V^{\dagger} \, ,
\end{equation}
and the unitary matrix $\V$ satisfies $\V = \V^T$. 
If $\V$ is a symmetric matrix, $(\V^{-1/2})$ is also a symmetric matrix. 
By performing a unitary transformation with $(\V^{-1/2})^*$, we find that $D^{\prime}(g) = (\V^{-1/2})^* D(g)  (\V^{-1/2})^T$ satisfy that
\begin{equation}
    \begin{aligned}
         D^{\prime}(g)^{*}& = \V^{-1/2} D(g)^*  (\V^{-1/2})^{\dagger}  \\
        & = (\V^{-1/2}) \V D(g) \V^{\dagger}  (\V^{-1/2})^{\dagger} \\
        & = (\V^{-1/2}) \V  (\V^{-1/2})^T D^{\prime}(g) (\V^{-1/2})^{*} \V^{\dagger}  (\V^{-1/2})^{\dagger} \\ 
        & =  D^{\prime}(g) \quad \forall g \in G
\label{real_irrep}
    \end{aligned}
\end{equation}
This implies that all elements of the real irrep can be made real by properly choosing a basis. $D(g)$ is a pesudo-real irrep, if $D(g)^* = \V D(g) \V^{\dagger}$, and $\V$ satisfies $ \V = -\V^T$.  As $\V$ is an anti-symmetric matrix, its dimension must be even, and consequently, the dimension of $D(g)$ must be even. 
(Suppose the dimension of $\V$ is odd, then $\V = - \V^T$ leads to $\det V = - \det V$ and hence $\det \V=0$, contradicts with the unitary condition of $\V$.)
Note that $\V^{1/2}$ is neither symmetric nor anti-symmetric. 
If $\V^{1/2}$ was symmetric or anti-symmetric, $\V = \V^{1/2} \V^{1/2}$ should be symmetric, which leads to a contradiction. 
Thus, we cannot choose a basis in which all elements of $D(g)$ are real as in Eq.~(\ref{real_irrep}), 
even though $D(g)$ is equivalent to its complex conjugate. 
If $D(g)$ is not equivalent to its complex conjugate, it is a complex irrep. Under any basis, some elements of $D(g)$ must be complex.

A direct sum of an irrep and its complex conjugate can be made real. Let $D_{ir}(g)$ be a complex or pseudo-real irrep, and let $D_{r}(g)$ be the direct sum of $D_{ir}(g)$ and its complex conjugate:
\begin{equation}
D_{r}(g)= \begin{pmatrix}
D_{ir}(g) & 0 \\
0 & D_{ir}(g)^*
\end{pmatrix} \, .
\end{equation}
Note that if $D_{ir}(g)$ is pseudo-real, $D_{r}(g)$ is a direct sum of two equivalent pseudo-real irreps.  
The matrix $D_r(g)$ satisfies that 
\begin{equation}
    D_{r}(g)^* = \sigma_x D_{r}(g) \sigma_x \, \forall g \in G \, ,
\end{equation}
and $\sigma_x$ is a symmetric matrix. 
Thus, with unitary transformation $ \sigma_x^{1/2}$, we can make $D_{r}(g)$ a real matrix as in Eq.~(\ref{real_irrep}).

We have shown that the representation matrices of a single complex irrep or a single pseudo-real irrep are necessarily complex in any basis, while the representation matrices of a conjugate pair of complex irreps or a pair of equivalent pseudo-real irreps can be made real. 
In summary, a real representation must be decomposed into real irreps, conjugate pairs of complex irreps, or pairs of equivalent pseudo-real irreps.

\subsection*{Irreducible representations of the little groups of space groups}

To study the irreps of a space group $P$, we first consider its translation subgroup $T$, given as 
\begin{equation}
    T = \brace{ \{ 1 | \v \} \big | \{ 1 | \v \} \in P } \, ,
\end{equation}
and irreps of $T$. 
Since $T$ is an Abliean group, it can only possess 1D irreps. 
An irrep ${\bm \Gamma}^{k}$ of $T$ is characterized by a vector $\k$ in the BZ, and the representation matrix $D_{\bm \Gamma^{k}}( \{1 | \v \}) = e^{i \k \cdot \v}$. 
Since $\k \cdot \v$ is a phase factor, two vectors $\k_1$ and $\k_2$ in the BZ are considered equivalent ($\k_1 \equiv \k_2$) if they satisfy the relation
\begin{equation}
\k_1 = \k_2 + \sum_{i=1}^3 m_i \b_i, \quad m_i \in \mathbb{Z},
\end{equation}
where $\b_i$ represents reciprocal lattice vectors, and $\b_i$ and the primitive lattice vectors $\a_i$ of $T$ satisfy $\a_i \cdot \b_j = 2\pi \delta_{i,j}$ for $i,j=1,2,3$. 

Next, let us consider the little group $P^{\k}$ of $\k$ and its allowable irreps $\rho^{\bm k}$. 
The little group $P^{\k}$ of $\k$ is defined as 
\begin{equation}
    P^{\k} = \brace{  \{\R | \v \} \big | \{\R | \v \} \in P , \, \R \k \equiv \k } \, .
\end{equation}
Based on this definition, the translation group $T$ is a subgroup of $P^{\k}$, and we can decompose $P^{\k}$ into cosets with respect to $T$ %
\begin{equation}
    P^{\k} = \{ R_1 | \v_1 \} T + \{ R_2 | \v_2 \} T  + \cdots +  \{ R_m | \v_m \} T\, .
\end{equation}
The point-group parts $R_1, R_2, \cdots, R_m$ must be distinct from 
one another %
and form a point group, which is referred to as the little co-group $\bar{P}^{\k}$.
An allowable irrep $\rho_{\k}$ of $P^{\k}$ satisfies 
$D_{\rho_{\k}} (\{ 1 | \v\}) = e^{ i \k \cdot \v} 1_{r \times r}$, where $r$ is the 
dimension of  $\rho_{\k}$. All the allowable irreps of $P^{\k}$ can be  
obtained from (projective) irreps of the little co-group $\bar{P}^{\k}$. To see this, let us consider the representation matrices of the coset representatives $\{ R_i | \v_i \}$ ($i = 1,2,\cdots,m$) and their multiplication 
\begin{equation}
    D_{\rho_{\k}}(\{ R_i | \v_i \} )  D_{\rho_{\k}}(\{ R_j | \v_j \} ) =  D_{\rho_{\k}}(\{ R_i R_j| R_i \v_j + \v_i \} ) \, .
\end{equation}
Suppose $R_i R_j$ is identical to $ R_{f(i,j)}$, where $f(i,j) \in \{ 1,2,\cdots,m\}$. Then, we have
\begin{equation}
\begin{aligned}
     & D_{\rho_{\k}}(\{ R_i | \v_i \} )  D_{\rho_{\k}}(\{ R_j | \v_j \} ) \\
     = &D_{\rho_{\k}}(\{ 1| R_i \v_j + \v_i -  \v_{f(i,j)} \}) D_{\rho_{\k}}(\{ R_{f(i,j)}| \v_{f(i,j)} \} ) \\
     = & e^{i \k \cdot \left( R_i \v_j + \v_i -  \v_{f(i,j)} \right)}  D_{\rho_{\k}}(\{ R_{f(i,j)}| \v_{f(i,j)} \} ) \, .
     \label{app_proj_1}
\end{aligned}
\end{equation}
Thus, the allowable irrep $ D_{\rho_{\k}}( \{ R_i| v_i \})$ can also be understood as a projective irrep of the point-group operation $R_i \in \bar{P}^{\k}$ with a factor system given by $e^{i\k \cdot \left( R_i \v_j + \v_i - \v_{f(i,j)} \right)}$. %
In the other way, all allowable irreps of the little group $P^{\k}$ can be obtained 
from the %
irrep of the little co-group $\bar{P}^{\k}$ with the projective phase given by Eq.~(\ref{app_proj_1}).

\section{Equivalence between \texorpdfstring{O($N$)}{O(N)} representations}
\label{app sec: equiv rep}
\label{app eq sub1}
\label{app eq sub2}
\label{app eq sub3}
\label{App_unit_cell}

In this section, we present some technique details for utilizing equivalence relationship between O($N$) representations. 

\subsection*{Choices of unit cells in different space groups}

We begin by discussing the restrictions imposed by the Bravais lattice setting on the lattice vector transformation matrix $V$.  
Subsequently, we present an algorithm for enumerating all possible choices of the unit cell for different space groups.

Suppose the lattice basis for a space group is $\a_i$ ($i=1,2,3$), and a transformation matrix $V$ satisfies that $V \a_i = \sum_{j=1}^3 \a_j M_{ji}$.
Then, on the basis $\a_i$'s, the matrix $V$ should be represented as $M$.
As discussed in Sec.~\CmdDistinctO, for any space group, $M$ must satisfy that $\det M = 1$ and $M_{ij} \in \mathbb{Z}$. 
The Bravais lattice should remain unchanged under the action of $M$.
It implies that the transformed unit cell satisfies the same restrictions regarding the interaxial angles $\alpha, \beta, \gamma$, the relative relationship of the lengths $a, b, c$ of the lattice vectors, and the choice of rotation axes as the original unit cell. 
The requirements for each lattice system are summarized below:
\begin{itemize}
    \item Triclinic lattice: The only requirements for $M$ are that $\det M = 1$ and $M_{ij} \in \mathbb{Z}$, as the interaxial angles $\alpha, \beta, \gamma$ and lattice lengths $a, b, c$ are all distinct.
    
    \item Monoclinic lattice: In this case, $\a_2$ represents the unique 2-fold rotation axis, and $\alpha = \gamma = \pi/2$. As a result, $M$ can be block diagonalized into a transformation $M_{xz} = \begin{pmatrix} M_{11} & M_{13} \\ M_{31} & M_{33} \\ \end{pmatrix}$ within the $\a_1, \a_3$ plane and a transformation along $\a_2$. $V \a_2 = M_{22}\a_2$ can be either $\a_2$ or $-\a_2$. As $ \beta$ does not necessarily equal $\pi/2$ and $ a$ does not necessarily equal $b$, they do not impose any restrictions on $M_{xz}$. 
    As $ \det M = 1$, if $M_{22} = 1$ ($-1$), $M_{xz}$ is an integer matrix with determinant $1$ ($-1$).

    \item Orthorhombic lattice: $\a_1, \a_2 , \a_3$ are all two-fold rotation axes. 
    Consequently, $V \a_i$ ($ i = 1,2,3$) can be any of $\pm \a_1, \pm \a_2 , \pm \a_3$. 
    The condition $\alpha = \beta = \gamma = \pi/2$ requires that $M$ be an orthogonal matrix. 
    Thus, $M$ must be a rotation in the octahedral point group $O$.

    \item Tetragonal lattice: The condition $\alpha = \beta = \gamma = \pi/2$ necessitates $M$ being an orthogonal matrix. Furthermore, $M \a_3$ should be parallel to $\a_3$, which is the unique 4-fold rotation axis. 
    $V \a_1$ can only be one of $\pm \a_1, \pm \a_2$. Therefore, $V$ (and $M$) must be a rotation in the dihedral group $D_4$.

    \item Hexagonal lattice: $\a_3$ represents the unique 6-fold or 3-fold rotation axis, and $\alpha = \gamma = \pi/2$. 
    Consequently, $V \a_3$ should be parallel to $\a_3$, and $V \a_1$ and $V \a_2$ should remain within the same plane as $\a_1$ and $\a_2$. 
    Additionally, $V \a_{1,2}$ still represents a lattice translation and can only be one of $\pm \a_1$, $\pm \a_2$, and $\pm (\a_1 + \a_2)$. 
    Therefore, $V$ must correspond to an operation in the $D_6$ group.

    \item Rhombohedral lattice: $a = b = c$, $ \alpha = \beta = \gamma$, $\a_1 + \a_2 + \a_3$ is the unique rotation axis. 
    $V \a_i$ ($ i = 1,2,3$) can only be one of $\pm \a_1, \pm \a_2, \pm \a_3$. Once $V \a_1$ is specified, $V \a_2$ and $V \a_3$ are also uniquely determined. Such $V$'s form the dihedral group $D_3$.

    \item For cubic lattice, $\a_1, \a_2 , \a_3$ are all four-fold rotation axes. 
    Consequently, $V \a_i$ ($ i = 1,2,3$) can be any of $\pm \a_1, \pm \a_2 , \pm \a_3$. 
    The condition $\alpha = \beta = \gamma = \pi/2$ requires $V$ to be an orthogonal matrix. 
    Thus, $V$ (and $M$) must be a rotation in the octahedral point group O.
\end{itemize}

In the following, we will introduce the algorithm for enumerating all possible choices of the unit cell for a given space group. 
We note that according to Eq.~\RefCoordGroup, the product of two allowable coordinate transformations is still allowable. 
Thus, all allowable coordinate transformations $\{V| \t \}$ form a group $P_C$, which generally has infinite order, and we only need to enumerate their generators. 

$P_C$ can be decomposed as $P_C = T_C \rtimes V_C$, where $T_C$ contains all allowable origin shift $\t$, and $V_C$ contains rotations of the lattice axes. 
We only need to enumerate elements in $T_C$ and the coset representatives of $P_C/T_C$.

Let us first consider $T_C$. 
If $\t$ is a lattice vector, it must produce an automorphism [Eq.~\RefCoordGroup], but it will not transform an irrep into a different one.
Consider an irrep $\rho_{\k} \uparrow P$ of space group $P$, which is induced by an irrep $\rho_{\k}$ of the little group $P^{\k}$ of vector $\k$. 
We investigate how irrep $\rho_{\k}$ of the little group is transformed under an origin shift $\t$, where $\t$ is a lattice vector.
Note that $\{1 | \t \} \in \rho_{\k}$ and $\rho_{\k}( \{1 | \t \}) = e^{i \k \cdot t} 1_{m_{\k} \times m_{\k}} $, where $m_{\k}$ is the dimension of $\rho_{\k}$.
According to Eq.~\RefTranRep, the transformed irrep $\rho_{\k}^{\prime}(g)$ ($g \in P^{\k}$) is given as 
\begin{equation} 
\begin{aligned} 
    \rho_{\k}^{\prime}(g) & = \rho_{\k}(\{1 | \t \} g \{1 | \t \}^{-1} ) \\ 
    & = \rho_{\k}(\{1 | \t \}) \rho_{\k}(g)  \rho_{\k}(\{1 | \t \}^{-1}) \\ 
    & =  e^{i \k \cdot t} 1_{m_{\k} \times m_{\k}} \rho_{\k}(g)   e^{-i \k \cdot t} 1_{m_{\k} \times m_{\k}} \\
    & = \rho_{\k}(g) \, .
\end{aligned}
\end{equation}
$\rho_{\k}^{\prime}$ is the same little group irrep as $\rho^{\k}$, and therefore, they induce the same irrep of the space group $P$. 
Thus, we only need to consider $\t$ within the primitive unit cell. 
According to Eq.~\RefCoordGroup, allowable $\t$ satisfies the condition:
\begin{equation}
    R_p \t - \t = \sum_{i = 1}^{3} m_i \a_i \, (m_i \in \mathbb{Z}) \, ,  \forall p \in P \, , 
\label{app_coord_group}
\end{equation}
where $R_p$ is the point-group part of operation $g$. 
It implies that the new origin must have all point-group symmetries of this space group, and $\t$ is solely determined by $R_g$.
In some groups, $\t$ can only take discrete groups, such as groups with inversion symmetry.
Eq.~(\ref{app_coord_group}) implies that the shifted origin should be an inversion-invariant point. 
In other groups, $\t$ can take continuous values.
For example, in group P1 (No.~1), $\t$ can be an arbitrary vector in the whole space since $R_p$ is always $1$. 
In such cases, we consider a subgroup $T_{C0}$ of $T_C$, in which $ R_p \t - \t = 0, \forall p \in P $. 
$T_{C0}$ contains allowable origin shift that can be continuously connected into an identity operation. 
If $\{1 | \t  \} \in T_{C0}$, $\{ 1  | \t \}^{-1} \{ \R | \v \} \{ 1 | \t \} = \{ \R | \v + (\R - 1)\t \}  =  \{ \R | \v \}$.
An origin shift in $T_{C0}$ will not transform a space-group operation or an irrep into a different one. 
Therefore, we only need to consider the effect of coset representatives of $T_C/ T_{C0}$ on the irreps. 
$T_C/T_{C0}$ only takes discrete values and only contains a finite number of $\t$ in the primitive unit cell. 
Thus, for each group, we can obtain a finite number of allowable $\t$, and these $\{1|\t\}$ serve as one part of generators of the allowable coordinate transformation.

To study $V_C$, we proceed as follows. 
In five lattice systems except for triclinic and monoclinic lattice, the number of choices of $V$ is finite due to the restrictions imposed by Bravais lattices, as discussed earlier in this section.
For each space group $P$ in these lattice systems and for each $V$, we perform a numerical search for a $\t$ within the unit cell that satisfies Eq.~\RefCoordGroup. 
If such a $\t_{V}$ exists, we include $\{V | \t_{V}\}$ into the coset representatives of $P_C/T_C$. 
For space groups ($P1$ and $P\bar{1}$) in triclinic lattices, all matrices $M \in SL_3(\mathbb{Z})$ are compatible with Bravais lattices ($M$ is the representation matrix of $V$ on the lattice basis). 
Using Eq.~\RefCoordGroup, one can check that in space groups $P1$ and $P\bar1$, for arbitrary $M \in SL_3 (\mathbb{Z})$ $\{M| \0\}$ is an allowable coordinate transformation, implying that the all the coset representatives of $P_C/T_C$ can always be chosen to have the form $\{M|\0\}$. 
Thus, we can choose generators (for example, see Ref.~\cite{conder92}) of the group $SL_3(\mathbb{Z})$ as coset representatives $P_C/T_C$.

We need to apply a more careful treatment for space groups in the monoclinic lattice systems.
As discussed earlier, due to the restriction on Bravais lattice, 
$M$ belongs to one of these two sets 
{\small
\begin{equation} 
    S_I  = \brace{ \begin{pmatrix}
    M_{11} & 0 & M_{13} \\
     0  &  1 & 0 \\
    M_{31} & 0 & M_{33}
  \end{pmatrix} \,\middle\vert\,
M_{xz} = \begin{pmatrix} 
  M_{11} & M_{13} \\
  M_{31} & M_{33} \\
\end{pmatrix}  \in SL_2(\mathbb{Z})}
\label{S_I}
\end{equation} }%
or 
\begin{equation} 
M_{\hx}(\pi) S_I \quad \text{with} \quad M_{\hx}(\pi) = 
\begin{pmatrix}
    1 & 0 & 0 \\
     0  &  -1 & 0 \\
    0 & 0 & -1
  \end{pmatrix} \, .
\end{equation} 
The generators of $SL_2(\mathbb{Z})$ are given as 
\begin{equation} 
  \begin{pmatrix} 
    1 & 1 \\
    0 & 1
  \end{pmatrix} ,  \,
  \begin{pmatrix} 
    0 & 1 \\
    -1 & 0
  \end{pmatrix} \, .
  \label{gen_SL2Z}
\end{equation} 
We note that $\{ M_{\hx}(\pi) | \0\}$ that reverses the $y,z$ directions is an allowable coordinate transformation for all space groups with Monoclinic lattice, and hence it serves as a coset representative of $P_C/T_C$.
While the number of transformation matrices in $S_I$ is infinite, we can generate them successively by the generators [Eq.~(\ref{gen_SL2Z})]. 
For each generated $M_I$ and the corresponding $V_I$, we perform a numerical search within the unit cell to find a $\t_{V}$ that satisfies Eq.~\RefCoordGroup.
We add $\{V_I | \t_{V_I} \}$ to the coset representatives of $P_C/T_C$ if such a $\t_{V_I}$ exists.  
This process continues until the newly generated $V_I$, which is determined as allowable, can be generated by those already in the set of coset representatives. 

In summary, the methods described above allow us to obtain the generators of allowable coordinate transformations for every space group.
These generators consist of generators for origin shifts and generators for changing the axes of the unit cell.

\subsection*{Example: type-V O(3) representations of space group \texorpdfstring{$P3$}{P3}}

As established in Sec.~\CmdDistinctO, O(3) representations of different types do not deform into each other under any coordinate transformation.
It enables us to study different types of representations separately. 
In this section, we will focus on the transformation of type-V O(3) representations for the space group $P3$ as an illustrative example.
As shown in Sec.~\RefSecExampleP, we need to consider six representations DT$i$DU$i\oplus$1, and P$i$PC$i\oplus$1 ($i=1,2,3$), where $1$ represents the identity irrep.
The explicit representation matrices are given in Eq.~\RefDTReP. 
The identity irrep is always invariant under any coordinate transformation. 
DT$i$DU$i$ (P$i$PC$i$) is the direct sum of DT$i$ (P$i$) and its complex conjugate.
Thus, it is sufficient to consider the transformation of the complex irreps DT$i$ and P$i$. 
The generators of allowable coordinate transformation consist of the generator of $D_6$ group, given as $\{ 6_{001}|\0 \}$, $\{2_{100} | \0\}$,as well as the generators of origin shift, given as $\{1|1/3,2/3, 0\}$, $\{1|2/3,1/3,0\}$, $\{1|0,0,z\}$. 

According to Eq.~\RefTranRep, the momentum $\k$ of an irrep will be transformed to $V^{-1}\k$ under the action of $\{V| \t\}$. 
The line DT ($\k = (0,0,v \pi)$) is invariant under the action of $V = 6_{001}$, $2_{100}$, $1$.
Similarly, the line P ($\k = (2\pi/3,2\pi/3,u \pi)$) is invariant under the action of $V = 2_{100}$ and $1$, and it is transformed to its TRS-partner line PC ($\k = (-2\pi/3,-2\pi/3,-v\pi)$) under $V = 6_{001}$.
Therefore, DT$i$'s and P$i$'s will not be transformed into each other.
In table~\ref{tab: trans C3z}, $\rho$ and $\sigma$ denote the irreps before and after the coordinate, respectively, and we show how $D_{\sigma} (\{ 3^+_{001}| \0 \})$ is determined by $\rho$ and coordinate transformation $\{  V | \t \}$. 
Only $\{2_{100} | \0\}$ has a non-trivial effect on DT$i$'s.
It transforms DT2 and DT3 into each other while leaving DT1 invariant. 
We have also shown in Sec.~\RefSecExampleP\ that the real representation induced by DT2 and DT3 should be put in the same class because they can be continuously deformed to each other. 
Thus, DT1DU1 belongs to one class, and DT2DU2 and DT3DU3 are in the other class. 
For the irreps on the line P, we observe that $D_{{\rm P}i}(C_{3z})$ transforms to $e^{i2\pi/3}D_{{\rm P}i}(C_{3z})$ and $e^{i4\pi/3}D_{{\rm P}i}(C_{3z})$ under the action of $\{1|2/3,1/3,0\}$ and $\{1|1/3,2/3,0\}$, respectively (see the fourth and fifth rows of Table~\ref{tab: trans C3z}). 
This implies that irrep P1 can be transformed into irreps P2 and P3, and hence these three irreps belong to the same class.

\begin{table}[h]
\caption[The transformation of $C_{3z}$ in the space group $P3$ (No.~143) under coordinate transformation.]{The transformation of $C_{3z}$ under coordinate transformation $\{ V| \t\}$. The irrep before the transformation is denoted as $\rho$ ($\rho$ = DT$i$, P$i$, $i=1,2,3$) with momentum $\k$, and $\sigma$ denotes the irrep after the action of $\{ V|\t\}$.
    }
    \begin{tabular}{cc|c}
        \hline \hline
        $ \{ V|\t\}$ & $ \{ V|\t\}\{ 3^+_{001}| \0 \} \{ V|\t\}^{-1}$ & $D_{\sigma}(C_{3z})$ \\ \hline
        $\{ 6_{001}|\0 \}$ & $ \{ 3^+_{001}| \0 \}$ &  $D_{\rho}(C_{3z})$\\
        $\{2_{100} | \0\}$ &  $ \{ 3^-_{001}| \0 \}$ & $\left( D_{\rho}(C_{3z}) \right)^{-1}$\\
        $\{1|1/3,2/3, 0\}$ & $ \{ 3^+_{001}| 1,1,0 \}$  &$e^{ik_1 + ik_2}D_{\rho}(C_{3z}) $\\
        $\{1|2/3,1/3,0\}$ & $ \{ 3^+_{001}| 1,0,0 \}$ & $e^{ik_1}D_{\rho}(C_{3z}) $ \\
        $\{1|0,0,z\}$ &  $\{ 3_{001}| \0 \}$ & $D_{\rho}(C_{3z})$\\
        \hline  \hline
    \end{tabular}
    \label{tab: trans C3z}
\end{table}

\subsection*{Details for the equivalence based on continuously connected wave vectors}

To fully utilize the third equivalence based on continuously connected wave vectors, we can follow three steps.
Firstly, if irreps are on the same high-symmetry line or plane or generic points, and they are labeled by the same symbol ({\it e.g.,} DT1 of $P3$ on different points on the DT line), they should be categorized into the same class. 
Secondly, a 2D real representation induced from non-HSP $\k$'s, which can be either $[\rho_{\k}\uparrow P]^{c}_1 \oplus [\rho_{\k}^* \uparrow P]^{c}_1$ or $[\rho_{\k}\uparrow P]^{r}_2$, should contain a pair of momenta $\pm \k$.
If $\pm \k$ belongs to the same connected piece of the fix-point manifold of $P^{\k}$, they can be continuously moved to each other while remaining on this manifold.
It is necessary to investigate whether this 2D real representation will be deformed into another one when the positions of $\pm \k$ are switched. For example, we find that ${\rm DT2DU2}$ of $P3$ can be deformed into ${\rm DT2DU3}$.
Thirdly, suppose that the fixed point of $P^{\k}$ with $\k$ being a non-HSP is a $n$-dimensional ($n > 0$) manifold. 
Momentum in this manifold can continuously move within this manifold along the $n$ independent tangent vectors and will return to its original position after completing a full winding around the first BZ.
We need to study whether a 2D real representation will be deformed to another one after it momenta $\pm \k$ go through the process together, with $\pm \k$ moving in opposite directions.

\subsection*{Algorithm of modulo equivalences}

To classify O($N$) representations into different classes by equivalence relationship, the following algorithm can be utilized.
This algorithm takes advantage of the transitivity property of the equivalence relation, and we only need to consider the generators of these equivalences.
We start by placing the inequivalent O($N$) representations in a given type into separate sets.
Then, we successively apply the generators of the second and third equivalences to these representations. 
If an O($N$) representation in one set can be transformed into a different representation in another set based on equivalence, we merge these two sets.
Repeat the above process iteratively for all the generators.
By following this algorithm, representations within the same set belong to the same class, indicating an equivalence, while those in different sets describe distinct SSGs.

\section{Electronic bands in SSG}
\label{App Sec: d_M}
\label{SNF}
\label{app sec: non-commuting SBZ}

\subsection*{\texorpdfstring{$d_{\rm SBZ}$}{dM} in commuting SBZ}

In this section, we study how SSG symmetries determine the dimension $d_{\rm SBZ}$ of the span of $\vS(\kk)$ in SBZ. 
We have shown how to determine $d_{\rm SBZ}$ for collinear SSGs (Sec.~\RefSecSpinTexture), and we show the details on non-collinear SSGs where unitary translation generators $\hat{t}_{1,2,3}$ all commute with each other.  
As established in Sec.~\RefSecSpinTexture, $d_{\rm SBZ}$ are only determined by spin-rotation translations, generated by $\hat{t}_{1,2,3}$, and spin-rotation $\mathcal{PT}$ operation, denoted as $\hat{\mathcal{P}}\hat{\T}$.
If $\hat{\mathcal{P}}\hat{\T}$ is present and $\boldsymbol{q}_{\mathcal{P}{\T}}=0$, $d_{\rm SBZ}=\mathrm{dim}[ \bigcap_{i=1}^3 \ker( \U_{t_i} - I) \bigcap \ker(\U_{\mathcal{PT}} + I )]$; $d_{\rm SBZ}=\mathrm{dim}[ \bigcap_{i=1}^3 \ker( \U_{t_i} - I)]$ otherwise.

First, we will prove that if any of $\hat{t}_{i}$'s are accompanied by non-trivial spin rotations, they must share the same rotation axis. Without loss of generality, suppose that $\U_{t_i} = U_{\n_i} (\theta_i)$ ($i = 1,2$) with $\theta_i  \in (0,2\pi)$. 
In SSGs with commuting SBZ, $[\hat{t}_1, \hat{t}_2] = 0$ requires that $[e^{i \frac{\theta_1}{2} \n_1 \cdot {\bm \sigma}},e^{i \frac{\theta_2}{2} \n_2 \cdot {\bm \sigma}}] = 0$. 
After some simple algebra, the equation is equivalent to $ (\n_1 \times \n_2  ) \cdot {\bm \sigma} = 0$, where ``$ \times$'' denotes the cross product. $\n_1$ should be parallel to $\n_2$.  
Thus, if any of $U_{t_i}$ does not equal identity, $\mathrm{dim}[ \bigcap_{i=1}^3 \ker( \U_{t_i} - I)]  = 1$.

Next, we investigate SSGs with $\mathcal{PT}$ symmetry and a non-trivial $U_{t_i}$. 
As the square of $\mathcal{P}\mathcal{T}$ operation equals the identity, $U_{\mathcal{PT}}^2 = I$, and $U_{\mathcal{PT}}$ can be either $I$ or a $\pi$-rotation $U_{\bm m}(\pi)$. 
The group structure constrains the relationship between $U_{\mathcal{PT}} $ and $U_{t_i} = U_{\n} (\theta_i)$. 
As the conjugate of $\hat{t_i}$ under $\hat{\mathcal{P}}\hat{T}$ is its inverse, 
\begin{equation}
    U_{\mathcal{PT}}^{-1} U_{\n} (\theta_i) U_{\mathcal{PT}} = U_{\n} (-\theta_i).
    \label{app eq: PT and t}
\end{equation}
Case (i): $U_{\mathcal{PT}} = I $. A non-zero $\theta_i$ must equal $\pi$. We notice that 
\begin{equation}
    i \sigma_y \mathcal{K} e^{i \frac{\pi}{2} \n \cdot {\bm \sigma} } \left( i \sigma_y \mathcal{K} \right)^{-1} =   -e^{-i \frac{\pi}{2} \n \cdot {\bm \sigma} }  \ .
\end{equation}
Thus, in such a case, the SBZ is nonsymmorphic ($\qq_{\mathcal{PT}} \neq 0$), and $d_{\rm SBZ}$ is only determined by $U_{t_i}$, and still equals $1$.   
Case (ii): $U_{\mathcal{PT}} = U_{\bm m}(\pi)$. 
Eq.~(\ref{app eq: PT and t}) implies that $U_{U_{\bm m}(\pi)\n}(\theta_i) = U_{\n}(-\theta_i)$. 
Therefore, $U_{\bm m}(\pi) \n$ can equal either $\n$ or $-\n$. 
If $U_{\bm m}(\pi) \n = -\n$, which means that ${\bm m} \cdot \n =0$  , $\theta_i$'s can take arbitrary values.  
One can verify that if ${\bm m} \cdot \n =0$, 
\begin{equation}
     i \sigma_y \mathcal{K}   e^{i \frac{\pi}{2} {\bm m} \cdot {\bm \sigma} } e^{i \frac{\theta_i}{2} \n \cdot {\bm \sigma} } e^{-i \frac{\pi}{2} {\bm m} \cdot {\bm \sigma} } \left( i \sigma_y \mathcal{K} \right)^{-1} =   e^{-i \frac{\theta_i}{2} \n \cdot {\bm \sigma} }  \ .
\end{equation}
Thus, in such a case, the SBZ is symmorphic.
$\mathcal{P}\T$ constraints $\S(\kk)$ in a 2D plane perpendicular to ${\bm m}$, ${t}_i$ constraint $\S(\kk)$ parallel to $\bm n$, which is a subspace of the 2D plane, resulting $d_{\rm SBZ} = 1$. 
If $U_{\bm m}(\pi) \n = \n$, which means that ${\bm m} = \n $, non-zero $\theta_i$'s can only be $\pi$.  
One can verify that 
\begin{equation}
    i \sigma_y \mathcal{K}  e^{i \frac{\pi}{2} {\n } \cdot {\bm \sigma} }
     e^{i \frac{\pi}{2} {\n } \cdot {\bm \sigma} } 
      e^{-i \frac{\pi}{2} {\n } \cdot {\bm \sigma} } 
    \left( i \sigma_y \mathcal{K} \right)^{-1} =   -e^{-i \frac{\pi}{2} \n \cdot {\bm \sigma} }  \ .
\end{equation}
Thus, the SBZ is nonsymmorphic with this $\U_{\mathcal{PT}}$, and $d_{\rm SBZ}$ is only determined by $U_{t_i}$ and equals $1$.

\begin{figure}[htb]
    \centering
    \includegraphics[width = 1 \linewidth]{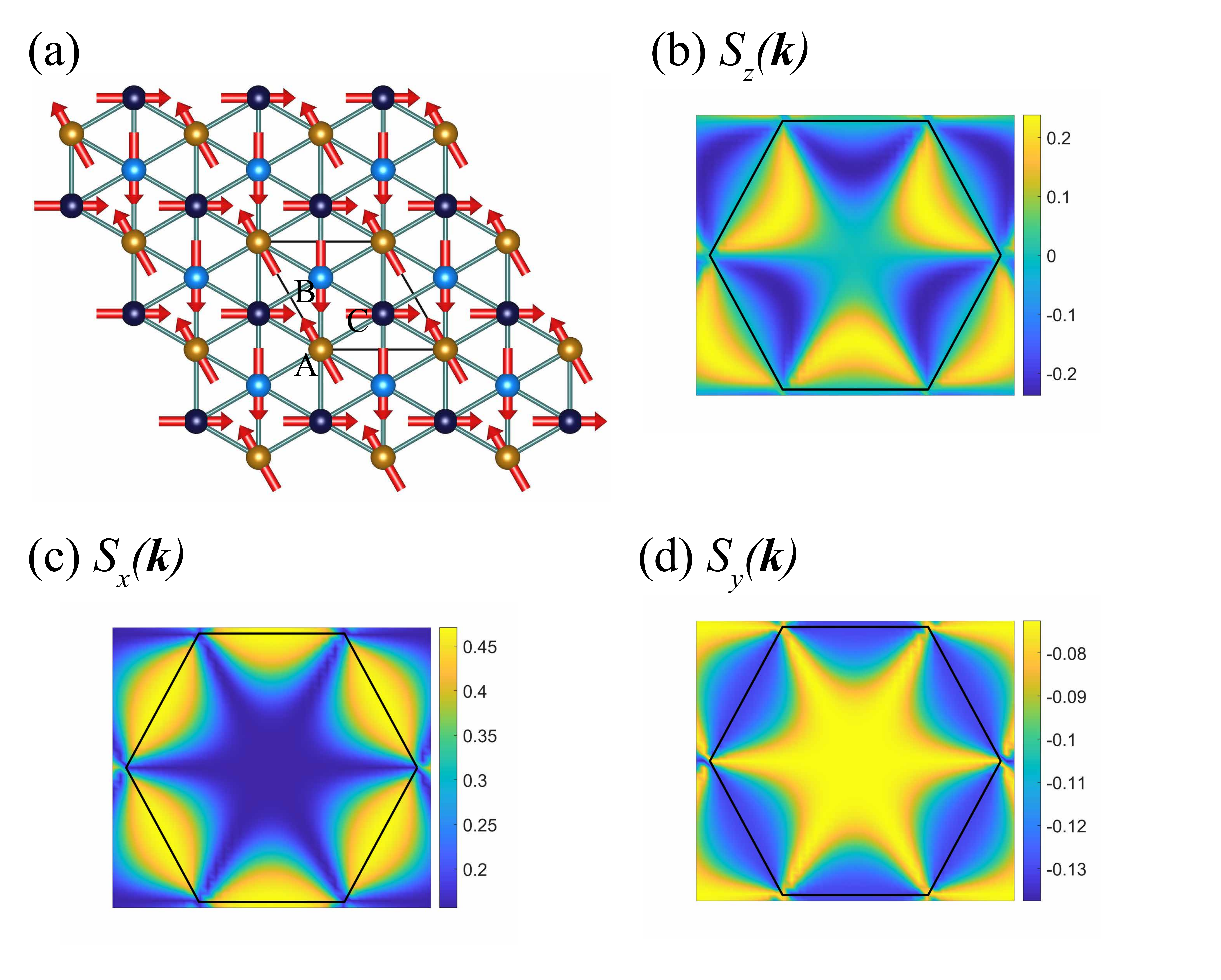}
    \caption{(a) Coplanar magnetic structure with SSG P143.1.1 GM1. (b)-(d) The $z$ [(b)], $x$ [(c)], $y$ [(d)] components of the spin expectation values of the Bloch states in the lowest band. The colors (see the color bar) represent the values. The hexagon is the boundary of the first BZ.}
    \label{fig: d_SBZ_P3_143}
\end{figure}

\blue{\subsection*{\texorpdfstring{A coplanar magnetic model exhibiting $d_{\rm SBZ} = 3$}{dM}}}
We consider a 2D coplanar magnetic structure, whose crystalline symmetry is described by the space group P3 (No.~143) with lattice constant $c = \infty$. 
Three atoms, which are labeled by A, B, and C in Fig.~\ref{fig: d_SBZ_P3_143}(a), exist in three $C_{3z}$ invariant points $(0,0,0)$, $(2/3,1/3,0)$, and $(1/3,2/3,0)$ in the unit cell, respectively.
The coplanar magnetic moments are localized around these atoms, given as  
\begin{equation}
    \S_A = (-\frac{1}{2},\frac{\sqrt{3}}{2},0), \, 
    \S_B = (0,-1,0), \, \S_C = (1,0,0) \, .
\end{equation}
The magnetic structure [Fig.~\ref{fig: d_SBZ_P3_143}(a)] remains invariant under operations $\{I|3_{001}^+| \0 \}$ and $\{I|1| \a_i \}$ ($i = 1,2$), and hence is described by the SSG P143.1.1 GM1.

We consider an electronic Hamiltonian in this magnetic structure, given as  
{
\begin{equation}
\begin{aligned}
        \H =  t \sum_{\langle \R ,\alpha ; \R^{\p} , \alpha^{\p} \rangle} c^{\dagger }_{ \R ,\alpha } \sigma_0 c_{\R^{\p}  , \alpha^{\p}}  +  J \sum_{\R ,\alpha} \S_{\alpha } \cdot {\bm s}_{\R, \alpha}  ,
\end{aligned}
\label{H_Kagome}
\end{equation}}%
with
\begin{equation*}
    {\bm s}_{\R,\alpha }^{\mu} = \frac{1}{2}c^{\dagger }_{ \R ,\alpha} \sigma_{\mu}  c_{ \R ,\alpha } \quad (\mu = x,y,z)
\end{equation*}
being the spin operator of the electrons.  
Here $c^{\dagger }_{ \R ,\alpha } = (c^{\dagger }_{ \R ,\alpha ,\uparrow}, c^{\dagger }_{ \R ,\alpha, \downarrow })$ ($\alpha=A,B,C$) is a two-component spin-1/2 electron creation operator in the unit cell $\R$, sub-lattice $\alpha $.
The first term of the Hamiltonian represents nearest-neighboring hopping, represented by the bonds in Fig.~\ref{fig: d_SBZ_P3_143} (a).
The hopping matrix is proportional to $\sigma_0$ because the SOC is considered negligible. 
The second term describes the on-site interaction between the conduction electron and localized magnetic moments $\S_{\R, \alpha}$.

We choose parameters as $t = 1$, $J = -1$, and study the spin texture $\vS(\k)$ in the momentum space. 
Since in this SSG, all translations are accompanied by the identity spin operations, the magnetic BZ and SBZ coincide. 
Figs.~\ref{fig: d_SBZ_P3_143}(b)-(d) show the expectation values $\vS(\k)$ of spin operators for the Bloch states in the lowest energy band. 
Although $\int_{\rm BZ} S_z(\k) d\k= 0 $, $S_z(\k)$ is non-zero at individual $\k$ [Fig.~\ref{fig: d_SBZ_P3_143}(b)]. 
For generic $\k$, $S_x(\k)$ and $S_y(\k)$ are also non-zero, leading to $d_{\rm SBZ} = d_{\rm BZ} = 3$, consistent with our theoretical analysis (Table IV in the main text).

\subsection*{Algorithm for determining whether an SBZ is symmorphic or nonsymmorphic}

In this section, we present the detailed algorithm for determining whether an SBZ of a collinear SSG with $[\hat{t}_i, \hat{t}_j] = 0 \, (i,j=1,2,3)$ is symmorphic or nonsymmorphic.
In Sec.~\RefSecPNSSGs, we have shown the SSG momentum $\kk$ is transformed to $s_gR_g \kk + \qq_g$ under a generic SSG operation $\hat{g}$, where $s_g = 1$ $(-1)$ if $g$ is unitary (anti-unitary), and have also explained how to determine the value of $\qq_g$.
To determine whether an SBZ is symmorphic or nonsymmorphic, we need to find out whether a $\bm \theta$ exists such that 
\begin{equation}
      -s_g{\bm \theta} + R_g {\bm \theta} \equiv   \qq_g
    \label{App_q_g}
\end{equation}
holds for all $g \in \mG$. 
Since operations in the unitary translation group $T_U$ leave $\kk$ invariant, we only need to consider the action of the coset representatives $g_1,g_2, \cdots, g_N$ of $\mG /T_U$.

To proceed, we write the matrix $R_g$ and vectors $\qq_g$ on the basis of reciprocal lattice vectors $\b_i$'s. 
$R_g \b_i$  can be written as $\sum_{j = 1}^3 \b_j M_{g;ji}$, where $M_{g;ji}$ are integers since the spatial operations must be compatible with the lattice vectors.
As shown in Sec.~\RefSecPNSSGs, the component of $\qq_g$ must be a half-integer on the basis of $\b_i$'s, 
We can construct a $3N \times 3$ matrices $A$ and a $3N \times 1$ vector $Q$ 
\begin{equation}
    C  = \begin{pmatrix}
     M_{g_1} - s_{g_1}I \\
     \vdots \\
      M_{g_N} - s{g_N}I 
    \end{pmatrix}, \quad 
    Q  = \begin{pmatrix}
     \qq_{g_1} \\
     \vdots \\
     \qq_{g_N}
    \end{pmatrix} \, . 
\end{equation}
With this construction, solving Eq.~(\ref{App_q_g}) is equivalent to finding a $\bm \theta$ such that $C {\bm \theta} - Q$ is an integer vector. 
A necessary condition of $C {\bm \theta} - Q$ being an integer vector is that $2 C {\bm \theta}$ is an integer vector because $2 Q$ is an integer matrix. 
To find all vectors $\bm{\theta}$ satisfying $2C \bm{\theta}$ being an integer vector, we can apply the Smith decomposition to the matrix $2C$, given as:
\begin{equation}
    2C = S A T \, ,
\end{equation}
where $S,A,T$ are $3N \times 3N$, $3N \times 3$, and $ 3 \times 3 $ integer matrices, respectively. 
Matrix $A$ is diagonal, given as
\begin{equation}
    A = \begin{pmatrix}
        \alpha_1 & 0 & 0 \\
        0 & \alpha_2 & 0 \\
        0 & 0 & \alpha_3 \\
        0 & 0 & 0 \\
        \vdots & \vdots & \vdots \\
         0 & 0 & 0
    \end{pmatrix} \, , 
\end{equation}
and the diagonal elements are non-negative integers that satisfy that $ \alpha_{i+1}$ being divisible by $\alpha_{i} $ $(i = 1,2)$. 
Some of $\alpha_i$ can be zero, and without loss of generality, we assume that $ \alpha_j > 0 \, ( j \leq r \leq 3)$ and $\alpha_j = 0 \, (j > r)$.  
$S$ and $T$ satisfy that $|\det (S)| = | \det (T)| = 1$, which means that $S^{-1}$ and $T^{-1}$ are also integer matrices. 
Thus, the condition for $2 C {\bm \theta} =S A T {\bm \theta}$ being an integer vector is equivalent to $ A T {\bm \theta} $ being an integer vector. 
If $ A T {\bm \theta} $ is an integer vector, $A T {\bm \theta}$ must be in the form 
\begin{equation}
    A T {\bm \theta} = \begin{pmatrix}
        m_1  \\
        \vdots \\
        m_r \\
        0 \\
        \vdots \\
        0
    \end{pmatrix}  \,  \text{with} \, m_i \in \mathbb{Z} \, (i \leq r).
    \label{eq: AT}
\end{equation}
We then need to determine whether an $A T \bm{\theta}$ of this form can satisfy $S (\frac{1}{2} A T \bm{\theta}) - Q$ to be an integer vector. 
Enumerating all integers $m_i$ is unnecessary because if $m_i$'s are all even, $S (\frac{1}{2} A T {\bm \theta}) $ must be an integer matrix.
Thus, we only need to test the cases $m_i \in \{0,1\}$. 
If there exist $m_1, \cdots, m_r$ that satisfy  $S (\frac{1}{2} A T {\bm \theta}) - Q$ to be an integer vector, the SBZ is symmorphic, and $ {\bm \theta}$ is given by  
\begin{equation}
    {\bm \theta} = T^{-1} \begin{pmatrix}
        m_1/\alpha_1 \\
        \vdots \\
        m_r/\alpha_r \\
        \vdots \\
        0 \\
    \end{pmatrix} \, .
\end{equation}
Otherwise, the SBZ is nonsymmorphic.

\subsection*{{Energy bands in non-commuting SBZ}}

In this section, we investigate the electron bands in the SSG N143.16.1 M1. 
This SSG can be generated by $\hat{t}_1 = \{ \sigma_y| 1 | \a_1\}$, $\hat{t}_2 = \{ \sigma_x| 1 | \a_2\}$, $\hat{t}_3 = \{ \sigma_0| 1 | \a_2\} $, and $\hat{C}_{3z} = \{ e^{i \frac{\pi}{3} \n \cdot {\bm \sigma}} | 3^+_{001} |  \0 \}$ with $\n = \frac{\sqrt{3}}{3} \left( {\bm e}_x +  {\bm e}_y +  {\bm e}_z \right)$. 
Note that in Sec.~\RefGenericH, $\hat{U}_{t_1}$, $\hat{U}_{t_2}$ are chosen as $i \sigma_y$, $i \sigma_z$.
Compared to that, we apply a gauge transformation, which can simplify the form of the transformation of SSG momentum. 
Bloch states are defined by using the eigenvalues of the commuting operators $\hat{t}_1$, $\hat{t}_2^2$, and $\hat{t}_3$, as shown in Eq.~\RefEqBlochNoncommuting.
Let us consider the action of $\hat{C}_{3z}$ on the Bloch states. 
In Sec.~\RefSecNoncommutingSBZ, we showed $\ket{\psi(\kk)}$ and $ \hat{t}_2\ket{\psi(\kk)}$ share the same energy $E(\kk)$,
Thus, their linear combination is also an eigenstate with energy $E(\kk)$.
Let $ \ket{\phi_{\pm}} = \ket{\psi(\kk)} \pm e^{-i (\tk_2 + \pi) /2}  \hat{t}_2 \ket{\psi(\kk)}$, and we will prove $ \hat{C}_{3z} \ket{\phi} $ is an Bloch state. 
We can verify that $ \hat{C}_{3z}^{-1} \hat{t}_1 \hat{C}_{3z} = i \hat{t}_1^{-1}\hat{t_2}^{-1} $, $ \hat{C}_{3z}^{-1} \hat{t}_2^2 \hat{C}_{3z} = \hat{t}_1^2 $, and $\hat{C}_{3z}^{-1} \hat{t}_1 \hat{C}_{3z} = \hat{t}_3$
Then, \begin{equation}
    \begin{aligned}
        & \hat{C}_{3z}^{-1} \hat{t}_1 \hat{C}_{3z}   \ket{\phi_{\pm}}  \\
        & = i \hat{t}_1^{-1} \left( \pm  e^{-i (\tk_2 + \pi)/2} \ket{\psi(\kk)} + \hat{t}_2 \hat{t}_2^{-2} \ket{\psi(\kk)} \right)\\
        & = i \hat{t}_1^{-1} \left( \pm  e^{-i (\tk_2 + \pi)/2} \ket{\psi(\kk)} + \hat{t}_2 e^{-i \tk_2} \ket{\psi(\kk)} \right) \\
        & = \pm  e^{-i (2\tk_1 + \tk_2)/2} \ket{\psi(\kk)} + \hat{t}_2 e^{-i \tk_2 - i \tk_1 - i \pi/2 } \ket{\psi(\kk)} \\
        & =  \pm e^{-i (2\tk_1 + \tk_2 )/2} \ket{\phi_{\pm }} \, .
    \end{aligned} 
\end{equation}
In this derivation, we utilize the fact that $\hat{t}_2 \ket{\psi(\kk)}$ is a Bloch state with SSG momentum $\kk + \frac{1}{2}\b_1$.
Similarly, we can show that $\hat{C}_{3z}^{-1} \hat{t}_2^2 \hat{C}_{3z} \ket{\phi_{\pm}} =  e^{2 \tk_1}  \ket{\phi_{\pm}}$ and  $\hat{C}_{3z}^{-1} \hat{t}_3 \hat{C}_{3z}  \ket{\phi_{\pm}} =  e^{ \tk_3}  \ket{\phi_{\pm}}$. 
Thus, the $\hat{C}_{3z} \ket{\phi_{+}}$ ($\ket{\phi_{-}}$) is a Bloch state with SSG momentum $ \kk^{\p} = \tk_1^{\p} \b_1 +\tk_2^{\p} \frac{1}{2} \b_2 + \tk_3^{\p} \b_3  $, where $\tk_1^{\p} = -(2\tk_1 + \tk_2 )/2 $ $\left(\tk_1^{\p} = -(2\tk_1 + \tk_2 + 2\pi)/2 \right) $, $\tk_2^{\p} = 2 \tk_1$, and $\tk_3^{\p} = 2 \tk_3$. The momenta after tranformation can also be written as $\kk^{\p}$ are $R_{\hz}({2 \pi}/{3}) \kk$ $\left( R_{\hz}({2 \pi}/{3}) \kk + \frac{1}{2} \b_1 \right)$.

\section{More discussions on the magnetic materials}
\label{Iden_SSG}
\label{app: material example}
\label{App Sec: InMnO3}
\label{app sec: SOC}

\subsection*{Identification of SSGs for magnetic materials}

In this section, we summarize how to find the SSG (or qSSG) of a crystalline magnetic structure.
We will first show an algorithm that can find all possible O(3) matrices $O$ that can produce the map 
\begin{equation} 
  O \S_{P_i} = \S_i  \quad (i = 1,2,\cdots,N) \, ,
  \label{U_operation}
\end{equation}
given a finite number of magnetic moments $\S_1, \cdots, \S_N$ and their permutation $\S_{P_1}, \cdots, \S_{P_N}$.
As discussed in Sec.~\RefClassGeneral, in the case that $\S_i$'s are non-co-planar, if $O$ exists, it must be unique.
For the coplanar ({\it e.g.,} along the $x,y$ plane) and collinear ({\it e.g.,} along the $z$ direction) cases, $O$ must be diagonal in the form ${\rm diag}\left( O_{xy},O_z \right)$.
Meanwhile, only $O_{xy}$ ($O_z$) influences the transformation in the coplanar (collinear) case, and if $O_{xy}$ ($O_z$) exists, it must be unique. 
Thus, we only need to find an O(2) (O(1)) matrix $O_{xy}$ ($O_z$) if $\S_i$'s are coplanar (collinear).

For the non-coplanar case, without loss of generality, we assume that $\S_1, \S_2, \S_3$ are linearly independent. 
The only $3 \times 3$ matrix $\V$ that can transform linearly independent vectors $\S_1, \S_2, \S_3$ to $\S_{P_1}, \S_{P_2}, \S_{P_3}$ is given by 
\begin{equation} 
  \V = \left(\S_{P_1}, \, \S_{P_2}, \, \S_{P_3}\right) \left(\S_1, \, \S_2, \, \S_3\right)^{-1} \, .
\end{equation} 
If $\V \in \mathrm{O(3)}$, {\it i.e.,} $\V \V^T = I$ and $\V \S_{P_i} = \S_i $ is satisfied for all $ i > 3$, we can choose the unique O(3) matrix $O  =\V$.
Otherwise, no O(3) matrix can satisfy Eq.~(\ref{U_operation}). 
For the coplanar case, without loss of generality, we assume that $\S_1, \S_2$ are linearly independent. 
The unique $2 \times 2$ matrix $\V$ that can transform linearly independent vectors $\S_1, \S_2$ to $\S_{P_1}, \S_{P_2}$ is given by 
\begin{equation} 
  \V_{xy} = \left(\S_{P_1} , \, \S_{P_2} \right)\left(\S_1, \, \S_2 \right)^{-1} \, ,
\end{equation} 
where $\S_i$ and $\S_{P_i} \, (i = 1,2)$ have zero $z$ component and are written as $2 \times 1$ column vector. 
If $\V_{xy} \in \mathrm{O(2)}$, and $\V_{xy} \S_{P_i} = \S_i $ is satisfied for all $ i > 2$, we can choose the unique O(2) matrix $O_{xy} =\V_{xy}$.
Otherwise, Eq.~(\ref{U_operation}) cannot be satisfied. 
For the collinear case, if $\S_{P_i} = \S_i $ ($\S_{P_i} = -\S_i $) is satisfied for all $ i $, $O_z = 1$ ($-1$).
Otherwise, Eq.~(\ref{U_operation}) cannot be satisfied.

Then, we present the detailed algorithm for finding the SSG of a magnetic structure.
Here, we use a non-coplanar magnetic structure as an example.
The algorithm for the coplanar (collinear) cases is very similar, and the only difference is that a spin operation is represented by an O(2) (O(1)) matrix but not an O(3) matrix. 
\begin{enumerate}%
    \item Find the space group $G_{\rm latt}$ of the crystalline structure without the magnetic moments. 
    Let $\a_{L1},\a_{L2},\a_{L3}$ be the lattice vectors of $G_{\rm latt}$, and $T_{\rm latt}$ be the translation subgroup of $G_{\rm latt}$.
    \item Find the lattice vectors $\a_{M1},\a_{M2},\a_{M3}$ of magnetic unit cell. 
    A finite number of magnetic moments $\S({\r_1}),\cdots, \S({\r_N})$ exist in a magnetic unit cell, and the magnetic moments satisfy $ \S(\r_i) = \S(\r_i + \a_{Mj}) \, (i \leq N, j \leq 3)$, and translations along the $\a_{Mi}$'s generates a subgroup $T_M$ of the translation group $T$.
    \item Some spatial rotation in $G_{\rm latt}$ is not compatible with $\a_{Mi}$'s, {\it i.e.,} cannot be written as an integer matrix in their lattice coordinate.
    Only spatial operations compatible with $\a_{Mi}$'s can belong to the parent space group of the SSG (see proof in the end of this section), and these spatial operations form a subgroup $P_{\rm latt}$ of $G_{\rm latt}$. 
    Let $p_1,p_2,\cdots, p_{m}$ be the coset representatives of the quotient group $P_{\rm latt}/T_{M}$. 
    \item After the spatial operation $p_j$, the magnetic moments in the position $\r_1, \cdots , \r_N$ are $\S({p_j^{-1} \r_1}),\cdots, \S({p_j^{-1} \r_N})$, which are a permutation of $\S({\r_1}),\cdots, \S({\r_N})$. 
    Using the algorithm introduced above, we can determine whether an O(3) matrix $O$ satisfying $O \S({g_j^{-1} \r_i}) = \S(\r_i) \, \forall i \leq N$ exists and find its value if $O$ exists. 
    Suppose for operation $p_{i_1},p_{i_2},\cdots,p_{i_n}$ ($n \leq m$), the O(3) matrix satisfying the condition exists and are given as $O_{p_{i_1}},\cdots,O_{p_{i_n}}$. 
    \item Up to now, we obtain the parent space group of $P$ the SSG, which are the union of these cosets  
    \begin{equation} 
    p_{i_1} T_M, \, p_{i_2} T_M , \, \cdots, \, p_{i_n} T_M \, .
    \end{equation} 
    The spin operations in each coset are the same, which are 
    \begin{equation} 
       O_{p_1}, O_{p_2}, \cdots ,O_{p_n} \, ,
    \end{equation}
    respectively. Note that some of $p_{i_1},p_{i_2},\cdots,p_{i_n}$ might be translation operations with a non-trivial spin operation.
\end{enumerate}

These spin operations, which are represented by O($N$) ($N=1,2,3$) matrices, form a representation of the parent space group.
When classifying SSGs, we categorize O($N$) representations into different types by the constituent irreps, and the nomenclature of SSGs is also based on the irreps.
Thus, to know the specific SSG that the O($N$) representation corresponds to, we should decompose the O($N$) representations into irreps.
To this end, we identify the momenta of the representation by studying the representation matrices of translations.
Let $\a_i$ ($i= 1,2,3$) be the lattice vectors of the parent space group, and $O_{\a_i}$'s be their representation matrices. 
$O_{\a_i}$'s must commute with each other, because $O_{\a_i} O_{\a_j}$ and $O_{\a_j} O_{\a_i}$ both equal the representation matrix of $\{ 1 |\a_i + \a_j\}$. 
Thus, $O_{\a_i}$'s ($i=1,2,3$) share a set of common eigenvectors $\ket{v_j}$ ($j=1,\cdots,N$): $ O_{\a_i} \ket{v_j} =e^{i  2 \pi \lambda_{ij}} \ket{v_j} $ $\left( \lambda_{ij} \in [0,1) \right)$.
The $j$th ($j \leq N$) momentum is given by $\k_j = \sum_{i = 1}^{3}\b_{i}  \lambda_{ij} $, where $ \b_{j}$ satisfying $\a_{i}  \cdot \b_{j}= 2 \pi \delta_{i,j}$ is the reciprocal lattice vector.
After knowing the momenta $\k_j$, we can reduce the O($N$) representations to the allowable representation $\rho_{\k_j}$ of the little group $P^{\k_j}$ of $\k_j$. 
Note that ``allowable'' means that translation $\{ 1  | \v \}$ is represented by $e^{i \k_j \cdot \v} I$.
Suppose that $T$ is the translation subgroup of the parent space group, and the little group of $P^{\k_j}$ consists of $ p^{\p}_{1} T, p^{\p}_{2} T, \cdots, p^{\p}_{{n^{\p}}} T$, and the O($N$) representation matrix of $p^{\p}_{i}$ is $O_{p^{\p}_{i}}$.
The representation matrix $D_{\rho_{\k_j}}(p^{\p}_{i})$ of $p^{\p}_{i}$ in $\rho_{\k_j}$ should be given by $Q_{\k_j} O_{p^{\p}_{i}} Q_{ \k_j}$, where $ Q_{ \k_j} = \sum_{i \leq N,\k_i = \k_j} \ket{v_i}\bra{v_i} $ is the projection operator to the eigenspace of the momentum $\k_j$.
Knowing the little group representation $\rho_{\k_j}$, we can easily reduce it to the irreps of $P^{\k}$ with the character theory.
Repeating this process for all $\k_j$, we can reduce the O($N$) representation to the irreps of the little groups and hence determine which SSG it corresponds to.

In the following, we will prove that the spatial-rotation part $R_g$ of an SSG operation $g \in \mG$ must be compatible with lattice vectors $\a_{Mi}$'s of pure translation subgroup $T_M = \brace{ \brace{I|1|\v} | \brace{I|1|\v} \in \mG }$.
If $ t =  \brace{I|1|\v} \in T_M$, $\v$ must be in the form $\sum_i^3 m_i \a_{Mi}$ with $m_i \in \mathbb{Z}$. 
The conjugate operation of $ t =  \brace{I|R|\v}$ under a generic SSG operation $g = \{X_gU_g | R_g| \v_g\}$ is $ g^{-1} t g =  \brace{I|1|R_g^{-1}\v}$, which still belongs to $T_M$. 
Thus, for any $R_g$, $R_g^{-1}\v$ must be in the form $\sum_i^3 m_i^{\p} \a_{Mi}$ with $m_i^{\p} \in \mathbb{Z}$, and it is the definition of rotation being compatible with lattice vectors.

\subsection*{Nonsymmorphic SBZ in \texorpdfstring{InMnO$_3$}{InMnO3}}

\begin{figure}[tb]
    \centering
   \includegraphics[width=1 \linewidth]{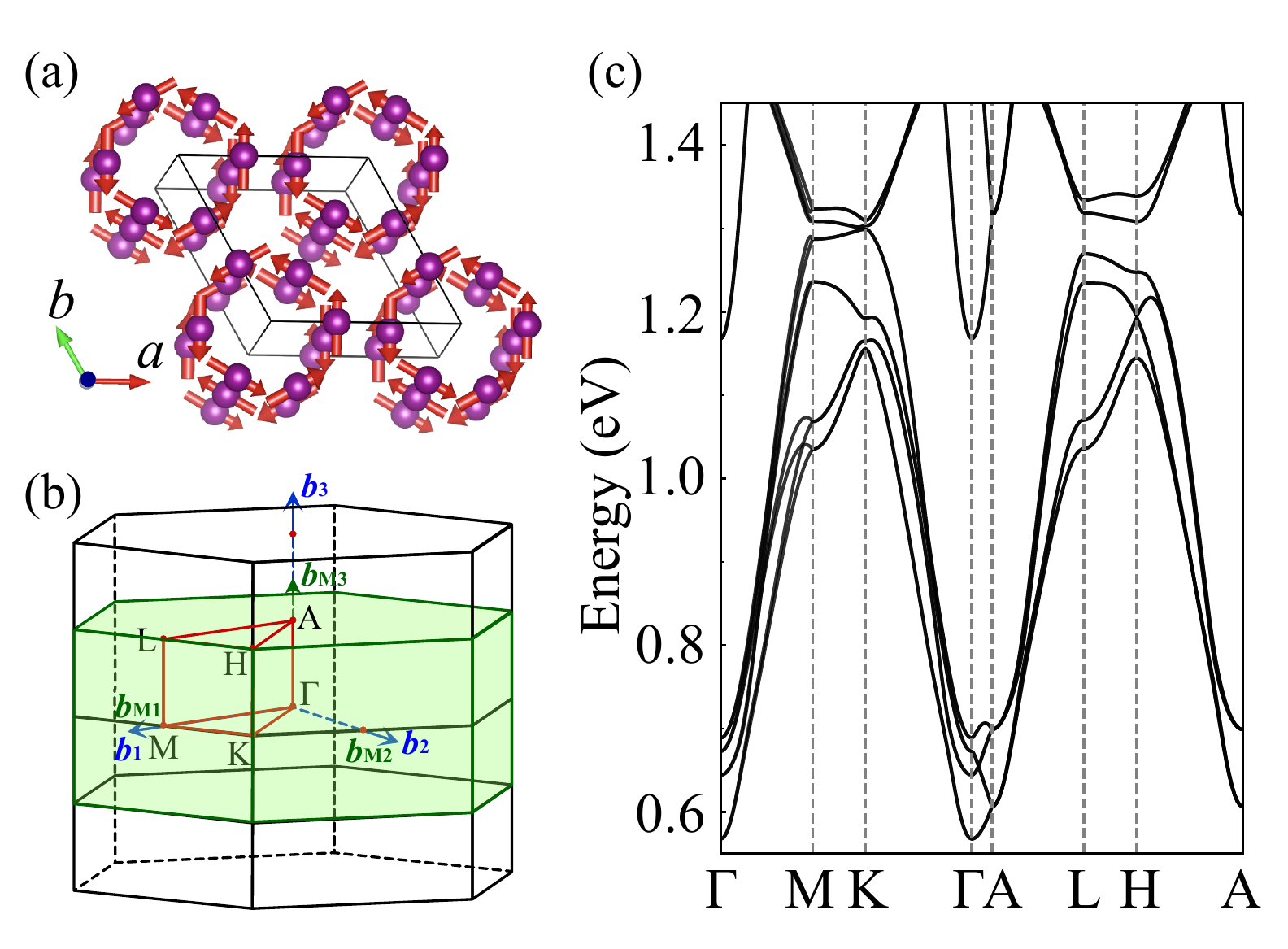}
    \caption[]{(a) The magnetic structure of InMnO$_3$ showing only the magnetic atoms (Mn). (b)The first BZs in the SSG reciprocal lattice $\b_i$'s (black lines) and MSG reciprocal lattice $\b_{Mi}$'s (green region).
    (c) The first-principle calculation energy bands.}
    \label{InMNO3_Fig}
\end{figure}

InMnO$_3$ (No.~1.525) has a hexagonal lattice structure. Let $\a_1,\a_2,\a_3$ be its lattice vectors along the three directions. 
In the magnetic phase, the spin moments lie in the $x,y$ plane~\cite{fabreges11}. 
In each layer along the $z$ direction, spins form a triangular antiferromagnetic structure, and spin moments related by lattice translation $\{ 1 | \a_3 \}$ have opposite directions. 
 The MSG of this structure is $P_{2c}31m^{\p}$ (No.~157.5.1288).
The magnetic unit cell is twice that in the paramagnetic phase, in which the MSG lattice vectors $\a_{M1,M2}= \a_{1,2}$, and $\a_{M3} = 2 \a_3$. 
Fig.~\ref{InMNO3_Fig}(c) shows the energy bands obtained from the first-principle calculation.
These bands are plotted in the BZ of the MSG, and the high symmetry points in MSG BZ are labeled in Fig.~\ref{InMNO3_Fig}(b). 
They exhibit the following features that MSG cannot explain:
(i) Along the line $\Gamma$-K-M, the bands are two-fold degenerate. However, the little co-group of the MSG along this line is the point group $m$, which only has two 1D coirreps.
(ii) At points A and H, the degree of degeneracy is either two or four. The little co-group of the MSG on A is $3m1^{\p}$ (the direct product point group $3m$ and $Z_2^T$ TRS group), and the little co-group on H is $3m$. Both of them only have two 1D coirreps and one 2D coirrep.
(iii) The bands are at least two-fold degenerate along the path A-L-H-A. However, the little co-groups on A-L and L-H-A are $m'$ (the $Z_2^T$ group generated by the joint operation of mirror and time-reversal) and $m$, respectively, either of which only has a 1D coirrep. 

\begin{table}[h]
    \caption[The generators of the SSG (P157.6.2 A3) of coplanar magnetic material InMnO$_3$.]{The generators of the SSG (P157.6.2 A3) of coplanar magnetic material InMnO$_3$.
    The last row is for the pure-spin-operation group $\mS$. 
    The last column shows how the SSG momentum is transformed under the SSG operations. In the forth row, $\n_{1\bar{1}0} = {\sqrt{3}}/{2}{\bm e}_x - 1/2{\bm e}_y$.
    }
    \begin{tabular}{c|ccc}
        \hline \hline
        SSG operation & \makecell[c]{Spin operations \\ on electrons} &  \makecell[c]{Transformation of \\  SSG momentum $\kk$}  \\   \hline
        $\{I |1 | \a_{1,2} \}$ &  $\sigma_0$ & $\kk$\\
        $\{U_{\hz} (\pi) |1 | \a_{3} \}$ &  $\sigma_z$ & $\kk$\\
        $\{ U_{\hz} (\frac{2\pi}{3})| 3^+_{001}| \0  \}$ & $-e^{i \frac{\pi}{3} \sigma_z }$ &  $R_{\hz} (\frac{2\pi}{3}) \kk$  \\
        $\{ U_{\hx}(\pi) | m_{1\bar{1}0} | \0 \}$ & $ \n_{ 1\bar{1}0 }  \cdot {\bm \sigma}$ & $ -R_{\n_{1\bar{1}0}}(\pi) \kk + \frac{1}{2}\b_3  $\\ 
        \hline
        $\{T U_{\hz}(\pi) | 1 | \0  \}$ & $ \sigma_x {\mathcal K}$& $-\kk  + \frac{1}{2}\b_3$\\ 
        \hline
    \end{tabular}
    \label{tab: InMnO3}
\end{table}

The SSG of this structure is P157.6.2 A3, whose parent space group is $P3m1$. 
The SSG shares the same lattice vectors $\a_i$'s as the space group in the paramagnetic phase.  
Table~\ref{tab: InMnO3} shows the generators of this SSG and how an SSG momentum $\kk$ is transformed under these operations. 
The mirror operation $\hat{M}_{\n_{1\bar{1}0}} = \{ \n_{1\bar{1}0} \cdot {\bm \sigma} | m_{1\bar{1}0}| \0 \}$ ($\n_{1\bar{1}0} = {\sqrt{3}}/{2}{\bm e}_x - 1/2{\bm e}_y $) acts a glide operation on the $\kk$. 
Therefore, the SBZ is nonsymmorphic.
Before investigating the energy bands in the SBZ, we study the relationship between SSG and crystal momentum.
Within our choice of spin rotation (see the first two rows of Table~\ref{tab: InMnO3}), translations $\{ 1 | \sum_{i=1}^3 m_i \a_{Mi} \}$ are always accompanied by $\sigma_0$ without an extra phase.
Thus, the crystal momentum $\k$ equals the SSG momentum $\kk$. 
On the other hand, the first SSG BZ is twice as large as the magnetic BZ, and $\b_3 = 2\b_{M3}$, indicating that SSG momenta $\kk$ and $\kk + 1/2\b_3$ are considered as equivalent in the magnetic BZ.
Then, we study the symmetries of different $\kk$. 
Let $\kk_{\rm \Gamma-K} = u \b_1 + u \b_2 $ with $u \in [0,1)$. 
$\hat{M}_{\n_{1\bar{1}0}} $ transforms $\kk_{ \rm \Gamma-K }$ to $\kk_{ \rm \Gamma-K}^{\p} = \kk_1 + \frac{1}{2} \b_1$, implying that $E_n(\kk_{\rm \Gamma-K}) = E_n(\kk_{\rm \Gamma-K}^{\p})$ in the SBZ. 
Similarly, $ \hat{M}_{\n_{1\bar{1}0}} \hat{C}_{3z}$ ($\hat{C}_{3z} = -\{e^{i \frac{\pi}{3} \sigma_z } |3^+_{001}| \0 \}$) transforms $\kk_{\rm K-M} = (1/3 +v) \b_1 + (1/3-2v) \b_2 $ with $v \in [0,1/6]$ to $\kk_{\rm K-M} + 1/2\b_3$. 
After folding the band, this symmetry explains the feature (i) in the magnetic BZ. 
Let $\kk_{\rm A} = \frac{1}{4}\b_3 = \frac{1}{2} \b_{M3}$, $ \kk_{\rm H} =  \frac{1}{3} (\b_1 + \b_2) + \frac{1}{4}\b_3 $. 
Both $\kk_{\rm A}$ and $ \kk_{\rm H}$ have the symmetries $\hat{\T} = \{ T U_{\hz}(\pi) | 1 | \0\}$ and $\hat{C}_{3z}$. 
The little co-groups of the SSG at $\kk_{\rm A,H}$ have one 1D coirrep ($D(\hat{C_{3z}}) = 1$) and one 2D coirrep ($D(\hat{C}_{3z}) = {\rm diag}( e^{i \frac{2\pi}{3}}, e^{i \frac{4\pi}{3}}))$.
Due to $\hat{M}_{\n_{1\bar{1}0}} $, $E_n(\kk_{\rm A,H} ) = E_n(\kk_{\rm A,H} + \frac{1}{2} \b_3)$. 
The feature (ii) in the magnetic BZ has been successfully explained. 
Let $\kk_{\rm A-L} = w \b_1 + \frac{1}{4} \b_3$ with $w \in [0,1)$. 
$\hat{\T} \hat{M}_{\n_{1\bar{1}0}} \hat{C}_{3z}$ transforms $\kk_{\rm A-L}$ to $\kk_{\rm A-L}^{\p} = w \b_1 - \frac{1}{4} \b_3$. 
Note that $\kk_{\rm A-L}$ and $\kk_{\rm A-L}^{\p}$ differ by $\frac{1}{2} \b_1$ and hence correspond to the same point in the magnetic BZ.
This explains the two-fold degeneracy on the line A-L in magnetic BZ. 
The two-fold-degeneracy on the L-H-A can be explained in the same manner as we adopted for the line $\rm \Gamma$-K-M.

\subsection*{Extra band degeneracy in non-coplanar \texorpdfstring{Mn$_3$Ge}{Mn3Ge}}

\begin{figure}[tb]
    \centering
   \includegraphics[width=1 \linewidth]{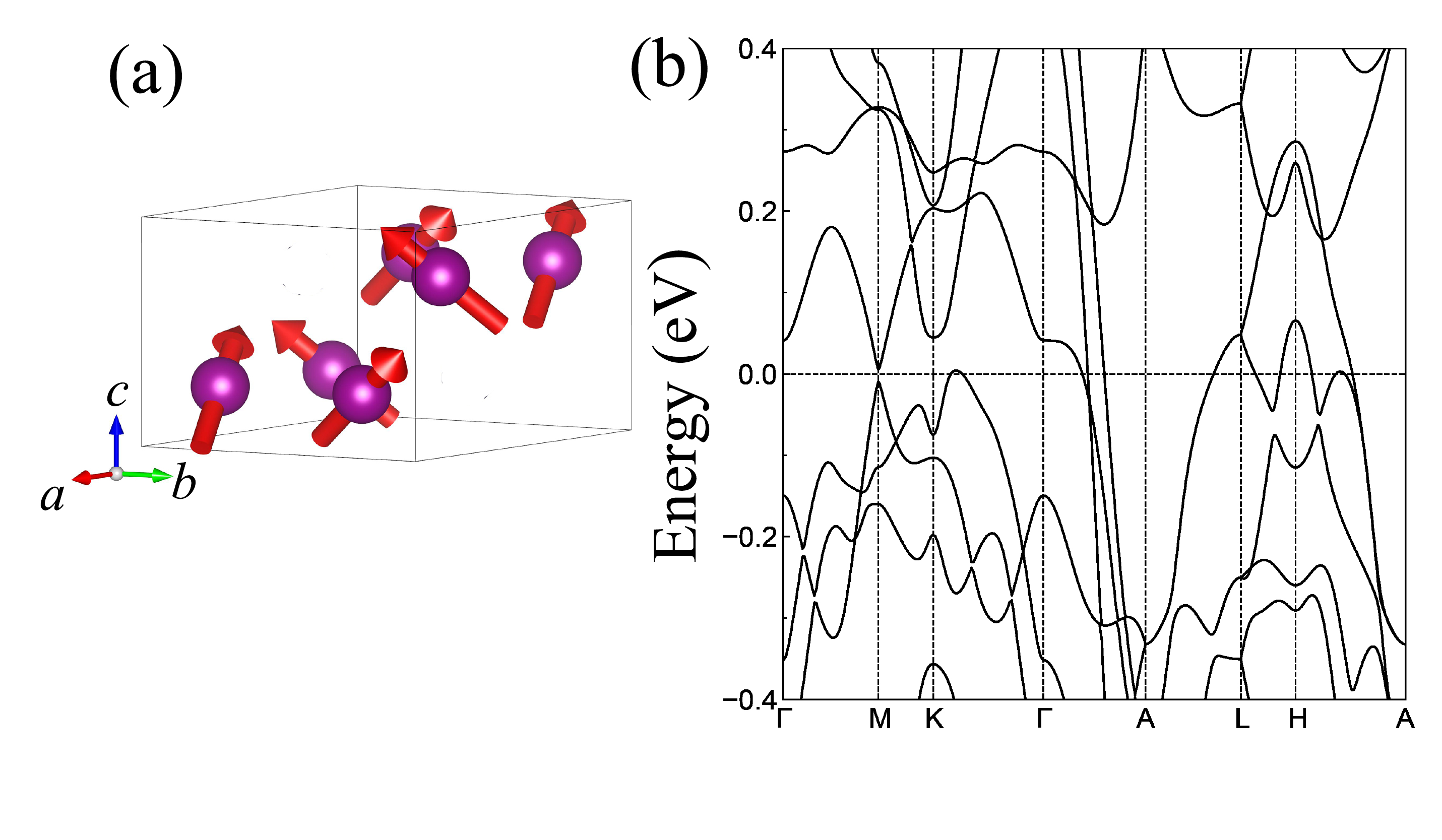}
    \caption[]{(a) The non-coplanar magnetic structure of Mn$_3$Ge showing only the magnetic atoms (Mn). (b) The first-principle calculation energy bands.}
    \label{Mn3Ge_N_Band}
\end{figure}

We use Mn$_3$Ge (No.~0.203) in a non-coplanar magnetic structure as an example to show that SSGs can explain the extra degeneracy of electron bands in non-coplanar magnetism. 
Under the ambient pressure, MnGe$_3$ has a coplanar triangular antiferromagnetic structure [see Figs.~7 (a) in the main text].
It was found that spin moments acquire an out-of-plane caning angle $\theta$ under high pressure, and $\theta$ gradually change with the increase of the pressure~\cite{Sukhanov18}, and the MSG of non-coplanar Mn$_3$Ge is $C2^{\p}/m^{\p}$ (No.~12.5.70). 
Since non-coplanar Structures with different $\theta$ ($\theta \neq 0$ or $\pi/2$) share similar symmetries, {\it i.e.,} are described by the same MSG and SSG, we study the case with $\theta = \pi/4$, without lost of generality. 
Fig.~\ref{Mn3Ge_N_Band} (b) shows the energy bands from the first-principle calculation with constrained spin directions.
On line A-L ($\k = (u, 0, \pi)$), the bands are always two-fold degenerate.
However, the little co-group in MSG on that line is $m'$, which only has a 1D coirrep.

The SSG of non-coplanar Mn$_3$Ge is N194.6.1 $\rm GM5^{+}$ with the parent space group being $P6_3/mmc$.
Its generators are given as $\hat{C}_{6z} = \{e^{i \frac{\pi}{3} \sigma_z}|6^+_{001} | 0, 0 , 1/2\} $. $\hat{\mathcal{P}} = \{ \sigma_0 | \bar{1} | \0 \}$, $\hat{M}_x \hat{\T} = \{ e^{i \pi \sigma_x}i \sigma_y \mathcal{K}|m_{100} | \0 \}$, and pure spatial translation $\{\sigma_0 | 1 | \a_i \}$ ($i=1,2,3$). 
The little co-group of the A-L line is generated by $ \hat{M}_y \hat{\T} = ( \hat{C}_{6z} )^3 \hat{M}_x \hat{\T}  = \{ i \sigma_z \mathcal{K} | m_{210}| 0,0,3/2 \}$ and $ \hat{M}_x \hat{\T} $. We find that $ (\hat{M}_y \hat{\T})^2 \ket{\psi(u,0,\pi)} =-\ket{\psi(u,0,\pi)} $, leading to the Kramer's degeneracy along line A-L.

\subsection*{\texorpdfstring{$B_{2g}\oplus B_{1u}$}{B2g+B1u} spin texture in \texorpdfstring{FePO$_4$}{FePO4}}

\begin{figure}[tb]
    \centering
   \includegraphics[width=\columnwidth]{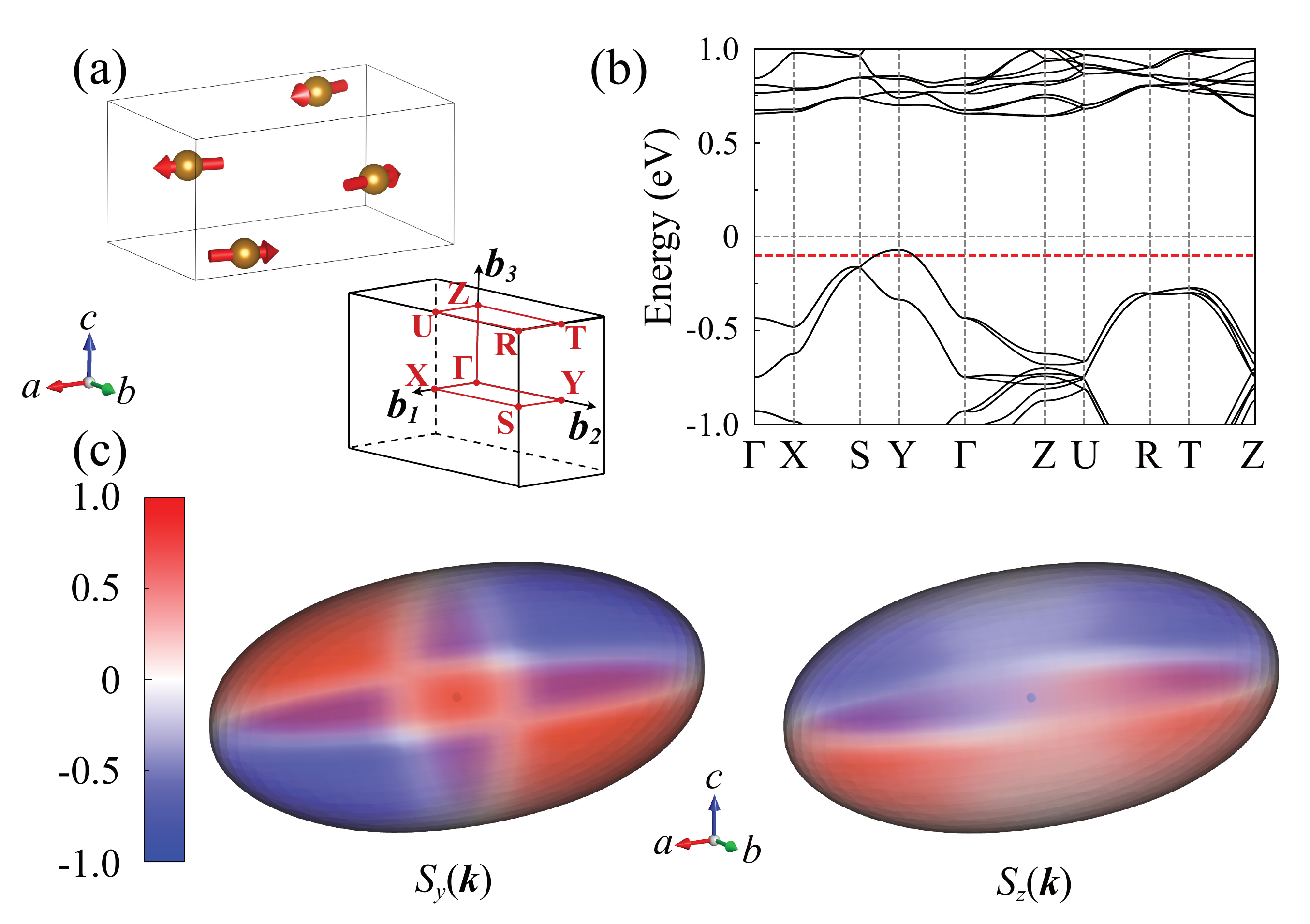}
    \caption[]{(a) Upper left: the magnetic structure of FePO$_4$ showing only the magnetic atoms (Fe). Bottom right: the BZ of FePO$_4$.  
    (b) The first-principle energy bands of FePO$_4$ along the path in the BZ shown in subfigure (a). (c) The spin texture $\vS(\k)$ on the Fermi surface centered at the Y point with hole doping  ($E = -0.08 \rm eV$). The $y$ and $z$ components ($S_y(\k)$ and $S_z(\k)$) are separately plotted.
    The color represents normalized $S_{\mu}(\k)$ ($\mu = y,z$), given as $S_{\mu}(\k)/\left(\max_{\k} |S_{\mu}(\k)|\right)$. $S_{\mu}(\k)$ forms an 1D real irrep of $mmm$.
    }
    \label{Fig: FePO4 all}
\end{figure}

The compound FePO$_4$ (No.~0.17) crystallizes in a primitive tetragonal lattice and exhibits a coplanar magnetic material described by MSG $P2_12_12_1$ (No.~19.1.119) \cite{rousse_magnetic_2003}. 
The spin moments are confined in $x,y$ plane and are shown in Fig.~\ref{Fig: FePO4 all} (a).
The SSG of FePO$_4$ is P62.3.24 $\rm GM3^{-} \oplus GM4^{+}$ with the parent space group being $Pnma$. 
Besides $\mS_{Z_2^T}$ [Eq.~\RefPSO] and lattice translations, the SSG is generated by $\mathcal{P} = \{ U_{\hy}(\pi) |\bar{1}| \0 \}$, $M_z = \{ U_{\hx}(\pi) | m_{001} | 1/2, 0 , 1/2 \} $, and $M_y = \{ I |m_{010} | 0 , 1/2 , 0 \}$.
The unitary lattice translations in this SSG are all accompanied by the identity spin rotation ($T_U = T_M$).
Hence, the SSG momentum and crystal momentum are equivalent, and the spin textures in the SBZ and magnetic BZ are the same.

Let us consider the spin texture $\vS(\k)$ by the method discussed in Sec.~\RefSecSpinTexture.
The spin-operation part of the $\mathcal{PT}$ operation is $T U_{\hx}(\pi)$, which flips the $S_x(\k)$ and leaves $S_{y,z}(\k)$ invariant.
Consequently, although spin moments are within the $x,y$ plane in real space, the spin texture $\vS(\k)$'s are confined to the $y,z$ plane in the $\k$ space. 
It is worth noting that this specific $\mathcal{PT}$ operation, which only forbids $S_z(\k)$ but not the entire $\vS(\k)$, does not exist in any MSG.
Therefore, the confinement of spin moments to two perpendicular planes in real and momentum space is generally impossible with only MSG symmetries.
$\vS(\k)$ respects a point group $\tilde{P} = \{s_g R_g | g \in \mG\}$, which is $mmm$ in this case.
The SSG requires that under $\mathcal{P}$, $\S_{y}(\k)$ is odd, $\S_z(\k)$ is even; under $M_z$ ($M_y$), $\S_{y,z}(\k)$ are both odd (even).
Thus, $S_{y}(\k)$ ($S_z(\k)$) forms a real 1D irrep of $mmm$, refered to as B$_{\rm 2g}$ (B$_{\rm 1u}$).   
The symmetry of $\vS(\k)$ of this material is also summarized in Sec.~\ref{sec:material-table}.
Then, we compare this analysis with the result of the first-principle calculation.

By the first-principle calculation, FePO$_4$ is identified as a band insulator [Fig.~\ref{Fig: FePO4 all} (b)].  
We study the spin texture $\vS(\k)$ on the Fermi surface ($E = -0.08 \rm eV$) with hole doping. 
The equal-energy surface is centered at Y point ($\k = (0,\pi,0)$) that is $mmm$-invariant. 
Although along the path S-Y-$\Gamma$ [Fig.~\ref{Fig: FePO4 all} (b)], the energy bands are double-degenerate, but this degeneracy does not exist in generic momentum as the square of the $\mathcal{PT}$ symmetry is equal to 1.
As a result, two independent hole-type Fermi surfaces exist.
The features of $\vS(\k)$ on two Fermi surfaces are similar, and we show one of them in Fig.~\ref{Fig: FePO4 all} (c). 
We verify that $S_x(\k) \equiv 0$ within the numerical accuracy. 
The red and blue colors represent the $S_{\mu}(\k)$ ($\mu = x,z$) is positive and negative, respectively, which are consistent with the prediction of SSGs.

\subsection*{Ths spin textures on the remaining Fermi surfaces of example materials} 
\begin{figure}[h]
    \centering
   \includegraphics[width=1 \linewidth]{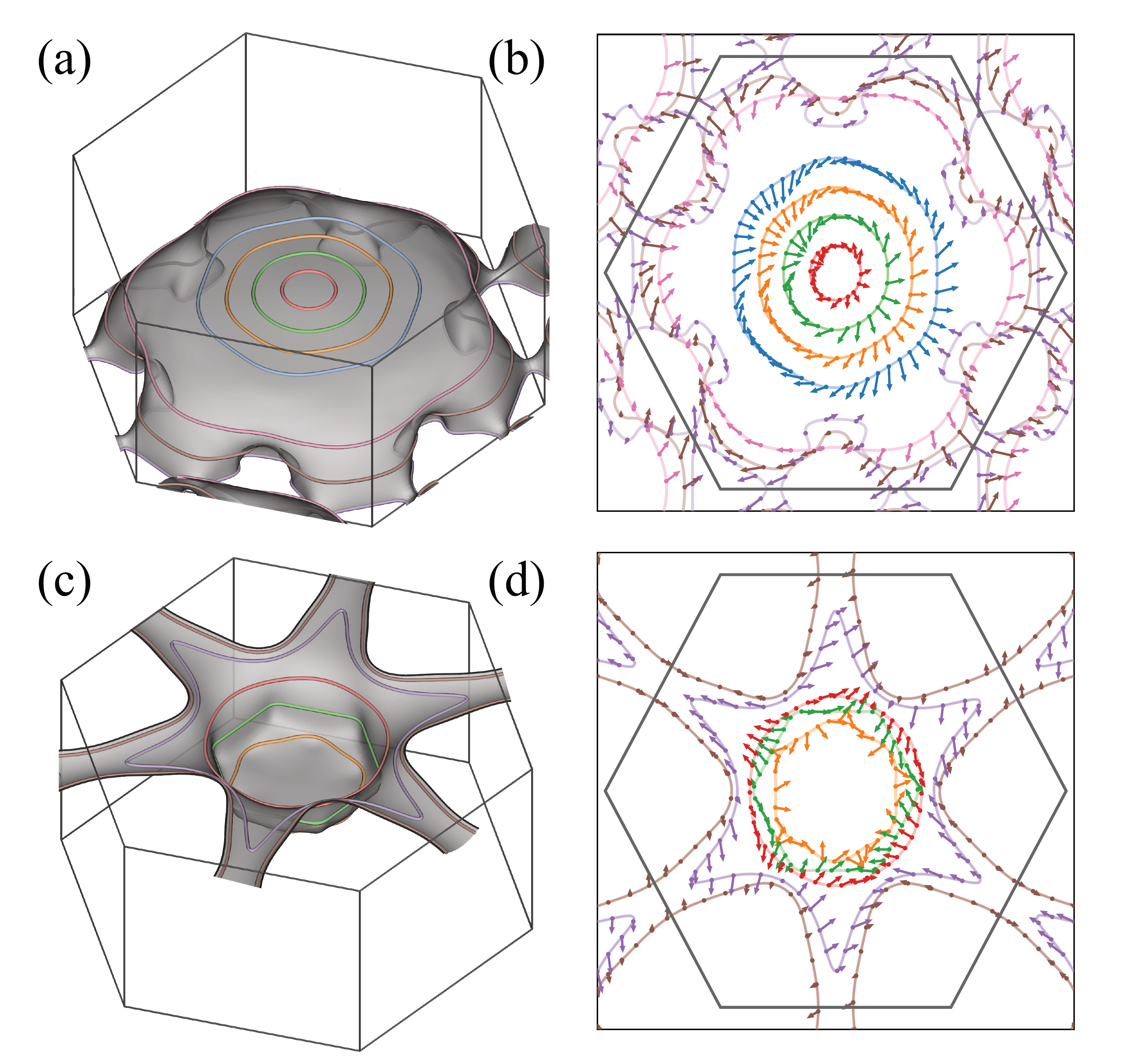}
    \caption[]{
    (a), (c) The second and third Fermi surfaces of Mn$_3$Ge.
    Different colors represent equal-$k_z$ lines.
    (b), (d) The spin texture $\vS(\k)$ on equal-$k_z$ lines labeled in the subfigures (a) and (c), respectively.
    Arrows represent the directions of $\vS(\k)$.}
    \label{Fig: app Mn3Ge}
\end{figure}

\begin{figure}[h]
    \centering
   \includegraphics[width=1 \linewidth]{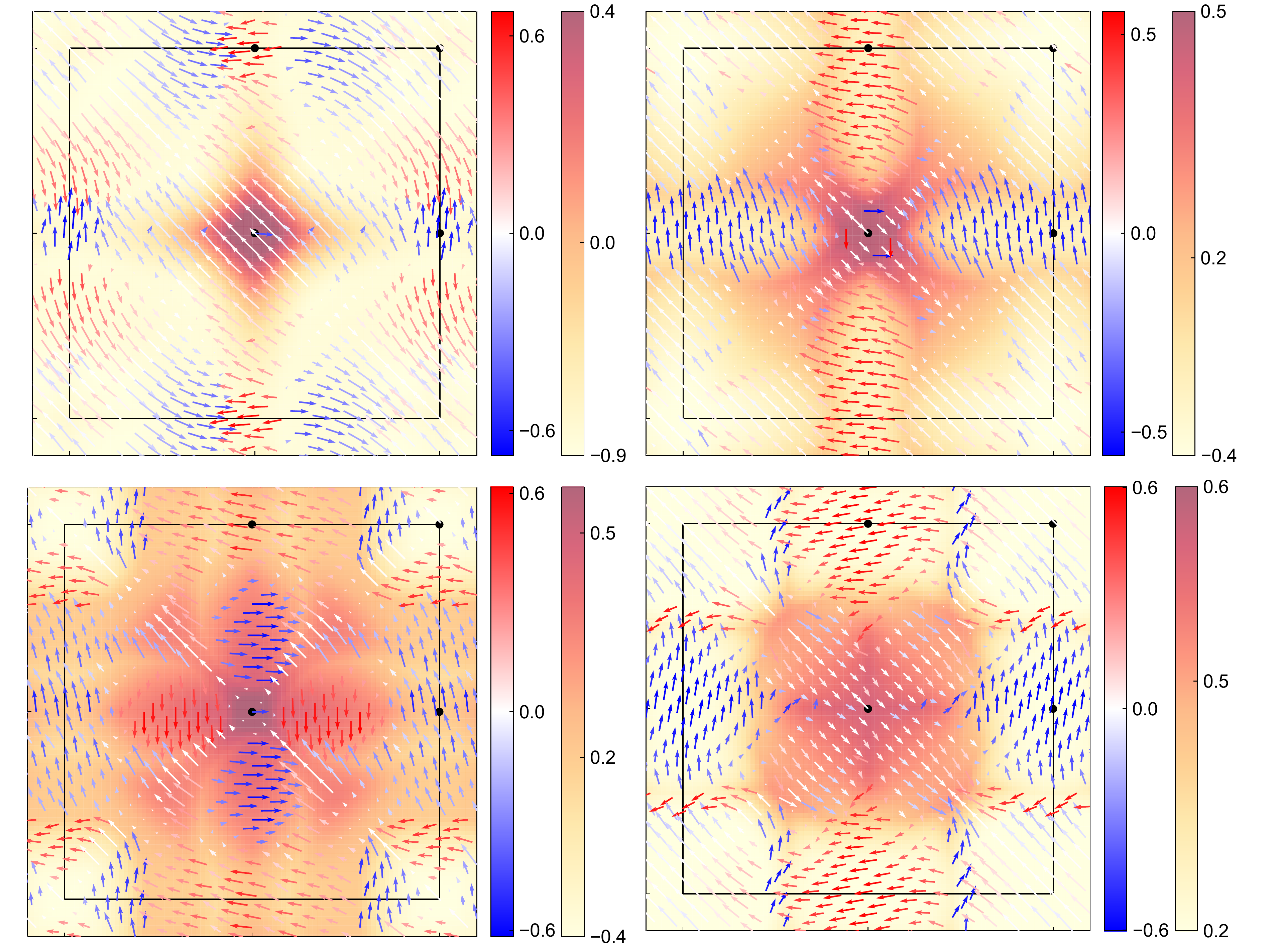}
    \caption[]{The spin texture $\vS(\k)$ on the $k_z = 0$ ($001$ plane) in the momentum space of different bands of Mn$_3$GaN.
    The background colors denote the values of $E(\k)$ (see the first color bar on the right).
    $\vS(\k)$'s are not confined to the $001$ plane, and the red or blue color (the second color bar on the right) represents the value of $S_z(\k)$.}
    \label{Fig: app Mn3GaN}
\end{figure}

\begin{figure}[h]
    \centering
   \includegraphics[width=1 \linewidth]{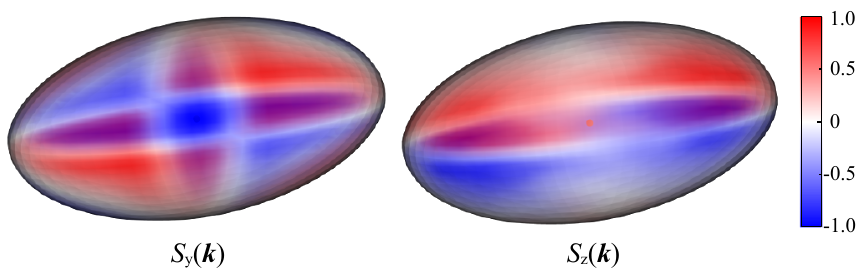}
    \caption[]{The spin texture $\vS(\k)$ on the second Fermi surface of FePO$_4$. The $y$ and $z$ components ($S_y(\k)$ and $S_z(\k)$) are plotted separately.
    The color represents normalized $S_{\mu}(\k)$ ($\mu = y,z$), given as $S_{\mu}(\k)/\left(\max_{\k} |S_{\mu}(\k)|\right)$. $S_{\mu}(\k)$ forms an 1D real irrep of $mmm$.}
    \label{Fig: app FePO4}
\end{figure}

In the main text and Appendix above, we showed the spin textures on one of the Fermi surfaces of Mn$_3$Ge, Mn$_3$GaN, and FePO$_4$.
In the subsection, we present the spin textures on the remaining Fermi surfaces of these materials (see Figs.~\ref{Fig: app Mn3Ge}, \ref{Fig: app Mn3GaN}, and \ref{Fig: app FePO4}).
One can verify that the symmetries of the spin textures are consistent with the SSG analysis in the main text.

\subsection*{The spin texture from first-principle calculation with SOC}
In the main text, we present the spin texture on the Fermi surface of Mn$_3$Ge obtained from first-principle calculation without SOC.
In the momentum space, the spin texture $\S(\k)$ is coplanar within the $x,y$ plane if the effect of SOC is not included.
We find that a tiny $|S_z(\k)| \approx 1/10 |S(\k)|$ is developed in the first-principle calculation with SOC, because $\mathcal{PT} = \{TU_{\hz}(\pi)|\bar{1}| 0, 0 , 0 \}$ is no longer an exact symmetry.
However, the spin texture of $(S_x(\k),S_y(\k))$ shows a similar vortex configuration with charge $S_V = 2$ (Fig.~\ref{Fig: app Mn3Ge SOC}), consistent with the analysis of SSG.

In the Appendix above, we study the spin textures of FePO$_4$. 
The SSG $\mathcal{PT}$ operation $\{ T U_{\hx}(\pi) | \bar1  | \0 \}$ require $\S(\k)$ to be confined within $y,z$ planes, as confirmed by the first-principle calculation without SOC. 
Similarly, $|S_x(\k)|  \approx 1/3 |S(\k)| $ is developed in the calculation with SOC, and $(S_y(\k),S_z(\k))$ shows a similar pattern as the case without SOC.
Note that the SOC-induced $|S_x(\k)|/|S(\k)|$ here is larger than SOC-induced $|S_z(\k)|/|S(\k)|$ observed in Mn$_3$Ge, which is explained as follows. 
Despite being in a coplanar magnetic structure, the spin moments of FePO$_4$ predominantly align along the $x$ direction (see Table~\ref{magnetic_moments}). 
If the $y$ components of magnetic moments were zero, the structure would be collinear anti-ferromagnetic, in which $|\vS(\k)| \equiv \0 $.
The canting towards the $y$ direction, which induces finite $S_y(\k)$ and $S_z\k)$, is a perturbative effect.
As a result, $|S_x(\k)|$ induced by the SOC, which is also a perturbative effect, is not negligible compared to $S_y(\k)$ and $S_z(\k)$.

\begin{figure}[h]
    \centering
   \includegraphics[width=1 \linewidth]{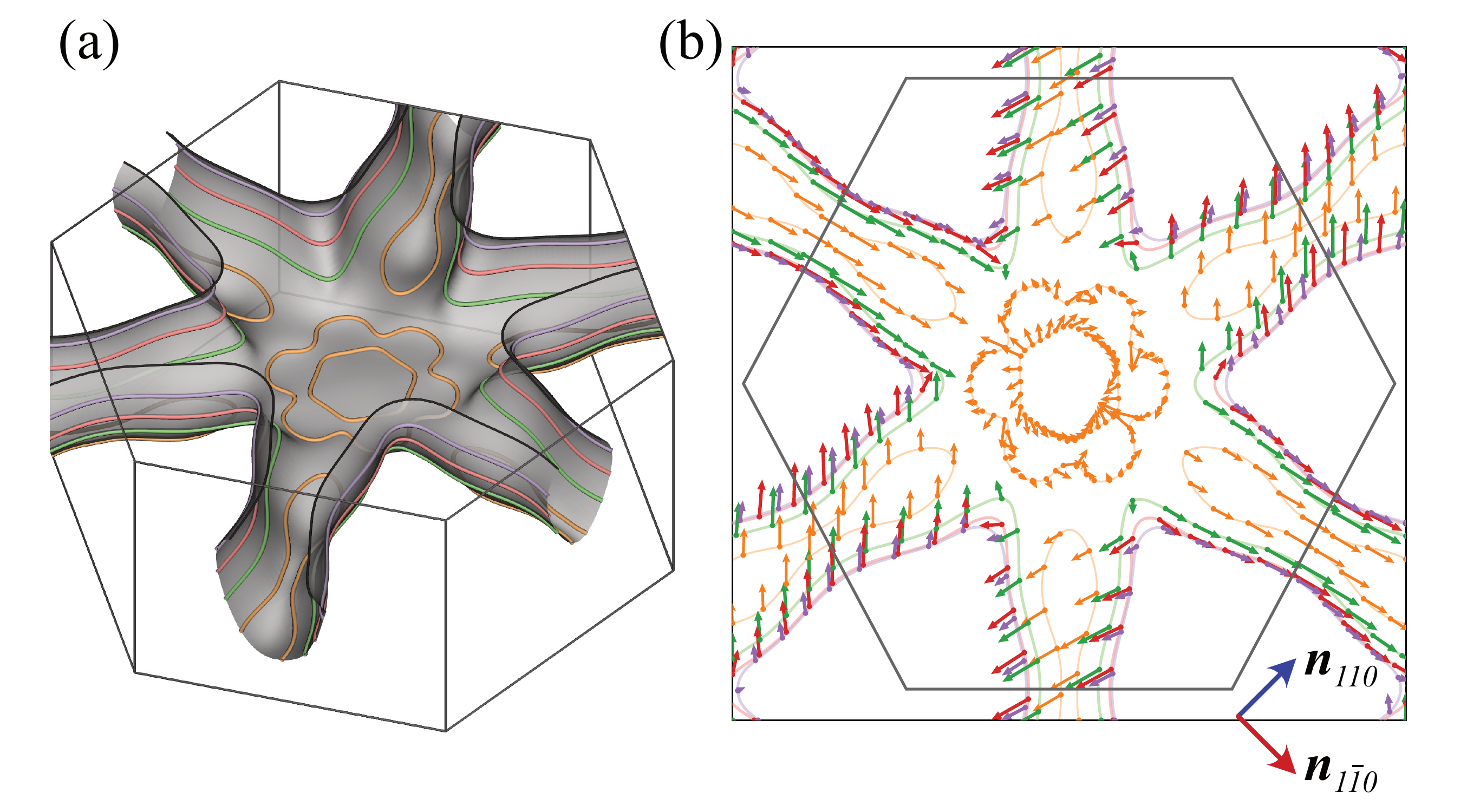}
    \caption[]{(a) The Fermi surface of Mn$_3$Ge obtained by the first-principle calculation with SOC. Different colors represent equal-$k_z$ lines.
    (b) The spin texture $\vS(\k)$ on equal-$k_z$ lines labeled in (a)}
    \label{Fig: app Mn3Ge SOC}
\end{figure}

\subsection*{Magnetic moments from first-principle calculations}

Table~\ref{magnetic_moments} compares the magnetic structures from DFT calculations to the experimental ones for all the seven materials discussed in this work.

\begin{table}[h]
    \caption[The magnetic moments of all the real materials studied in the main text and Appendices.]{
        The magnetic moments of all the real materials studied in the main text and Appendices. The column ``DFT'' (``Exp'') lists the magnetic moments obtained from the first-principle calculation (experimental measurement) with the unit being $\mu_B$. We only list magnetic moments of inequivalent atoms. 
    }
    \def\arraystretch{1.2}
    \begin{tabular}{p{54pt}<{\centering} | c | rrr | rrr}
        \hline \hline
            & \multirow{2}{*}{atom} & \multicolumn{3}{c|}{DFT} & \multicolumn{3}{c}{Exp} \\
        \cline{3-8}
            &      &  $M_x$ & $M_y$ & $M_z$  &  $M_x$ & $M_y$ & $M_z$ \\
        \hline
        \href{http://webbdcrista1.ehu.es/magndata/index.php?this_label=1.519}{CoSO$_4$} & Co1 &  0.00  &  2.46 & -0.76 & \multirow{2}{*}{0.00} & \multirow{2}{*}{2.99} & \multirow{2}{*}{-1.39} \\
        \cline{1-5}
        CoSO$_4$ (constrained) &  Co1 & 0.00  &  2.34 & -1.09 \\
        \hline
        \href{http://webbdcrista1.ehu.es/magndata/index.php?this_label=1.557}{FeGe$_2$} & Fe1 &  1.52  &  0.00 &  0.00 &  1.21 & 0.00 & 0.00 \\
        \hline

        \href{http://webbdcrista1.ehu.es/magndata/index.php?this_label=0.377}{Mn$_3$Ge (coplanar)} & Mn1 &  2.278  &  1.316 &  0.000 & 2.29 & 1.33 & 0.00 \\ 
        & Mn2& -2.278  &  1.315 &  0.000 &  -2.29 &  1.33 & 0.00\\
        \hline
        \href{http://webbdcrista1.ehu.es/magndata/index.php?this_label=0.177}{Mn$_3$GaN} & Mn1 &  0.000  &  1.819 & -1.819 & 0.00 & 0.83 & -0.83 \\
        \hline
        \href{http://webbdcrista1.ehu.es/magndata/index.php?this_label=1.525}{InMnO$_3$} & Mn1 &  0.000  &  3.389 &  0.000 & 0.00 & 3.25 & 0.00 \\
        & Mn2 &  0.000  &  3.389 &  0.000 &  0.00 & 3.25 & 0.00   \\        \hline
       \href{http://webbdcrista1.ehu.es/magndata/index.php?this_label=0.203}{Mn$_3$Ge (non-coplanar)} (constrained) & Mn1 &  0.000  &  1.981 &  1.643  & 0.00 & 1.30 & 1.10\\
              &  Mn2 &  1.716  & -0.991 &  1.643  & 1.12 & -0.65 & 1.10 \\      
              \hline
               \href{http://webbdcrista1.ehu.es/magndata/index.php?this_label=0.17}{FePO$_4$} & Fe1   & 3.85  &  0.86 &  0.00 & 4.06 & 0.90 & 0.00\\
        \hline       
              \hline
    \end{tabular}
    \label{magnetic_moments}
\end{table}

\begin{figure*}[tb]
    \centering
   \includegraphics[width=0.8 \linewidth]{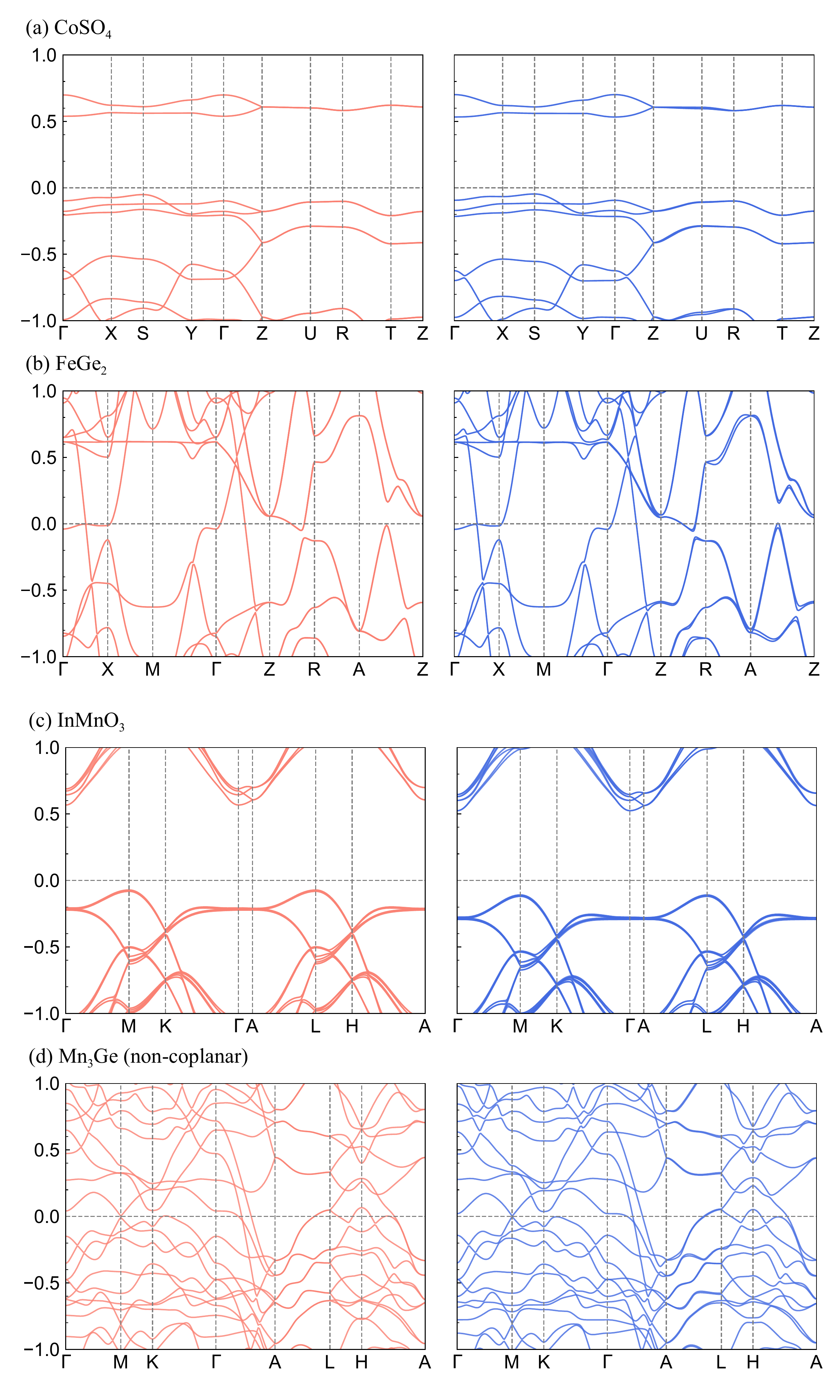}
    \caption[]{Band structures for CoSO$_4$, FeGe$_2$, InMnO$_3$, and Mn$_3$Ge (non-coplanar). Bands in the left and right columns are obtained without and with SOC, respectively. }
    \label{Fig: SOC 1}
\end{figure*}

\begin{figure*}[tb]
    \centering
   \includegraphics[width=0.8 \linewidth]{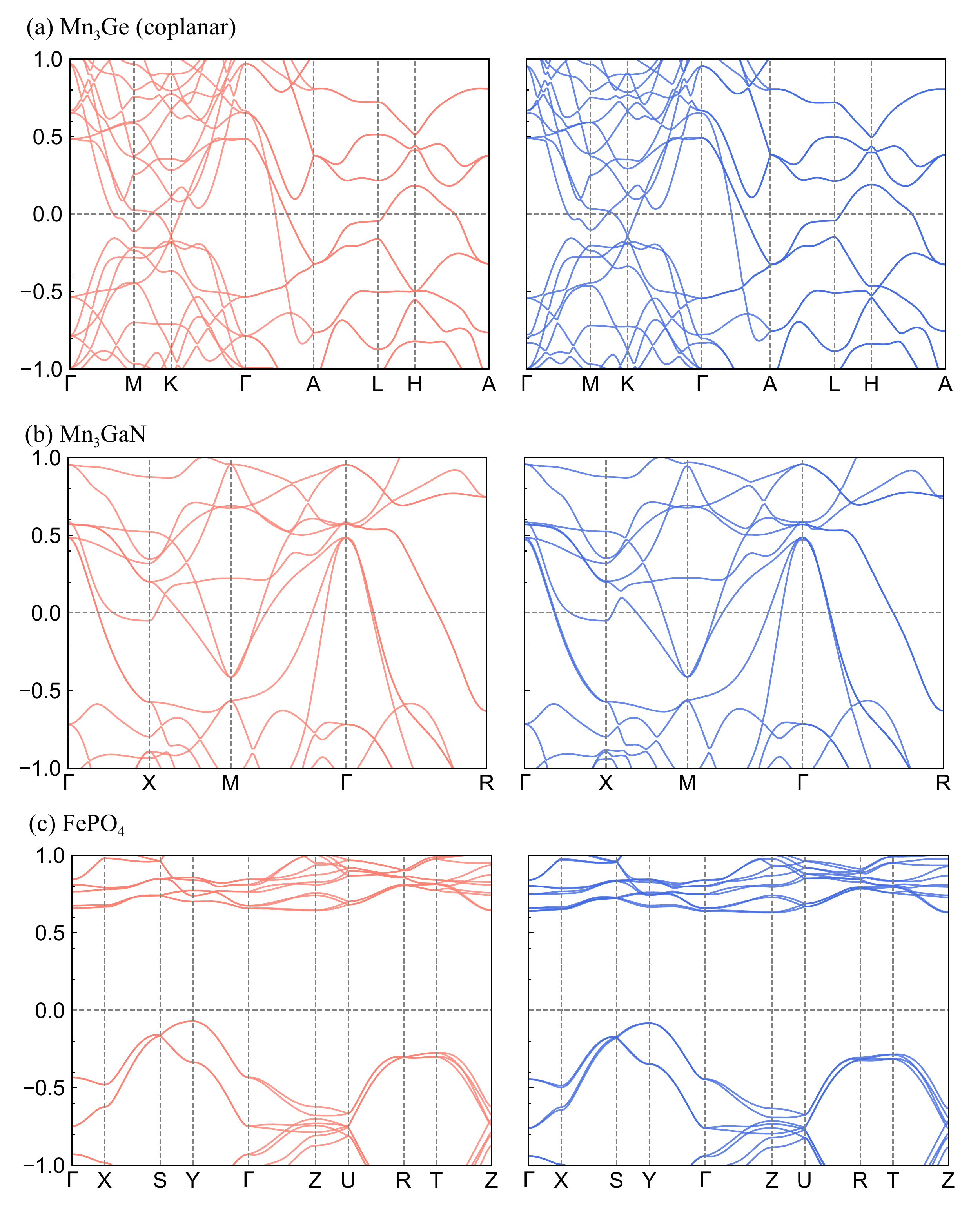}
    \caption[]{Band structures for Mn$_3$Ge (co-planar), Mn$_3$GaN, and FePO$_4$. Bands in the left and right columns are obtained without and with SOC, respectively. }
    \label{Fig: SOC 2}
\end{figure*}

We also perform first-principle calculations with the SOC effect for all the materials discussed in this work. 
Figs.~\ref{Fig: SOC 1} and \ref{Fig: SOC 2} compare the energy bands without and with SOC for all the materials discussed in this work. 

\section{Dirac theory for the 3D \texorpdfstring{$\mathbb{Z}_2$}{Z2} TI}
\label{App Sec: Dirac TI}

In this section, we discuss how the surface Dirac cone of the 3D $\mathbb{Z}_2$ TI emerges by the Dirac theory.
Let us consider $y=N_y$ surface of the 3D TI, and its Hamiltonian can be obtained by coupling helical edge modes of the 2D $\mathbb{Z}_2$ TI in different layers:
\begin{equation}
    \H_s (k_x,k_z)= \begin{pmatrix}
        v_F k_x\sigma_z & V_s(k_z) \\
         V_s(k_z)^{\dagger} & -v_F k_x\sigma_z
    \end{pmatrix} \, ,
\end{equation}
where $\pm v_F k_x\sigma_z $ describe the band dispersion of the helical edge states on $z = n$ and $z = n + 1/2$ ($n \in \mathbb{Z}$) planes, and $V_s(k_z)$ describe the hopping between them. 
This surface and $H_s$ should respect the symmetries $\hat{G}_x =  \{ I |  m_{100} | 0,0,1/2 \} $  and  $\hat{M}_z\hat{T} = \{ \T |  m_{001} | \0 \}$. 
The $G_x$ and $ M_z T $ in this basis are given as 
\begin{equation}
    D_s(G_x) = \begin{pmatrix}
        0 &  e^{ik_z} \\
        1  & 0
    \end{pmatrix} \otimes \sigma_0  , \, 
 D_s(M_z \T) = \begin{pmatrix}
        1 &  0 \\
        0  & e^{ik_z}
    \end{pmatrix} \otimes i \sigma_y \mathcal{K} \,.
\end{equation}
On the $G_x$-and-$M_zT$-invariant line $k_x = 0$,  $\H_s (0,k_z)$ satisfies
\begin{equation}
    D_s(g)^{-1} \H_s (0,k_z) D_s(g) = \H_s (0,k_z) \quad \text{for } g = G_x, M_z \T \, .
\end{equation}
After some algebra, we find that 
\begin{equation}
    V_s(k_z) = f(k_z) e^{-i k_z /2} \sigma_0 \, ,
\end{equation}
where $f(k_z) \in \mathbb{R}$. 
The eigenvalues of $\H_s(k_x,k_z)$ are given as 
\begin{equation}
    E_s(k_x,k_z) = \sqrt{ v_F^2k_x^2 + f(k_z)^2}
\end{equation}
As $V_s(0)$ should equal $V(2 \pi)$, $f(0) = -f(2\pi)$, which means that a $k_0$ exists such that $V_s(k_0) = 0$. The Dirac point should be located at $(k_x,k_z) = (0,k_0)$ or $(\pi,k_0)$.

\onecolumngrid

\clearpage

\section{Table of SSGs }
\label{sec:SSG-table}

The following tables present all spin space groups (SSGs). 
The SSGs ($\mG$) for collinear, coplanar, and non-coplanar magnetic structures can be represented by O(1), O(2), and O(3) representations of the parent space groups, respectively.
For each parent space group, SSGs for the collinear, co-planar, and non-coplanar magnetic structures are listed in the three adjacent tables.
The $230 \times 3$ tables are arranged in ascending order based on the indices of their parent space groups.
Each row in the column ``SSG" provides the name of an SSG, in the form $\alpha \mathcal{I.J.K}$ $\rho$, where $\alpha$ indicates collinear ($\alpha = \rm L$), coplanar ($\alpha = \rm P$), or non-coplanar magnetism ($\alpha = \rm N$), $\mathcal{I}=1\cdots 230$ is the index of the parent space group $P$, $\mathcal{J}=1\cdots 16$ is the type of the representation, $\mathcal{K}$ is the additional numbering of the SSG for given $\alpha \mathcal{I.J}$, and $\rho$ specifies the constituent irreps forming the O($N$) representation (see Sec.~\RefSumClass). 
The explicit representation matrices of these irreps can be accessed on the \href{https://www.cryst.ehu.es/cgi-bin/cryst/programs/representations.pl?tipogrupo=dbg}{Bilbao Crystallographic Server}~\cite{Elcoro17}.
In cases where several inequivalent O($N$) representations correspond to the same SSG, we provide one of them in the table and consistently use the chosen one in the nomenclature.
The remaining O($N$) representations describing the same SSG can be found in the text files~\cite{data}.

The additional columns of the tables are the properties of electronic states determined by the SSG and its representation theory.
For non-collinear SSGs, if all operations in the unitary translation subgroup $T_U$ commute with each other, the SSG possesses a commuting SSG Brillouin zone (SBZ); it possesses a non-commuting SBZ otherwise.
We indicate a commuting SBZ with "$\surd$" and a non-commuting SBZ with "$\times$" in the "Commute" column.
In cases of commuting SBZs, the column ``$d_{\rm SBZ}$'' (``$d_{\rm BZ}$'') lists the dimension $d_{\rm SBZ}$ ($d_{\rm BZ}$) of the span of the spin texture $\tS(\kk)$ ($\vS(\k)$) in the SBZ (magnetic BZ). 
Commuting SBZs can be further classified into symmorphic (``$\surd$'' in the column ``Symmorphic SBZ'') and nonsymmorphic (``$\times$'') depending on how SSG momentum $\kk$ is transformed under SSG operations.
In a symmorphic SBZ, the momentum transforms normally under the SSG operations, {\it i.e.,} $\kk \to \kk^{\prime} = s_g R_g \kk$, with $R_g$ being the point group part of SSG operation $g$ and $s_g$ being $1$ ($-1$) for unitary (anti-unitary) $g$.
The symmetry of the spin texture $\tS(\kk)$ in the SBZ can be characterized by a $d_{\rm SBZ}$-dimensional representation $\tilde{\rho}$ of a point group $\tilde{P} = \{s_g R_g | g\in \mG\}$.
It is also notable that in the cases of symmorphic SBZs, $d_{\rm BZ} = d_{\rm SBZ}$, and $\vS(\k)$ realizes the same representation $\tilde{\rho}$ of $\tilde{P}$ as $\tS(\k)$ (see Sec.~\RefSecSpinTexture).
The point group $\tilde{P}$ and its representation $\tilde{\rho}$, expressed as a direct sum of the irreps of $\tilde{P}$, are shown in the columns ``Sym of $\tS(\kk)$'' and ``representations'', respectively.
We adopted the notations of irreps of crystalline point groups in Ref.~\cite{bradley10}, where the explicit forms and characters of these irreps can also be accessed.
Conversely, in a nonsymmorphic SBZ, $\kk$ transforms to $s_g \left(R_g \kk + \qq_g\right)$, where $\qq_g$ is not a reciprocal lattice.
Characterizing the symmetry of $\tS(\kk)$ in a nonsymmorphic SBZ is left for future study.
For collinear SSGs, electronic Bloch states are effectively described by the linear representations of single-valued grey space groups. 
It is unnecessary to introduce the concepts of non-commuting or nonsymmorphic SBZ. 
For collinear SSGs, $d_{\rm BZ} \leq 1$ because the electronic Hamiltonian can be diagonalized into spin-up and spin-down parts.

\section{Table of published magnetic materials}
\label{app sec ssg materials}
We identify and show the SSGs of 1595 published experimental magnetic structures in the \href{http://webbdcrista1.ehu.es/magndata/}{MAGNDATA} database~\cite{gallego_magndata_2016,gallego_magndata_2016-1} on the Bilbao Crystallographic Server in the following tables. 
The SSGs of collinear, coplanar, and non-coplanar magnetic structures are displayed in three separate tables.
Within each table, the materials are further divided into two groups: light-element materials, where all consistent elements are from the first four periods of the periodic table, and heavy-element materials.
The column ``BCSID'' contains the labels of magnetic structures in the database.
The column ``MSG'' lists the magnetic space group of each magnetic material, using the Opechowski-Guccione notation~\cite{opechowski95}.
``PSG'' denotes the parent space group of the SSG. 
The column ``Trans from parent structure'' shows the transformation matrix and the origin shift from the paramagnetic structure's cell to the conventional cell of the parent space group.
The choice of paramagnetic structures' cells can be found in the magnetic crystallographic information files in the \href{http://webbdcrista1.ehu.es/magndata/}{MAGNDATA} database.
The column ``Rep'' lists the O($N$) ($N= 1,2,3$) representation realized by the transformation of the spin moments under operations of the parent space group.  
The column ``SSG'' lists the names of the SSGs. 
Note that inequivalent O($N$) representations can describe the same SSGs; hence, the representation in the "Rep" column may differ from the representation in the name of an SSG.

For the convenience of future studies, we prepare additional tables listing all the materials with non-commuting or nonsymmorphic SBZs.
We also include tables for materials exhibiting non-zero spin texture $\tS(\kk)$ ($d_{\rm SBZ} \neq 0$) with the symmorphic SBZ. 
The columns ``$d_{\rm SBZ}$'', ``Sym of $\tS(\kk)$'', and ``Representation'' denote the same meaning as explained for the tables of SSGs.
If the irreps listed in the "representation" column are all trivial irreps, we classify the spin texture as trivial; otherwise, it is classified as non-trivial.

\blue{\section{Notations of SSGs in the related works}}
We note that related works on SSGs were carried out by Chen Fang's~\cite{fang2023_ssg} group and Qi-Hang Liu's group~\cite{liu2023_ssg}. 
In this section, we explain the notations of SSGs they adopted. 
Their notations of SSGs are related to their classification approach, which is akin to Livtin's (also discussed in Sec. VI in the main text).
Litvin's approach involves the introduction of a normal subgroup $P_0=\{ \{R_g | \v_g \} | X_g U_g = I, g\in \mG \}$ of the parent space group $P$, consisting of pure spatial operations that leave the magnetic structure unchanged. 
Additionally, a supergroup $\mathcal{B} = \{ X_g U_g| g\in \mG \}$ of the pure-spin-operation group $\mS$ is introduced, containing the spin operation parts of all SSG operations. 
$P$ can be decomposed as cosets of $P_0$ 
\begin{equation}
    P = p_1 P_0 + p_2P_0 + \cdots + p_n P_0\ .
\end{equation}
Distinct SSGs are defined based on distinct $P$, $P_0$ and $\mathcal{B}$, and the isomorphisms between $P/P_0$ and $\mathcal{B}/\mS$, modulo the equivalence induced by coordinate transformation.
The group $P_0$ has a translation subgroup $T_M = \left\{ \{ 1| \v_g \} | X_g U_g = I, R_g = 1,  g \in \mG \right\}$ consisting of pure-translation operations, which is also a subgroup of the translation subgroup $ T = \left\{ \{ 1| \v_g \} | R_g = 1,  g \in \mG \right\} $ of $P$. 
    
In Ref.~\cite{fang2023_ssg}, the name of an SSG is given by four indices and a letter $N_{\rm SG}.I_{k}.I_{t}.N_{\rm rep}.\mathcal{\alpha}$.
Here $N_{\rm SG}$ is the group index for the parent space group $P$; $I_k = |T|/|T_M|$, equalling the ratio between the size of the magnetic supercell and the size of the SSG unit cell.
The index $I_t$ represents the ratio of distinct point-group operations in $P$ to $P_0$.
The last index $N_{\rm rep}$ specifies the SSGs with the same first three indices, which characterize distinct $\mathcal{B}$'s and distinct isomorphisms between $P/P_0$ and $\mathcal{B}/\mS$.
The letter $\alpha = L$ for collinear SSGs, and $\alpha = P$ for coplanar SSGs, while $\alpha$ is omitted for non-coplanar SSGs. 

In Ref.~\cite{liu2023_ssg}, the name of an SSG is given by four indices $N_{\rm SG2}.N_{\rm SG}.I_{k}.m$.
Similarly, $N_{\rm SG}$ is the group index for the parent space group, and $I_k =  |T|/|T_M|$.  
The index $N_{\rm SG}$ is the group index of space group $P_0$ defined before.
The last index $m$ specifies the SSGs with the same first three indices. 
In their nomenclature, they do not specify whether an SSG is for collinear, coplanar, or non-coplanar magnetic structures.

To illustrate the correspondence between their SSG notations and ours, we consider SSG N143.5.1 DT1DU1 (in our notation). 
As also discussed in Sec.~II C in the main text, N143.5.1 DT1DU1 with different momenta $(0,0,u\pi)$ [$u \in (0,1)$] on the line DT1 have different sizes of magnetic supercells, leading to distinct SSGs in their classification. 
For $\frac{1}{2}u = p/q$ with $p,q$ being coprime integers, the lattice vectors of pure translation group $T_0$ of this SSG are $\a_1,\a_2$ and $q \a_3$ (see Sec.~II C4 in the main text).
Thus, the index $I_k = q$.
Since the $C_{3z}$ operation in this SSG is accompanied by the identity spin operation, independent of $u$. 
The subgroup $P_0$ is given as $T_0 + T_0 C_{3z} + T_0 C_{3z}^2$, isomorphic to the space group $P3$ (No.~143). 
Thus, the index $I_t = 1$, and the group index $N_{\rm SG2} = 143$.
Note that the above analysis can only specify the first three indices in their notations, and the fourth index is determined by explicitly considering the spin operations of the generators of the space group.
In the case of $u = 2/3$, $I_k = 3$, we find that it corresponds to the SSG 143.3.1.1 in Ref.~\cite{fang2023_ssg}, and corresponds to 143.143.3.1 in Ref.~\cite{liu2023_ssg}.
In the case of $u = 1/2$, $I_k = 4$, it corresponds to the SSG 143.4.1.1 and 143.143.4.1 in these two works, respectively. 
If $u = 1/6$, $I_k = 12$, it corresponds to the SSG 143.12.1.1 in Ref.~\cite{fang2023_ssg}. 
Such an SSG does not exist in Ref.~\cite{liu2023_ssg} since they only consider the case with $I_k \leq 8$.
The SSG N143.5.1 DT1DU1 with $\frac{1}{2}u = 1/p$ and $p > 12 $ is only studied in our classification, but not studied in either of their works, because Ref.~\cite{fang2023_ssg} only consider the case $I_k \leq 12$.

We also map an SSG they study to the notation in our classification (see the related technique in Appendix~\ref{Iden_SSG}). 
In Ref.~\cite{fang2023_ssg}, the authors study non-coplanar SSG 83.2.4.1 (in their notation) with parent space group $P4/m$ (No.~83). The SSG is generated by 
\begin{equation} 
  \{I | m_{001}| {\bm 0} \}, \{I | 4_{001}^+| {\bm 0} \},  \{T \cdot U_{\hat{z}} (\pi) | 1| 001 \},   \{I | 1| 010 \}, \text{ and } \{I | 1| 100 \} \, .
\end{equation}
Since the corresponding O(3) matrices of translation $\{1|\a_i \}$ (i = 1,2,3) are $I, I$, and ${\rm diag}(1,1,-1)$, respectively. 
Thus, the momenta of the O(3) representation are $\k_1 = \0$ (GM point), $\k_2 = \0$ (GM point), and $\k_3 = (0,0,\pi)$ (Z point).
Meanwhile, the representation matrices of $\{m_{001}| {\bm 0} \}$ and $\{4_{001}^+| {\bm 0} \}$ are both identity matrices. 
Therefore, this SSG's corresponding O(3) representation is $\rm GM1^+ \oplus GM1^+ \oplus Z1^+$. 
Since Z1$^+$ is a 1D real irrep, this SSG belongs to type II and is named N83.2.7 Z1$^+$. 
Note that the trivial irrep $ \rm GM1^+$ is omitted in our notation. 

In Ref.~\cite{liu2023_ssg}, the authors study SSG 13.54.1.2 (in their notation) with parent space group $Pcca$ (No.~54). The SSG is generated by which is generated by 
\begin{equation} 
  \{U_{\hz}(\pi) | 2_{010}| 0 0 \frac{1}{2} \}, \{I | 2_{001}| 0 0 \frac{1}{2} \},  \{I | \bar{1}| \0 \}, \{I | 4_{001}^+| {\bm 0} \},  \{I | 1| \a_i \} (i = 1,2,3) \, .
\end{equation}
Since the translations are all accompanied by the identity matrices, the three momenta of the O(3) representations are all $\0$ (GM point). 
The representation matrices for other operations are diagonal, and hence the O(3) representation is given by the direct sum of three 1D irreps. 
By studying the explicit forms of spin operations, we find that the O(3) representation is  $\rm GM1^+\oplus \rm GM2^+ \oplus GM2^+$, and the SSG is named N54.3.12 $\rm \rm GM2^+ \oplus GM2^+$ in our classification.

\listoftables

\setcounter{table}{0}
\renewcommand{\thetable}{S\arabic{table}}
\renewcommand{\arraystretch}{1.3}
\LTcapwidth=\textwidth

\input{table_dim_all}
\begin{longtable}{cp{3cm}<{\centering}|cp{3.4cm}<{\centering}cp{1.8cm}c}
    \caption{SSGs of collinear magnetic structures.} \\
    \hline \hline
        BCSID  & Chemical  formula   & MSG &  \makecell[c]{Trans from \\ parent structure}  & PSG & Rep           & SSG              \\
        \hline \endfirsthead 
    \caption*{SSGs of collinear magnetic structures (continued).} \\
     \hline \hline
        BCSID  & Chemical  formula   & MSG &  \makecell[c]{Trans from \\ parent structure}  & PSG & Rep           & SSG              \\
        \hline \endhead 
    \multicolumn{2}{c|}{  light-element materials } & &  & & \\ \hline
    0.4 & NiCr$_{2}$O$_{4}$ & \makecell[l]{No.~70.4.619 \\ $ Fd^{\p}d^{\p}d$} & $  a,b,c;0,0,0  $ & \makecell[l]{No.~141 \\ $ I4_1/amd$} & $\rm GM1^+$ & \makecell[l]{L141.1.1 \\ $\rm GM1^+$} \\ \hline 
    0.5 & Cr$_{2}$S$_{3}$ & \makecell[l]{No.~2.1.4 \\ $ P\bar{1}$} & $  a,b,c;0,0,0  $ & \makecell[l]{No.~148 \\ $ R\bar{3}$} & $\rm GM1^+$ & \makecell[l]{L148.1.1 \\ $\rm GM1^+$} \\ \hline 
    0.15 & MnF$_{2}$ & \makecell[l]{No.~136.5.1156 \\ $ P4_2^{\p}/mnm^{\p}$} & $  a,b,c;0,0,0  $ & \makecell[l]{No.~136 \\ $ P4_2/mnm$} & $\rm GM4^+$ & \makecell[l]{L136.2.6 \\ $\rm GM4^+$} \\ \hline 
    0.19 & MnTiO$_{3}$ & \makecell[l]{No.~148.3.1249 \\ $ R\bar{3}^{\p}$} & $  a,b,c;0,0,0  $ & \makecell[l]{No.~148 \\ $ R\bar{3}$} & $\rm GM1^-$ & \makecell[l]{L148.2.1 \\ $\rm GM1^-$} \\ \hline 
    0.23 & Ca$_{3}$Mn$_{2}$O$_{7}$ & \makecell[l]{No.~36.3.251 \\ $ Cm^{\p}c2_1^{\p}$} & $  a,b,c;0,0,0  $ & \makecell[l]{No.~36 \\ $ Cmc2_1$} & $\rm GM2$ & \makecell[l]{L36.2.1 \\ $\rm GM2$} \\ \hline 
    0.24 & LiMnPO$_{4}$ & \makecell[l]{No.~62.9.510 \\ $ Pn^{\p}m^{\p}a^{\p}$} & $  a,b,c;0,0,0  $ & \makecell[l]{No.~62 \\ $ Pnma$} & $\rm GM3^-$ & \makecell[l]{L62.2.5 \\ $\rm GM3^-$} \\ \hline 
    0.35 & Cu$_{2}$OSeO$_{3}$ & \makecell[l]{No.~146.1.1240 \\ $ R3$} & $  a,b,c;0,0,0  $ & \makecell[l]{No.~198 \\ $ P2_13$} & $\rm GM1$ & \makecell[l]{L198.1.1 \\ $\rm GM1$} \\
\hline 
    \hline 
    \end{longtable}

    \LTcapwidth=\textwidth

    \begin{longtable}{cp{3cm}<{\centering}|cp{3.2cm}<{\centering}cp{2.8cm}c}
        \caption{SSGs of coplanar magnetic structures.} \\
        \hline \hline
        BCSID  & Chemical  formula   & MSG &  \makecell[c]{Trans from \\ parent structure}  & PSG & Rep           & SSG              \\ \endfirsthead
        \caption*{SSGs of coplanar magnetic structures (continued).} \\
        \hline \hline
        BCSID  & Chemical  formula   & MSG &  \makecell[c]{Trans from \\ parent structure}  & PSG & Rep           & SSG              \\ 
        \hline \endhead
    \multicolumn{2}{c|}{  light-element materials } & &  & & \\ \hline
    0.7 & ScMnO$_{3}$ & \makecell[l]{No.~185.5.1433 \\ $ P6_3c^{\p}m^{\p}$} & $  a,b,c;0,0,0  $ & \makecell[l]{No.~185 \\ $ P6_3cm$} & $\rm GM6$ & \makecell[l]{P185.6.2 \\ $\rm GM6$} \\ \hline 
    0.8 & ScMnO$_{3}$ & \makecell[l]{No.~173.1.1360 \\ $ P6_3$} & $  a,b,c;0,0,0  $ & \makecell[l]{No.~185 \\ $ P6_3cm$} & $\rm GM6$ & \makecell[l]{P185.6.2 \\ $\rm GM6$} \\ \hline 
    0.17 & FePO$_{4}$ & \makecell[l]{No.~19.1.119 \\ $ P2_12_12_1$} & $  a,b,c;0,0,0  $ & \makecell[l]{No.~62 \\ $ Pnma$} & $\rm GM3^-\oplus GM4^+$ & \makecell[l]{P62.3.24 \\ $\rm GM3^-\oplus GM4^+$} \\ \hline 
    0.28 & LiFeSi$_{2}$O$_{6}$ & \makecell[l]{No.~14.4.89 \\ $ P2_1/c^{\p}$} & $  a,b,c;0,0,0  $ & \makecell[l]{No.~14 \\ $ P2_1/c$} & $\rm GM1^-\oplus GM2^-$ & \makecell[l]{P14.3.3 \\ $\rm GM1^-\oplus GM2^-$} \\ \hline 
    0.36 & NiF$_{2}$ & \makecell[l]{No.~58.6.476 \\ $ Pnn^{\p}m^{\p}$} & $  a,b,c;0,0,0  $ & \makecell[l]{No.~136 \\ $ P4_2/mnm$} & $\rm GM4^+$ & \makecell[l]{P136.2.6 \\ $\rm GM4^+$} \\ \hline 
    0.40 & Mn$_{2}$O$_{3}$-alpha & \makecell[l]{No.~61.1.497 \\ $ Pbca$} & $  a,b,c;0,0,0  $ & \makecell[l]{No.~61 \\ $ Pbca$} & $\rm GM2^+\oplus GM3^+$ & \makecell[l]{P61.3.5 \\ $\rm GM2^+\oplus GM3^+$} \\ \hline 
    0.41 & Mn$_{2}$O$_{3}$-alpha & \makecell[l]{No.~61.1.497 \\ $ Pbca$} & $  a,b,c;0,0,0  $ & \makecell[l]{No.~61 \\ $ Pbca$} & $\rm GM2^+\oplus GM3^+$ & \makecell[l]{P61.3.5 \\ $\rm GM2^+\oplus GM3^+$} \\
\hline 
    \hline 
    \end{longtable}
    
    \begin{longtable}{cp{3.2cm}<{\centering}|cp{2.8cm}<{\centering}cp{4cm}c}
        \caption{SSGs of non-coplanar magnetic structures.} \\
        \hline \hline
        BCSID  & Chemical  formula   & MSG &  \makecell[c]{Trans from \\ parent structure}  & PSG & Rep           & SSG              \\ \endfirsthead
            \caption*{SSGs of non-coplanar magnetic structures (continued).} \\
        \hline \hline
        BCSID  & Chemical  formula   & MSG &  \makecell[c]{Trans from \\ parent structure}  & PSG & Rep           & SSG              \\ 
        \hline \endhead
        \hline
    \multicolumn{2}{c|}{  light-element materials } & &  & & \\ \hline
    0.60 & $\rm [NH_{2}(CH_{3})_{2}]_n$ $\rm [Fe^{III}Fe^{II}(HCOO)_{6}]_n $ & \makecell[l]{No.~167.5.1338 \\ $ R\bar{3}c^{\p}$} & $  a,b,c;0,0,0  $ & \makecell[l]{No.~167 \\ $ R\bar{3}c$} & $\rm GM3^+$ & \makecell[l]{N167.6.1 \\ $\rm GM3^+$} \\ \hline 
    0.64 & MnV$_{2}$O$_{4}$ & \makecell[l]{No.~88.1.742 \\ $ I4_1/a$} & $  a,b,c;0,0,0  $ & \makecell[l]{No.~88 \\ $ I4_1/a$} & $\rm GM3^+GM4^+$ & \makecell[l]{N88.4.1 \\ $\rm GM3^+GM4^+$} \\ \hline 
    0.70 & Na$_{3}$Co(CO$_{3}$)$_{2}$Cl & \makecell[l]{No.~203.1.1527 \\ $ Fd\bar{3}$} & $  a,b,c;0,0,0  $ & \makecell[l]{No.~203 \\ $ Fd\bar{3}$} & $\rm GM4^+$ & \makecell[l]{N203.15.1 \\ $\rm GM4^+$} \\ \hline 
    0.78 & NiN$_{2}$O$_{6}$ & \makecell[l]{No.~148.1.1247 \\ $ R\bar{3}$} & $  a,b,c;0,0,0  $ & \makecell[l]{No.~148 \\ $ R\bar{3}$} & $\rm T1^+\oplus T2^+T3^+$ & \makecell[l]{N148.10.6 \\ $\rm T1^+\oplus T2^+T3^+$} \\ \hline 
    0.96 & CoSO$_{4}$ & \makecell[l]{No.~62.1.502 \\ $ Pnma$} & $  a,b,c;0,0,0  $ & \makecell[l]{No.~62 \\ $ Pnma$} & $\rm GM2^+\oplus GM3^+\oplus GM4^+$ & \makecell[l]{N62.9.42 \\ $\rm GM2^+\oplus GM3^+\oplus GM4^+$} \\ \hline 
    0.102 & Mn$_{2}$GeO$_{4}$ & \makecell[l]{No.~62.1.502 \\ $ Pnma$} & $  a,b,c;0,0,0  $ & \makecell[l]{No.~62 \\ $ Pnma$} & $\rm GM2^+\oplus GM3^+\oplus GM4^+$ & \makecell[l]{N62.9.42 \\ $\rm GM2^+\oplus GM3^+\oplus GM4^+$} \\ \hline 
    0.103 & Mn$_{2}$GeO$_{4}$ & \makecell[l]{No.~14.1.86 \\ $ P2_1/c$} & $  -b,-c,a;0,0,0  $ & \makecell[l]{No.~14 \\ $ P2_1/c$} & $\rm GM2^+\oplus GM2^+$ & \makecell[l]{N14.3.5 \\ $\rm GM2^+\oplus GM2^+$} \\
\hline 
    \hline 
    \end{longtable}

\begin{longtable}{cp{4cm}<{\centering}|p{3cm}<{\centering}p{4cm}<{\centering}}
\caption{Coplanar magnetic materials with non-commuting SBZs.} \\
\hline \hline
BCSID  & Chemical  formula   & PSG & SSG \\ \hline \endfirsthead
\caption*{Coplanar magnetic materials with non-commuting SBZs (continued).} \\
\hline \hline
BCSID  & Chemical  formula   & PSG & SSG \\ \hline \endhead
\multicolumn{2}{c|}{  light-element materials } & &   \\ \hline
2.31 & Mn$_{3}$ZnN & \makecell[c]{No.~123 \\ $ P4/mmm$} & \makecell[c]{P123.3.151 \\ $\rm M1^+\oplus Z1^+$} \\ \hline 
\multicolumn{2}{c|}{  heavy-element materials } & &   \\ \hline
2.2 & Sr$_{2}$F$_{2}$Fe$_{2}$OS$_{2}$ & \makecell[c]{No.~12 \\ $ C2/m$} & \makecell[c]{P12.7.1 \\ $\rm V1^+$} \\ \hline 
2.6 & Nd$_{2}$CuO$_{4}$ & \makecell[c]{No.~139 \\ $ I4/mmm$} & \makecell[c]{P139.7.1 \\ $\rm X1^+$} \\ \hline 
2.7 & Sm$_{2}$CuO$_{4}$ & \makecell[c]{No.~139 \\ $ I4/mmm$} & \makecell[c]{P139.7.1 \\ $\rm X1^+$} \\ \hline 
2.13 & UP & \makecell[c]{No.~139 \\ $ I4/mmm$} & \makecell[c]{P139.7.1 \\ $\rm X1^+$} \\ \hline 
2.14 & NdMg & \makecell[c]{No.~123 \\ $ P4/mmm$} & \makecell[c]{P123.7.3 \\ $\rm X1^+$} \\
\hline 
\hline 
\end{longtable}

\begin{longtable}{cp{4cm}<{\centering}|p{3cm}<{\centering}p{4cm}<{\centering}}
\caption{Non-coplanar magnetic materials with non-commuting SBZs.} \\
\hline \hline
BCSID  & Chemical  formula   & PSG & SSG \\ \hline \endfirsthead
\caption*{Non-coplanar magnetic materials with non-commuting SBZs (continued).} \\
\hline \hline
BCSID  & Chemical  formula   & PSG & SSG \\ \hline \endhead
\multicolumn{2}{c|}{  light-element materials } & &   \\ \hline
3.4 & MgCr$_{2}$O$_{4}$ & \makecell[c]{No.~119 \\ $ I\bar{4}m2$} & \makecell[c]{N111.13.28 \\ $\rm M2\oplus X2$} \\ \hline 
3.24 & CaFe$_{3}$Ti$_{4}$O$_{12}$ & \makecell[c]{No.~148 \\ $ R\bar{3}$} & \makecell[c]{N148.16.2 \\ $\rm F1^+$} \\ \hline 
\multicolumn{2}{c|}{  heavy-element materials } & &   \\ \hline
3.2 & UO$_{2}$ & \makecell[c]{No.~225 \\ $ Fm\bar{3}m$} & \makecell[c]{N225.16.1 \\ $\rm X1^+$} \\ \hline 
3.6 & DyCu & \makecell[c]{No.~221 \\ $ Pm\bar{3}m$} & \makecell[c]{N221.16.1 \\ $\rm M1^+$} \\ \hline 
3.7 & NpBi & \makecell[c]{No.~225 \\ $ Fm\bar{3}m$} & \makecell[c]{N225.16.1 \\ $\rm X1^+$} \\ \hline 
3.8 & NdZn & \makecell[c]{No.~221 \\ $ Pm\bar{3}m$} & \makecell[c]{N221.16.5 \\ $\rm X1^+$} \\ \hline 
3.12 & USb & \makecell[c]{No.~225 \\ $ Fm\bar{3}m$} & \makecell[c]{N225.16.1 \\ $\rm X1^+$} \\ \hline 
3.18 & HoRh & \makecell[c]{No.~221 \\ $ Pm\bar{3}m$} & \makecell[c]{N221.16.5 \\ $\rm X1^+$} \\ \hline 
3.21 & TmGa$_{3}$ & \makecell[c]{No.~221 \\ $ Pm\bar{3}m$} & \makecell[c]{N221.16.1 \\ $\rm M1^+$} \\ \hline 
3.22 & Eu$_{3}$PbO & \makecell[c]{No.~221 \\ $ Pm\bar{3}m$} & \makecell[c]{N221.16.5 \\ $\rm X1^+$} \\ \hline 
\hline 
\end{longtable}

\begin{longtable}{cp{4cm}<{\centering}|p{3cm}<{\centering}p{4cm}<{\centering}}
\caption{Coplanar magnetic materials with non-symmorphic SBZs.} \\
\hline \hline
BCSID  & Chemical  formula   & PSG & SSG \\ \hline \endfirsthead
\caption*{Coplanar magnetic materials with non-symmorphic SBZs (continued).} \\
\hline \hline
BCSID  & Chemical  formula   & PSG & SSG \\ \hline \endhead
\multicolumn{2}{c|}{  light-element materials } & &   \\ \hline
1.0.33 & FeF$_{3}$ & \makecell[c]{No.~191 \\ $ P6/mmm$} & \makecell[c]{P191.6.5 \\ $\rm A5^+$} \\ \hline 
1.2 & CuSe$_{2}$O$_{5}$ & \makecell[c]{No.~15 \\ $ C2/c$} & \makecell[c]{P15.3.15 \\ $\rm Y1^+\oplus Y2^+$} \\ \hline 
1.25 & KFe$_{3}$(OH)$_{6}$(SO$_{4}$)$_{2}$ & \makecell[c]{No.~166 \\ $ R\bar{3}m$} & \makecell[c]{P166.6.3 \\ $\rm T3^+$} \\ \hline 
1.39 & LiFeGe$_{2}$O$_{6}$ & \makecell[c]{No.~14 \\ $ P2_1/c$} & \makecell[c]{P14.3.12 \\ $\rm Y1^+\oplus Y2^+$} \\ \hline 
1.88 & Mn$_{5}$Si$_{3}$ & \makecell[c]{No.~63 \\ $ Cmcm$} & \makecell[c]{P63.3.87 \\ $\rm Y1^+\oplus Y2^+$} \\ \hline 
1.117 & NaFePO$_{4}$ & \makecell[c]{No.~11 \\ $ P2_1/m$} & \makecell[c]{P11.3.13 \\ $\rm B1^+\oplus B1^+$} \\ \hline 
1.126 & NaCoSO$_{4}$F & \makecell[c]{No.~15 \\ $ C2/c$} & \makecell[c]{P15.3.15 \\ $\rm Y1^+\oplus Y2^+$} \\
\hline 
\hline 
\end{longtable}

\begin{longtable}{cp{4cm}<{\centering}|p{3cm}<{\centering}p{4cm}<{\centering}}
\caption{Non-coplanar magnetic materials with non-symmorphic SBZs.} \\
\hline \hline
BCSID  & Chemical  formula   & PSG & SSG \\ \hline \endfirsthead
\caption*{Non-coplanar magnetic materials with non-symmorphic SBZs (continued).} \\
\hline \hline
BCSID  & Chemical  formula   & PSG & SSG \\ \hline \endhead
\multicolumn{2}{c|}{  light-element materials } & &   \\ \hline
2.61 & Fe$_{3}$F$_{8}$(H$_{2}$O)$_{2}$ & \makecell[c]{No.~12 \\ $ C2/m$} & \makecell[c]{N12.9.70 \\ $\rm GM2^+\oplus A1^+\oplus A1^+$} \\ \hline 
\multicolumn{2}{c|}{  heavy-element materials } & &   \\ \hline
2.38 & Pb$_{2}$MnWO$_{6}$ & \makecell[c]{No.~7 \\ $ Pc$} & \makecell[c]{N7.9.10 \\ $\rm GM2\oplus Y1\oplus Y2$} \\ \hline 
2.63 & DyCrO$_{3}$ & \makecell[c]{No.~11 \\ $ P2_1/m$} & \makecell[c]{N11.9.29 \\ $\rm GM2^+\oplus B1^+\oplus B1^+$} \\ \hline 
2.64 & DyCrO$_{3}$ & \makecell[c]{No.~11 \\ $ P2_1/m$} & \makecell[c]{N11.9.29 \\ $\rm GM2^+\oplus B1^+\oplus B1^+$} \\ \hline 
2.75 & Sr$_{2}$Fe$_{3}$S$_{2}$O$_{3}$ & \makecell[c]{No.~10 \\ $ P2/m$} & \makecell[c]{N10.9.173 \\ $\rm B1^+\oplus B1^+\oplus Z2^+$} \\ \hline 
\hline 
\end{longtable}

    \begin{longtable}{cp{2.6cm}<{\centering}|p{2cm}<{\centering}p{2.4cm}|ccc}
        \caption{Collinear magnetic materials with non-zero spin texture $\vS(\k)$.} \\
        \hline \hline
        BCSID  & Chemical  formula   & PSG & \makecell[c]{SSG} &  $d_M$   & Sym of $\vS(\k)$ & Representation           \\ \hline \endfirsthead
        \caption*{Collinear magnetic materials with non-zero spin texture $\vS(\k)$ (continued).} \\
        \hline \hline
        BCSID  & Chemical  formula   & PSG & \makecell[c]{SSG} &  $d_M$   & Sym of $\vS(\k)$ & Representation           \\ \hline \endhead
    \multicolumn{2}{c|}{  light-element materials } & &  & & \\ \hline
    0.4 & NiCr$_{2}$O$_{4}$ & $ I4_1/amd$ & L141.1.1 $\rm GM1^+$ & 1 & $4/mmm$  &  $\rm A1_g$ \\ \hline 
    0.5 & Cr$_{2}$S$_{3}$ & $ R\bar{3}$ & L148.1.1 $\rm GM1^+$ & 1 & $\bar{3}$  &  $\rm A_g$ \\ \hline 
    0.15 & MnF$_{2}$ & $ P4_2/mnm$ & L136.2.6 $\rm GM4^+$ & 1 & $4/mmm$  &  $\rm B2_g$ \\ \hline 
    0.23 & Ca$_{3}$Mn$_{2}$O$_{7}$ & $ Cmc2_1$ & L36.2.1 $\rm GM2$ & 1 & $mmm$  &  $\rm B1_g$ \\ \hline 
    0.35 & Cu$_{2}$OSeO$_{3}$ & $ P2_13$ & L198.1.1 $\rm GM1$ & 1 & $m\bar{3}$  &  $\rm A_g$ \\ \hline 
    0.38 & GaFeO$_{3}$ & $ Pna2_1$ & L33.1.1 $\rm GM1$ & 1 & $mmm$  &  $\rm A_g$ \\ \hline 
    0.50 & MnTiO$_{3}$ & $ R3c$ & L161.2.1 $\rm GM2$ & 1 & $\bar{3}m$  &  $\rm A2_g$ \\
\hline 
    \hline 
    \end{longtable}
    
    \begin{longtable}{cp{2.6cm}<{\centering}|p{2cm}<{\centering}p{3.6cm}|ccc}
        \caption{Coplanar magnetic materials with non-zero spin texture $\tS(\kk)$.} \\
        \hline \hline
        BCSID  & Chemical  formula   & PSG & \makecell[c]{SSG} &  $d_M$   & Sym of $\tS(\kk)$ & Representation           \\ \hline \endfirsthead
                \caption*{Coplanar magnetic materials with non-zero spin texture $\tS(\kk)$ (continued).} \\
        \hline \hline
        BCSID  & Chemical  formula   & PSG & \makecell[c]{SSG} &  $d_M$   & Sym of $\tS(\kk)$ & Representation           \\ \hline \endhead
    \multicolumn{2}{c|}{  light-element materials } & &  & & \\ \hline
    0.7 & ScMnO$_{3}$ & $ P6_3cm$ & P185.6.2 $\rm GM6$ & 3 & $6/mmm$  &  $\rm A1_u \oplus E1_g$ \\ \hline 
    0.8 & ScMnO$_{3}$ & $ P6_3cm$ & P185.6.2 $\rm GM6$ & 3 & $6/mmm$  &  $\rm A1_u \oplus E1_g$ \\ \hline 
    0.17 & FePO$_{4}$ & $ Pnma$ & P62.3.24 $\rm GM3^-\oplus GM4^+$ & 2 & $mmm$  &  $\rm B1_u \oplus B2_g$ \\ \hline 
    0.36 & NiF$_{2}$ & $ P4_2/mnm$ & P136.2.6 $\rm GM4^+$ & 2 & $4/mmm$  &  $\rm A1_g \oplus B2_g$ \\ \hline 
    0.40 & Mn$_{2}$O$_{3}$-alpha & $ Pbca$ & P61.3.5 $\rm GM2^+\oplus GM3^+$ & 2 & $mmm$  &  $\rm B1_g \oplus B3_g$ \\ \hline 
    0.41 & Mn$_{2}$O$_{3}$-alpha & $ Pbca$ & P61.3.5 $\rm GM2^+\oplus GM3^+$ & 2 & $mmm$  &  $\rm B1_g \oplus B3_g$ \\ \hline 
    0.69 & Co$_{4}$(OH)$_{2}$ (C$_{10}$H$_{16}$O$_{4}$)$_{3}$ & $ P2_1/c$ & P14.2.2 $\rm GM2^+$ & 2 & $2/m$  &  $\rm A_g \oplus B_g$ \\
\hline 
    \hline 
    \end{longtable}
    
    \begin{longtable}{cp{2.8cm}<{\centering}|p{2cm}<{\centering}p{5cm}|ccc}
        \caption{Non-coplanar magnetic materials with non-zero spin texture $\tS(\kk)$.} \\
        \hline \hline
        BCSID  & Chemical  formula   & PSG & \makecell[c]{SSG} &  $d_M$   & Sym of $\tS(\kk)$ &  Representation         \\ \hline \endfirsthead
                \caption*{Non-coplanar magnetic materials with non-zero spin texture $\tS(\kk)$ (continued).} \\
        \hline \hline
        BCSID  & Chemical  formula   & PSG & \makecell[c]{SSG} &  $d_M$   & Sym of $\tS(\kk)$ &  Representation         \\ \hline \endhead
    \multicolumn{2}{c|}{  light-element materials } & &  & & \\ \hline
    0.60 & $\rm [NH_2(CH_3)_2]_n$ $\rm [Fe^{III}Fe^{II}(HCOO)_6]_n $ & $ R\bar{3}c$ & N167.6.1 $\rm GM3^+$ & 3 & $\bar{3}m$  &  $\rm A1_g \oplus E_g$ \\ \hline 
    0.64 & MnV$_{2}$O$_{4}$ & $ I4_1/a$ & N88.4.1 $\rm GM3^+GM4^+$ & 3 & $4/m$  &  $\rm A_g \oplus \prescript{2}{}E_g \oplus \prescript{1}{}E_g$ \\ \hline 
    0.70 & Na$_{3}$Co(CO$_{3}$)$_{2}$Cl & $ Fd\bar{3}$ & N203.15.1 $\rm GM4^+$ & 3 & $m\bar{3}$  &  $\rm T_g$ \\ \hline 
    0.96 & CoSO$_{4}$ & $ Pnma$ & N62.9.42 $\rm GM2^+\oplus GM3^+\oplus GM4^+$ & 3 & $mmm$  &  $\rm B1_g \oplus B3_g \oplus B2_g$ \\ \hline 
    0.102 & Mn$_{2}$GeO$_{4}$ & $ Pnma$ & N62.9.42 $\rm GM2^+\oplus GM3^+\oplus GM4^+$ & 3 & $mmm$  &  $\rm B1_g \oplus B3_g \oplus B2_g$ \\ \hline 
    0.103 & Mn$_{2}$GeO$_{4}$ & $ P2_1/c$ & N14.3.5 $\rm GM2^+\oplus GM2^+$ & 3 & $2/m$  &  $\rm A_g \oplus B_g \oplus B_g$ \\ \hline 
    0.135 & Ni$_{3}$B$_{7}$O$_{13}$Br & $ Pca2_1$ & N29.3.2 $\rm GM2\oplus GM3$ & 3 & $mm2$  &  $\rm A1 \oplus A2 \oplus B2$ \\
\hline 
    \hline 
    \end{longtable}

\label{sec:material-table}

\clearpage

\end{document}